\documentclass[fleqn,useAMS,usenatbib]{mn2e}
\usepackage{subfiles}
\usepackage{amsmath}	
\DeclareMathOperator\arctanh{arctanh}
\usepackage{amssymb}
\usepackage[english]{babel}
\usepackage{color}
\usepackage{fontawesome}
\usepackage{natbib}
\usepackage{graphics}
\usepackage{graphicx}
\usepackage{hyperref}
\usepackage[capitalise]{cleveref}
\usepackage{pifont}
\usepackage{multirow}
\usepackage{titlesec}
\usepackage{orcidlink}
\usepackage{adjustbox}

\usepackage{tabularx}
\usepackage{multirow}
\usepackage{boldline}

\def\simlt{\lower.5ex\hbox{$\; \buildrel < \over \sim \;$}}
\def\simgt{\lower.5ex\hbox{$\; \buildrel > \over \sim \;$}}

\defcitealias{Etherington2022}{E22}

\newcommand*\tens[1]{\ensuremath{\mathsf{#1}}}
\newcommand{\xmark}{\text{\ding{55}}}

\title[Scanning For DM Subhalos With Strong Lensing]
{Scanning For Dark Matter Subhalos\\in \textit{Hubble Space Telescope} Imaging of 54 Strong Lenses}
\author[Nightingale et al.]
{\parbox{\textwidth}{James W.\ Nightingale$^{1,2}$\orcidlink{0000-0002-8987-7401}\thanks{e-mail: james.w.nightingale@durham.ac.uk},
Qiuhan He$^{2}$\orcidlink{0000-0003-3672-9365},
Xiaoyue Cao$^{4, 5, 3}$,
Aristeidis \\Amvrosiadis\orcidlink{0000-0002-4465-1564}$^{2}$, 
Amy Etherington$^{1}$,
Carlos S.\ Frenk\orcidlink{0000-0002-2338-716X}$^{2}$,
Richard G.\ Hayes$^{1}$,
Andrew Robertson\orcidlink{0000-0002-0086-0524}$^{6}$,
Shaun Cole\orcidlink{0000-0002-5954-7903}$^{2}$, 
Samuel Lange$^{2}$,
Ran Li$^{3, 5, 4}$  \&
Richard Massey\orcidlink{0000-0002-6085-3780}$^{1,2}$ \\
}\\
$^{1}$Department of Physics, Centre for Extragalactic Astronomy, Durham University, South Rd, Durham, DH1 3LE, UK \\
$^{2}$Department of Physics, Institute for Computational Cosmology, Durham University, South Rd, Durham DH1 3LE, UK \\
$^{3}$National Astronomical Observatories, Chinese Academy of Sciences, 20A Datun Road, Chaoyang District, Beijing 100012, China \\
$^{4}$School of Astronomy and Space Science, University of Chinese Academy of Sciences, Beijing 100049, China \\
$^{5}$Institute for Frontiers in Astronomy and Astrophysics, Beijing Normal University, Beijing 102206, China \\
$^{6}$Jet Propulsion Laboratory, California Institute of Technology, 4800 Oak Grove Drive, Pasadena, CA 91109, USA \\
}

\begin{document}

\bibliographystyle{mn2e}
\bibpunct{(}{)}{;}{a}{}{;}
\date{\today}
\pagerange{\pageref{firstpage}--\pageref{lastpage}} 
\pubyear{2018}
\maketitle
\label{firstpage}

\begin{abstract}

The cold dark matter (DM) model predicts that every galaxy contains thousands of DM subhalos; almost all other DM models include a physical process that smooths away the subhalos. The subhalos are invisible, but could be detected via strong gravitational lensing, if they lie on the line of sight to a multiply-imaged background source, and perturb its apparent shape. We present a predominantly automated strong lens analysis framework, and scan for DM subhalos in Hubble Space Telescope imaging of 54 strong lenses. We identify five DM subhalo candidates, including two especially compelling candidates (one previously known in SLACS0946+1006) where a subhalo is favoured after all of our tests for systematics. We find that the detectability of subhalos depends upon the assumed parametric form for the lens galaxy's mass distribution, especially its degree of azimuthal freedom. Using separate components for dark matter and stellar mass reveals two DM subhalo candidates and removes four false-positives compared to the single power-law mass model that is common in the literature. We identify 45 lenses {\it without} substructures, the number of which is key to statistical tests able to rule out models of e.g.\ warm or self-interacting DM. Our full analysis results are available at \url{https://github.com/Jammy2211/autolens_subhalo}.

\end{abstract}


\section{Introduction}\label{Intro}

The nature by which cold dark matter (CDM) leads to the formation of the large-scale structure of the Universe, the `cosmic web', has been modelled in incredible detail by state-of-the-art cosmological $N$-body simulations \citep{Springel2005}. The picture of hierarchical growth has been established, where density peaks of CDM within the Universe’s initial density field collapse to form self-bound virialized halos. The lowest mass halos form first, and successively merge to form higher mass halos, a process that occurs over the full range of halo masses in a self-similar manner. In conjunction with a cosmological constant, $\Lambda$, this process describes structure formation in our concordance cosmological model, $\Lambda$CDM, which on large scales has now made numerous testable predictions which have shown remarkable agreement with observations, such as the clustering of galaxies \citep{Hildebrandt2017} and the growth of baryon acoustic oscillations \citep{Anderson2014}.

A key prediction of $\Lambda$CDM on smaller scales is the hierarchy of subhalos within each dark matter (DM) halo \citep{Diemand2008, Springel2008}. This states that orbiting within every DM halo are many lower mass satellite halos that it has previously accreted. This hierarchy extends on, with DM halos hosting subhalos that themselves host subhalos \citep{Diemand2007}. CDM thus predicts an abundance of low-mass ($10^{-3}$M$_{\rm \odot}$ to $10^{8}$M$_{\rm \odot}$) halos throughout the Universe. The majority of such halos are completely dark, as radiation from the ultraviolet background reheats the inter-galactic medium and prevents gas from cooling and forming stars \citep{Sawala2016, Benitez-Llambay2020}. Owing to this lack of luminous emission, DM halos below masses of $10^{\rm 8.5}$\,M$_{\rm \odot}$ are yet to be observed, with the lowest mass DM halos known being those of Milky Way dwarf galaxies \citep{Belokurov2014}. Observing completely dark halos below masses of $10^{\rm 8.5}$\,M$_{\rm \odot}$ would provide evidence in favour of $\Lambda$CDM on scales smaller than previously tested. However, if one could definitively show their absence, it would indicate that a different model for the DM particle is needed, for example warmer flavours \citep{Bode2001}. This would then disfavour a Weakly Interacting Massive Particle (WIMP) from being the DM, and would instead point to alternatives which change the relativistic properties of DM in the early universe, so as to suppress halo formation at low masses (e.g. the sterile neutrino \citealt{Shi1999}).

Strong gravitational lensing, where a background source is multiply imaged by a foreground deflector galaxy, provides a means to detect dark matter subhalos that do not emit light. When an extended source galaxy is lensed, light rays emanating from different regions of the source trace through (and are lensed by) different regions of the lens. The observed, distorted shape thus contains a high resolution imprint of the distribution of mass in the lens. If a DM subhalo is along any line of sight, it will perturb the image in a unique and observable way. This technique has provided multiple detections of DM subhalos \citep{Vegetti2010, Vegetti2012, Vegetti2014, Hezaveh2016} as well as non-detections that further constrain the subhalo mass function \citep{Ritondale2019a}. These observations have been translated into constraints on sterile neutrino cosmologies \citep{Vegetti2018, Enzi2021}. The technique also recently led to the discovery of an ultramassive black hole \citep{Nightingale2023}.

Much effort has gone into understanding which DM subhalos this technique can detect. Sensitivity mapping has shown that \textit{Hubble} Space Telescope (HST) imaging can detect subhalos of mass $\sim 10^{\rm 9.0}$\,M$_{\rm \odot}$, whereas higher resolution very long baseline interferometry probes masses as low as $\sim 10^{\rm 6.0}$\,M$_{\rm \odot}$ \citep{McKean2015, Li2016b, Despali2018, Despali2022}. These studies assume DM substructures lie on a mass-concentration relation (e.g.\ \citealt{Ludlow2016}). Instead, \citet{Amorisco2022} performed sensitivity mapping over the scatter in this relation and showed that DM halos $\sim 0.5$ dex lower in mass become detectable when they have a higher than average concentration. Furthermore, for DM cosmologies with a cut-off mass (e.g.\ around $\sim 10^{\rm 8.5}$\,M$_{\rm \odot}$ for warmer DM with a sterile neutrino) high concentration halos below this cut-off do not exist and therefore do not become detectable -- amplifying the contrast between the expected number of detections in CDM and warmer models. DM substructures in the lens galaxy and line-of-sight objects at a different redshift to the lens are both detectable \citep{Li2016a, Despali2018, He2022b, Despali2022, Amorisco2022}, with  their relative contributions depending on the redshifts of the lens and the source. If subhalos within the lens galaxy are detected, then interpreting them in terms of DM models is subject to uncertainties due to galaxy formation, for example reductions in subhalo mass by tidal stripping or stellar feedback \citep{Despali2017}. Line-of-sight objects are unaffected by this.

Subhalo analysis comprises two parts: (i) confirming that the inclusion of a parametric DM subhalo is favoured when fitting the lens data and; (ii) for the detection to be reproduced by a non-parametric model which adds corrections to the gravitational potential on top of the best-fit mass model \citep{Koopmans2005, Suyu2010, Vegetti2009, Ritondale2019a, Vernardos2022}. The latter, often called the `potential corrections', requires that the non-parametric model of the convergence resembles a \textit{local} over density of mass; the expected signal of a DM subhalo. However, the correction often produces non-zero convergence on larger \textit{global} scales, due to systematics associated with the assumed mass model being too simple (e.g. \citealt{Ritondale2019a}). In this scenario, a DM subhalo candidate is rejected, irrespective of how much the parametric model favours the DM subhalo. Early implementations of the potential corrections relied on some level of human input to choose aspects like the regularization \citep{Koopmans2005, Vegetti2009}, whereas \citet{Vernardos2022} recently placed the method in a Bayesian framework. This work does not use the potential corrections and therefore cannot make a definitive claim as to whether any subhalo detection is genuine or not. Our focus is to understand how different lens model assumptions impact whether a parametric DM subhalo is favoured.

This work presents a predominantly automated search for subhalos in strong lenses using the open-source strong lens modelling software {\tt PyAutoLens}\footnote{\url{https://github.com/Jammy2211/PyAutoLens}} \citep{Nightingale2018, Nightingale2021}. The software approaches lens modeling using the same Bayesian framework as the methods of \citet{Vegetti2009, Hezaveh2016} but differs in many aspects of its implementation (e.g. the source reconstruction). We scan for subhalos in a sample of $54$ strong lenses from the Strong Lens Advanced Camera for Surveys (SLACS) survey \citep{Bolton2008a} and BOSS GALaxy-Ly$\alpha$ EmitteR sYstems (BELLS-GALLERY) sample \citep{Shu2016}. This sample includes $10$ lenses analysed by \citet{Vegetti2010} and \citet{Vegetti2014} and $16$ of the systems analysed by \citet{Ritondale2019a}. We therefore perform DM subhalo detection in $28$ objects never previously analysed. Our results build on \citet{Etherington2022}, who performed automated lens modeling with {\tt PyAutoLens} in a sample of $59$ strong lenses from the SLACS and BELLS-GALLERY samples and investigated the redshift evolution of the lens galaxy mass distributions \citep{Etherington2023}.

After an initial analysis of the $54$ strong lenses we focus on `false positive' detections. Here, a lens model including a DM subhalo is favoured at $> 3\sigma$ over a model without a DM subhalo, but more detailed investigation led us to conclude the result is spurious. This has been seen in previous studies and attributed to inflexibility of mass models to fit the complex distribution in real galaxies \citep{Hsueh2016, Hsueh2017, Hsueh2018, He2023}. To mitigate false positives, previous studies have employed strict criteria for a DM subhalo detection, for example requiring that the Bayesian evidence of the lens model with a DM subhalo is favoured at $5\sigma$ \citep{Vegetti2014} or $10\sigma$ \citep{Ritondale2019a}. They are also flagged by the potential corrections technique discussed previously. \textbf{Our results do not imply that any previous DM subhalo detections are false positives}. Instead, we reproduce false positive signals found in previous studies (which are typically below the 5$\sigma$ or 10$\sigma$ threshold these studies used) and quantify which deficiencies in the strong lens model are the cause, in order to outline where improvements should be made in the future. 

We place an emphasis on understanding what impact changing the lens galaxy mass model has on the final DM subhalo inference. We scan for DM subhalos assuming a total of five different mass model parameterizations from the literature \citep{Chu2013, Tessore2015, Nightingale2019, Oriordan2020}. We quantify whether fitting more complex models leads one to favour or reject a DM subhalo, when fitting a simpler model either did or did not. This is only possible because our analysis is predominantly automated, and therefore straightforward to repeat with a variety of model assumptions. Our large sample of 54~lenses yields the first quantitative study of how different types of model complexity impact subhalo detectability.

This paper is structured as follows.
In \S\ref{Data}, we describe the HST imaging data.
In \S\ref{Method}, we describe {\tt PyAutoLens} and our substructure detection pipelines.
In \S\ref{ResultsData} we show results for fits to HST strong lenses.
In \S\ref{Discussion}, we discuss the implications of our measurements, and we give a summary in \S\ref{Summary}.
In \cref{ResultSim}, we show our substructure detection method works on simulated images.
{We assume a Planck 2015 cosmology throughout \citep{PlanckCollaboration2015a}}.
The analysis scripts and data used in this work are publically available at \url{https://github.com/Jammy2211/autolens_subhalo}.

\section{Hubble Space Telescope Data}\label{Data}

In this work we fit HST imaging of 54 strong lenses from the SLACS \citep{Bolton2008a} and BELLS-GALLERY \citep{Shu2016} samples. Full details of these datasets and their data reduction are given in \citet{Bolton2008a} and \citet{Shu2016}. SLACS data are observed in the HST Advanced Camera for Surveys F814W band and BELLS-GALLERY the HST Wide Field Camera 3 F606W band. \citet{Etherington2022} describe post processing steps which remove contaminating foreground light (e.g. of stars and line-of-sight galaxies) via a graphical user interface (GUI). We only use the gold sample presented in \citet{Etherington2022}, which removes five lenses where a poor lens light subtraction would negatively impact the quality of the lens model.
\section{Method}\label{Method}

\begin{figure*}
\centering
\includegraphics[width=0.24\textwidth]{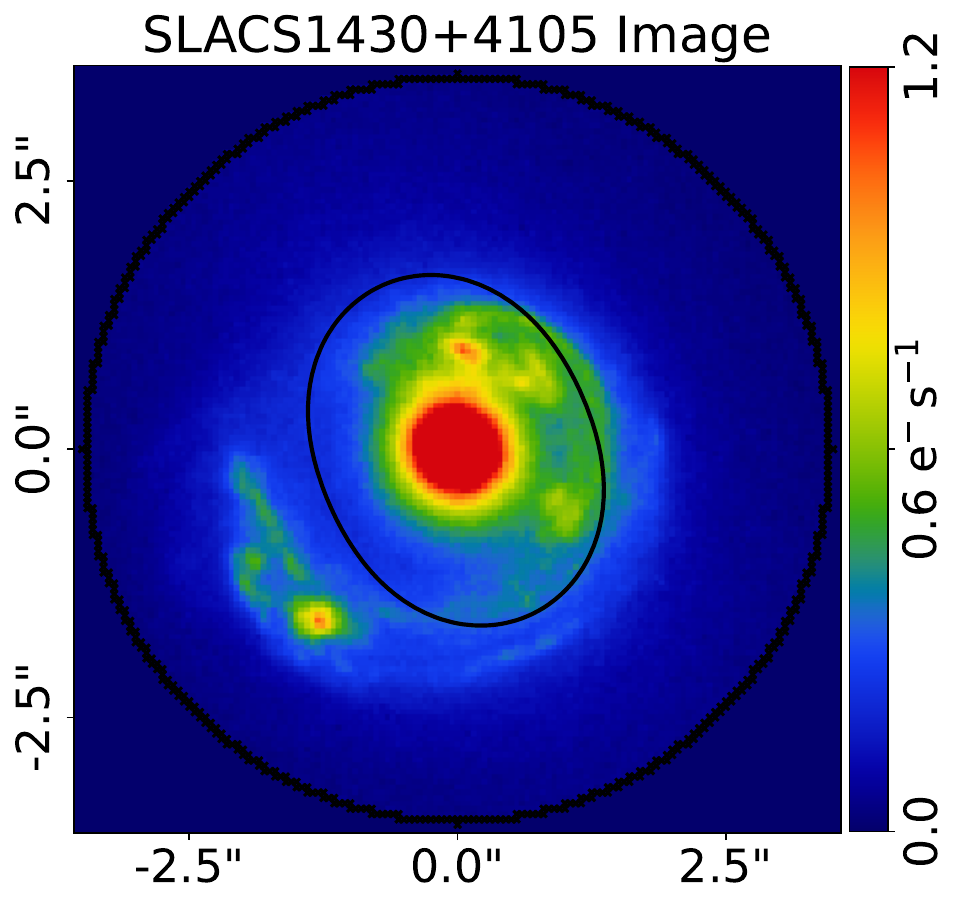}
\includegraphics[width=0.24\textwidth]{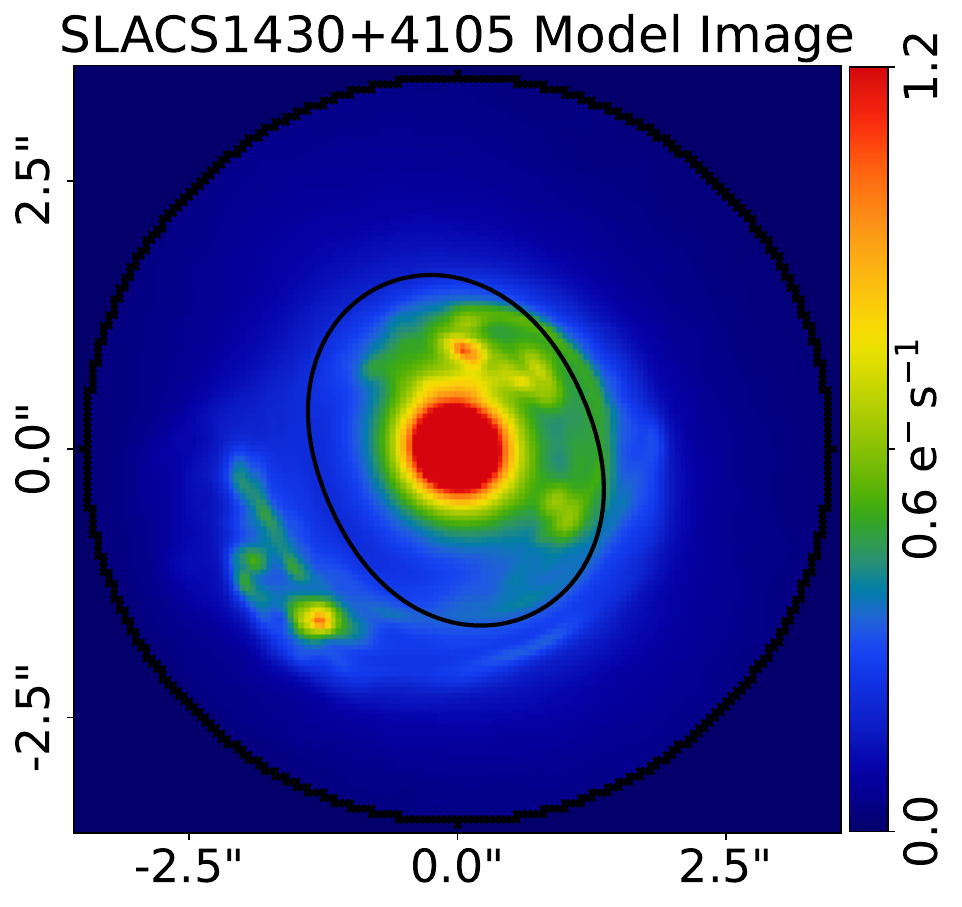}
\includegraphics[width=0.24\textwidth]{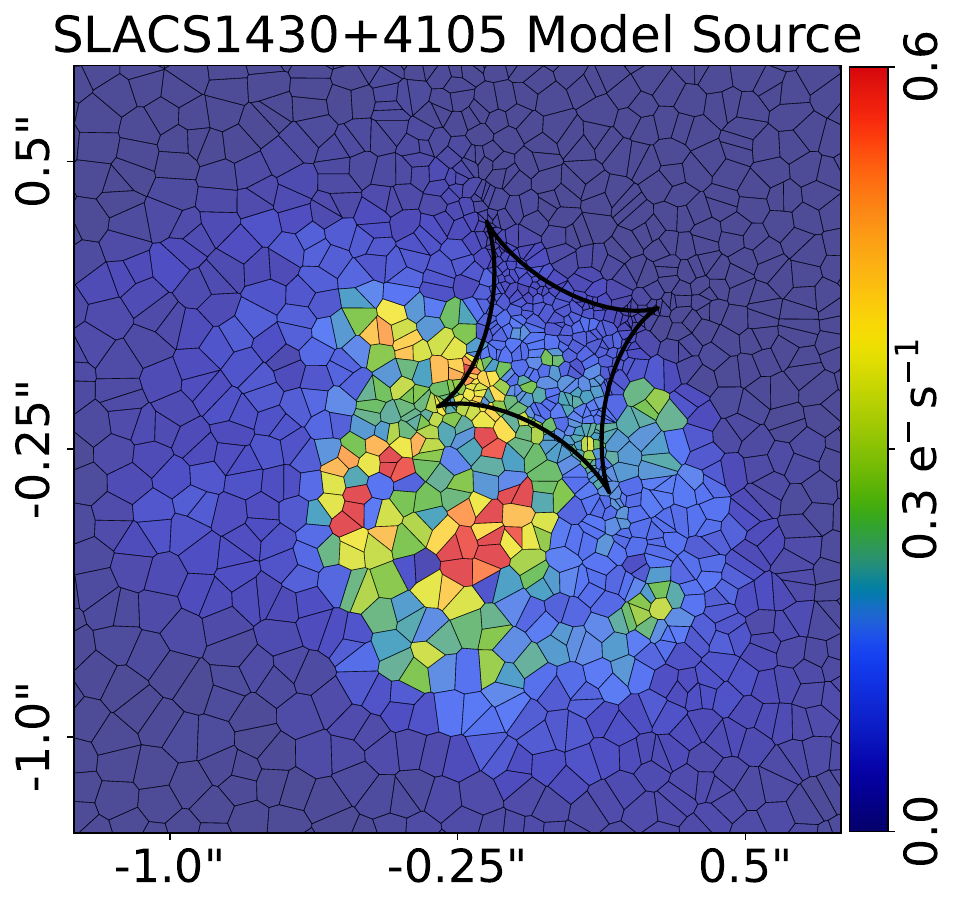}
\includegraphics[width=0.24\textwidth]{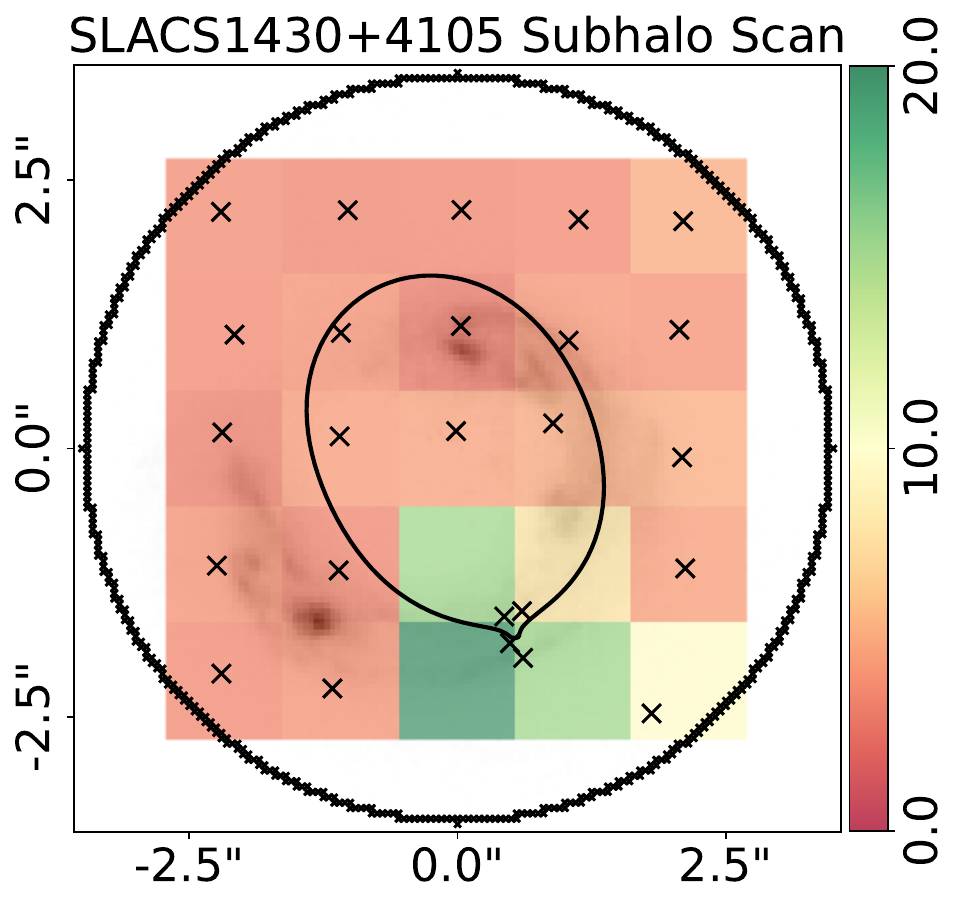}
\caption{
A concise visual overview of the {\tt PyAutoLens} analysis performed in this study, shown for an example strong lens SLACS1430+4105. The images from left to right are: (i) the observed HST imaging data of SLACS1430+4105; (ii) the model lens light and lensed source inferred from a {\tt PyAutoLens} model-fit; (iii) the inferred model source in the source-plane, which is reconstructed using a Voronoi mesh and; (iv) the subhalo scanning results, where the colorbar shows the log Bayesian evidence increase for lens models including a subhalo within 2D segments of the image-plane. The black lines show the tangential critical curves and caustics and black crosses in the subhalo scanning results show the inferred subhalo $(x^{\rm sub},\,y^{\rm sub})$ coordinates for each model including a subhalo. The example subhalo scan of SLACS1430+4105 favours a DM subhalo at  $(x^{\rm sub},\,y^{\rm sub}) \sim (0.2\arcsec,-1.2\arcsec)$, however systematic tests of the lens mass model will reveal this is a false positive.
}
\label{figure:Overview}
\end{figure*}

\subsection{Overview}\label{Overview}

We perform lens modelling using the open-source software {\tt PyAutoLens}, which is described in \citet{Nightingale2015, Nightingale2018, Nightingale2021} and builds on the methods of \citet[][WD03 hereafter]{Warren2003}; \citet{Suyu2006, Vegetti2009}. We compose pipelines which perform predominantly automated lens modelling using the probabilistic programming language {\tt PyAutoFit}\footnote{\url{https://github.com/rhayes777/PyAutoFit}} \citep{Nightingale2021a}, a spin-off project of {\tt PyAutoLens} which generalizes the methods used to model strong lenses into an accessible statistics library.

A concise visual overview of the {\tt PyAutoLens} analysis performed in this work is shown in \cref{figure:Overview}. Given an observed image of a strong lens the analysis: (i) defines a $3.5\arcsec$ circular mask within which the lens model is fitted (this mask extends beyond the lensed source in order to better constrain the lens light model); (ii) uses a model containing light and mass profiles for the lens to produce model images of the lens galaxy and lensed source, which are convolved with the instrumental Point Spread Function (PSF) and compared to the data; (iii) reconstructs the source galaxy in the source-plane using a Voronoi mesh and; (iv) produces a subhalo scanning map indicating how much a lens model with a DM subhalo at a specific location in the image-plane increases the Bayesian evidence compared to a lens model without a DM subhalo.

We now describe each step in more detail. The following link (\url{https://github.com/Jammy2211/autolens_likelihood_function}) contains Jupyter notebooks providing a visual step-by-step guide of the {\tt PyAutoLens} likelihood function used in this work. 

\subsection{Light Profiles}\label{Profiles}

Light and mass profile quantities are computed using elliptical coordinates $\xi = \sqrt{{x}^2 + y^2/q^2}$, with minor to major axis-ratio $q$ and position angle $\phi$ defined counter clockwise from the positive x-axis. For model-fitting, these are parameterized as two components of ellipticity
\begin{equation}
\epsilon_{1} =\frac{1-q}{1+q} \sin 2\phi, \, \, 
\epsilon_{2} =\frac{1-q}{1+q} \cos 2\phi \, .    
\label{eq: ellip}
\end{equation}
Light profiles are modelled using one or more elliptical S\'ersic profiles
\begin{equation}
\label{eqn:Sersic}
I_{\rm  Ser} (\xi) = I \exp \bigg\{ -k \bigg[ \bigg( \frac{\xi} R \bigg)^{\frac{1}{n}} - 1 \bigg] \bigg\} ,
\end{equation}
which have up to seven free parameters: $(x,y)$, the light centre in arc-seconds, $(\epsilon_{\rm 1}, \epsilon_{\rm 2})$ the elliptical components, $I$, the intensity in electrons per second at the effective radius $R$ in arc-seconds and $n$, the S\'ersic index. $k$ is not a free parameter, but is instead a function of $n$ \citep{Ciotti1999}. This study assumes a model with two S\'ersic profiles which have the same centre, with each individual profile’s intensities evaluated and summed. Parameters are given the superscripts `bulge' and `disk', which are used to distinguish which component of the lens galaxy they are modelling, for example the S\'ersic index of the bulge component is $n^{\rm bulge}$.

\subsection{Mass Profiles}\label{MassProfiles}

\subsubsection{Dark Matter Subhalos}

Dark matter subhalos (superscript `sub') are modelled as a spherical Navarro-Frenk-White (NFW) profile. The NFW represents the universal density profile predicted for dark matter halos by cosmological N-body simulations \citep{Zhao1996, Navarro1996, Navarro1997},  and with a volume mass density given by
\begin{equation}
    \rho = \frac{\rho_{\rm s}^{\rm dark}}{(r/r_{\rm s}^{\rm dark}) (1 + r/r_{\rm s}^{\rm dark})^2}.
    \label{eqn:MassNFW}
\end{equation}
The halo normalization is given by $\rho_{\rm s}^{\rm sub}$ and the scale radius in arc-seconds by $r_{\rm s}^{\rm sub}$. The dark matter normalization is parameterized using $M_{\rm 200}^{\rm sub}$ (the enclosed mass in solar masses at the radius $r_{200}$ within which the average density is 200 times the critical density of the Universe) as a free parameter. The scale radius is set via $M_{\rm 200}^{\rm sub}$ using the mean of the mass-concentration relation of \citet{Ludlow2016}. The convergence is given by
\begin{equation}
\label{eqn:DMkap}
\kappa (\xi) = 2 \kappa^{\rm  sub} \frac{1 - \mathcal{F}(\xi)}{\xi^2 – 1} ,
\end{equation}
where
\begin{equation}
\label{eqn:DMFunc}
\mathcal{F}(\xi) = \left\{
  \begin{array}{lr}
    \frac{1}{\sqrt{\xi^2 - 1}} \arctan \sqrt{\xi^2 - 1} & : \xi > 1 \\
    \frac{1}{\sqrt{1 - \xi^2}} \arctanh \sqrt{1 - \xi^2} & : \xi < 1 \\
     1 & : \xi = 1 
  \end{array}
\right.
\end{equation}
and $\kappa^{\rm sub}$ is related to the lens halo normalization by $\kappa^{\rm sub} = \rho_{\rm s} r_{\rm s} / \Sigma_{\rm  cr}$ and $\Sigma_{\rm cr}$ is the critical surface density. The lens and source redshifts are used to perform unit conversions, for example to calculate $M_{\rm 200}^{\rm sub}$ in solar masses. All DM subhalos are assumed to be at the lens galaxy redshift.

\subsubsection{Elliptical Power-Law}

For the lens mass model we assume an elliptical power-law (PL) density profile representing the total mass of the lens (e.g. star and dark matter) of form
\begin{equation}
\label{eqn:SPLEkap}
\kappa (\xi) = \frac{(3 - \gamma^{\rm mass})}{1 + q^{\rm mass}} \bigg( \frac{\theta^{\rm mass}_{\rm E}}{\xi} \bigg)^{\gamma^{\rm mass} - 1} ,
\end{equation}
where parameters associated with the lens mass profile have superscript `mass'. $\theta^{\rm mass}_{\rm E}$ is the model Einstein radius in arc-seconds. The power-law density slope is $\gamma^{\rm mass}$, and setting $\gamma^{\rm mass} = 2$ gives the singular isothermal ellipsoid (SIE) model. Deflection angles for the power-law are computed via an implementation of the method of \citet{Tessore2015} in {\tt PyAutoLens}.

\subsubsection{Broken Power-Law}\label{ModelBPL}

We also fit the elliptical broken power law (BPL) profile \citep{Oriordan2019, Oriordan2020, Oriordan2021}, again representing the total mass of the lens, with convergence
\begin{equation}\label{equ:power_law}
    \kappa{\left(r\right)}=\left\{
    \begin{array}{ll}
    \kappa^{\rm mass}_{\rm E}\left(r^{\rm mass}_{\rm b} / r \right)^{t^{\rm mass}_1}, & r \leq r^{\rm mass}_{\rm b} \\
    \kappa^{\rm mass}_{\rm E}\left(r^{\rm mass}_{\rm b} / r \right)^{t^{\rm mass}_2}, & r > r^{\rm mass}_{\rm b}
    \end{array}
    \right.,
\end{equation}
where $r^{\rm mass}_{\rm b}$ is the break radius in arc-seconds, $\kappa^{\rm mass}_{\rm E}$ is the convergence at the break radius, $t^{\rm mass}_1$ is the inner slope and $t^{\rm mass}_2$ is the outer slope.

\subsubsection{Power-Law With Internal Multipoles}\label{ModelMultipole}

We fit an extension to the PL profile which includes multipole-like terms describing internal angular structure in its mass distribution, by extending the parameterization given by \citet{Chu2013}. This model captures smooth deviations from ellipticity in the mass distribution. The functional form of the convergence is 
\begin{equation}
    \kappa(r, \phi) = \frac{1}{2} \left(\frac{\theta_{\rm E}^{\rm mass}}{r}\right)^{\gamma^{\rm mass} - 1} k^{\rm mass}_m \, \cos(m(\phi - \phi^{\rm mass}_m)) \, ,
\end{equation}
where we express the convergence in polar coordinates, with $r$ in arc-secconds. $m$ is the multipole order and $m=4$. $k^{\rm mass}_m$ is the multipole strength and $\phi^{\rm mass}_m$ its orientation angle, which is defined counter clockwise from the positive x-axis. The multipole $\theta_{\rm E}^{\rm mass}$ and $\gamma^{\rm mass}$ values are fixed to that of the underlying PL. We parameterize $k^{\rm mass}_m$ and $\phi^{\rm mass}_m$ as multipole components $(\epsilon_{\rm 1}^{\rm mp},\ \epsilon_{\rm 2}^{\rm mp})$ which are given by
\begin{equation}
    \label{eq:shear}
    \phi^{\rm mass}_m = \arctan{\frac{\epsilon_{\rm 2}^{\rm mp}}{\epsilon_{\rm 1}^{\rm mp}}}, \, \,
    k^{\rm mass}_m = \sqrt{{\epsilon_{\rm 1}^{\rm mp}}^2 + {\epsilon_{\rm 2}^{\rm mp}}^2} \, .
\end{equation}

\subsubsection{Stellar and Dark Matter Mass}\label{ModelDecomp}

We fit decomposed mass models for the lens, which decompose its mass into its stellar and dark components (in contrast to the PL models above). The stellar mass is modelled as a sum of S\'ersic profiles which are tied to those of the light. The S\'ersic profile given by \cref{eqn:Sersic} is used to give the light matter surface density profile
\begin{equation}
\label{eqn:Sersickap}
\kappa_{Ser} (\xi) = \Psi \bigg[\frac{q  \xi}{R}\bigg]^{\Gamma} I_{Ser} (\xi_l) \, \, ,
\end{equation}
where $\Psi$ gives the mass-to-light ratio and $\Gamma$ folds a radial dependence into the conversion of mass to light. A constant mass-to-light ratio is given for $\Gamma = 0$. This work assumes there are two light profile components (denoted the bulge and disk) which assume independent values of $\Psi$ and $\Gamma$. We therefore do not assume that mass fully traces light. Deflection angles for this profile are computed via an adapted implementation of the method of \citet{Oguri2021}, which decomposes the convergence profile into multiple cored steep elliptical profiles and efficiently computes the deflection angles from each.

The dark matter component of the lens galaxy's host halo is given by an elliptical NFW profile, whose parameters have superscript `dark'. This is again parameterized with $M_{\rm 200}^{\rm dark}$ as a free parameter and a scale radius set via the mean of the mass-concentration relation of \citet{Ludlow2016}. The convergence is given by \cref{eqn:DMFunc}.

\begin{figure*}
\centering
\includegraphics[width=0.19\textwidth]{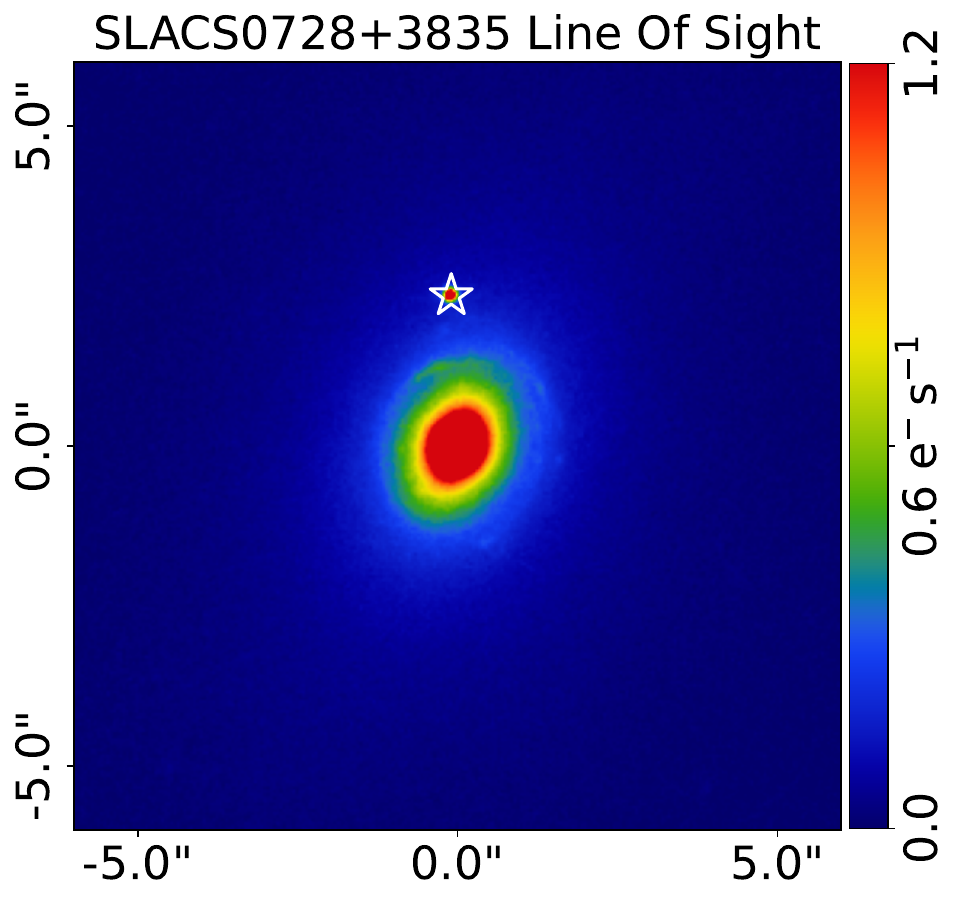}
\includegraphics[width=0.19\textwidth]{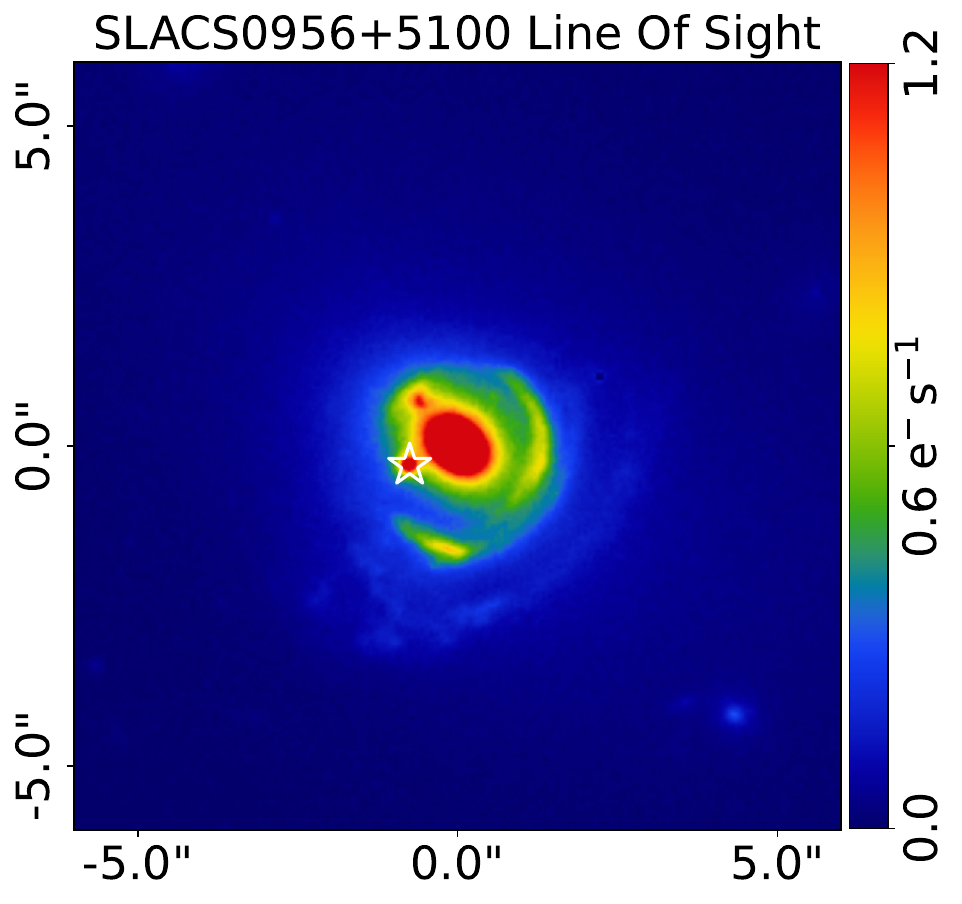}
\includegraphics[width=0.19\textwidth]{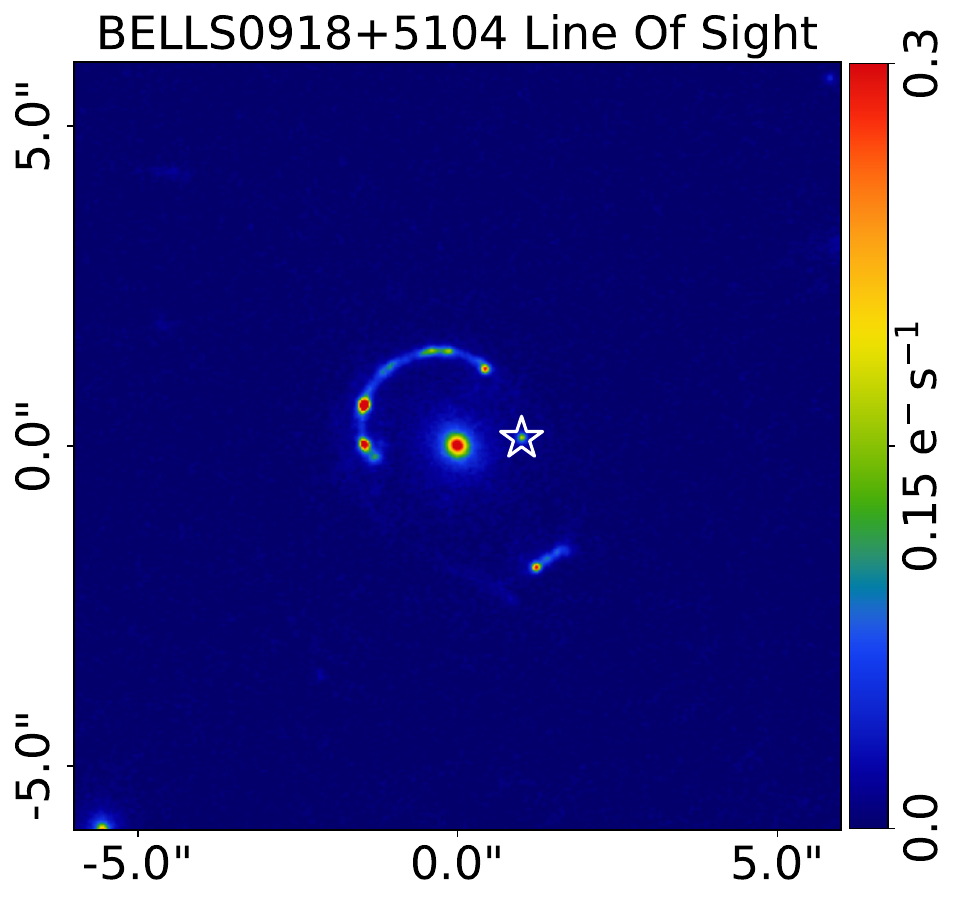}
\includegraphics[width=0.19\textwidth]{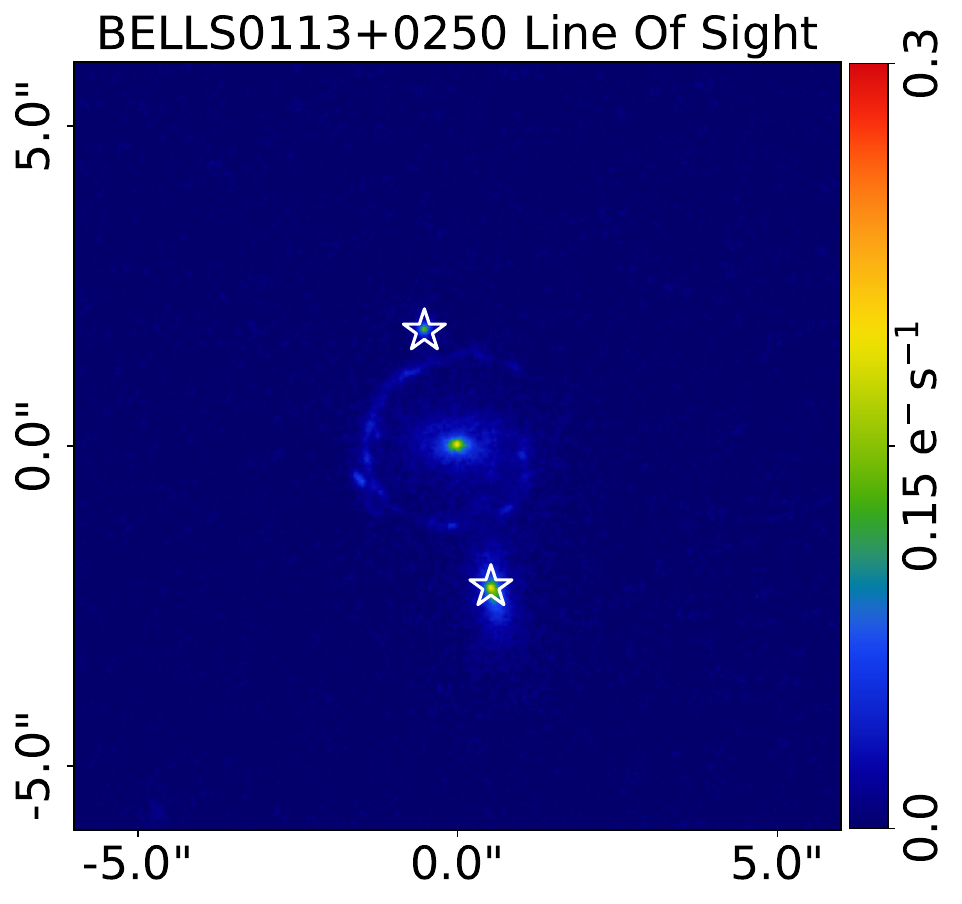}
\includegraphics[width=0.19\textwidth]{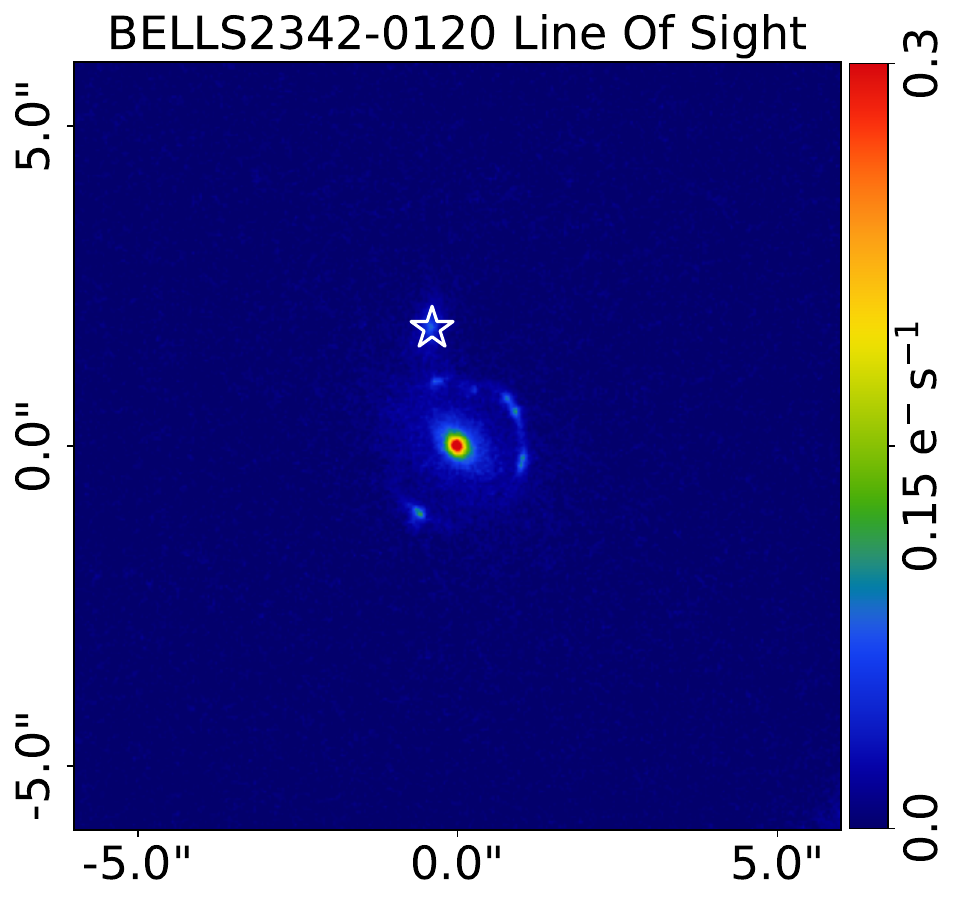}
\caption{
The five lenses with line-of-sight galaxies that are closest to the centre of their lens galaxy, which should have an impact on the inferred lens model. For two lenses, SLACS0956+5100 and BELLS0918+5104, the line-of-sight galaxy is within the Einstein radius, whereas for SLACS0728+3835, BELLS0113+0250 and BELLS2342-0120 the galaxy(s) are slightly outside the Einstein radius. Black stars mark the centre of each line-of-sight galaxy, which are modeled with SIS mass profiles fixed to these centres.
}
\label{figure:FPClump0}
\end{figure*}

\subsubsection{Line-Of-Sight Galaxies}\label{ModelLos}

Nearby line-of-sight galaxies may be included as spherical isothermal spheres (SISs), corresponding to an SIE where $(\epsilon^{\rm mass}_{1},\ \epsilon^{\rm mass}_{2}) = (0, 0)$. To decide whether to include line-of-sight galaxies in the mass model we use a GUI, where a user looks at $10\arcsec$ cut-outs of each lens and clicks on up to two galaxies nearby to add to the mass model. Galaxies are selected subjectively based on their proximity and size. Each galaxy is then included as an SIS, the centre of which is fixed to the galaxy's brightest pixel and with a redshift that is the same as the lens galaxy. The prior on $\theta^{\rm mass}_{\rm E}$ for each SIS is a uniform prior from $0.0\arcsec$ to $0.5\arcsec$. For the majority of line-of-sight galaxies a value of $\theta^{\rm mass}_{\rm E} = 0.5\arcsec$ is significantly above the mass one would estimate based on its luminosity. This is an intentional choice not to use more informative priors, so that we can investigate how line-of-sight galaxies change the DM inference with maximal freedom. 

\cref{figure:FPClump0} shows the five lenses with line-of-sight galaxies closest to the lens galaxy centre, which are all within $2.0\arcsec$ of it. These objects are close enough to the lensed source that we anticipate they will impact the inferred lens model. For the lenses SLACS0956+5100 and BELLS0918+5104 the line-of-sight galaxy is within the Einstein radius, whereas for SLACS0728+3835, BELLS0113+0250 and BELLS2342-0120 the galaxy(s) are slightly outside the Einstein radius. Models including line-of-sight galaxies for the remaining $49$ lenses are performed, noting the galaxies are typically much further (e.g. over 4.0$\arcsec$) from the lens centre.

\subsubsection{External Shear}

An external shear (superscript `ext') field is included and parameterized as two elliptical components $(\gamma_{\rm 1}^{\rm ext}, \gamma_{\rm 2}^{\rm ext})$. The shear magnitude, $\gamma^{\rm  ext}$, and orientation measured counter-clockwise from north, $\phi^{\rm  ext}$, are given by
\begin{equation}
    \label{eq:shear}
    \gamma^{\rm ext} = \sqrt{\gamma_{\rm 1}^{\rm ext^{2}}+\gamma_{\rm 2}^{\rm ext^{2}}}, \, \,
    \tan{2\phi^{\rm ext}} = \frac{\gamma_{\rm 2}^{\rm ext}}{\gamma_{\rm 1}^{\rm ext}}.
\end{equation}
The deflection angles due to the external shear are computed analytically. Every mass model above is combined with an external shear. A recent study by \citet{Etherington2023a} suggests that this external shear component is representing missing complexity in the lens mass distribution, as opposed to line-of-sight galaxies.

\subsection{Source Model}

After subtracting the foreground lens emission and ray-tracing coordinates to the source-plane via the mass model, the source is reconstructed in the source-plane using an adaptive mesh which accounts for irregular or asymmetric source morphologies. We use a Voronoi mesh with natural neighbour interpolation \citep{Sibson1981} and in \cref{ResultSim} we compare DM subhalo results assuming different source reconstruction methods.

\subsubsection{Mesh Centres}

The method first determines the centres of the $I$ Voronoi source pixels. Initial fits overlay a rectangular Cartesian grid of shape $(y_{\rm pix}, x_{\rm pix})$ over the image-plane, which extends to and from the mask edges (e.g. from $-3.5\arcsec$ to $3.5\arcsec$ for the mask shown in \cref{figure:Overview}). $y_{\rm pix}$ and $x_{\rm pix}$ are the height and width of this grid in pixels and are treated as free parameters. All coordinates on this uniform grid which fall within the mask are retained and traced to the source-plane via the mass model (pixels outside the mask are discarded). These coordinates, $M_i$, are used as the centre of the Voronoi cells, which therefore trace the mass model magnification\footnote{This corresponds to \texttt{PyAutoLens}'s \texttt{VoronoiNNMagnification} mesh object}.

Subsequent fits adapt the mesh centres $M_i$ to the source's unlensed morphology. This uses a previous model of the lensed source emission, $\Xi_{j}$, which is used to compute the weights
\begin{equation}
\label{eqn:KMeans}
W_j = (\frac{\Xi_{j} - \min{\Xi}} {\max{\Xi_{j}} - \min{\Xi}}) + W_{\rm floor} + \max{\Xi_{j}}. 
\end{equation}
The first term on the right hand side runs from zero to one, where values closer to one correspond to the lensed source's brightest pixels. $W_{\rm floor}$ controls how much weight is given to the source's brightest pixels and is a free parameter in certain fits. $W$ is passed to a weighted KMeans clustering algorithm \citep{scikit-learn} to determine image-plane coordinates which are traced to the source-plane. The KMeans assumes $N_{\rm pix}$ source pixels, which is treated as a free parameter in certain fits. This scheme adapts to the lensed source emission. \footnote{This corresponds to \texttt{PyAutoLens}'s  \texttt{VoronoiNNBrightnessImage} mesh objects}.

\subsubsection{Mapping Matrix}

The reconstruction computes the linear superposition of PSF-smeared source pixel images which best fits the observed image. This uses the mapping matrix $f_{ij}$, which maps the $j$-th pixel of each lensed image to each source pixel $i$, giving a total of $J$ lensed image pixels and $I$ source pixels. When constructing $f_{ij}$ we apply image-plane subgridding of degree $4 \times 4$, meaning that $16 \times J$ sub-pixels are fractionally mapped to source pixels with a weighting of $\frac{1}{16}$, removing aliasing effects \citep{Nightingale2015}. 

Each image sub-pixel is mapped to multiple Voronoi source pixels weighted via interpolation. We use Voronoi natural neighbor interpolation via Sibson's technique \citep{Sibson1981}. For every sub-pixel, $j$, the method considers a new polygon that adding this point to the Voronoi mesh computed from $M_i$ would create. The new polygon captures some of the area that was previously covered by its neighbors, which the method computes and uses to compute the interpolation weights in $f_{ij}$ as
\begin{equation}
\label{eqn:NaturalNeighbor}
w = f_{ij} = \frac{1}{\sum_{i=1}^{K} A_{\rm capture}} \sum_{k=1}^{K} A_{\rm capture} \, z_{k} \, ,
\end{equation}
where $K$ is the number of neighbors of a given Voronoi cell $i$. \footnote{More details about the natural neighbor interpolation technique can be found at \url{https://gwlucastrig.github.io/TinfourDocs/NaturalNeighborTinfourAlgorithm/index.html}}.

\subsubsection{Regularization}

Performing an inversion using \cref{eqn:NaturalNeighbor} by itself is ill-posed, therefore to avoid over-fitting noise the solution is regularized using a linear regularization matrix $\tens{H}$ described by WD03. The matrix $\tens{H}$ applies a prior on the source reconstruction, penalizing solutions where the difference in reconstructed flux of neighboring Voronoi source pixels is large. Initial fits use gradient regularization (see WD03) adapted to a Voronoi mesh (see \citealt{Nightingale2015}) \footnote{This corresponds to the \texttt{PyAutoLens} regularization scheme \texttt{Constant}}. DM subhalo results use a scheme which adapts the degree of smoothing to the reconstructed source's luminous emission and interpolates values at a cross of surrounding points \footnote{This corresponds to the \texttt{PyAutoLens} regularization scheme \texttt{AdaptiveBrightnessSplit}}. The formalism for the calculation of these regularization matrices $\tens{H}$ is given in \cref{RegFormal}.

\subsubsection{Variance Scaling}\label{VarScale}

Lens galaxies can have complex morphologies which leave significant central residuals after subtraction via multiple S\'ersic profiles, which the source reconstruction will attempt to fit. We mitigate this by allowing the method to increase the variances (the noise value in each image pixel) at the centre of an image. First, we estimate the fractional contribution in each pixel $j$ from the lens light
\begin{equation}
\Omega_{\rm  Lens,j} = \frac {\L{\rm _j}}{T{\rm _j} + \omega_{\rm  Lens}} ,
\label{eqn:FracLens}
\end{equation}
where $\L{\rm _j}$ and $T{\rm j}$ are estimates of the lens light emission and total emission from a previous lens model and $\omega_{\rm  Lens}$ is a free parameter. Values of $\L{\rm _j}$ or $T{\rm j}$ less than $1\%$ their maximum value are rounded up to this value to ensure no values are negative. $\Omega_{\rm Lens}$ is divided by its maximum value such that it ranges between values just above 0 and 1. Initial fits which do not have $\L{\rm _j}$ and $T{\rm j}$ vectors use $\sigma_{j}^{\rm obs}$, the observed image statistical uncertainties. This contribution map is used to scale the noise in lens light dominated pixels as 
\begin{eqnarray}
\label{eqn:NoiseMap}
 \sigma_{j} = \omega_{\rm  Lens} (\sigma_{j}^{\rm obs} \, \Omega_{\rm  Lens,j})^{\omega_{\rm  Lens2}} ,
\end{eqnarray}
where $\omega_{\rm  Lens}$ and $\omega_{\rm  Lens2}$ are free parameters.

\subsubsection{Inversion}

Following the formalism of WD03, we define the data vector $\vec{D}_{i} = \sum_{j=1}^{J}f_{ij}(d_{j} - b_{j})/(\sigma_{j})^2$ and curvature matrix $F_{ik} = \sum_{j=1}^{J}f_{ij}f_{kj}/(\sigma_{j})^2$, where $d_{j}$ are the observed image flux values and $ b_{j}$ are the model lens light values. The source pixel surface brightnesses are given by $s = [F + H]^{-1} \vec{D}$ which are solved via a linear inversion that minimizes
\begin{equation}
\label{eqn:ChiSquared}
\chi^2 + G_{\rm L} = \sum_{j=1}^{J} \bigg[ \frac{(\sum_{\rm  i=1}^{I} s_{i} f_{ij}) + b_{j} - d_{j}}{\sigma_{j} } \bigg]^2 + s^{T}Hs \, .
\end{equation}
The term $\sum_{i=1}^{I} s_{i} f_{ij}$ maps the reconstructed source back to the image-plane for comparison with the observed data and $G_{\rm L} = s^{T}Hs$ is a regularization term.

The degree of smoothing is chosen objectively using the Bayesian formalism introduced by \citet{Suyu2006}. The likelihood function is taken from \citet{Dye2008a} and is given by
\begin{eqnarray}
\label{eqn:evidence2}
-2 \,{  \mathrm{ln}} \, \mathcal{Z} &=& \chi^2 + s^{T}Hs
+{ \mathrm{ln}} \, \left[ { \mathrm{det}} (F+H)\right]
-{ \mathrm{ln}} \, \left[ { \mathrm{det}} (H)\right]
\nonumber \\
& &
+ \sum_{j=1}^{J}
{ \mathrm{ln}} \left[2\pi (\sigma{_j})^2 \right]  \, .
\end{eqnarray}

The step-by-step Jupyter notebooks linked to in \cref{Overview} describe how the different terms in this likelihood function compare and ranks different source reconstructions, allowing one to objectively determine the lens model that provides the best fit to the data in a Bayesian context.

\subsection{Non-linear Search}

We use the nested sampling algorithm {\tt dynesty} \citep{Speagle2020} to fit every lens model. We use the static sampler with random walk nested sampling, which tests revealed gave faster and more reliable lens model fits. 


\subsection{Lens Modeling Pipelines}\label{SLaM}

{\renewcommand{\arraystretch}{1.83}
\begin{table*}
\centering
    \begin{tabularx}{\textwidth}{@{}*{6}{c}{X}}
        \hlineB{1}
        \textbf{Pipeline} & \textbf{Phase} & \textbf{Galaxy Component} & \textbf{Model} & \textbf{Varied} & \textbf{Prior info} & \textbf{Phase Description}\\
        \hlineB{1}
        \multirow{6}{*}{\shortstack[l]{Source \\ Parametric}} & \textbf{SP}$^{1}$ & Lens light & S\'ersic + Exp &  & - & Fit only the lens light model and subtract it from the data image. \\
        \cline{2-7}
        &\multirow{2}{*}{\textbf{SP}$^{2}$} & Lens mass & SIE + shear &  & - &\multirow{2}{\hsize}{Fit mass model and source light using lens subtracted image from \textbf{SP}$^{1}$.}  \\
        & & Source light & S\'ersic &  & - \\
        \cline{2-7}
        &\multirow{3}{*}{\textbf{SP}$^{3}$} & Lens light & S\'ersic + Exp &  & -  & \multirow{3}{\hsize}{Refit the lens light model with default priors and fit the mass and source models with priors informed from \textbf{SP}$^{2}$.}\\
        & & Lens mass & SIE + shear &  & \textbf{SP}$^{2}$\\
        & & Source light & S\'ersic &  & \textbf{SP}$^{2}$ \\
        \cline{1-7}
        \multirow{12}{*}{\shortstack[l]{Source \\ Inversion}}& \multirow{3}{*}{\textbf{SI$^1$}} & Lens light & S\'ersic + Exp & \checkmark & \textbf{SP}$^3$ & \multirow{3}{\hsize}{Fix lens light and mass parameters from \textbf{SP}$^{3}$ and fit magnification adaptive Voronoi mesh and constant regularisation parameters.}\\
        & & Lens mass & SIE + shear & \checkmark & \textbf{SP}$^3$ \\
        & & Source light & Voronoi Magnification &  & - \\
        \cline{2-7}
        & \multirow{3}{*}{\textbf{SI}$^2$} & Lens light & S\'ersic + Exp & \checkmark & \textbf{SP}$^3$  & \multirow{3}{\hsize}{Refine the lens mass model parameters, keeping lens light and source parameters fixed to those from previous phases.} \\
        & & Lens mass & SIE + shear &  & \textbf{SP}$^3$  \\
         & & Source light & Voronoi Magnification & \checkmark & \textbf{SI}$^1$   \\
        \cline{2-7}
        & \multirow{3}{*}{\textbf{SI}$^3$} & Lens light & S\'ersic + Exp & \checkmark & \textbf{SP}$^3$  & \multirow{3}{\hsize}{Fit brightness adaptive Voronoi mesh and luminosity adaptive regularisation. Lens parameters fixed from \textbf{SP}$^{3}$.}\\
        & & Lens mass & SIE + shear & \checkmark & \textbf{SP}$^3$ \\
        & & Source light &Voronoi Brightness&  & -  \\
        \cline{2-7}
        & \multirow{3}{*}{\textbf{SI}$^4$} & Lens light & S\'ersic + Exp & \checkmark & \textbf{SP}$^3$ & \multirow{3}{\hsize}{Refine lens mass model parameters using Voronoi Brightness grid. Fix lens light and source parameters to previous phases.} \\
        & & Lens mass & SIE + shear &  & \textbf{SI}$^2$   \\
        & & Source light & Voronoi Brightness & \checkmark & \textbf{SI}$^3$   \\
        \hlineB{1}
        \multirow{3}{*}{\shortstack[l]{Light \\ Parametric}} & \multirow{3}{*}{\textbf{LP}$^1$} & Lens light & S\'ersic + S\'ersic &  & - & \multirow{3}{\hsize}{Fit lens light parameters with broad uniform priors. Lens mass and source parameters fixed from \textbf{SI}$^4$.}\\
        & & Lens mass & SIE + shear & \checkmark & \textbf{SI}$^4$  \\
         & & Source light & Voronoi Brightness & & \textbf{SI}$^4$  \\
        \hlineB{1}
        \multirow{3}{*}{\shortstack[l]{Mass \\ Total}} & \multirow{3}{*}{\textbf{MT}$^1$} & Lens light & S\'ersic + S\'ersic & \checkmark & \textbf{LP}$^1$ & \multirow{3}{\hsize}{Fit the lens mass parameters, with subset of priors informed from \textbf{SI}$^4$. Lens and source light are fixed from \textbf{LP}$^1$ and \textbf{SI}$^4$.}\\
        & & Lens mass & See \cref{Mass} &  & \textbf{SI}$^{4}$  \\
        & & Source light & Voronoi Brightness &  & \textbf{LP}$^1$  \\
        \hlineB{1}
        \multirow{9}{*}{\shortstack[l]{Subhalo}} & \multirow{3}{*}{\textbf{SH}$^1$} & Lens light & S\'ersic + S\'ersic & \checkmark & \textbf{LP}$^1$ & \multirow{3}{\hsize}{Fit the lens mass parameters, with priors informed from \textbf{MT}$^1$. Lens and source light are fixed from \textbf{LP}$^1$ and \textbf{SI}$^4$.}\\
        & & Lens mass & See \cref{Mass} &  & \textbf{MT}$^{1}$  \\
        & & Source light & Voronoi Brightness &  & \textbf{MT}$^1$  \\
        \cline{2-7}
         & \multirow{3}{*}{\textbf{SH}$^2$} & Lens light & S\'ersic + S\'ersic & \checkmark & \textbf{LP}$^1$ & \multirow{3}{\hsize}{Performs grid search of DM subhhalos (see \cref{SubScan}).}\\
        & & Lens mass & See \cref{Mass} + Subhalo &  & \textbf{MT}$^{1}$  \\
        & & Source light & Voronoi Brightness &  & \textbf{MT}$^1$  \\
        \cline{2-7}
         & \multirow{3}{*}{\textbf{SH}$^3$} & Lens light & S\'ersic + S\'ersic & \checkmark & \textbf{LP}$^1$ & \multirow{3}{\hsize}{Fits for DM subhalo using priors based on \textbf{SH}$^2$. Bayesian evidence compared to \textbf{SH}$^1$ for DM subhalo inference.}\\
        & & Lens mass & See \cref{Mass} + Subhalo &  & \textbf{MT}$^{1}$  \\
        & & Source light & Voronoi Brightness &  & \textbf{MT}$^1$  \\
        \cline{2-7}
        \hlineB{1}
    \end{tabularx}
    \caption{Composition of the Source, Light and Mass (SLaM) pipelines that make up our uniform analysis. This is an adaptation and extension of the table presented by \citet{Etherington2022}, who used the SLaM pipelines to fit HST imaging of strong lenses. After the phases
    \textbf{SI}$^4$, \textbf{LP}$^1$ and \textbf{MT}$^1$ the best-fit model images (the $\Xi_{j}$ and $\L{\rm _j}$) are updated for the adaptive mesh, regularization and variance scaling features. Their associated parameters are refit via a standalone \texttt{dynesty} fit using a fixed lens model.} 
    \label{table: pipelines}
\end{table*}}

The models of lens mass, lens light and source light are complex and their parameter spaces are highly dimensional. Without human intervention or careful set up, the model-fitting algorithm (e.g. {\tt dynesty}) may converge very slowly to the global maximum a posteriori solution or falsely converge on a local maximum. \texttt{PyAutoLens} therefore breaks the fit into a sequence of simpler fits. Using the probabilistic programming language {\tt PyAutoFit}\footnote{\url{https://github.com/rhayes777/PyAutoFit}}, we fit a series of parametric lens models with growing complexity. Fits to simpler model parameterizations provide information which initialises subsequent fits to the next more complex model. We use the Source, Light and Mass (SLaM) pipelines described by \citet[][hereafter E22]{Etherington2022}, \citet{Cao2021} and \citet{He2023}. \cref{table: pipelines} provides a step-by-step overview of the pipelines used in this work. The SLaM pipelines are available at \url{https://github.com/Jammy2211/autolens_workspace}. 

An overview of the SLaM pipelines is as follows:

\begin{itemize}
    \item \textbf{Source Pipelines:} Initializes the Voronoi mesh source model by inferring a robust lens light subtraction (using a double S\'ersic model) and total mass model (using an SIE plus shear). The initial stages of this pipeline fit the source using a parametric S\'ersic profile and perform the variance scaling described in \cref{VarScale}.
    
    \item \textbf{Light Pipeline:} Uses fixed values of the mass and source parameters corresponding to the maximum likelihood model of the Source pipeline. This is the first time the lens light is fitted for simultaneously with a Voronoi mesh source instead of Sersic profile. The lens mass is therefore again described by an SIE plus shear. The only free parameters in this pipeline are those of a double S\'ersic lens light model and $\omega_{\rm  Lens}$ which controls the magnitude of variance scaling. The maximum likelihood lens light subtracted image inferred by this pipeline is output for use by additional fits investigating lens modelling systematics.
    
    \item \textbf{Mass Pipeline:} Fits a PL, BPL, PL with multipoles, decomposed mass model or PL with line-of-sight gaalxies, which are all more complex than the SIE fitted previously. The lens light is fixed to the maximum likelihood model of the Light pipeline.

    \item \textbf{Subhalo Pipeline:} Determines the increase in log Bayesian evidence when a DM subhalo is included in the lens model, which is described next in \cref{SubScan}.
\end{itemize}

The SLaM pipelines use prior passing (see E22) to initialize the regions of parameter space that \texttt{dynesty} will search in later \texttt{dynesty} fits, based on the results of earlier fits. The priors of every lens model fitted in this work can be found at \url{https://github.com/Jammy2211/autolens_subhalo}. Priors are set up carefully to ensure they are sufficiently broad to not omit viable lens model solutions. 

\subsection{Subhalo Scanning}\label{SubScan}

Including a subhalo in the mass model produces a complex and multimodal parameter space that a nested sampler like \texttt{dynesty} may struggle to sample efficiently and robustly. We tested a model-fitting approach which simply adds a subhalo to a lens model 
assuming broad uniform priors on the subhalo's image-plane position ($x^{\rm sub}$, $y^{\rm sub}$) and mass $M_{\rm 200}^{\rm sub}$. However, these fits did not always reliably infer the subhalo's input properties on simulated datasets. 

We instead perform a grid-search of \texttt{dynesty} searches, where each grid-search cell places uniform priors on the image-plane ($x^{\rm sub}$, $y^{\rm sub}$) position of the subhalo, spatially confining it to a small 2D square segment of the image-plane. We perform $25$ model-fits (corresponding to a $5 \times 5$ grid in the image-plane), where the size of the box containing this grid is chosen via visual inspection of each lens. An example subhalo scan is shown in \cref{figure:Overview}. This removes the multi-modality in the parameter space created by the subhalo model, simplifying it such that the global maxima solution in parameter space is reliably inferred. For each grid-cell, log uniform priors with masses between $10^{6}$M$_{\rm \odot}$ - $10^{12}$M$_{\rm \odot}$ are assumed for $M^{\rm sub}_{\rm 200}$. We always assume the subhalo is at the same redshift as the lens galaxy (e.g. single plane lensing). 

Once a grid search is complete, a final non-linear search is performed which provides accurate constraints on the subhalo mass $M^{\rm sub}_{\rm 200}$ and image-plane coordinates ($x^{\rm sub}$, $y^{\rm sub}$). The subhalo centre's priors are set via prior passing, using the highest evidence model of the grid search (the lens model parameters also use this result). This prior allows for a wider range of subhalo centres than the uniform priors defining each 2D grid-cell, but is centred on the highest evidence grid search model, ensuring \texttt{dynesty} sampling remains reliable. The subhalo retains its log uniform prior on $M^{\rm sub}_{\rm 200}$ with masses between $10^{6}$M$_{\rm \odot}$ - $10^{12}$M$_{\rm \odot}$, to avoid overly tight priors reducing the inferred error. \texttt{dynesty} settings are adjusted to sample parameter space more thoroughly at the expense of longer computational run-time. 

We quantify whether models including a subhalo are favoured using the Bayesian evidence, $\mathcal{Z}$, of the lens models with and without a DM subhalo. The evidence is the integral of the likelihood over the prior and therefore naturally includes a penalty term for including too much complexity in a model. $\mathcal{Z}$ is inferred by \texttt{dynesty} (see equation 2 of \citealt{Speagle2020}) and therefore available for every fit performed in this work. We define the log evidence difference in favour of the lens model with a DM subhalo as
\begin{equation}
\label{eqn:Evidence}
\Delta\,\mathrm{ln}\,\mathcal{Z} = \mathrm{ln}\,\mathcal{Z}_{\rm sub} - \mathrm{ln}\,\mathcal{Z}_{\rm none},
\end{equation}
where $\mathrm{ln}\,\mathcal{Z}_{\rm sub}$ is the Bayesian evidence inferred by the fit after the subhalo scanning grid search and $\mathrm{ln}\,\mathcal{Z}_{\rm none}$ is the evidence of the lens model without a subhalo before the grid search. Superscripts are added to $\Delta\,\mathrm{ln}\,\mathcal{Z}$ to denote model-fits which make different assumptions, for example $\Delta\,\mathrm{ln}\,\mathcal{Z}^{\rm Base}$ denotes the increase in log evidence for the baseline lens model with a subhalo assuming a PL mass model, double S\'ersic lens light model and where the source is reconstructed on a Voronoi mesh. An increase of $\Delta\,\mathrm{ln}\,\mathcal{Z} = 4.5$ for one model over another corresponds to odds of 90:1 in favour of that model;  a $\rm 3\sigma$ preference. An increase of $\Delta \ln \mathcal{Z} = 12.5$ corresponds to a $\rm 5\sigma$ preference. Our criteria for a candidate subhalo detection is that we infer $\Delta\,\mathrm{ln}\,\mathcal{Z} > 10$. The subhalo scanning analysis is the same as that used in \citet{He2023}, who modeled strong lenses simulated via cosmological simulations with {\tt PyAutoLens}.
        
\section{Results}\label{ResultsData}

We now present the results of subhalo scanning different datasets. In \cref{SimSummary} we give a concise summary of fits to simulated lens datasets which are described fully in \cref{ResultSim}.We use these results as a starting point to investigate false positives DM subhalo detections due to lens modeling systematics.

\begin{table*}
\Large
\begin{adjustbox}{max width=\textwidth}
\begin{tabular}{ l | l | l | l | l | l | l | l | l | l | l | l | l | l | l} 
\multicolumn{1}{p{2.0cm}|}{Lens Name} 
& \multicolumn{1}{p{1.2cm}|}{$\Delta\,\mathrm{ln}$ \, $\mathcal{Z}^{\rm Base}$}  
& \multicolumn{1}{p{2.0cm}|}{$\log_{10}$ $[M^{\rm sub}_{\rm 200}$\,/\,M$_{\rm \odot}]$} 
& \multicolumn{1}{p{1.3cm}|}{$\Delta\,\mathrm{ln}$ \,$\mathcal{Z}^{\rm Light}$}  
& \multicolumn{1}{p{1.3cm}|}{Light $\Delta\,\mathrm{ln}\,\mathcal{Z}$ Decrease?}  
& \multicolumn{1}{p{1.3cm}|}{$\Delta\,\mathrm{ln}$ \, $\mathcal{Z}^{\rm Source}$}  
& \multicolumn{1}{p{1.3cm}|}{Source  $\Delta\,\mathrm{ln}\,\mathcal{Z}$ Decrease?}  
& \multicolumn{1}{p{1.5cm}|}{$\Delta\,\mathrm{ln}$ \, $\mathcal{Z}^{\rm BPL}$} 
& \multicolumn{1}{p{1.5cm}|}{$\Delta\,\mathrm{ln}$ \, $\mathcal{Z}^{\rm Multipole}$} 
& \multicolumn{1}{p{1.5cm}|}{$\Delta\,\mathrm{ln}$ \, $\mathcal{Z}^{\rm Decomp}$} 
& \multicolumn{1}{p{1.3cm}|}{Mass $\Delta\,\mathrm{ln}\,\mathcal{Z}$ Decrease?}  
& \multicolumn{1}{p{1.5cm}|}{$\Delta\,\mathrm{ln}$ \, $\mathcal{Z}^{\rm Los}$} 
& \multicolumn{1}{p{1.3cm}|}{Los $\Delta\,\mathrm{ln}\,\mathcal{Z}$ Decrease?}  
& \multicolumn{1}{p{1.5cm}|}{$\Delta\,\mathrm{ln}$ \, $\mathcal{Z}^{\rm Final}$} 
& \multicolumn{1}{p{1.2cm}}{Category} \\ 
\hline
& & & & & & & & & & & & & & \\[-6pt]
SLACS2341+0000 & $\textbf{157.51}$ & $11.98^{+0.02}_{-0.21}$ & $\textbf{32.7}$ & $\checkmark$ & $\textbf{24.91}$ & \xmark & $\textbf{25.94}$ & $\textbf{18.46}$ & 8.52 & $\checkmark$ & Demag & \xmark & \textbf{11.61} & ND / Los \\[2pt]
SLACS1432+6317 & $\textbf{79.34}$ & $11.96^{+0.04}_{-0.30}$ & 8.65 & $\checkmark$ & -0.23 & \xmark & Demag & $\textbf{10.87}^{*}$ & 1.83 & \xmark & $\textbf{15.24}^{*}$ & \xmark & 1.82 & ND \\[2pt]
SLACS0946+1006 & $\textbf{52.86}$ & $11.91^{+0.09}_{-0.55}$ & $\textbf{75.76}^{*}$ & \xmark & $\textbf{72.36}$ & \xmark & $\textbf{74.27}$ & $\textbf{76.81}$ & $\textbf{22.52}$ & $\checkmark$ & $\textbf{29.24}$ & $\checkmark$ & \textbf{22.51} & Cand \\[2pt]
SLACS0956+5100 & $\textbf{40.11}$ & $11.77^{+0.23}_{-0.58}$ & $\textbf{18.04}$ & $\checkmark$ & $\textbf{23.35}$ & \xmark & $\textbf{12.07}$ & $\textbf{11.52}$ & $\textbf{10.78}$ & $\checkmark$ & $\textbf{31.37}$ & \xmark & \textbf{10.77} & ND / Los \\[2pt]
SLACS1020+1122 & $\textbf{38.71}$ & $11.93^{+0.07}_{-0.43}$ & 5.36 & $\checkmark$ & 2.1 & \xmark & 3.23 & 0.99 & 7.81 & \xmark & 4.42 & \xmark & 0.99 & ND \\[2pt]
SLACS1250+0523 & $\textbf{30.87}$ & $10.68^{+0.84}_{-0.33}$ & 8.23 & $\checkmark$ & $\textbf{13.41}$ & \xmark & $\textbf{18.68}$ & -2.09 & $\textbf{17.4}$ & $\checkmark$ & 8.17 & \xmark & 3.16 & ND / FP-PL \\[2pt]
SLACS1032+5322 & $\textbf{27.79}$ & $11.96^{+0.04}_{-0.28}$ & -0.49 & $\checkmark$ & 0.81 & \xmark & Demag & -1.68 & 6.26 & \xmark & -0.08 & \xmark & 2.82 & ND \\[2pt]
SLACS0959+0410 & $\textbf{21.39}$ & $11.91^{+0.09}_{-0.43}$ & $\textbf{49.66}^{*}$ & \xmark & $\textbf{19.62}$ & $\checkmark$ & Demag & 5.95 & $\textbf{28.12}$ & \xmark & 3.37 & $\checkmark$ &  -24.90 & ND / FP-PL \\[2pt]
SLACS0029-0055 & $\textbf{20.58}$ & $11.12^{+0.50}_{-0.52}$ & $\textbf{33.92}^{*}$ & \xmark & 4.82 & $\checkmark$ & $\textbf{18.75}^{*}$ & 7.22 & $\textbf{35.15}^{*}$ & \xmark & 1.36 & \xmark & \textbf{21.69} & Cand / Decomp \\[2pt]
SLACS1023+4230 & $\textbf{19.73}$ & $11.14^{+0.44}_{-0.60}$ & 4.32 & $\checkmark$ & 4.82 & \xmark & 1.26 & -1.31 & 0.72 & \xmark & 1.5 & \xmark & -1.30 & ND \\[2pt]
SLACS1143-0144 & $\textbf{19.17}$ & $11.94^{+0.06}_{-0.50}$ & 5.35 & $\checkmark$ & 2.37 & \xmark & 0.65 & -0.45 & $\textbf{17.23}^{*}$ & \xmark & $\textbf{13.29}^{*}$ & \xmark & -0.44 & ND \\[2pt]
SLACS0157-0056 & $\textbf{16.38}$ & $11.52^{+0.48}_{-5.40}$ & $\textbf{17.65}$ & \xmark & -0.57 & $\checkmark$ & -0.29 & -0.93 & 1.06 & \xmark & -0.37 & \xmark & -0.57 & ND \\[2pt]
SLACS1451-0239 & $\textbf{13.78}$ & $8.39^{+2.45}_{-2.36}$ & -1.09 & $\checkmark$ & -0.44 & \xmark & -0.67 & 2.43 & 2.98 & \xmark & -0.22 & \xmark & -0.22 & ND \\[2pt]
SLACS1430+4105 & $\textbf{12.0}$ & $10.80^{+0.50}_{-0.69}$ & $\textbf{11.14}$ & \xmark & $\textbf{13.4}$ & \xmark & 5.25 & $\textbf{14.06}$ & 6.54 & \xmark & $\textbf{15.72}$ & \xmark & 6.53 & ND / FP-PL \\[2pt]
SLACS0903+4116 & 9.72 & $11.89^{+0.11}_{-0.76}$ & $\textbf{15.35}$ & \xmark & 3.91 & $\checkmark$ & 3.79 & 1.05 & 9.13 & \xmark & 2.16 & \xmark & 3.90 & ND \\[2pt]
SLACS2303+1422 & 9.66 & $11.10^{+0.47}_{-0.79}$ & 1.1 & \xmark & 2.14 & \xmark & 2.82 & -1.6 & 1.73 & \xmark & 8.06 & \xmark & 3.80 & ND \\[2pt]
SLACS1213+6708 & 8.73 & $11.88^{+0.12}_{-1.10}$ & 1.1 & \xmark & 3.49 & \xmark & 0.21 & 0.25 & 1.52 & \xmark & 1.18 & \xmark & 1.52 & ND \\[2pt]
SLACS1630+4520 & 8.45 & $10.89^{+0.45}_{-0.64}$ & 5.41 & \xmark & -0.58 & \xmark & -0.43 & Demag & -1.66 & \xmark & 7.15 & \xmark & -1.66 & ND \\[2pt]
SLACS0822+2652 & 7.94 & $11.87^{+0.13}_{-0.92}$ & 2.67 & \xmark & -1.25 & \xmark & 2.1 & 1.46 & 2.19 & \xmark & -1.92 & \xmark & -1.38 & ND \\[2pt]
SLACS1029+0420 & 4.03 & $11.75^{+0.25}_{-3.47}$ & -1.28 & \xmark & 2.17 & \xmark & Demag & 3.71 & $\textbf{22.57}^{*}$ & \xmark & 1.71 & \xmark & \textbf{10.57} & Cand / Decomp \\[2pt]
SLACS2300+0022 & 4.49 & $9.41^{+2.57}_{-3.40}$ & -1.7 & \xmark & -0.46 & \xmark & -0.29 & -0.47 & 2.23 & \xmark & 1.47 & \xmark & -0.46 & ND \\[2pt]
SLACS1205+4910 & 4.12 & $6.92^{+1.42}_{-0.91}$ & 8.97 & \xmark & 2.14 & \xmark & 0.63 & 2.25 & $\textbf{31.15}^{*}$ & \xmark & -2.38 & \xmark & 1.87 & ND \\[2pt]
SLACS0912+0029 & 3.92 & $11.89^{+0.11}_{-1.05}$ & 3.1 & \xmark & 1.19 & \xmark & 2.36 & 1.49 & 5.33 & \xmark & 0.33 & \xmark & 2.07 & ND \\[2pt]
SLACS1402+6321 & 3.41 & $9.74^{+1.36}_{-3.73}$ & -0.14 & \xmark & -0.53 & \xmark & -2.01 & -1.03 & 4.64 & \xmark & -0.7 & \xmark & 0.01 & ND \\[2pt]
SLACS0252+0039 & 3.32 & $10.71^{+0.69}_{-1.76}$ & -1.04 & \xmark & 6.53 & \xmark & -0.89 & $\textbf{17.66}^{*}$ & 5.26 & \xmark & 5.89 & \xmark & -0.89 & ND \\[2pt]
SLACS1218+0830 & 3.25 & $11.06^{+0.92}_{-5.04}$ & -0.12 & \xmark & 0.24 & \xmark & -1.35 & 0.16 & 2.42 & \xmark & -0.92 & \xmark & 0.23 & ND \\[2pt]
SLACS1525+3327 & 3.15 & $10.94^{+1.02}_{-4.81}$ & 1.34 & \xmark & 0.24 & \xmark & Demag & Demag & Demag & \xmark & Demag & \xmark & 0.24 & ND \\[2pt]
SLACS1627-0053 & 2.52 & $11.37^{+0.58}_{-5.12}$ & 2.51 & \xmark & 7.9 & \xmark & 3.25 & 8.18 & 5.27 & \xmark & 5.93 & \xmark & 7.89 & ND \\[2pt]
SLACS0008-0004 & 2.51 & $8.83^{+2.94}_{-2.82}$ & 1.59 & \xmark & -1.44 & \xmark & -1.43 & -0.38 & -0.46 & \xmark & -0.28 & \xmark & -1.43 & ND \\[2pt]
SLACS1420+6019 & 1.92 & $10.81^{+1.06}_{-1.96}$ & 0.64 & \xmark & 3.11 & \xmark & 0.6 & -1.56 & 2.79 & \xmark & 2.94 & \xmark & 2.79 & ND \\[2pt]
SLACS0330-0020 & 1.19 & $10.83^{+0.36}_{-4.66}$ & -1.44 & \xmark & -0.88 & \xmark & 2.29 & $\textbf{11.04}^{*}$ & 8.12 & \xmark & 3.69 & \xmark & 0.01 & ND \\[2pt]
SLACS1142+1001 & 0.71 & $11.41^{+0.59}_{-5.17}$ & -0.42 & \xmark & 0.31 & \xmark & 0.78 & 0.53 & $\textbf{10.47}^{*}$ & \xmark & 0.64 & \xmark & 6.00 & ND \\[2pt]
SLACS0936+0913 & 0.33 & $10.30^{+1.56}_{-4.26}$ & -2.2 & \xmark & -1.12 & \xmark & -0.8 & -2.01 & 0.25 & \xmark & 0.33 & \xmark & -1.12 & ND \\[2pt]
SLACS2238-0754 & -3.36 & $8.72^{+2.91}_{-2.70}$ & -0.1 & \xmark & 0.78 & \xmark & -2.39 & $\textbf{17.66}^{*}$ & 8.53 & \xmark & 0.62 & \xmark & -0.22 & ND \\[2pt]
SLACS0216-0813 & -1.43 & $11.31^{+0.68}_{-5.23}$ & -3.4 & \xmark & -0.65 & \xmark & 0.55 & -1.41 & -0.12 & \xmark & -1.04 & \xmark & -0.11 & ND \\[2pt]
SLACS0737+3216 & -1.34 & $9.43^{+2.51}_{-3.38}$ & 1.21 & \xmark & 0.33 & \xmark & -1.99 & -0.74 & 0.99 & \xmark & -0.52 & \xmark & 0.99 & ND \\[2pt]
SLACS0728+3835 & -4.17 & $8.71^{+0.28}_{-0.27}$ & -6.58 & \xmark & 2.92 & \xmark & Demag & $\textbf{16.18}^{*}$ & 1.81 & \xmark & -10.47 & $\checkmark$ & 0.25 & ND / FP-Los \\[2pt]
\end{tabular}
\end{adjustbox}
\caption{
The log Bayesian evidence increase $\Delta\,\mathrm{ln}\,\mathcal{Z}$ of subhalo scanning for the SLACS sample. Column 1 gives the lens name. Column 2 shows $\Delta\,\mathrm{ln}\,\mathcal{Z}^{\rm Base}$ for the baseline model, which assumes a PL plus shear lens mass model, a two S\'ersic lens light subtraction, and a Voronoi mesh source reconstruction. Column 3 gives the corresponding inferred DM subhalo mass at $3\sigma$ confidence intervals. Column 4-13 show the results of different systematic tests. Column 4 shows $\Delta\,\mathrm{ln}\,\mathcal{Z}^{\rm Light}$ values, where lens light cleaned data is fitted (see \cref{sec:lens_light_results}). Column 5 indicates if $\Delta\,\mathrm{ln}\,\mathcal{Z}^{\rm Base} -  \Delta\,\mathrm{ln}\,\mathcal{Z}^{\rm Light} > 10$ (\textit{e.g. does improving the lens light subtraction decrease} $\Delta\,\mathrm{ln}\,\mathcal{Z}$?). Column 6 shows $\Delta\,\mathrm{ln}\,\mathcal{Z}^{\rm Source}$ values, where source only masked data is fitted (see \cref{sec:source_only_results}). Column 7 indicates if $\Delta\,\mathrm{ln}\,\mathcal{Z}^{\rm Light} -  \Delta\,\mathrm{ln}\,\mathcal{Z}^{\rm Source} > 10$ (\textit{e.g. does improving the source reconstruction decrease }$\Delta\,\mathrm{ln}\,\mathcal{Z}$?). Columns 8-10 show fits assuming different mass models: a broken power-law ($\Delta\,\mathrm{ln}\,\mathcal{L}^{\rm BPL}$), a PL with internal multipoles ($\Delta\,\mathrm{ln}\,\mathcal{L}^{\rm Multipole}$) and a decomposed mass model ($\Delta\,\mathrm{ln}\,\mathcal{L}^{\rm Decomp}$). Column 11 indicates whether $\Delta\,\mathrm{ln}\,\mathcal{Z}^{\rm Source} -  \max(\Delta\,\mathrm{ln}\,\mathcal{Z}^{\rm BPL}, \Delta\,\mathrm{ln}\,\mathcal{Z}^{\rm Multipole}, \Delta\,\mathrm{ln}\,\mathcal{Z}^{\rm Decomp}) > 10$ \textit{(e.g. does assuming any of the three more complex mass models decrease} $\Delta\,\mathrm{ln}\,\mathcal{Z}$?). Column 12 shows $\Delta\,\mathrm{ln}\,\mathcal{L}^{\rm Los}$ values, which fit a PL plus shear mass model that includes line-of-sight galaxies. Column 13 indicates if $\Delta\,\mathrm{ln}\,\mathcal{Z}^{\rm Source} -  \Delta\,\mathrm{ln}\,\mathcal{Z}^{\rm Los} > 10$ (\textit{e.g. does including line-of-sight galaxies decrease} $\Delta\,\mathrm{ln}\,\mathcal{Z}$?). Bold text indicates any fit above the detection threshold of $\Delta\,\mathrm{ln}\,\mathcal{Z} = 10.0$.  Asterix symbols indicate that $\Delta\,\mathrm{ln}\,\mathcal{Z}$ increases by over $10$ for any of the four comparisons shown in the table (\textit{e.g. does improving the lens light, source, mass model or including line-of-sight galaxies increase} $\Delta\,\mathrm{ln}\,\mathcal{Z}$?). Column 13 shows $\Delta\,\mathrm{ln}\,\mathcal{Z}^{\rm Final}$, the highest evidence of any mass model with a DM subhalo (corresponding to fits given in columns 5, 7, 8, 9, 11) minus the highest evidence of any mass model without a DM subhalo. The final column categorizes each DM subhalo result: \textit{ND} for non-detection, \textit{Cand} for candidate, \textit{FP-PL} for power-law mass false positive and \textit{FP-Los} for line-of-sight false positive. Entries labeled `Demag' are solutions where the fit inferred an unphysical solution (see \cref{Mass})}.
\label{table:DetectSLACS}
\end{table*}

\begin{figure*}
\centering
\includegraphics[width=0.162\textwidth]{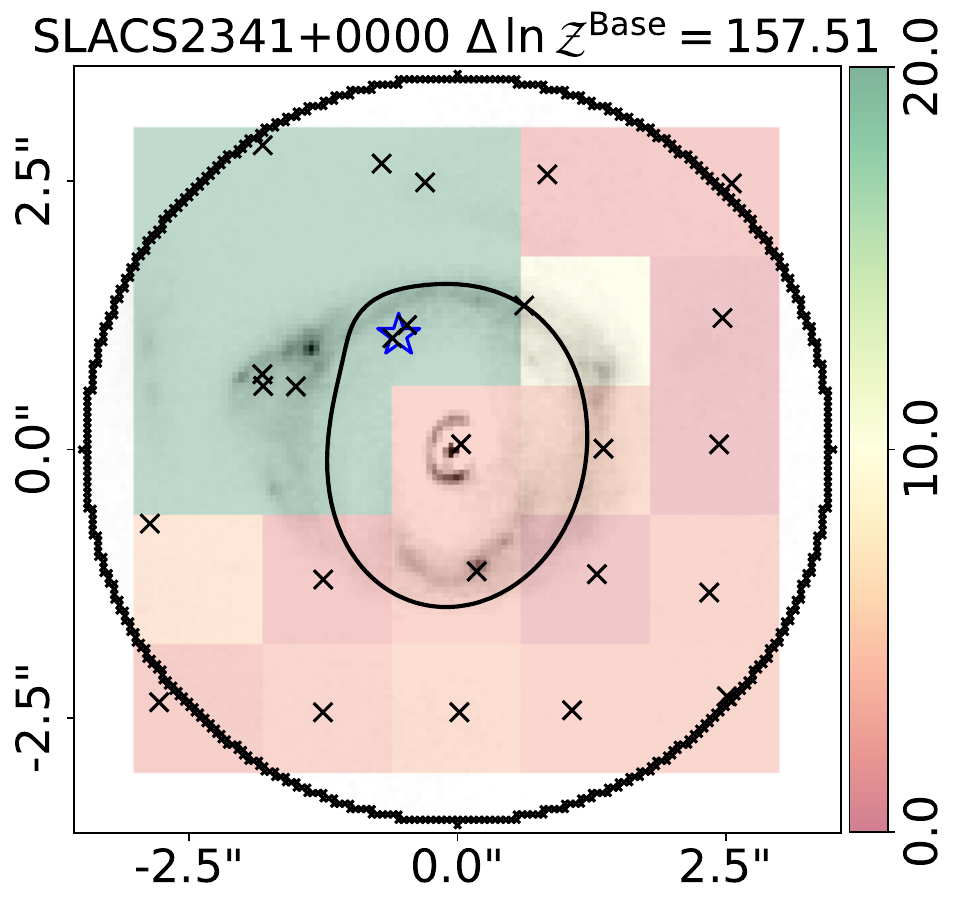}
\includegraphics[width=0.162\textwidth]{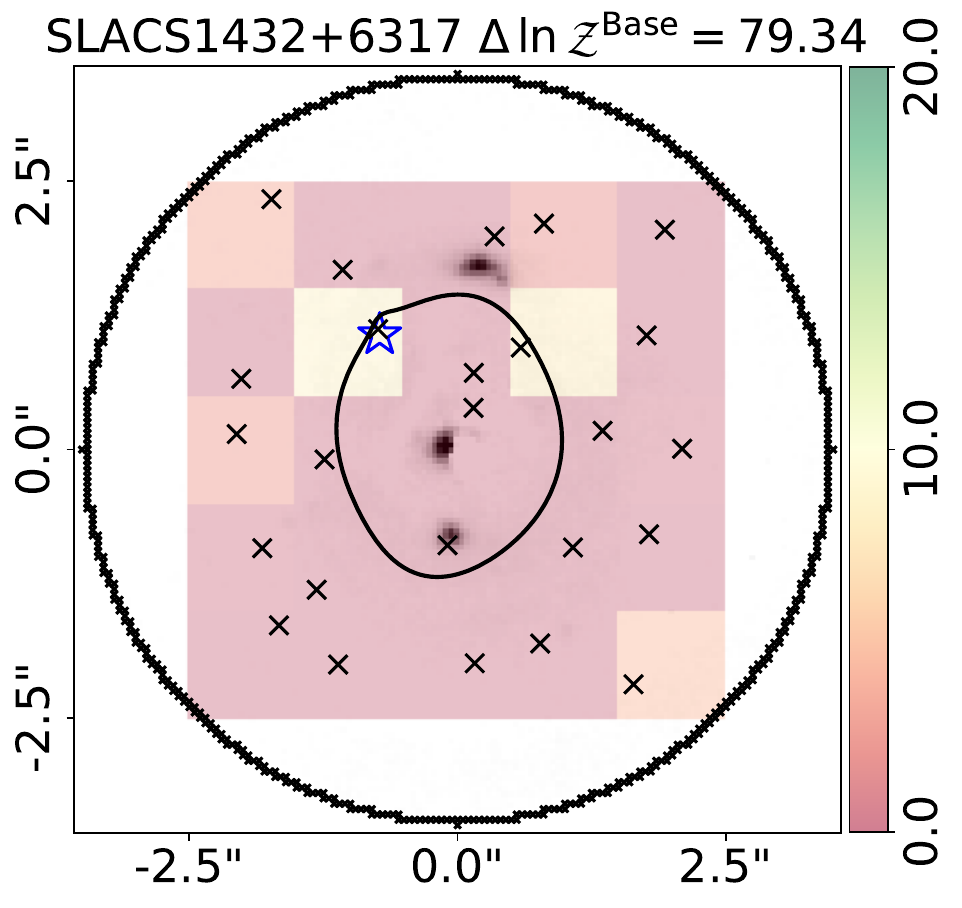}
\includegraphics[width=0.162\textwidth]{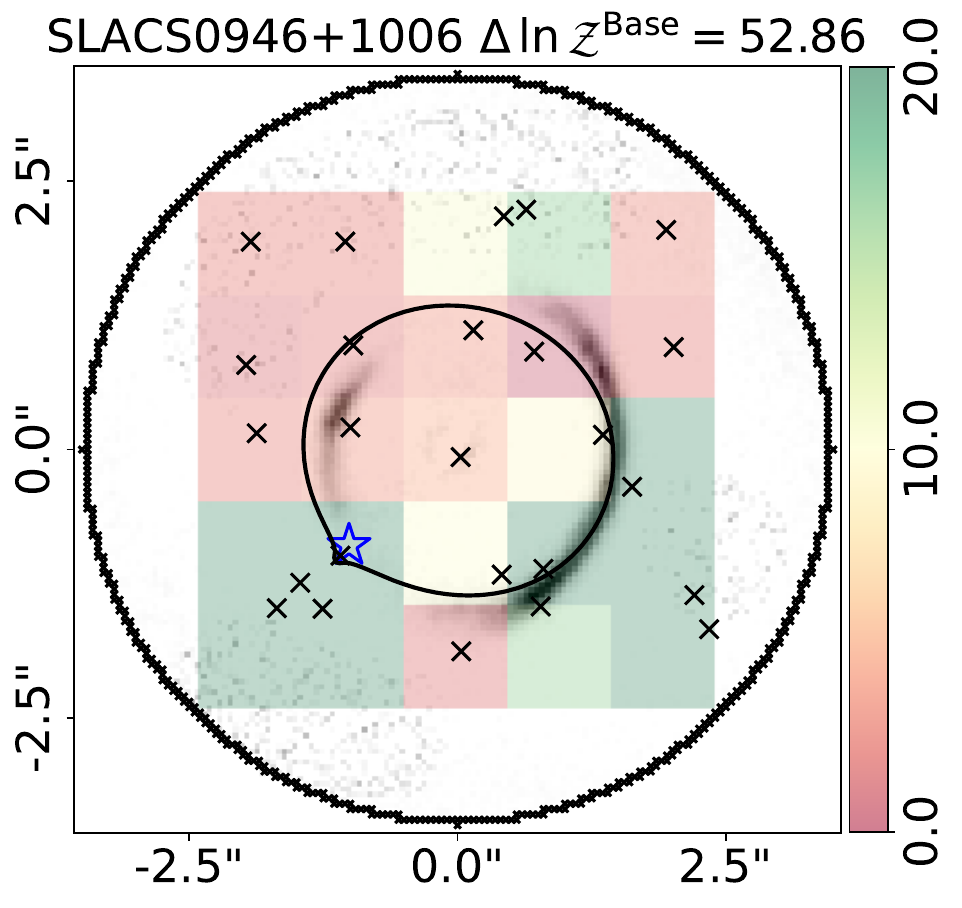}
\includegraphics[width=0.162\textwidth]{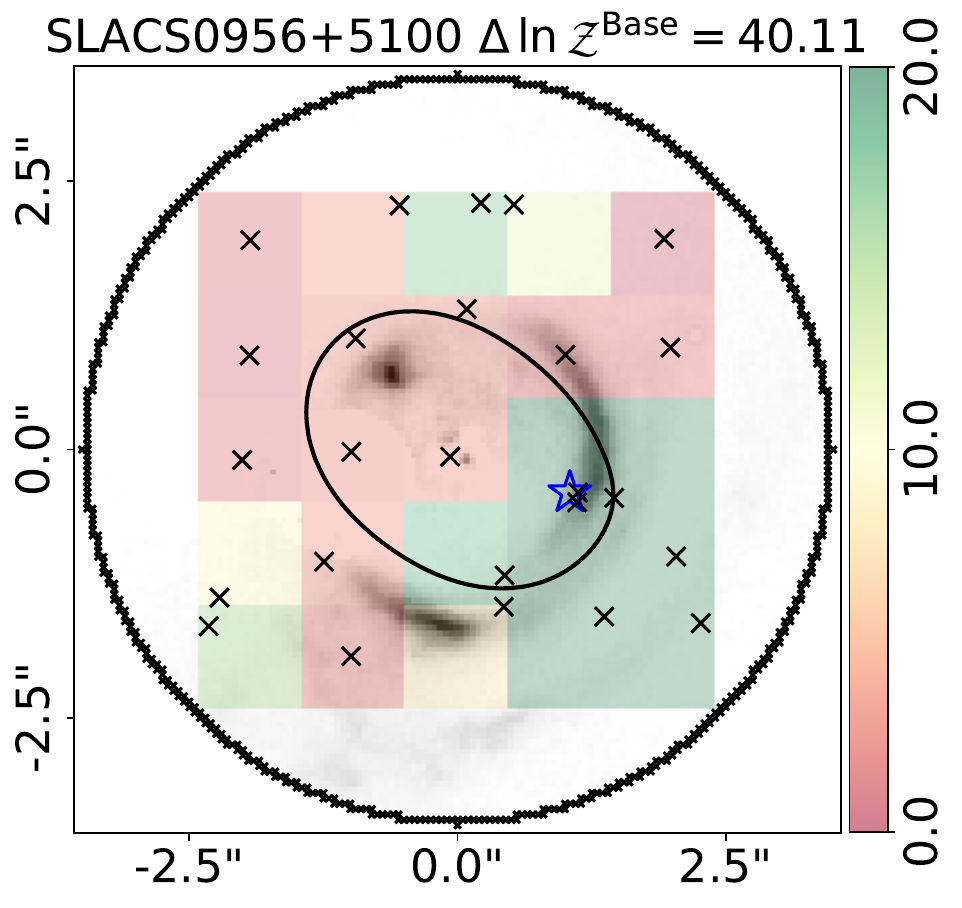}
\includegraphics[width=0.162\textwidth]{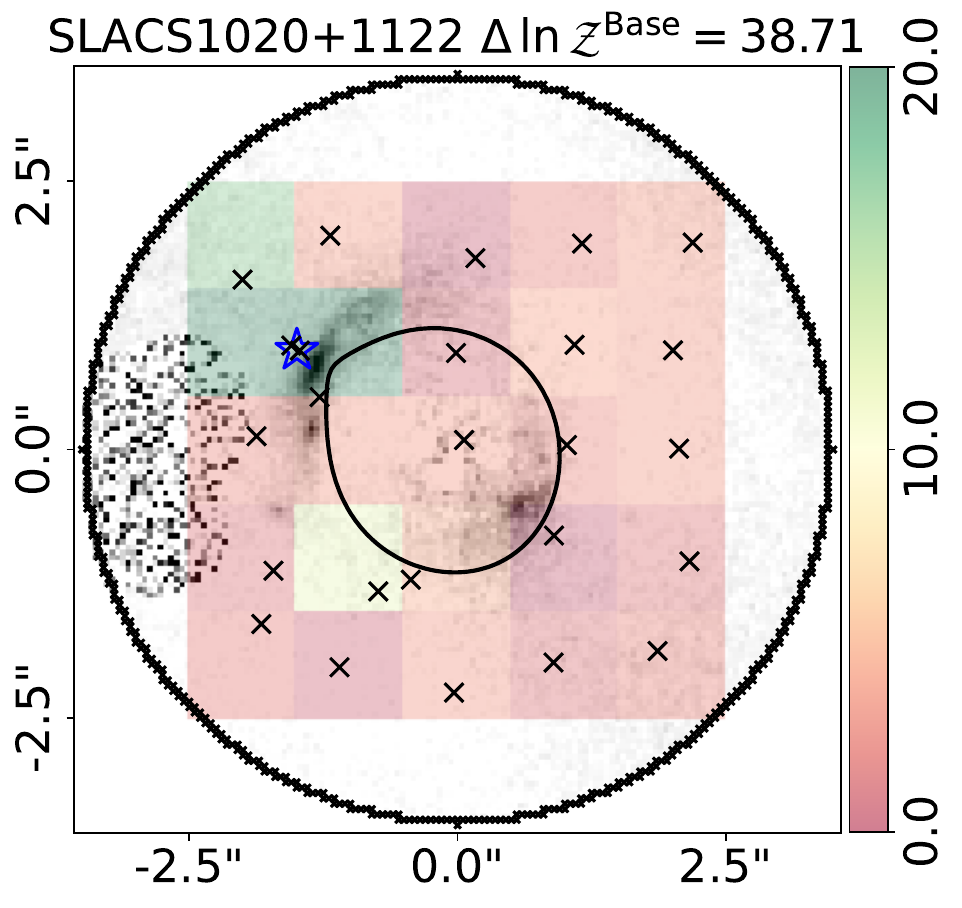}
\includegraphics[width=0.162\textwidth]{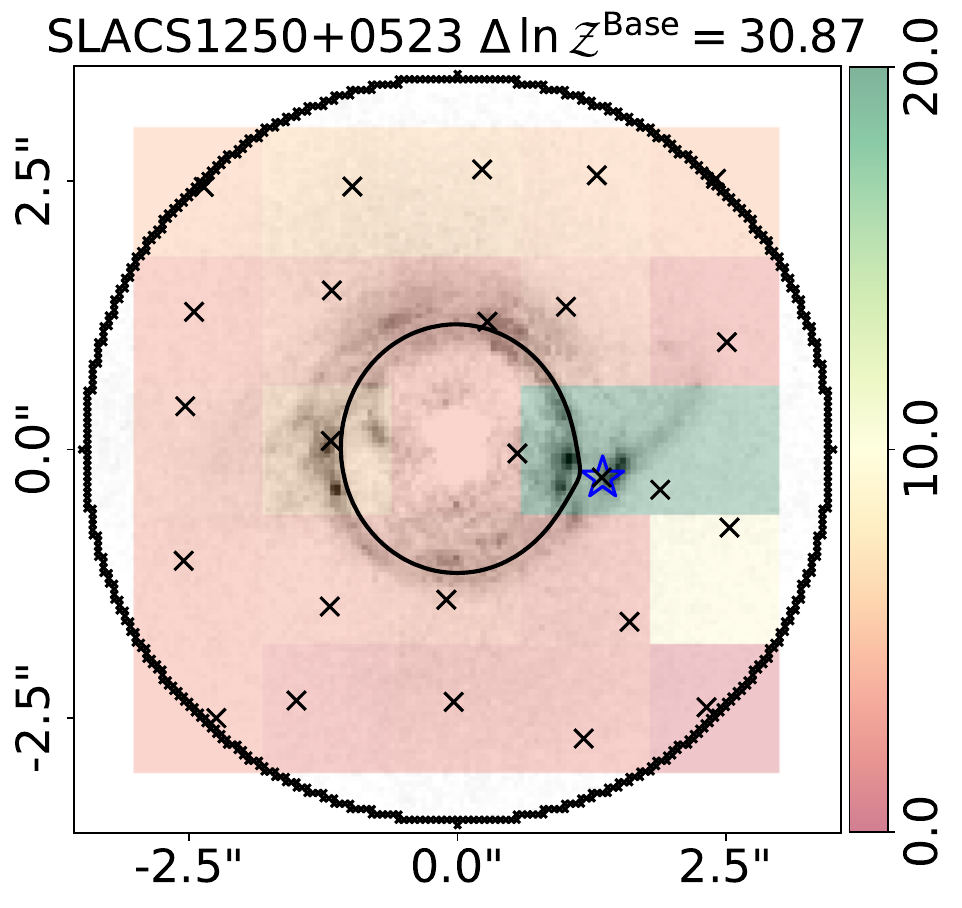}
\includegraphics[width=0.162\textwidth]{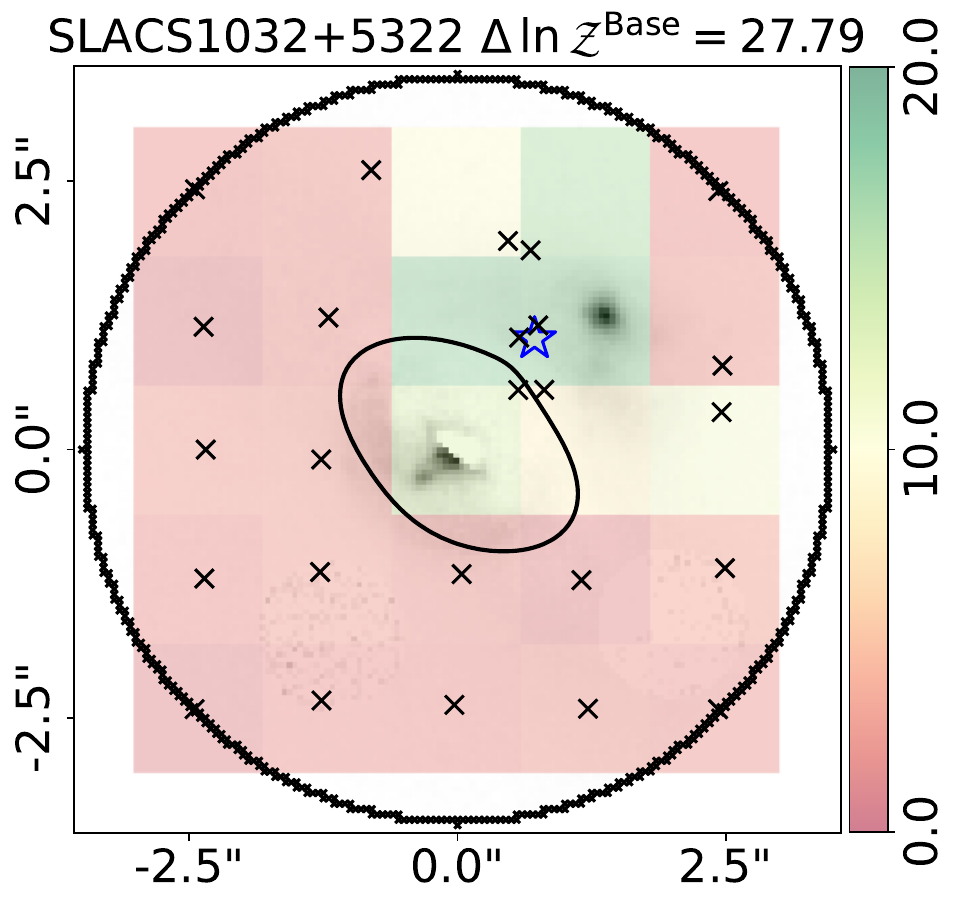}
\includegraphics[width=0.162\textwidth]{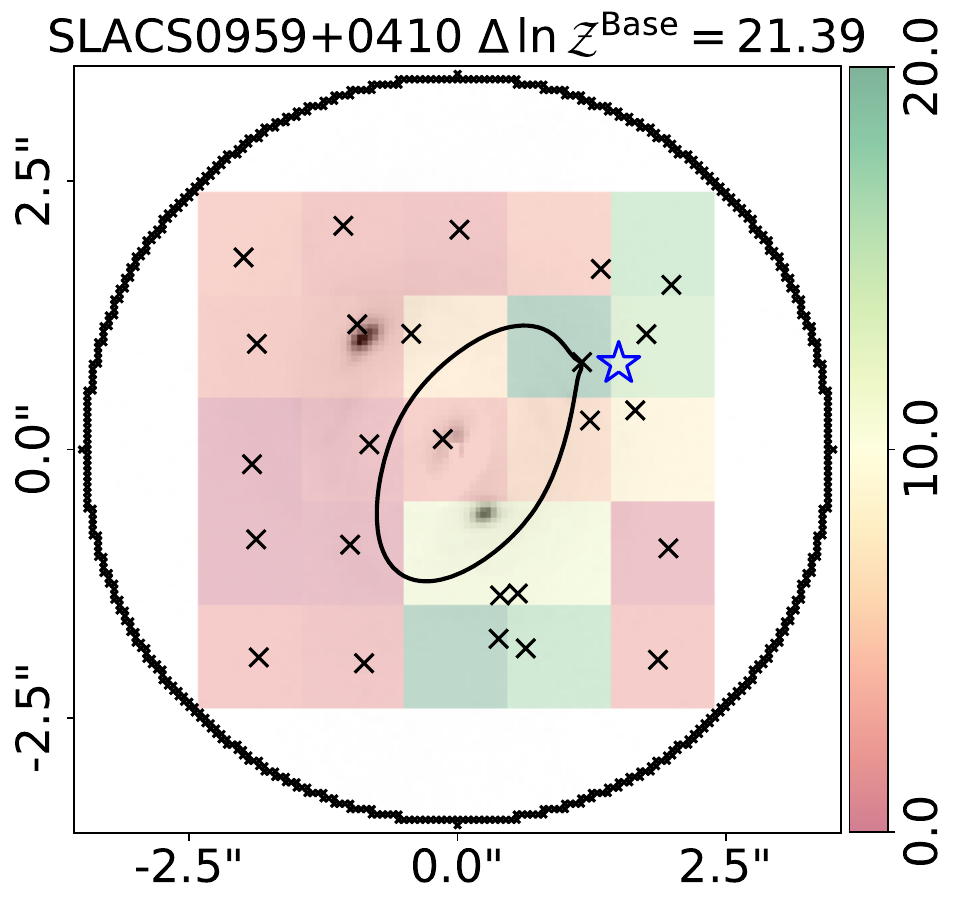}
\includegraphics[width=0.162\textwidth]{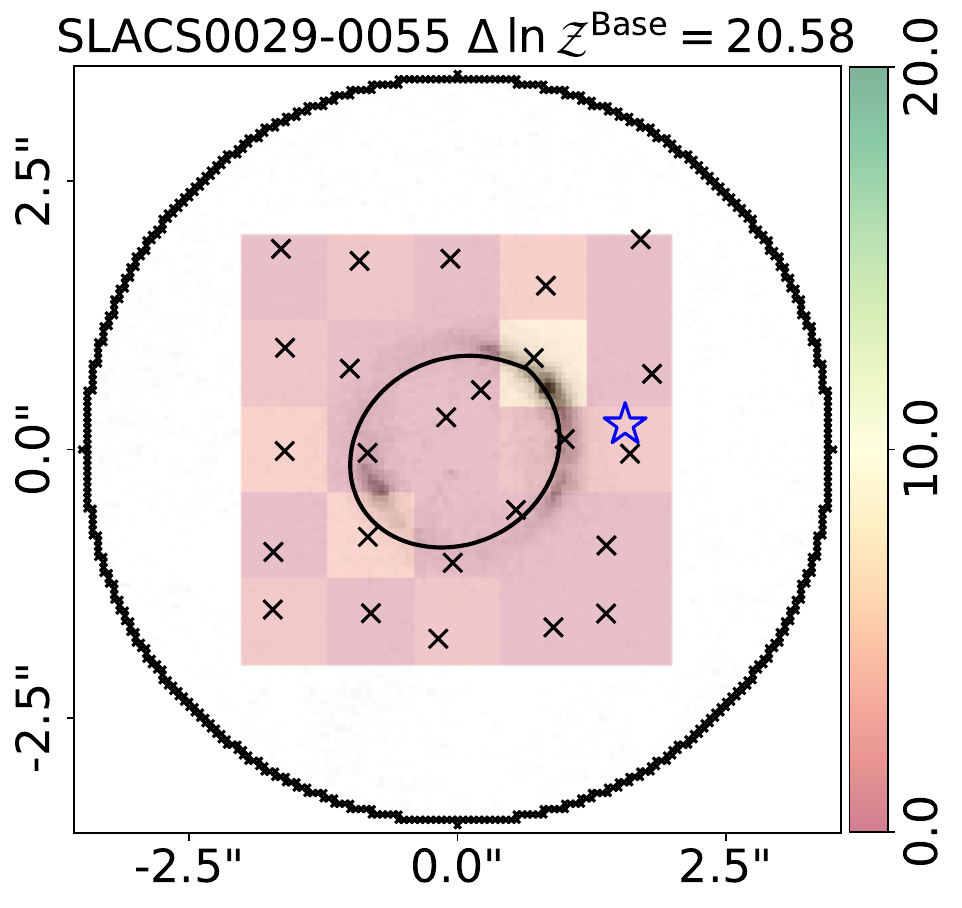}
\includegraphics[width=0.162\textwidth]{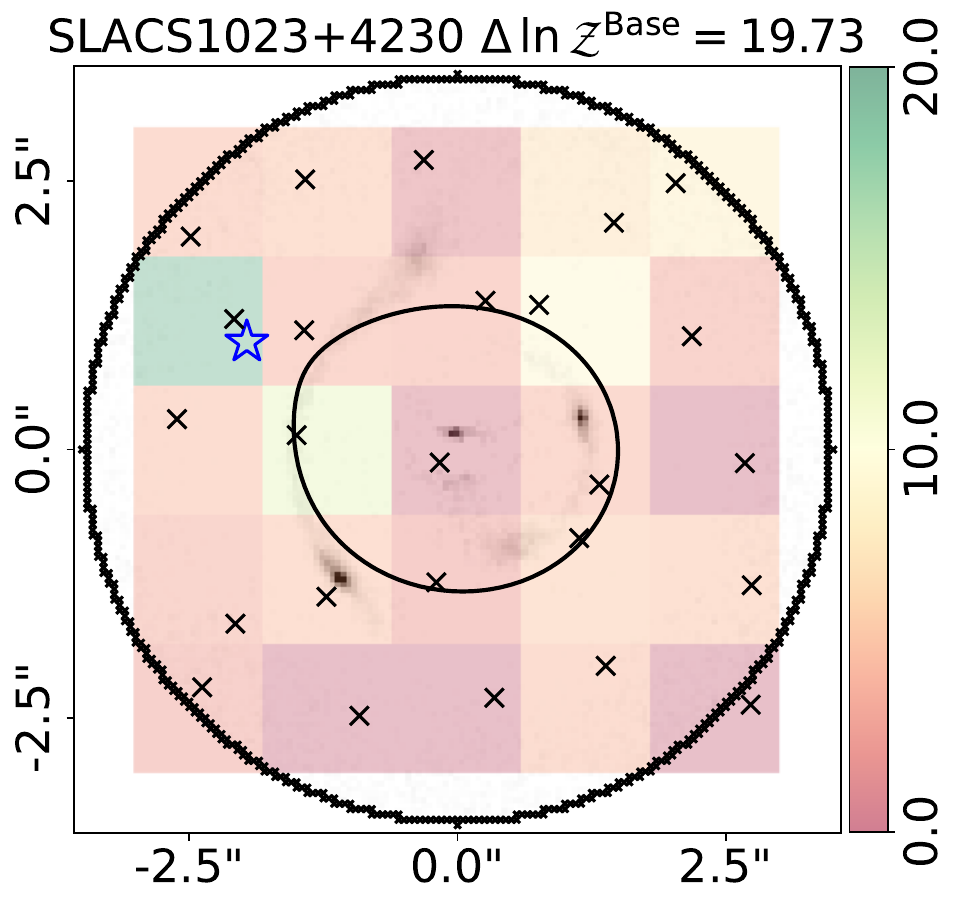}
\includegraphics[width=0.162\textwidth]{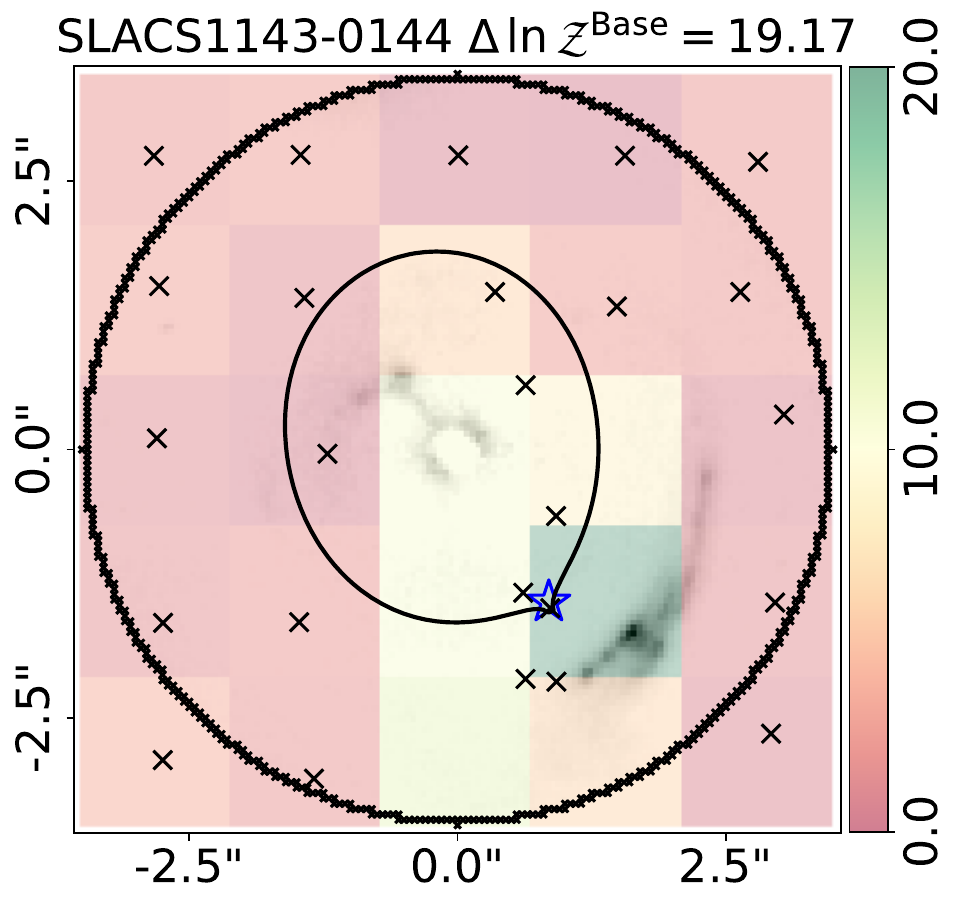}
\includegraphics[width=0.162\textwidth]{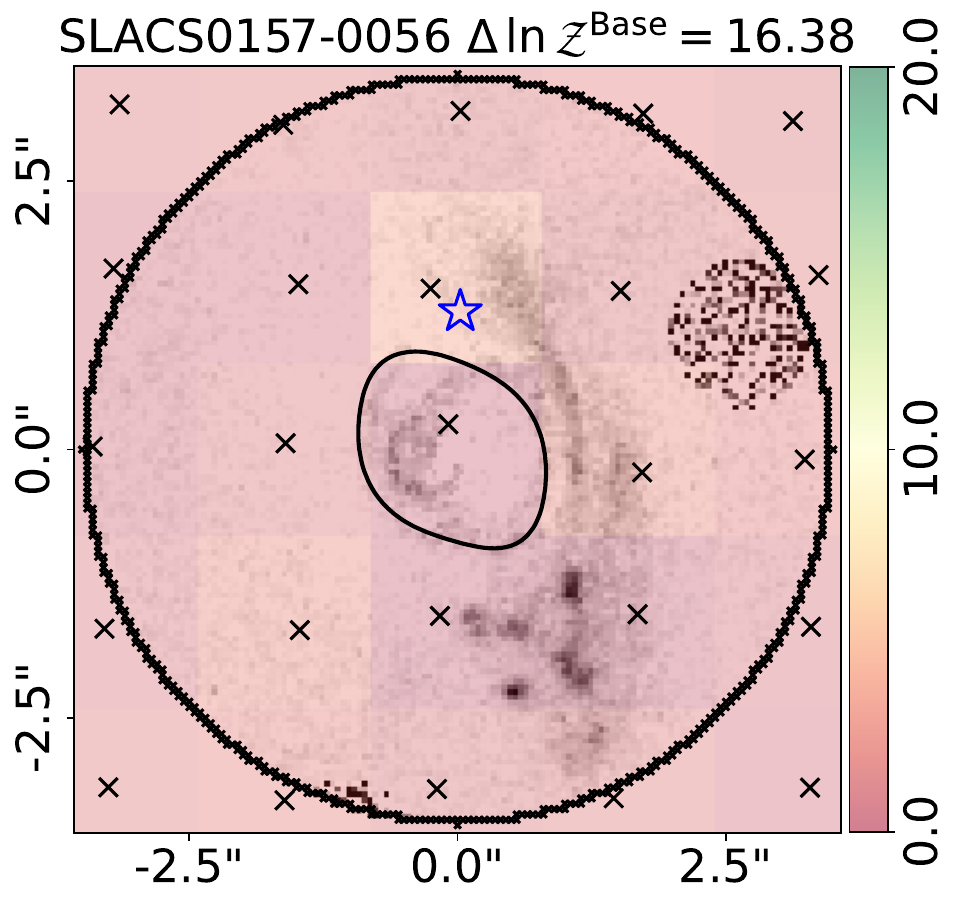}
\includegraphics[width=0.162\textwidth]{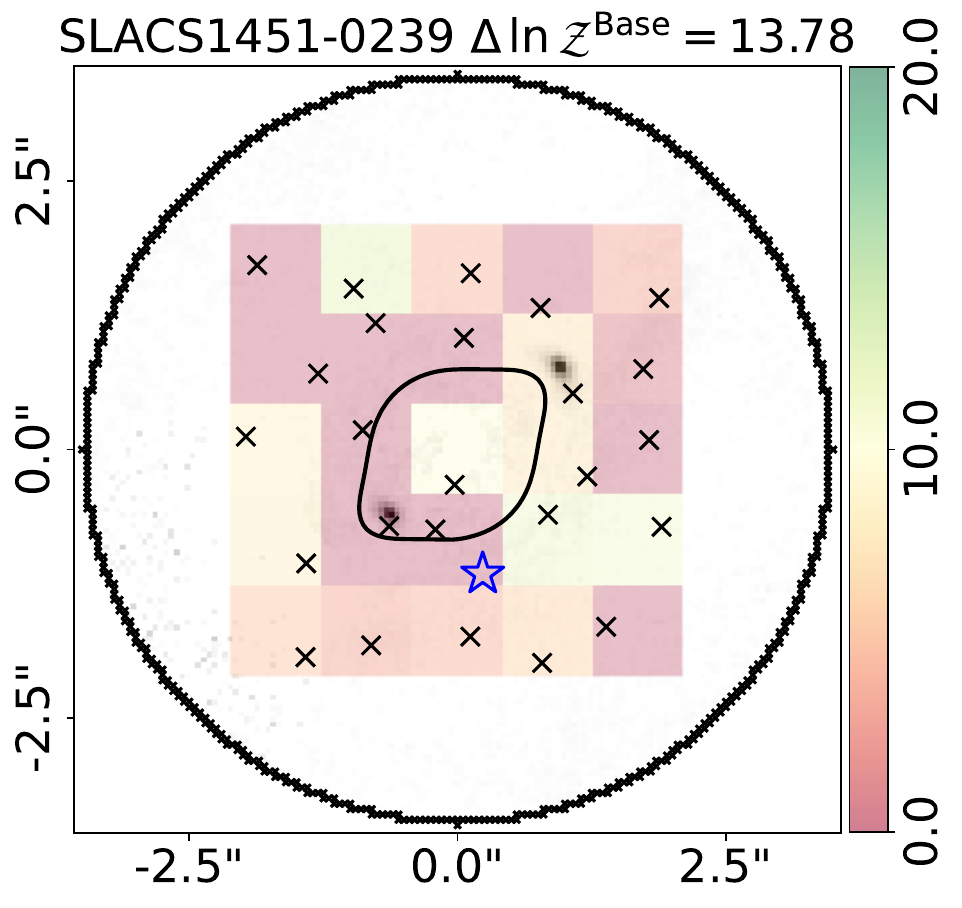}
\includegraphics[width=0.162\textwidth]{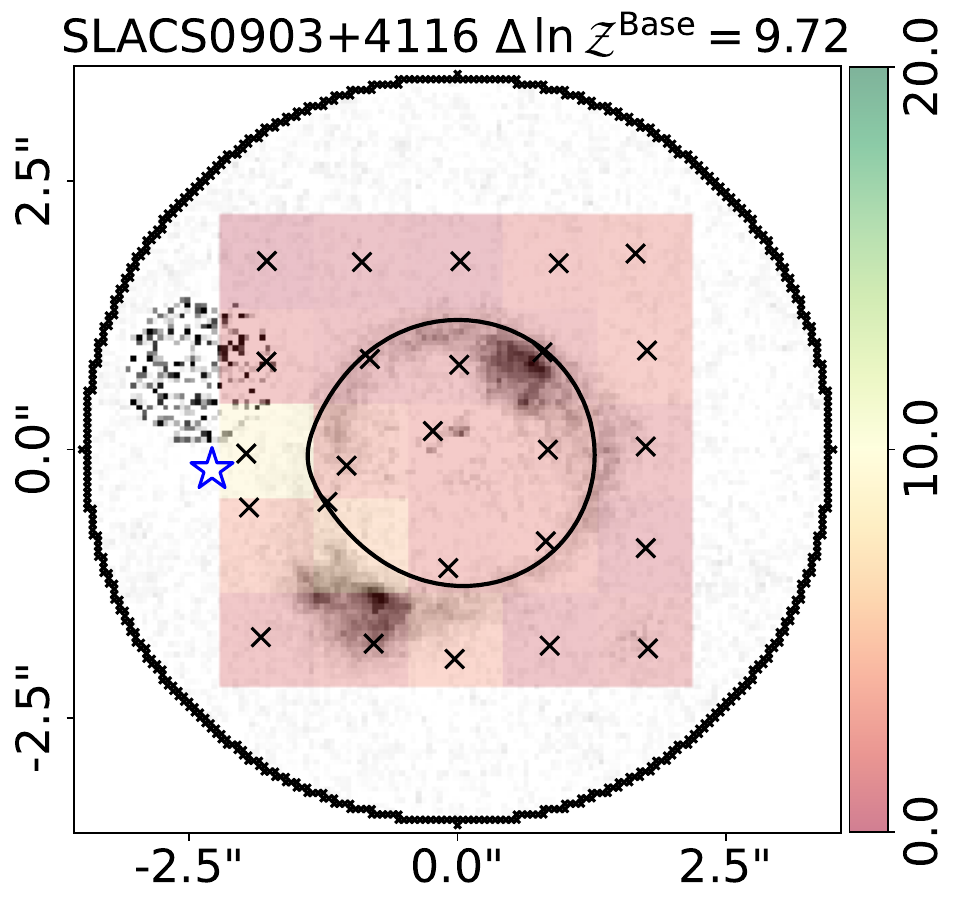}
\includegraphics[width=0.162\textwidth]{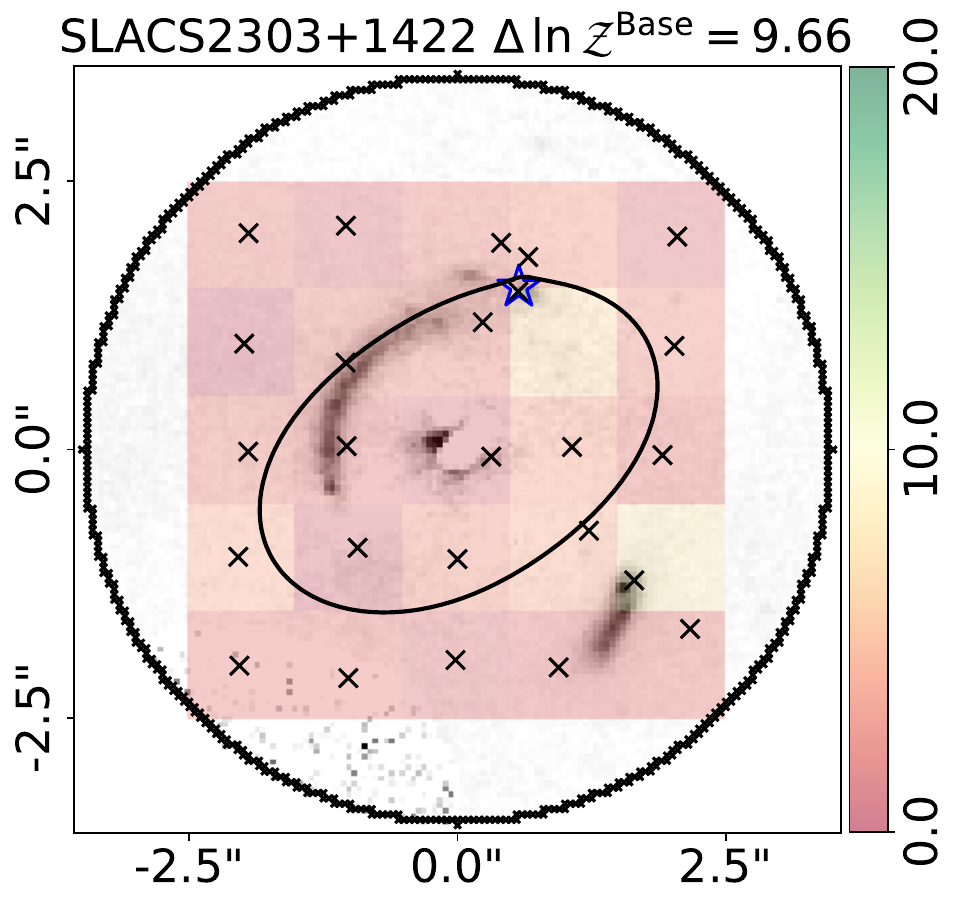}
\includegraphics[width=0.162\textwidth]{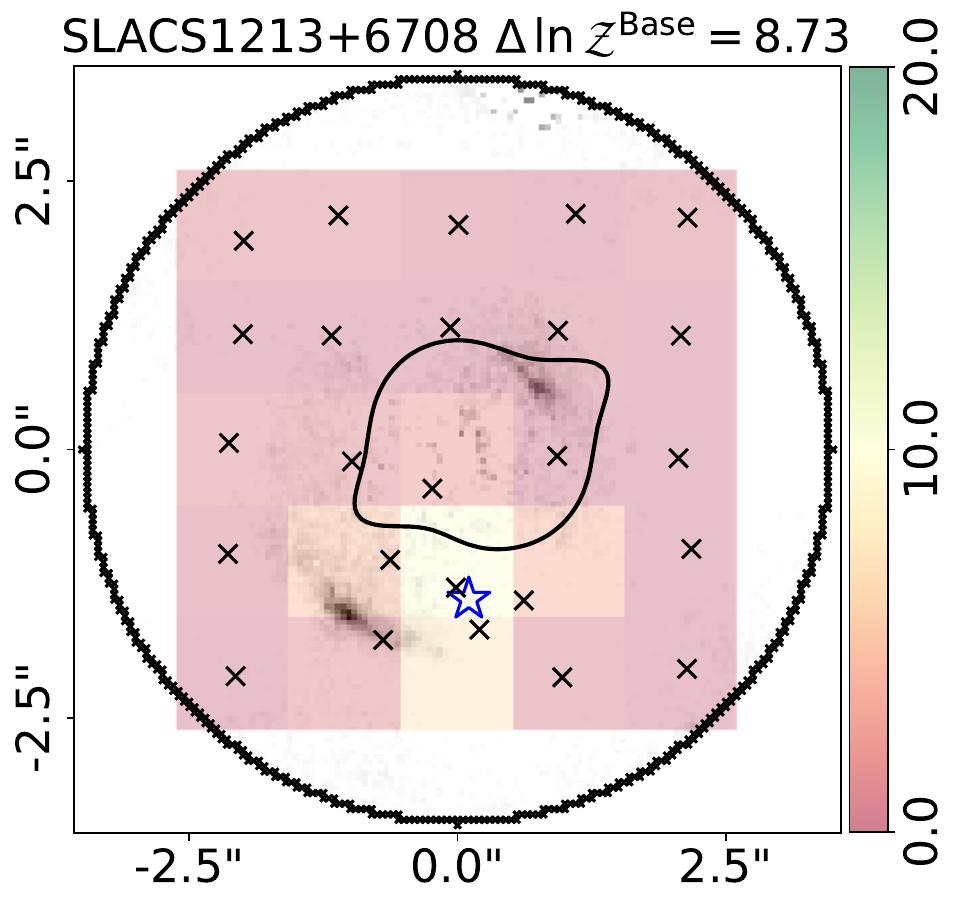}
\includegraphics[width=0.162\textwidth]{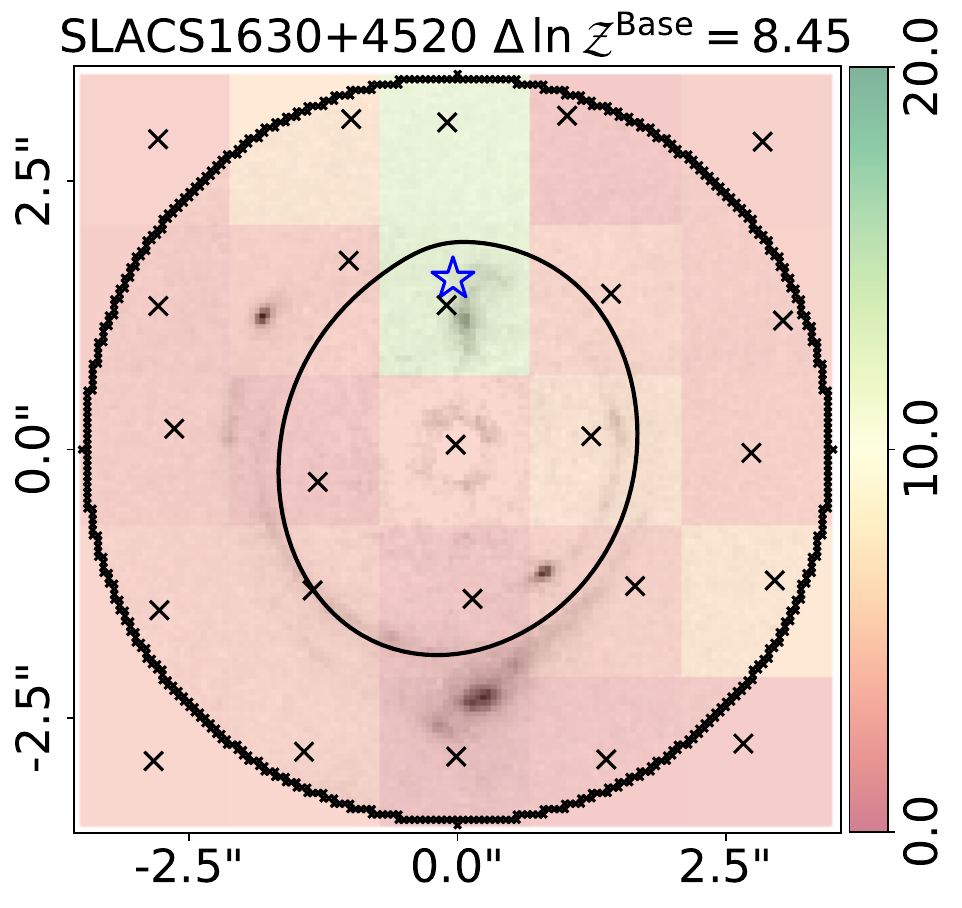}
\includegraphics[width=0.162\textwidth]{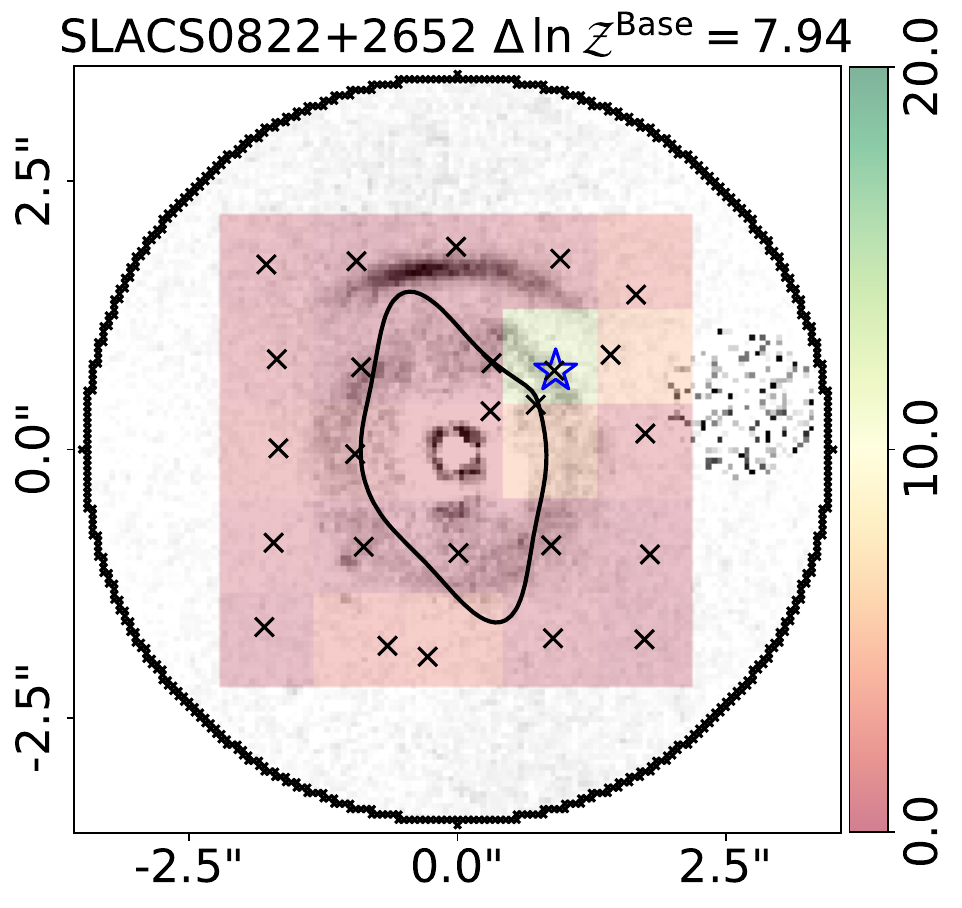}
\includegraphics[width=0.162\textwidth]{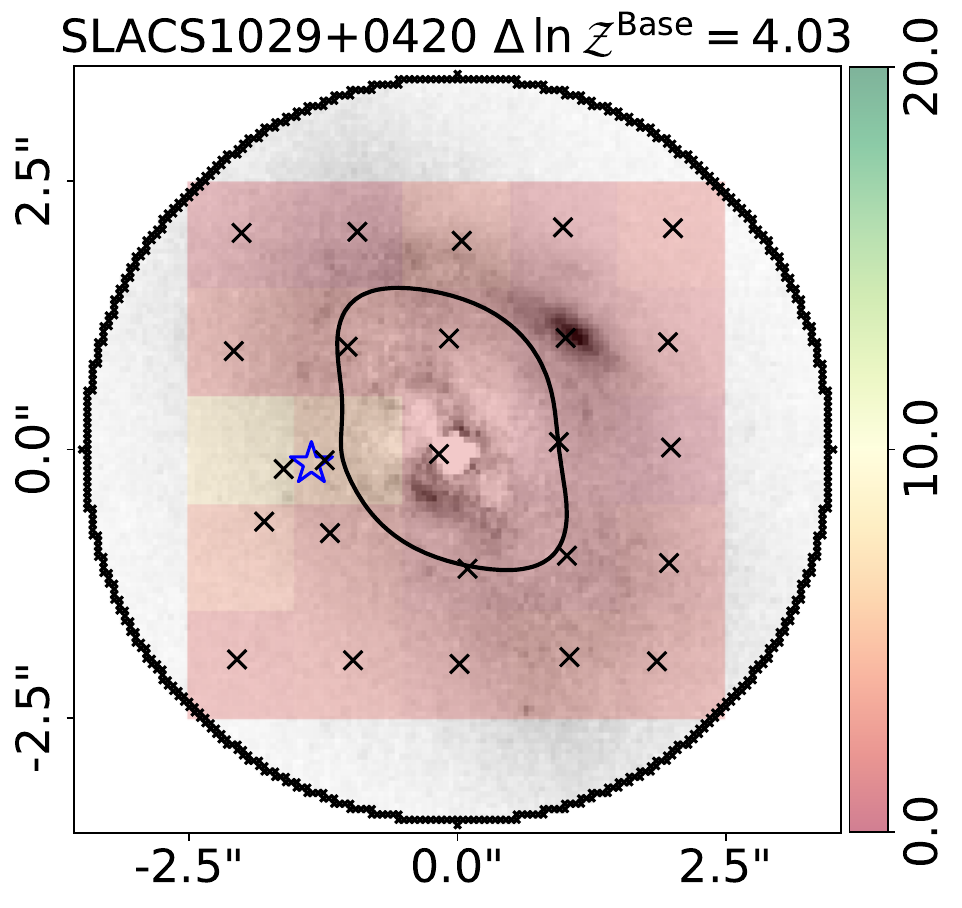}
\includegraphics[width=0.162\textwidth]{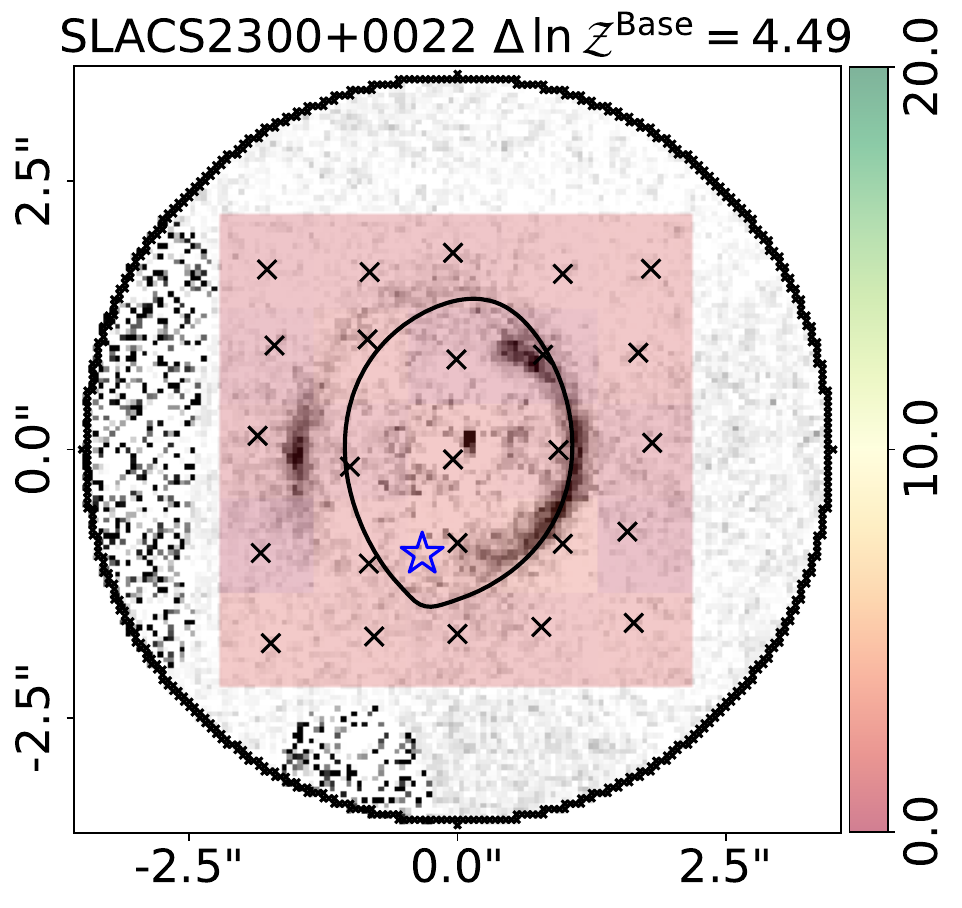}
\includegraphics[width=0.162\textwidth]{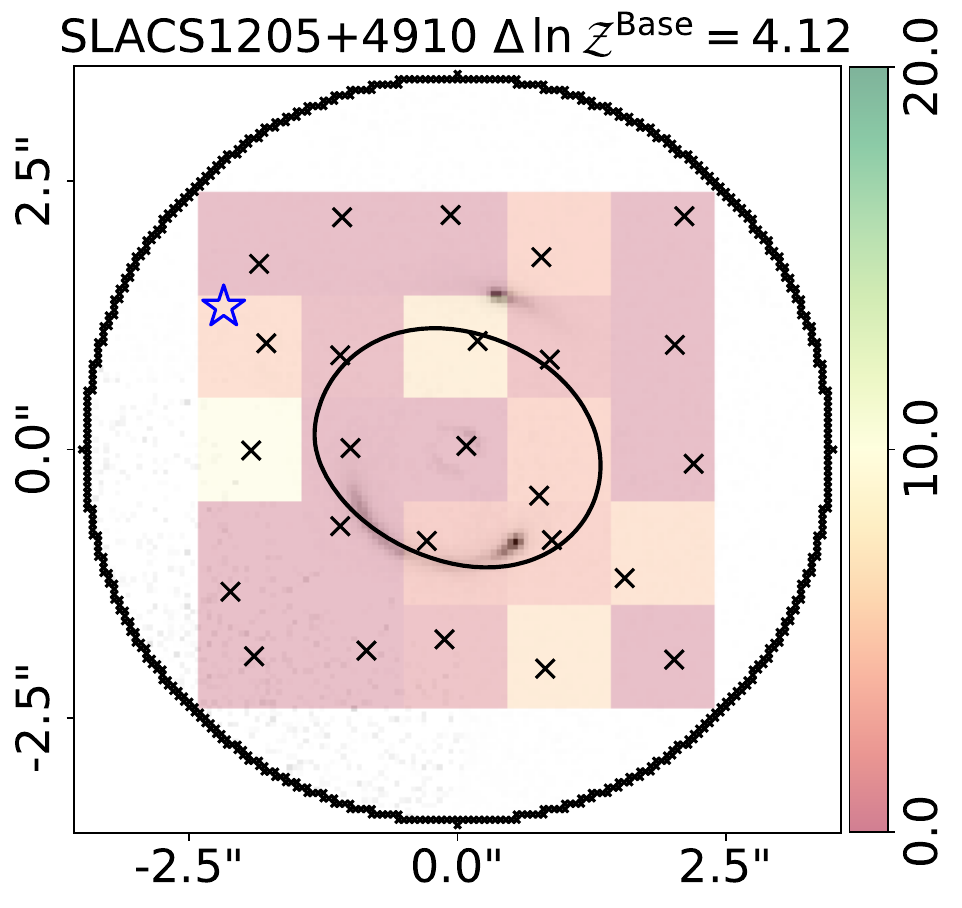}
\includegraphics[width=0.162\textwidth]{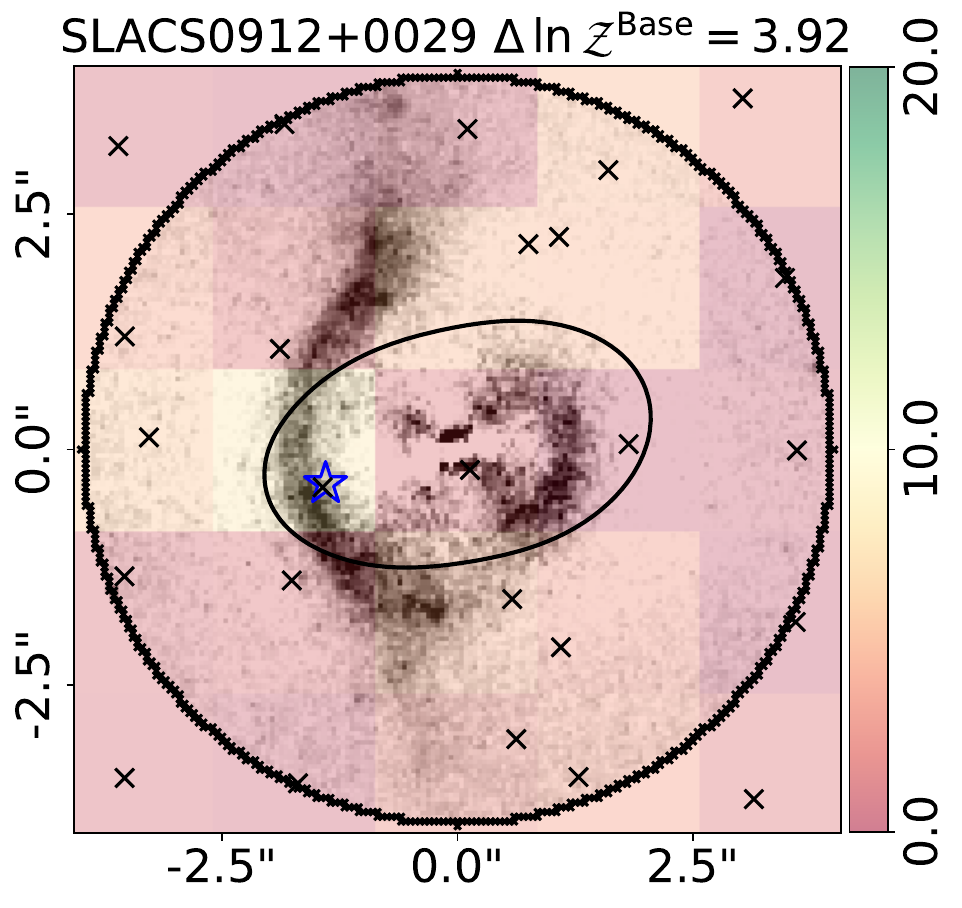}
\includegraphics[width=0.162\textwidth]{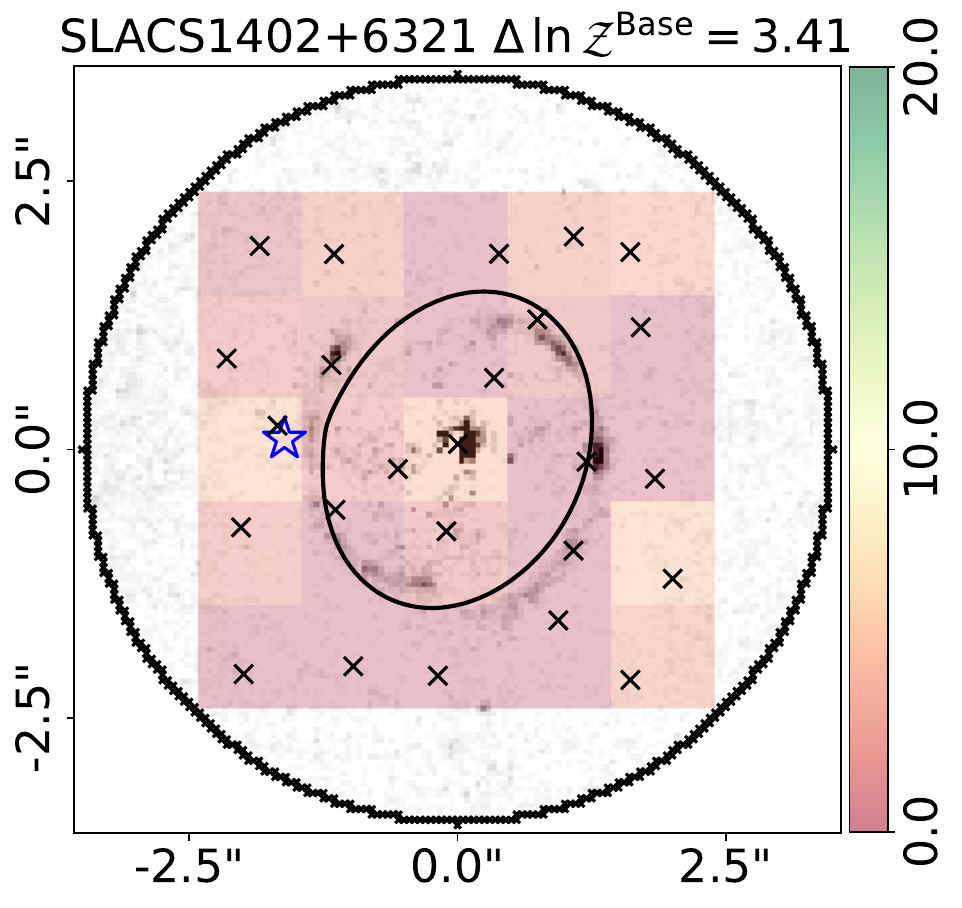}
\includegraphics[width=0.162\textwidth]{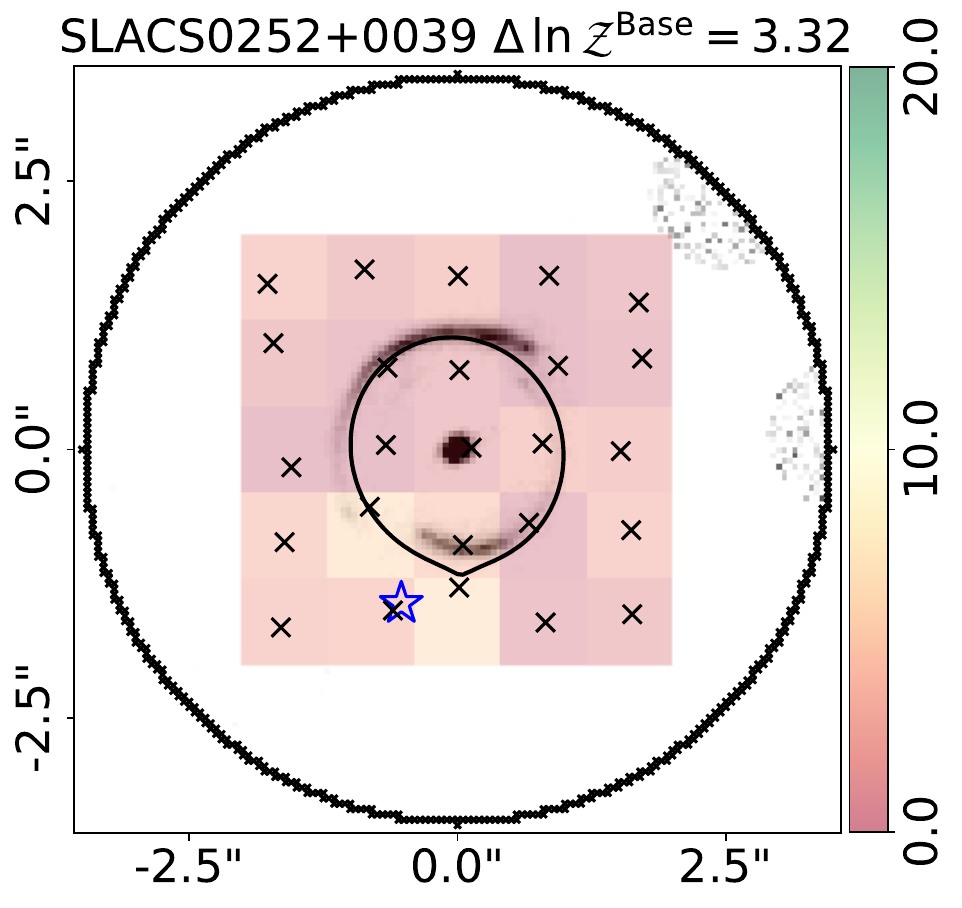}
\includegraphics[width=0.162\textwidth]{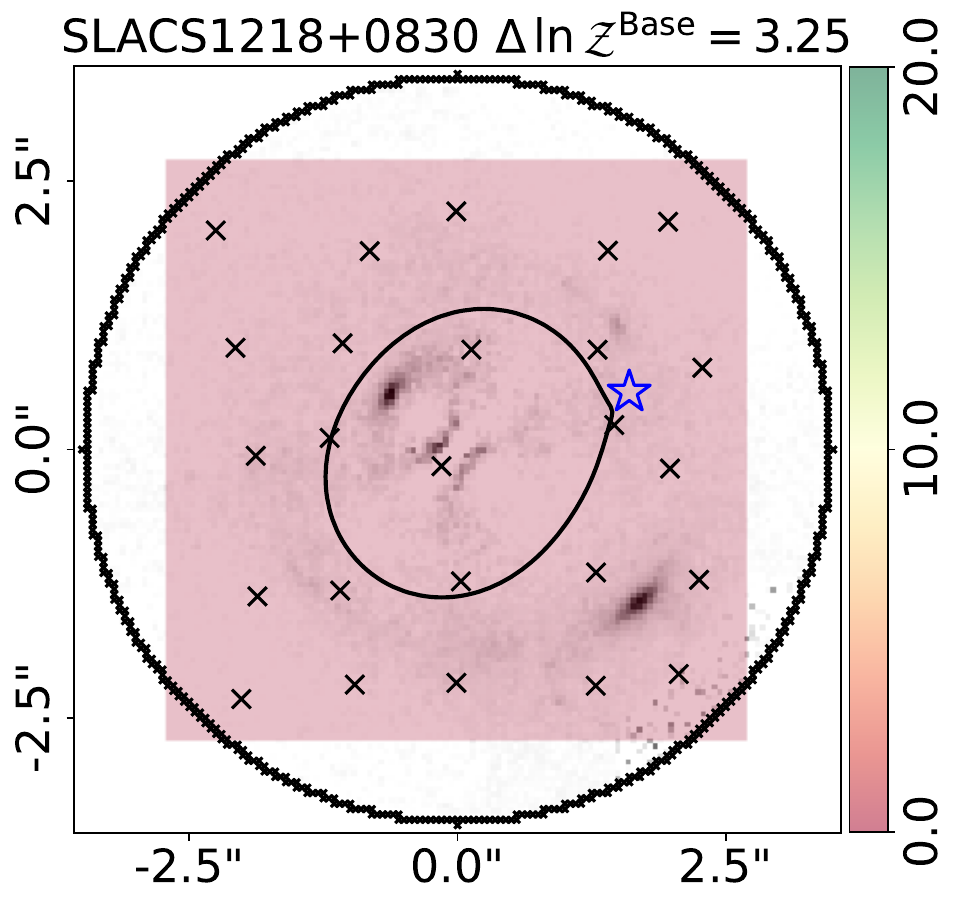}
\includegraphics[width=0.162\textwidth]{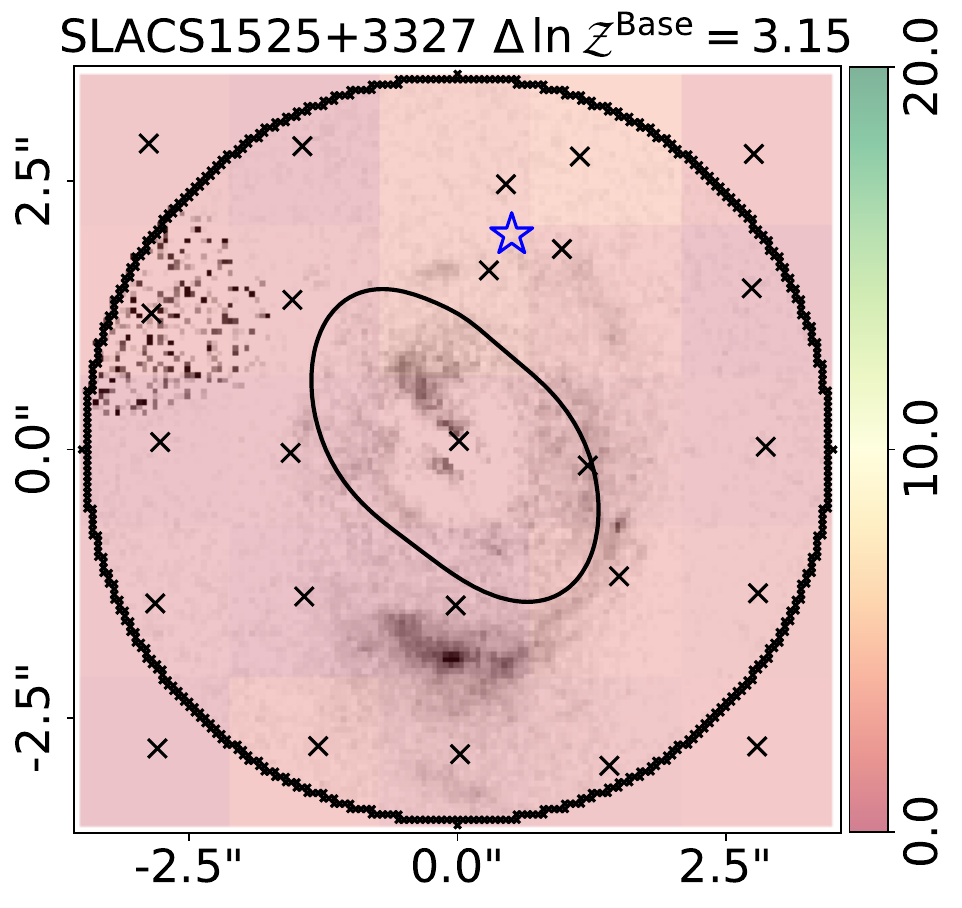}
\includegraphics[width=0.162\textwidth]{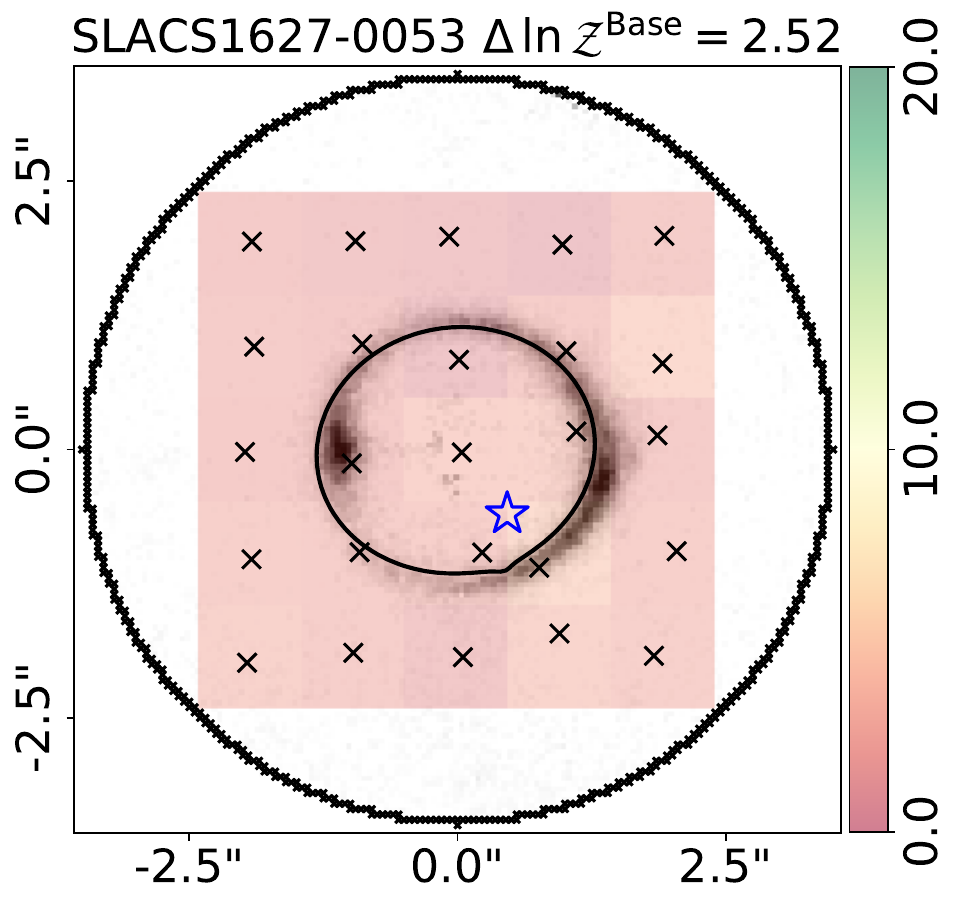}
\includegraphics[width=0.162\textwidth]{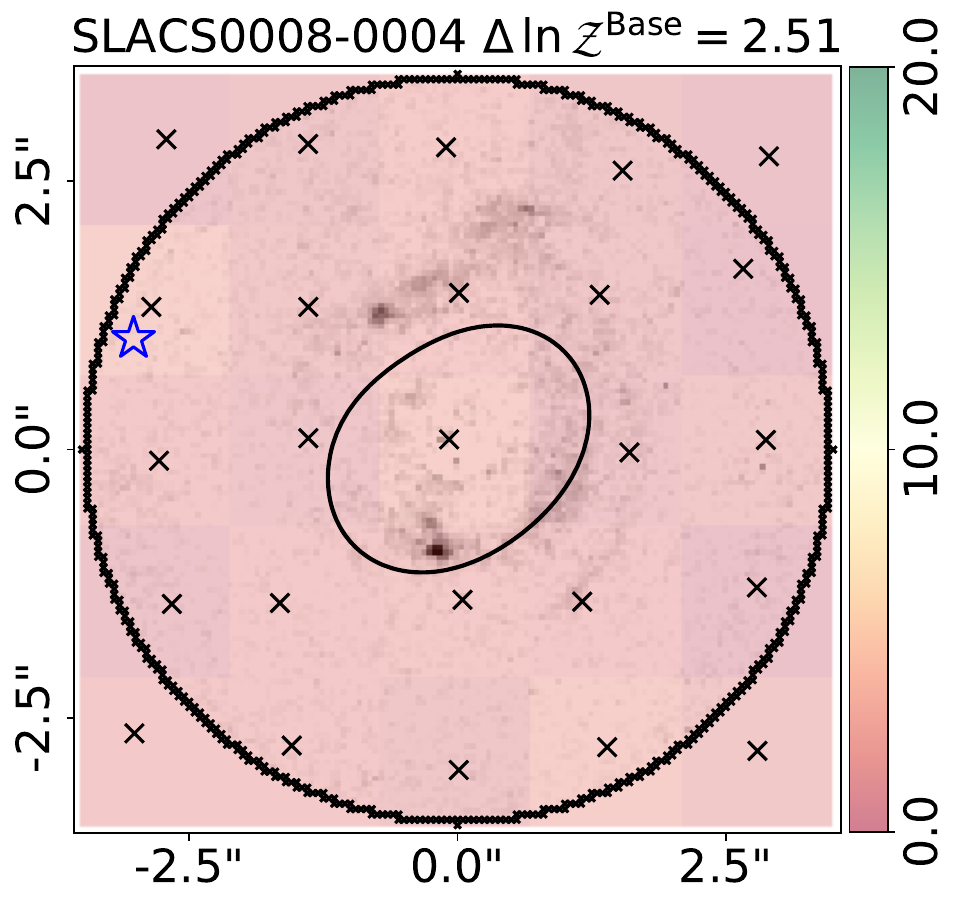}
\includegraphics[width=0.162\textwidth]{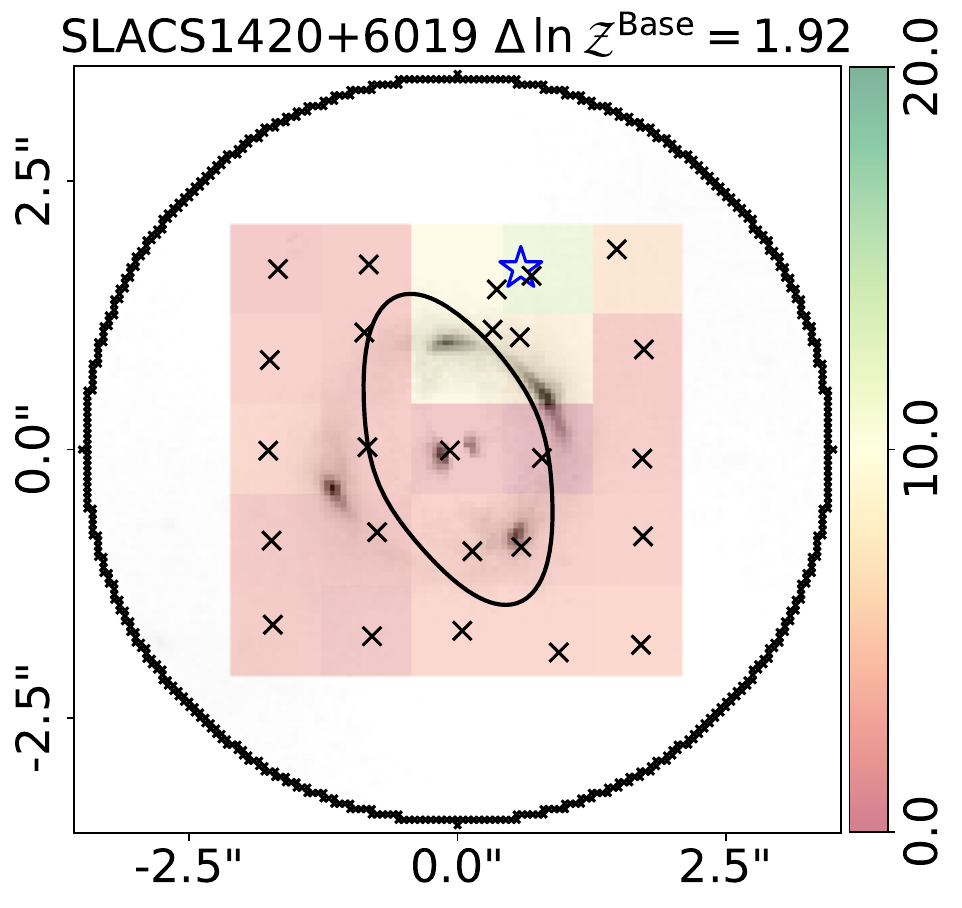}
\includegraphics[width=0.162\textwidth]{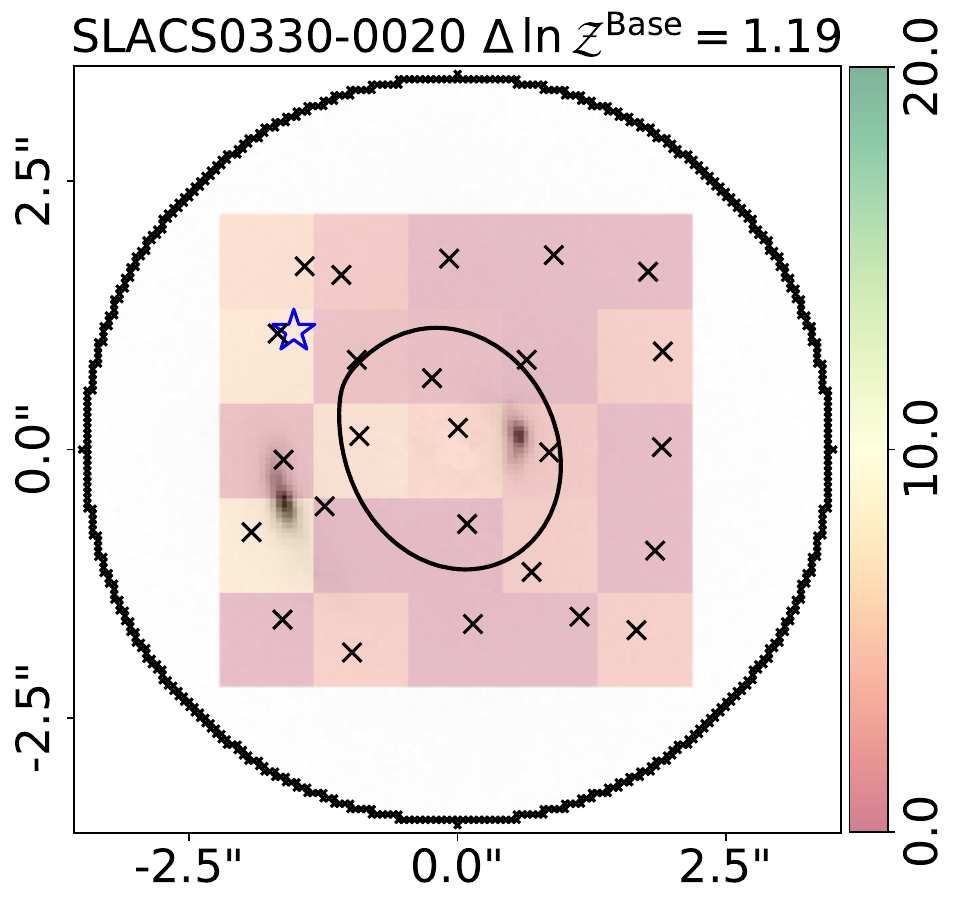}
\includegraphics[width=0.162\textwidth]{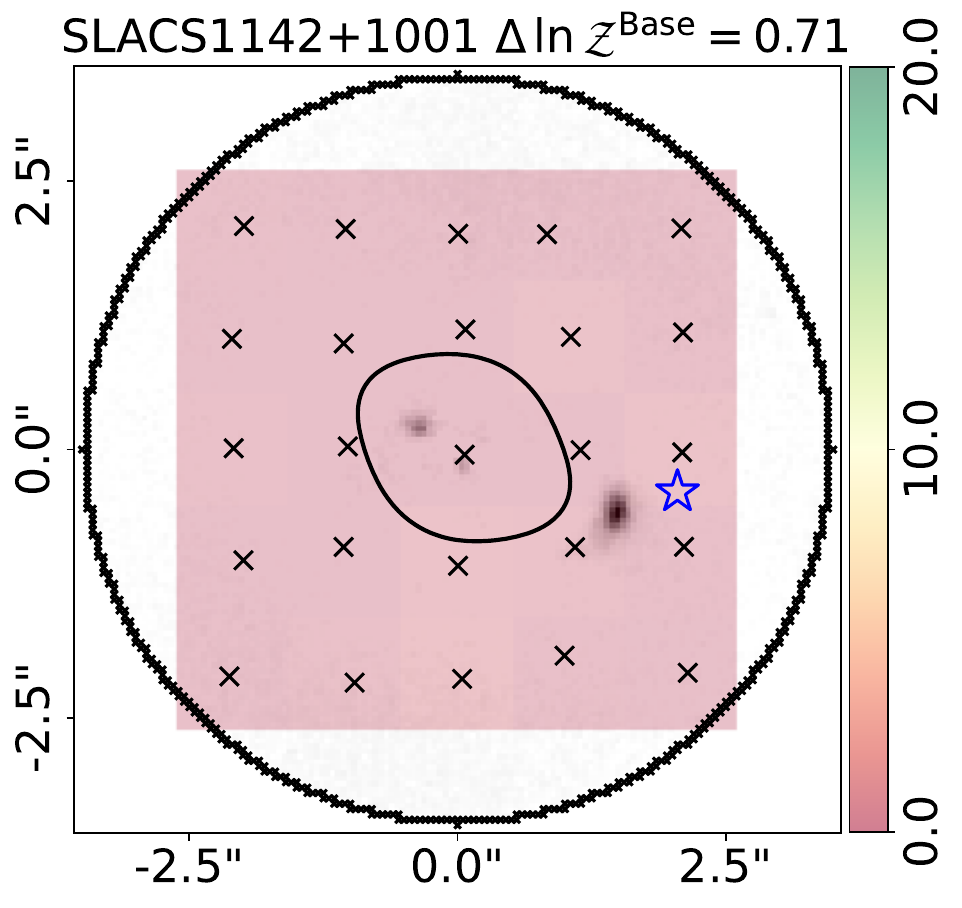}
\includegraphics[width=0.162\textwidth]{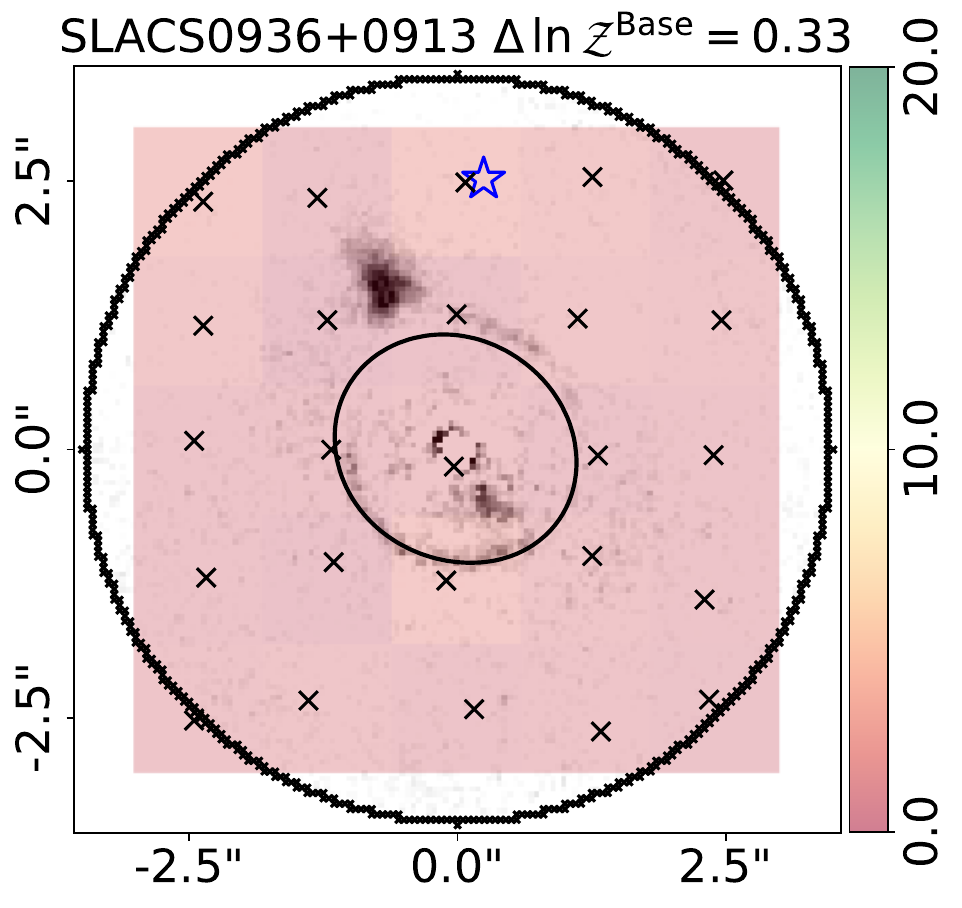}
\includegraphics[width=0.162\textwidth]{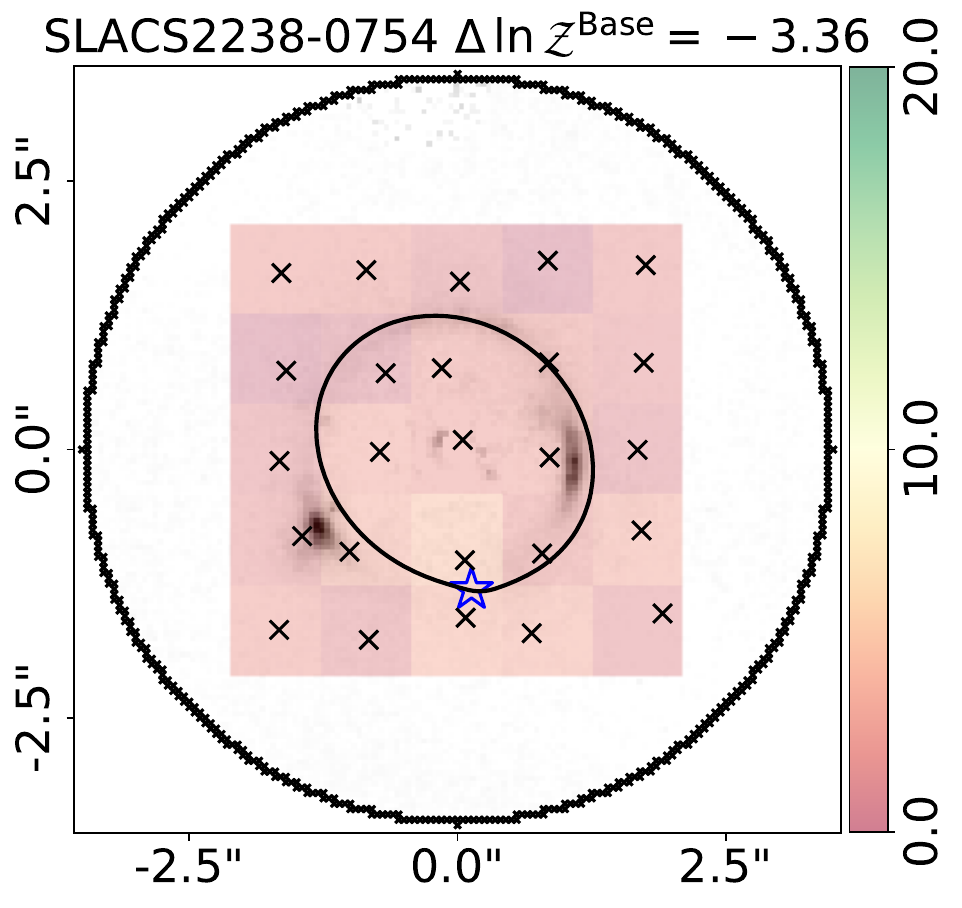}
\includegraphics[width=0.162\textwidth]{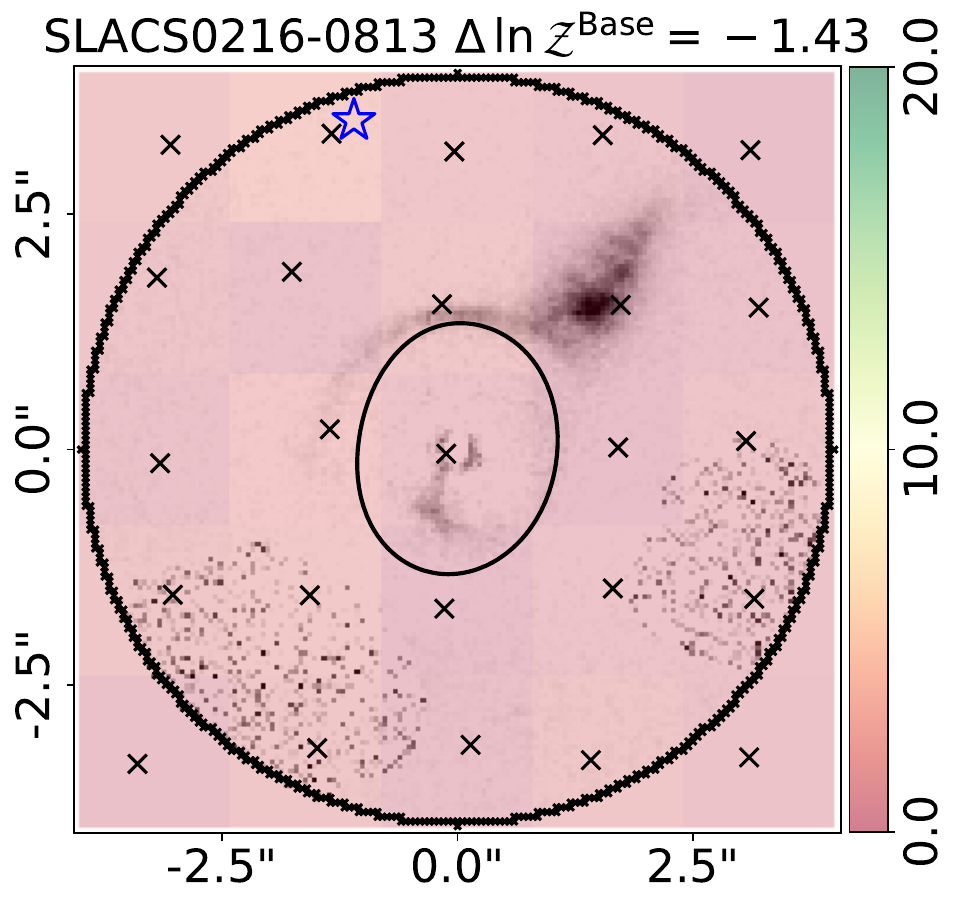}
\includegraphics[width=0.162\textwidth]{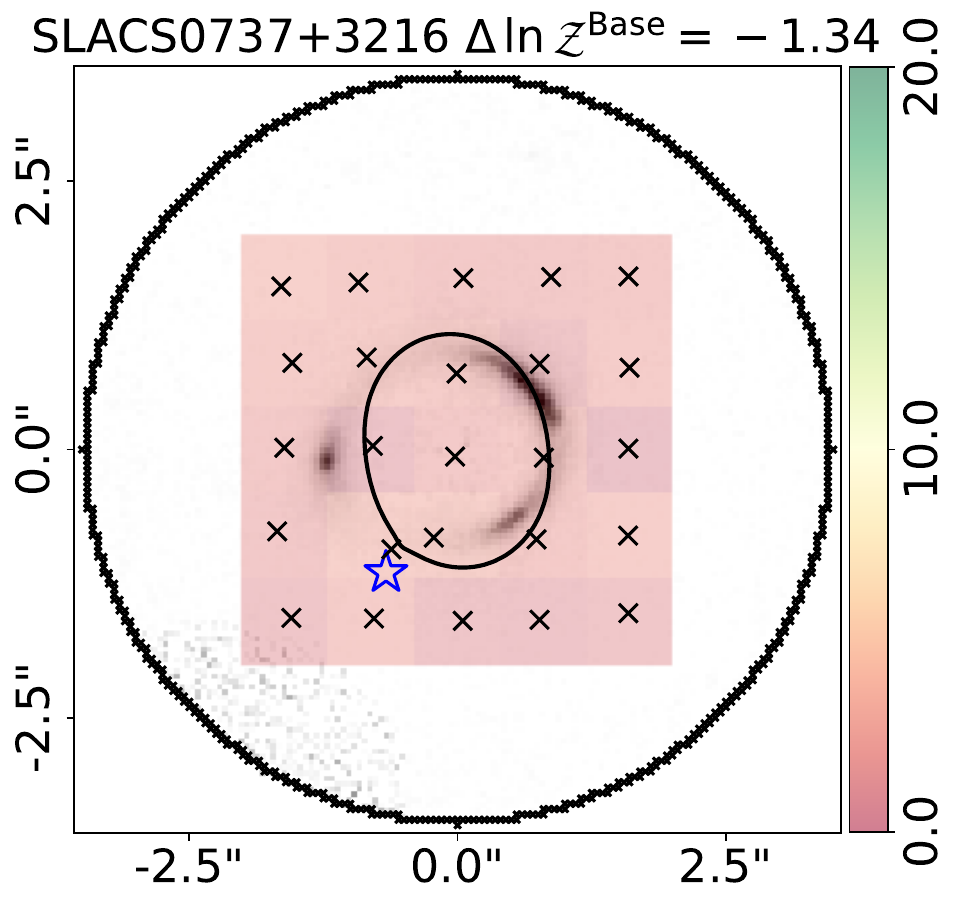}
\includegraphics[width=0.162\textwidth]{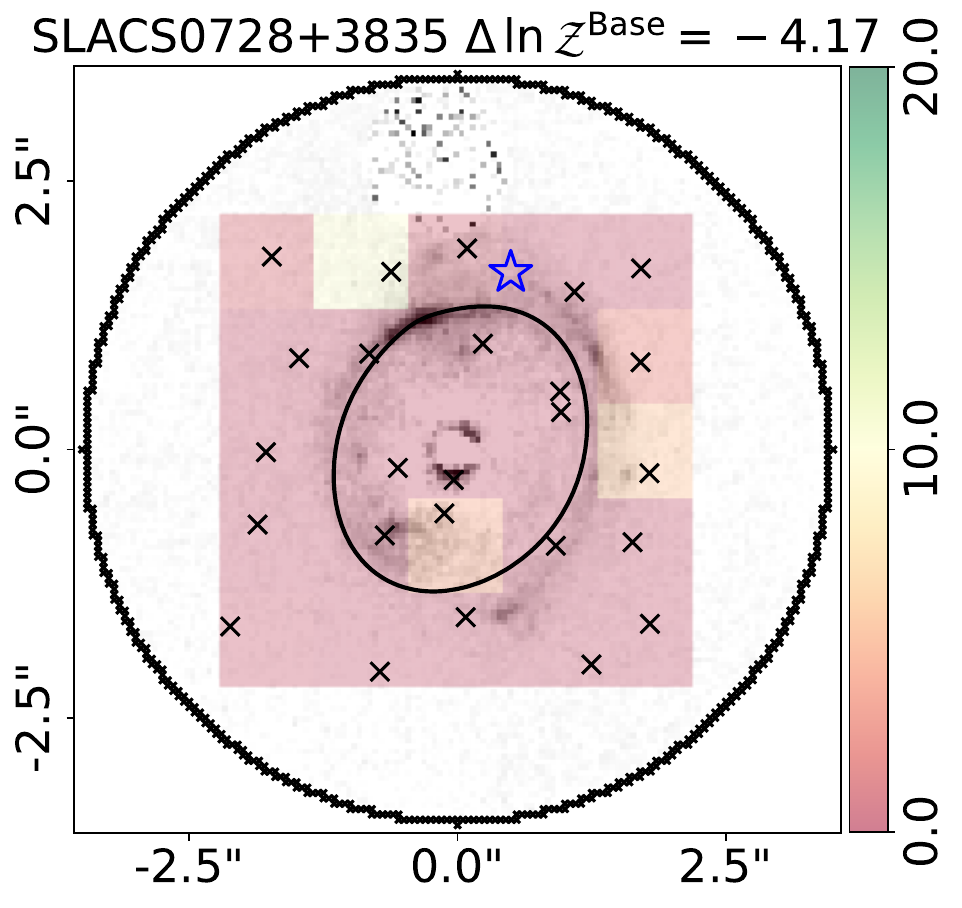}
\caption{
The subhalo scanning results of all SLACS lenses. Each scan uses a $5 \times 5$ grid of {\tt dynesty} searches fitting a PL plus shear model with a subhalo whose (x, y) coordinates are confined to a 2D grid-cell via uniform priors (see \cref{SubScan}). Fits are shown in descending order of highest log evidence increase. The background images show the lens subtracted data, where the lens subtraction uses a double S\'ersic fit. In \cref{ResultsData} we demonstrate that the majority of these DM subhalo candidates are false positives. The green/red grid shows the increase of log evidences of each subhalo scan grid cell. The $\Delta\,\mathrm{ln}\,\mathcal{Z}^{\rm Base}$ values in each plot title correspond to the $\textbf{SH3}$ fit performed after the subhalo grid search where the subhalo position priors extend beyond a small 2D cell. This $\Delta\,\mathrm{ln}\,\mathcal{Z}^{\rm Base}$ value is sometimes much higher than the inferred grid search values and is why certain lenses show predominantly red or yellow cells but have high overall $\Delta\,\mathrm{ln}\,\mathcal{Z}^{\rm Base}$ values in the title (e.g. SLACS1432+6316). The blue star shows the DM subhalo's maximum a posteriori inferred location for this fit. The colorbar ranges between 0 and 20 so that candidate DM subhalos ($\Delta\,\mathrm{ln}\,\mathcal{Z}^{\rm Base} > 10$ are colored green whereas non-candidates are red. Certain lenses (e.g. SLACS1020+1122 on its left-hand side) show patches of noise, which are a result of emission from a foreground galaxy or star being removed via the noise-scaling GUI described in \citep{Etherington2022}.
The lens SLACS1430+4105, shown in \cref{figure:Overview}, is omitted from this figure.
} 
\label{figure:DetectSlacs}
\end{figure*}

\subsection{Subhalo Scanning on Simulated Data}\label{SimSummary}

In \cref{ResultSim} we simulate and fit a sample of 16 strong lenses, in four groups of lenses with the same lens and source galaxies but with DM subhalos of masses $10^{10.5}$M$_{\odot}$, $10^{10.0}$M$_{\odot}$, $10^{9.5}$M$_{\odot}$ or no subhalo. The simulated lenses are idealized, because their lens light (double Sersic) and mass (PL plus shear) are simulated using the same model assumed to fit the data. Cautioning that these conclusions only hold in this idealized scenario, a summary is as follows:

\begin{itemize}
    
    \item For 7 out of the 8 datasets containing a $10^{10.5}$M$_{\odot}$ or $10^{10.0}$M$_{\odot}$ DM subhalo the analysis successfully detects the DM subhalo. 
    
    \item For 2 out of the 4 datasets containing a $10^{9.5}$M$_{\odot}$ DM subhalo the analysis successfully detects the input DM subhalo. For the two datasets where the input DM subhalo is not detected we attribute this to the data not being sensitive enough.
    
    
    \item For all four datasets not containing a DM subhalo, we correctly disfavor a DM subhalo provided the source reconstruction has sufficiently high resolution.

    \item Our subhalo inference does not depend on the source reconstruction assumptions (e.g. it is insensitive to using a different regularization scheme).
    
    \item The lens mass model is degenerate with the DM subhalo, whereby the inferred mass model changes its inferred parameters to `absorb' some of the DM subhalo signal.
    
\end{itemize}

False positive DM subhalo detections were not seen for the mock lenses (provided the source reconstruction was high enough resolution). This procedure therefore verifies that for our analysis of HST imaging of real lenses, false positives are because the lens model assumptions are not robust (or it is a geniune DM subhalo detection).

\subsection{Subhalo Scanning on HST data with Simple Models}\label{FitBase}

We now present subhalo scanning of HST imaging of 54 strong lenses from the SLACS \citep{Bolton2008a} and BELLS-GALLERY \citep{Shu2016} samples. Results for each sample are given separately, because the compact nature of BELLS-GALLERY sources changes their sensitivity to DM subhalos \citep{Despali2022}. We first present results for our simplest baseline lens model, which assumes two S\'ersic profiles with the same centres for the lens light, a power-law plus external shear mass model and Voronoi mesh source reconstruction. All fits adopt a $3.5\arcsec$ circular mask. 

Column 2 of \cref{table:DetectSLACS} lists $\Delta\,\mathrm{ln}\,\mathcal{Z}^{\rm Base}$, the log evidence increase for a model including a subhalo for the 37 SLACS lenses. 14 out of 37 lenses favour the inclusion of a DM subhalo and meet our criterion of $\Delta\,\mathrm{ln}\,\mathcal{Z}^{\rm Base} > 10$. \cref{figure:DetectSlacs} shows the corresponding subhalo grid search results for these objects, where from the top left rightwards and then downwards lenses are plotted in descending order of $\Delta\,\mathrm{ln}\,\mathcal{Z}^{\rm Base}$. The lens SLACS2341+0000 infers the highest value, $\Delta\,\mathrm{ln}\,\mathcal{Z}^{\rm Base} = 157.51$. 24 lenses are non-detections with $\Delta\,\mathrm{ln}\,\mathcal{Z}^{\rm Base} < 10$. Column 3 of \cref{table:DetectSLACS} shows the inferred subhalo masses $M^{\rm sub}_{\rm 200}$\,\,M$_{\rm \odot}$, which span $10^{8.39}$\,\,M$_{\rm \odot}$ and $10^{11.98}$\,\,M$_{\rm \odot}$ for models where $\Delta\,\mathrm{ln}\,\mathcal{Z}^{\rm Base} > 10$.

\begin{figure*}
\centering
\includegraphics[width=0.16\textwidth]{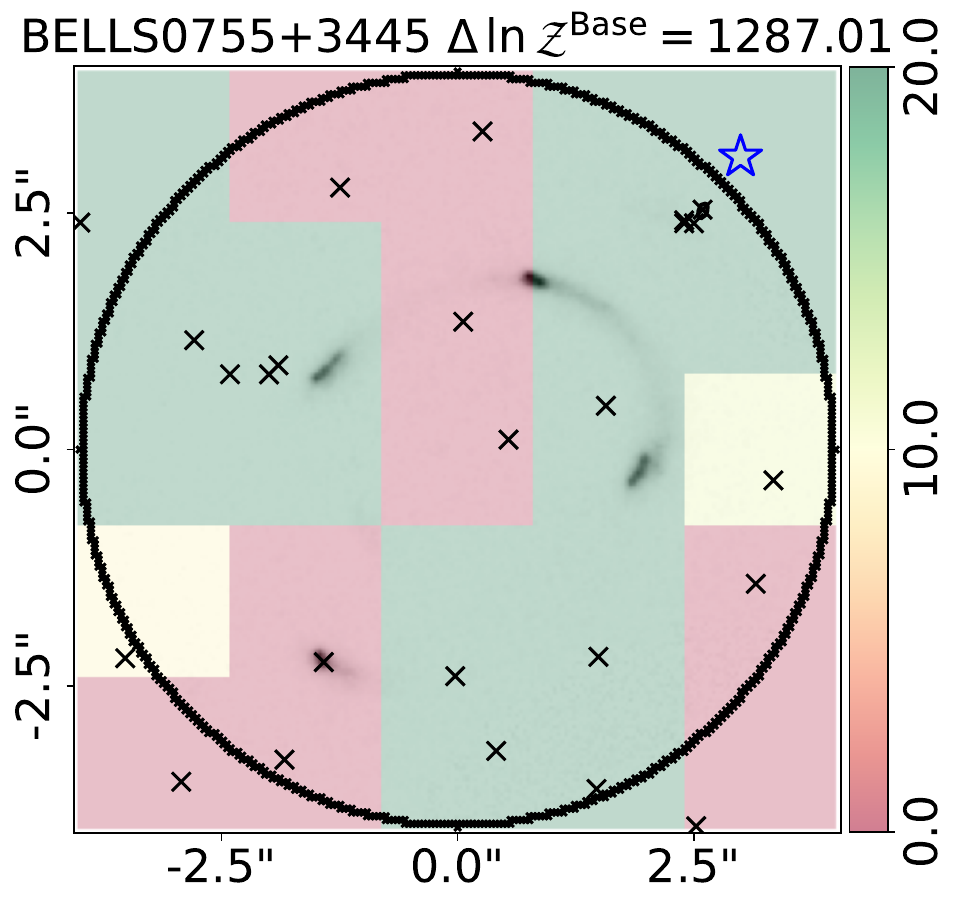}
\includegraphics[width=0.16\textwidth]{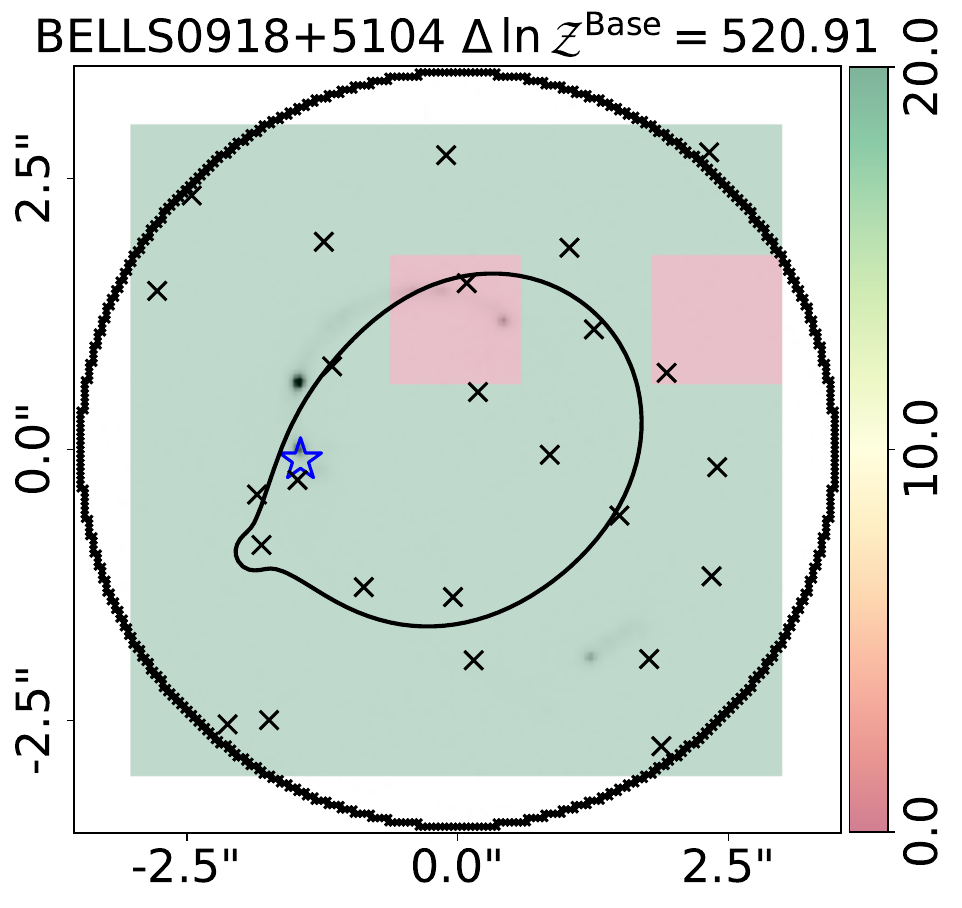}
\includegraphics[width=0.16\textwidth]{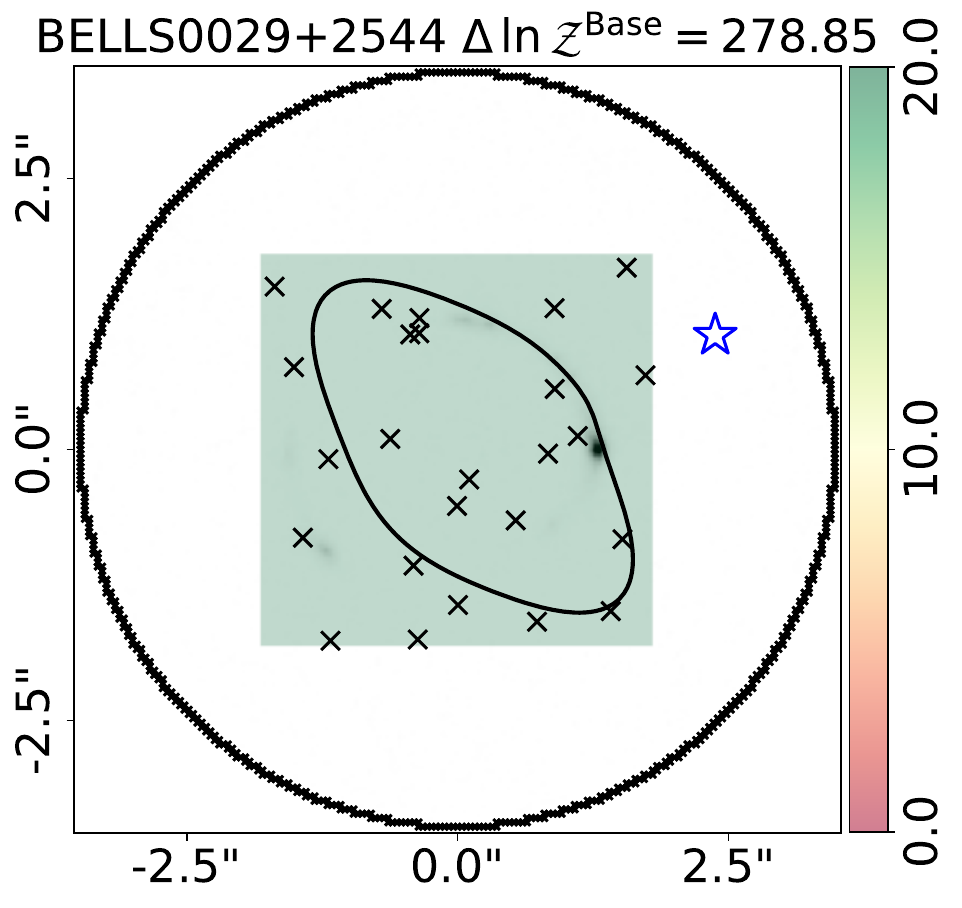}
\includegraphics[width=0.16\textwidth]{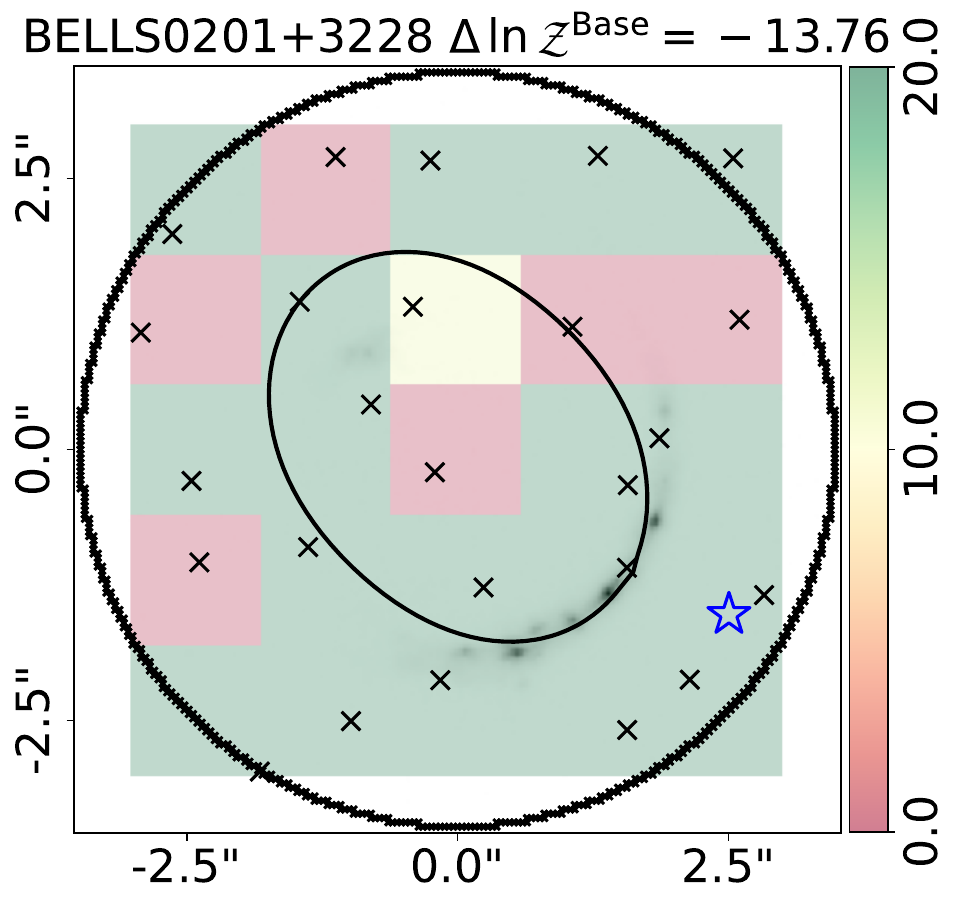}
\includegraphics[width=0.16\textwidth]{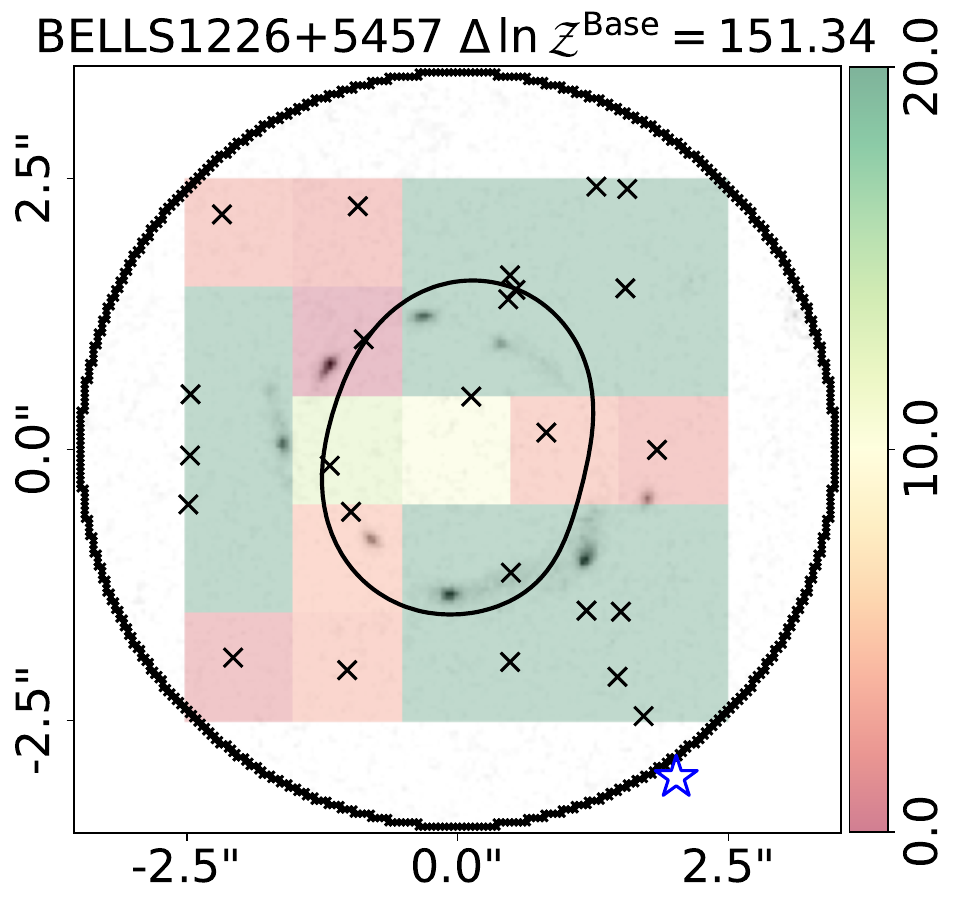}
\includegraphics[width=0.16\textwidth]{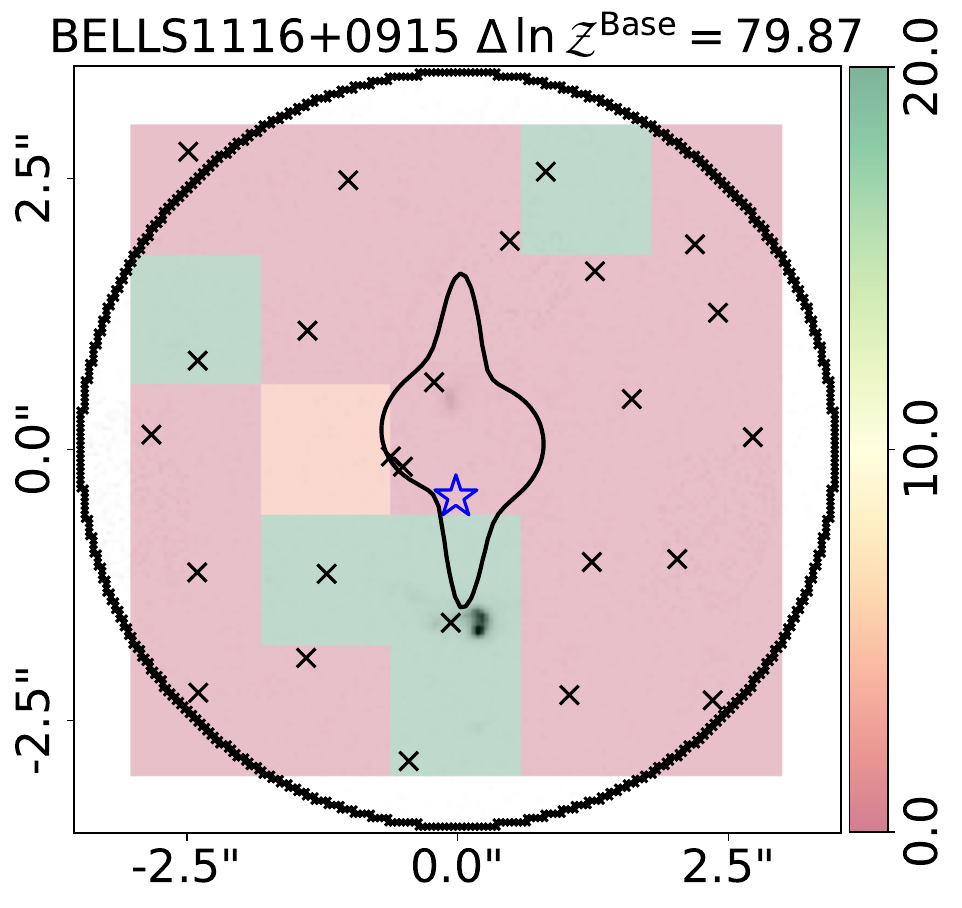}
\includegraphics[width=0.16\textwidth]{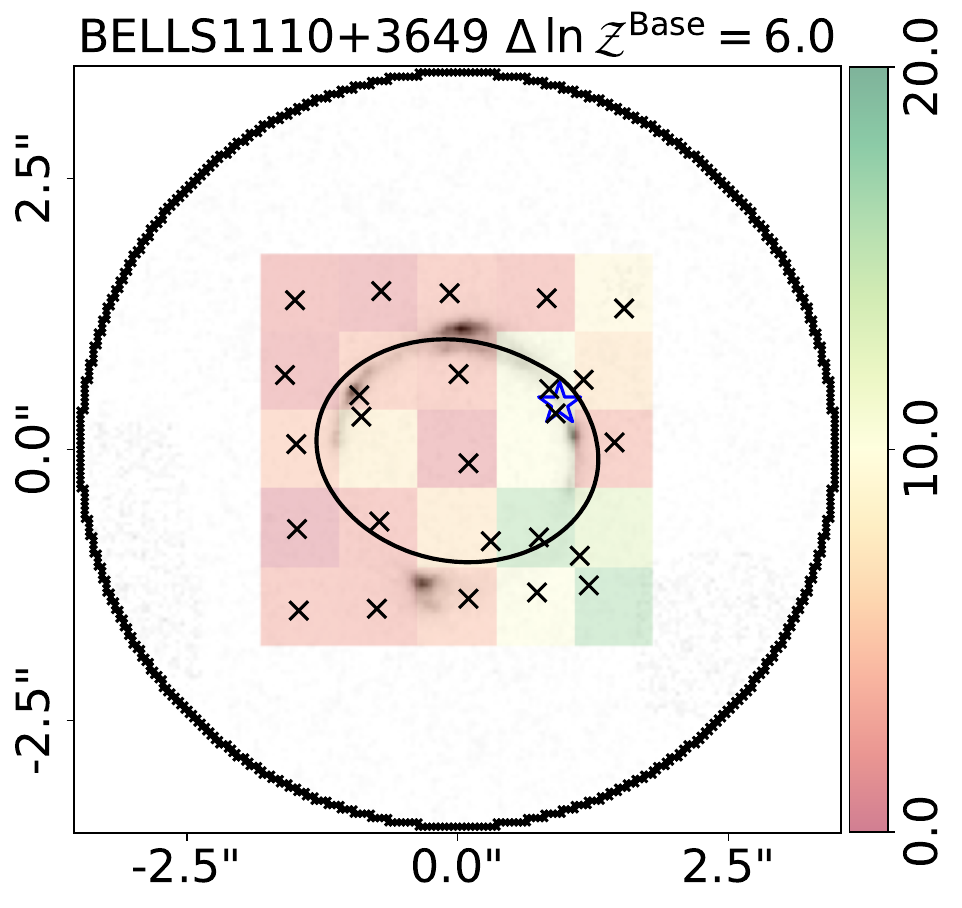}
\includegraphics[width=0.16\textwidth]{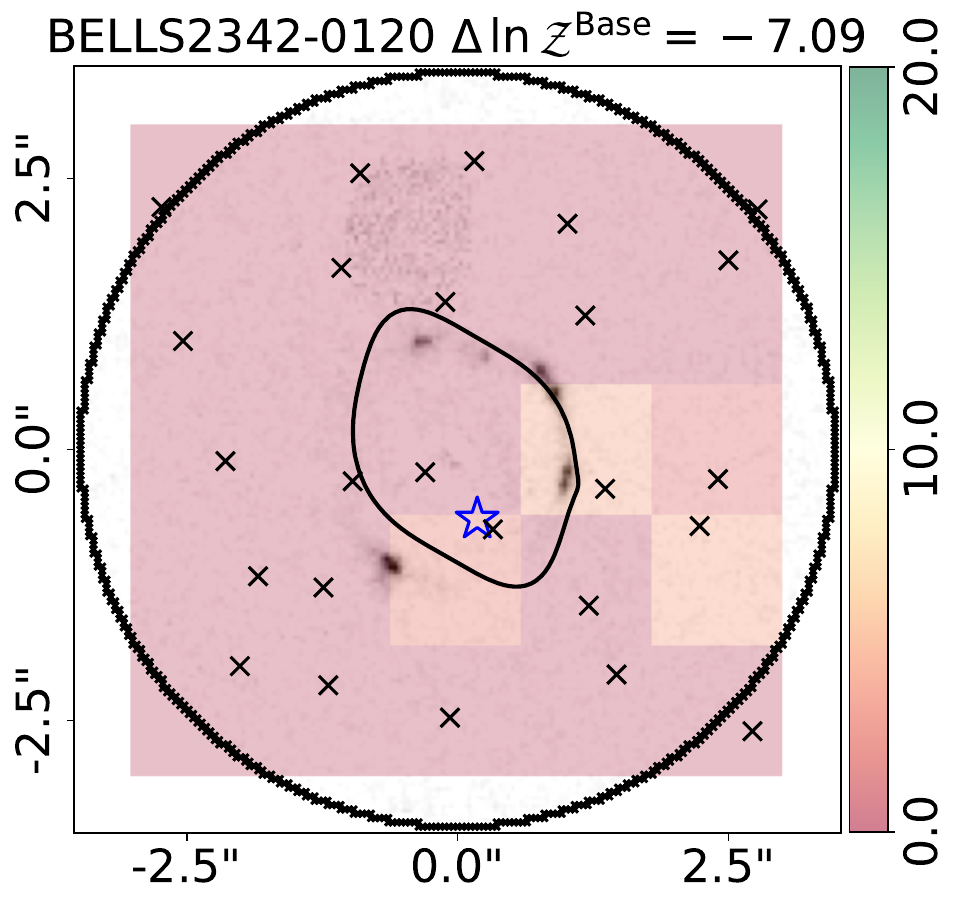}
\includegraphics[width=0.16\textwidth]{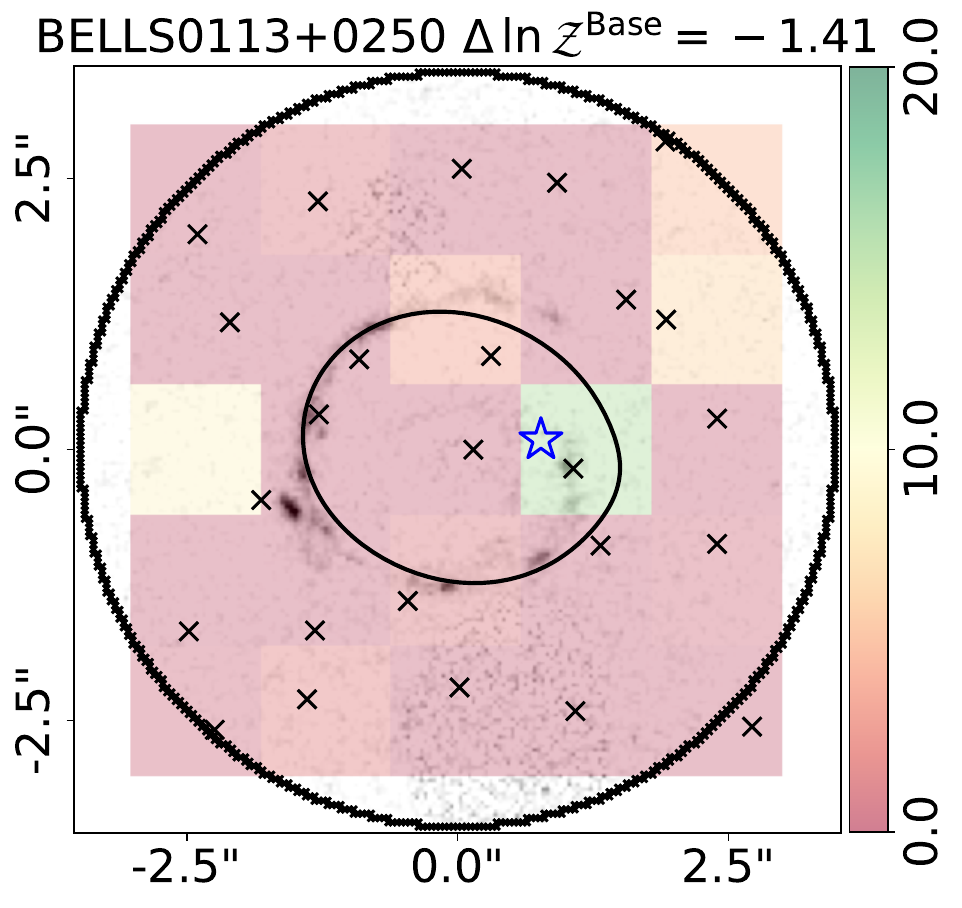}
\includegraphics[width=0.16\textwidth]{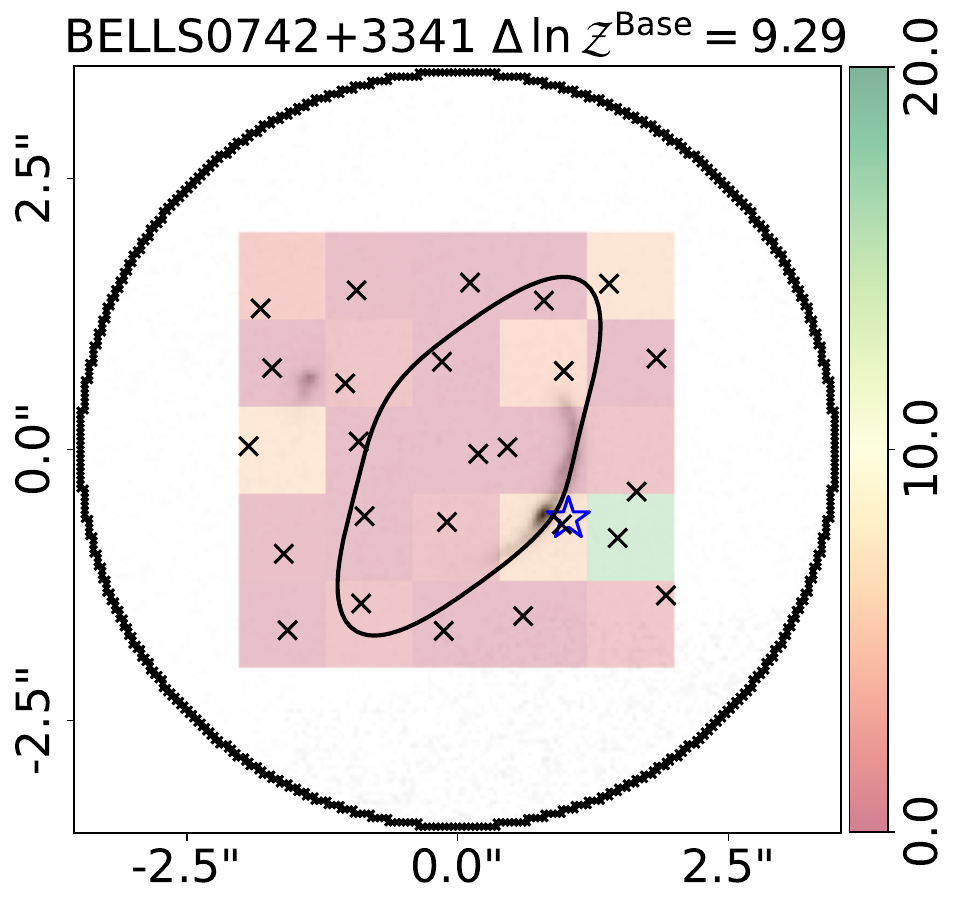}
\includegraphics[width=0.16\textwidth]{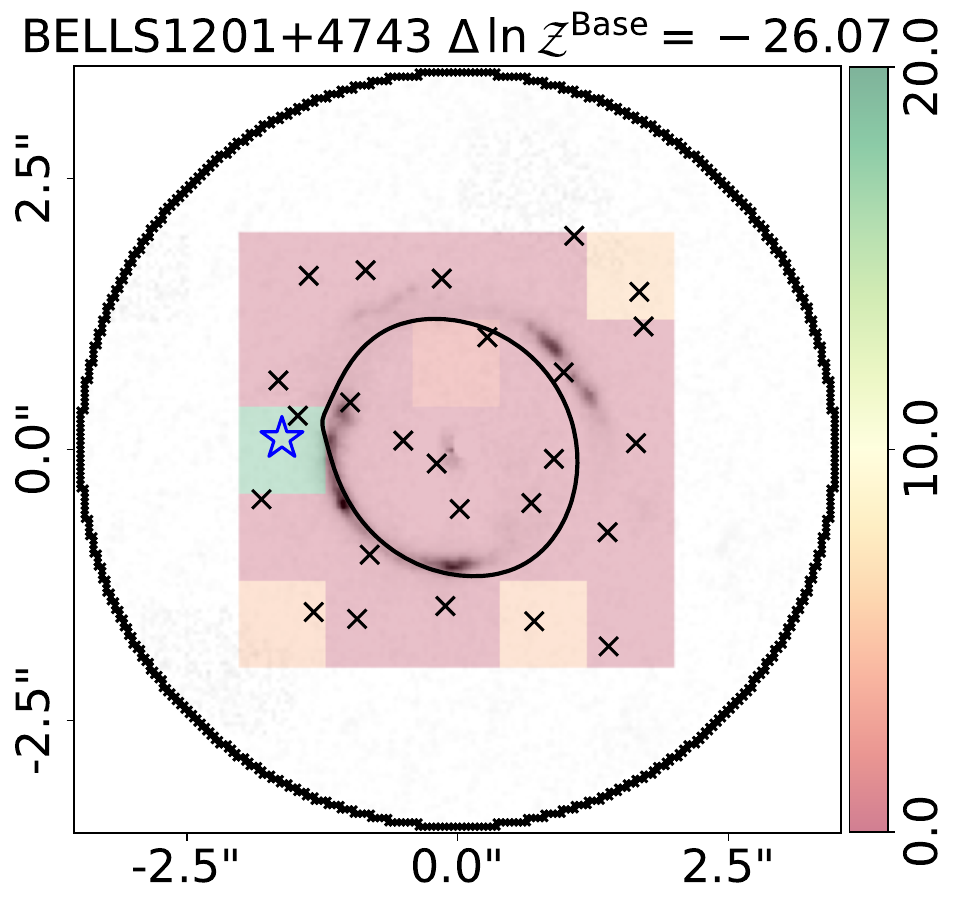}
\includegraphics[width=0.16\textwidth]{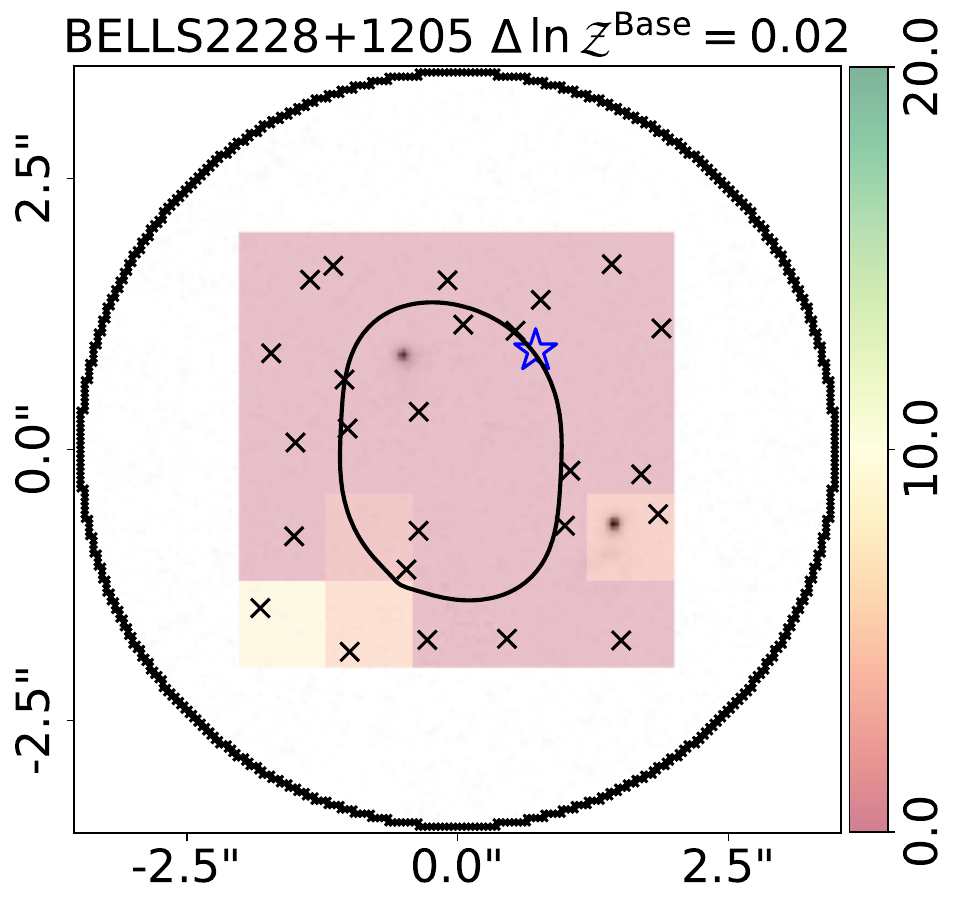}
\caption{
As \cref{figure:DetectSlacs}, but for the BELLS-GALLERY sample instead of SLACS.
}
\label{figure:DetectSBells}
\end{figure*}

\begin{table*}
\begin{adjustbox}{max width=\textwidth}
\Large
\begin{tabular}{ l | l | l | l | l | l | l | l | l | l | l | l | l | l | l} 
\multicolumn{1}{p{2.0cm}|}{Lens Name} 
& \multicolumn{1}{p{1.2cm}|}{$\Delta\,\mathrm{ln}$ \, $\mathcal{Z}^{\rm Base}$}  
& \multicolumn{1}{p{2.0cm}|}{$\log_{10}$ $[M^{\rm sub}_{\rm 200}$\,/\,M$_{\rm \odot}]$} 
& \multicolumn{1}{p{1.3cm}|}{$\Delta\,\mathrm{ln}$ \,$\mathcal{Z}^{\rm Light}$}  
& \multicolumn{1}{p{1.3cm}|}{Light $\Delta\,\mathrm{ln}\,\mathcal{Z}$ Decrease?}  
& \multicolumn{1}{p{1.3cm}|}{$\Delta\,\mathrm{ln}$ \, $\mathcal{Z}^{\rm Source}$}  
& \multicolumn{1}{p{1.3cm}|}{Source  $\Delta\,\mathrm{ln}\,\mathcal{Z}$ Decrease?}  
& \multicolumn{1}{p{1.5cm}|}{$\Delta\,\mathrm{ln}$ \, $\mathcal{Z}^{\rm BPL}$} 
& \multicolumn{1}{p{1.5cm}|}{$\Delta\,\mathrm{ln}$ \, $\mathcal{Z}^{\rm Multipole}$} 
& \multicolumn{1}{p{1.5cm}|}{$\Delta\,\mathrm{ln}$ \, $\mathcal{Z}^{\rm Decomp}$} 
& \multicolumn{1}{p{1.3cm}|}{Mass $\Delta\,\mathrm{ln}\,\mathcal{Z}$ Decrease?}  
& \multicolumn{1}{p{1.5cm}|}{$\Delta\,\mathrm{ln}$ \, $\mathcal{Z}^{\rm Los}$} 
& \multicolumn{1}{p{1.3cm}|}{Los $\Delta\,\mathrm{ln}\,\mathcal{Z}$ Decrease?}  
& \multicolumn{1}{p{1.5cm}|}{$\Delta\,\mathrm{ln}$ \, $\mathcal{Z}^{\rm Final}$} 
& \multicolumn{1}{p{1.2cm}}{Category} \\  \hline
& & & & & & & & & & & & & & \\[-6pt]
BELLS0755+3445 & $\textbf{1287.01}$ & $11.98^{+0.02}_{-0.10}$ & $\textbf{1360.14}^{*}$ & \xmark & $\textbf{1268.78}$ & $\checkmark$ & $\textbf{1155.44}$ & $\textbf{860.44}$ & $\textbf{889.39}$ & $\checkmark$ & $\textbf{184.49}$ & $\checkmark$ &  1155.43 & X \\[2pt]
BELLS0918+5104 & $\textbf{520.91}$ & $9.31^{+0.01}_{-0.01}$ & $\textbf{323.52}$ & $\checkmark$ & $\textbf{417.56}^{*}$ & \xmark & $\textbf{250.44}$ & $\textbf{248.04}$ & $\textbf{193.93}$ & $\checkmark$ & $\textbf{247.81}$ & $\checkmark$ & 248.04 & X \\[2pt]
BELLS0029+2544 & $\textbf{278.85}$ & $10.99^{+0.12}_{-0.04}$ & -164.99 & $\checkmark$ & $\textbf{157.26}^{*}$ & \xmark & -1269.31 & $\textbf{10.78}$ & -35.81 & $\checkmark$ & -37.86 & $\checkmark$ & 24.74 & X \\[2pt]
BELLS0201+3228 & -13.76 & $8.30^{+0.01}_{-0.01}$ & $-3.1^{*}$ & \xmark & -62.65 & $\checkmark$ & $-52.48^{*}$ & $\textbf{38.39}^{*}$ & $1.37^{*}$ & \xmark & $-44.7^{*}$ & \xmark & -52.47 & X \\[2pt]
BELLS1226+5457 & $\textbf{151.34}$ & $11.95^{+0.05}_{-0.24}$ & $\textbf{96.47}$ & $\checkmark$ & $\textbf{105.9}$ & \xmark & $\textbf{97.29}$ & $\textbf{72.0}$ & $\textbf{97.33}$ & $\checkmark$ & $\textbf{22.46}$ & $\checkmark$ & \textbf{79.08} & Cand \\[2pt]
BELLS1116+0915 & $\textbf{110.36}$ & $6.68^{+1.19}_{-0.66}$ & 1.6 & $\checkmark$ & -0.36 & \xmark & None & 1.32 & 0.9 & \xmark & -1.47 & \xmark & -0.35 & ND \\[2pt]
BELLS1110+2808 & $\textbf{11.14}$ & $11.82^{+0.17}_{-3.90}$ & 4.37 & \xmark & 3.71 & \xmark & 0.32 & 1.03 & 3.98 & \xmark & -0.19 & \xmark & 0.32 & ND \\[2pt]
BELLS0742+3341 & 9.29 & $10.39^{+0.04}_{-0.19}$ & 7.59 & \xmark & -3.59 & $\checkmark$ & -2.58 & -1.83 & 1.42 & \xmark & -1.03 & \xmark & -3.58 & ND \\[2pt]
BELLS1110+3649 & 6.0 & $9.77^{+0.43}_{-0.55}$ & $\textbf{21.3}^{*}$ & \xmark & $\textbf{12.65}$ & \xmark & $\textbf{20.26}$ & $\textbf{14.99}$ & 3.05 & \xmark & $\textbf{10.82}$ & \xmark & 3.04 & ND / FP-PL \\[2pt]
BELLS0237-0641 & 5.64 & $6.91^{+2.20}_{-0.38}$ & -8.41 & $\checkmark$ & -0.96 & \xmark & 0.75 & 3.74 & 2.1 & \xmark & 2.1 & \xmark & -0.95 & ND \\[2pt]
BELLS0856+2010 & 5.41 & $7.85^{+2.33}_{-1.08}$ & 9.14 & \xmark & -1.77 & $\checkmark$ & Demag & -0.11 & 1.57 & \xmark & -0.09 & \xmark & 1.53 & ND \\[2pt]
BELLS1201+4743 & 1.25 &  $8.65^{0.96}_{-1.23}$  & 3.11 & \xmark & $\textbf{20.36}^{*}$ & \xmark & $\textbf{16.52}$ & 3.59 & 3.69 & $\checkmark$ & $\textbf{25.58}$ & \xmark & \textbf{18.11} & Cand \\[2pt]
BELLS2228+1205 & 0.02 & $10.27^{+0.27}_{-0.05}$ & 1.46 & \xmark & 2.23 & \xmark & -2.0 & -1.4 & 2.38 & \xmark & -2.0 & \xmark & 2.22 & ND \\[2pt]
BELLS0113+0250 & -1.41 & $10.71^{+0.60}_{-0.05}$ & $\textbf{29.99}^{*}$ & \xmark & 8.55 & $\checkmark$ & $\textbf{11.37}$ & -7.81 & -1.0 & $\checkmark$ & -0.17 & \xmark & 8.55 & ND \\[2pt]
BELLS2342-0120 & -7.09 & $8.20^{+2.59}_{-0.48}$ & $\textbf{17.97}^{*}$ & \xmark & $\textbf{16.61}$ & \xmark & $\textbf{13.19}$ & 5.62 & $\textbf{23.69}$ & $\checkmark$ & 2.81 & $\checkmark$ & \textbf{13.18} & ND / FP-Los \\[2pt]
BELLS1141+2216 & -18.2 & $9.41^{+0.19}_{-0.32}$ & $-2.13^{*}$ & \xmark & 3.86 & \xmark & 0.73 & 2.64 & 3.88 & \xmark & $\textbf{10.08}$ & \xmark & 0.72 & ND \\[2pt]
\end{tabular}
\end{adjustbox}
\caption{
As \cref{table:DetectSLACS}, but for the BELLS-GALLERY sample instead of SLACS.
}
\label{table:DetectBells}
\end{table*}

Column 2 of Table \ref{table:DetectBells} lists $\Delta\,\mathrm{ln}\,\mathcal{Z}^{\rm Base}$ for all 16 BELLS-GALLERY lenses and \cref{figure:DetectSBells} shows the corresponding subhalo scanning results. 7 out of 16 lenses meet our criterion of producing $\Delta\,\mathrm{ln}\,\mathcal{Z}^{\rm Base} > 10$. Four lenses give $\Delta\,\mathrm{ln}\,\mathcal{Z}^{\rm Base} > 100$. Nine lenses are non-detections with $\Delta\,\mathrm{ln}\,\mathcal{Z}^{\rm Base} < 10$. \cref{table:DetectBells} also shows the inferred subhalo masses $M^{\rm sub}_{\rm 200}$, which again span $10^{8.3}$\,\,M$_{\rm \odot}$ and $10^{11.98}$\,\,M$_{\rm \odot}$ for models where $\Delta\,\mathrm{ln}\,\mathcal{Z}^{\rm Base} > 10$.

\subsection{Subhalo Scanning With Different Lens Light Subtraction}\label{sec:lens_light_results}

\begin{figure}
\centering
\includegraphics[width=0.235\textwidth]{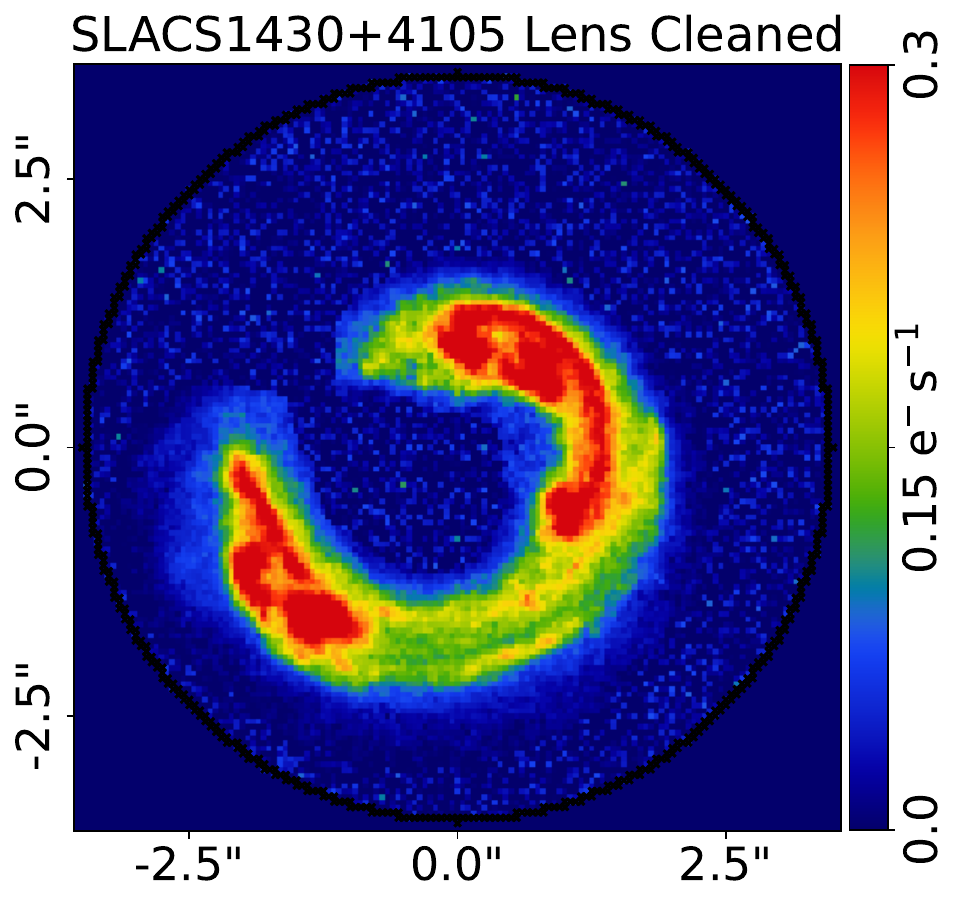}
\includegraphics[width=0.235\textwidth]{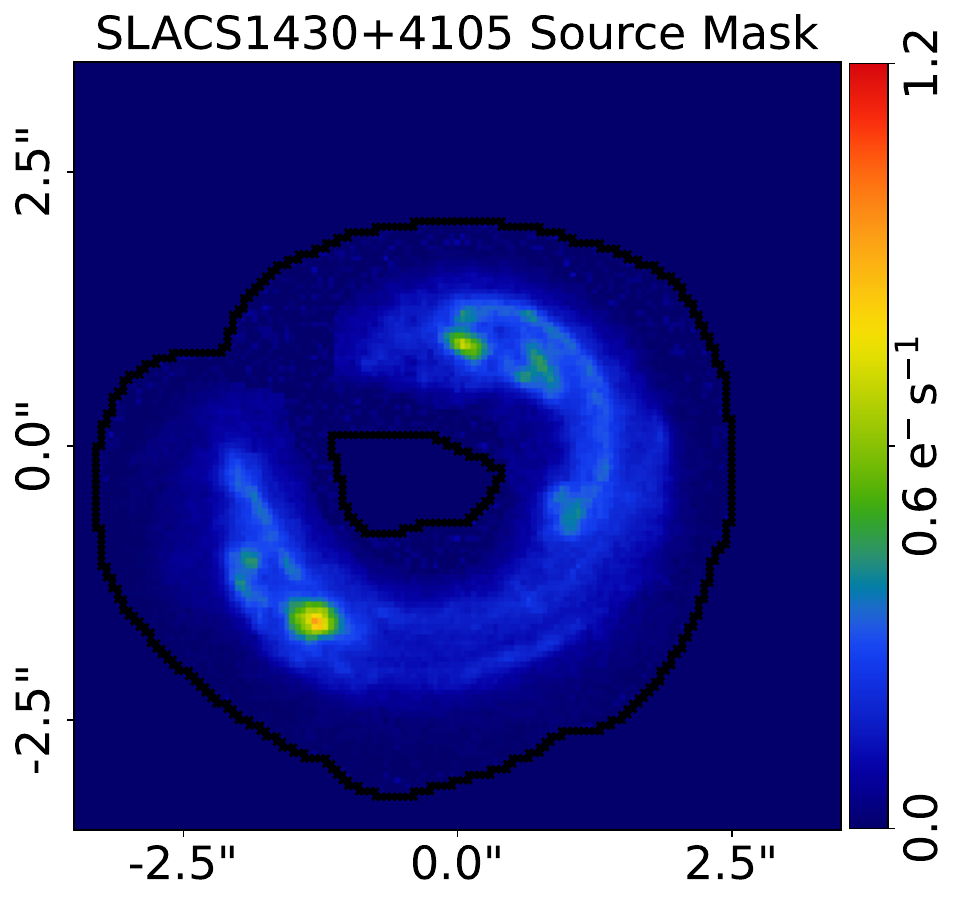}

\caption{
Illustration of the data augmentation schemes used to test for different systematics in the DM subhalo analysis, using the example lens SLACS1430+4105 (the image with the default $3.5\arcsec$ mask is shown in \cref{figure:Overview}). The left panel illustrates lens light cleaned data, where image regions containing predominantly signal from the lens light have their signal replaced with Gaussian noise, and their corresponding variances increased to arbitrarily large values (meaning the pixels do not contribute to the likelihood). These image-pixels are still ray traced to the source plane and reconstructed by the Voronoi mesh. Lens light cleaned data isolates whether lens light residuals are a systematic on the subhalo results (by comparing to fits using the default masks) and it uses evidence increases denoted $\Delta\,\mathrm{ln}\,\mathcal{Z}^{\rm Light}$. The right panel illustrates a source-only mask where the removed image-pixels are not ray-traced to the source-plane at all. The source reconstruction therefore dedicates more Voronoi cells to image pixels containing predominately the lensed source. Source-only masks isolate whether insufficient source resolution impacts the subhalo result (by comparing to fits using lens light-cleaned data) and evidence increases are denoted $\Delta\,\mathrm{ln}\,\mathcal{Z}^{\rm Source}$. 
}
\label{figure:MaskScaled}
\end{figure}

\begin{figure*}
\centering
\includegraphics[width=0.24\textwidth]{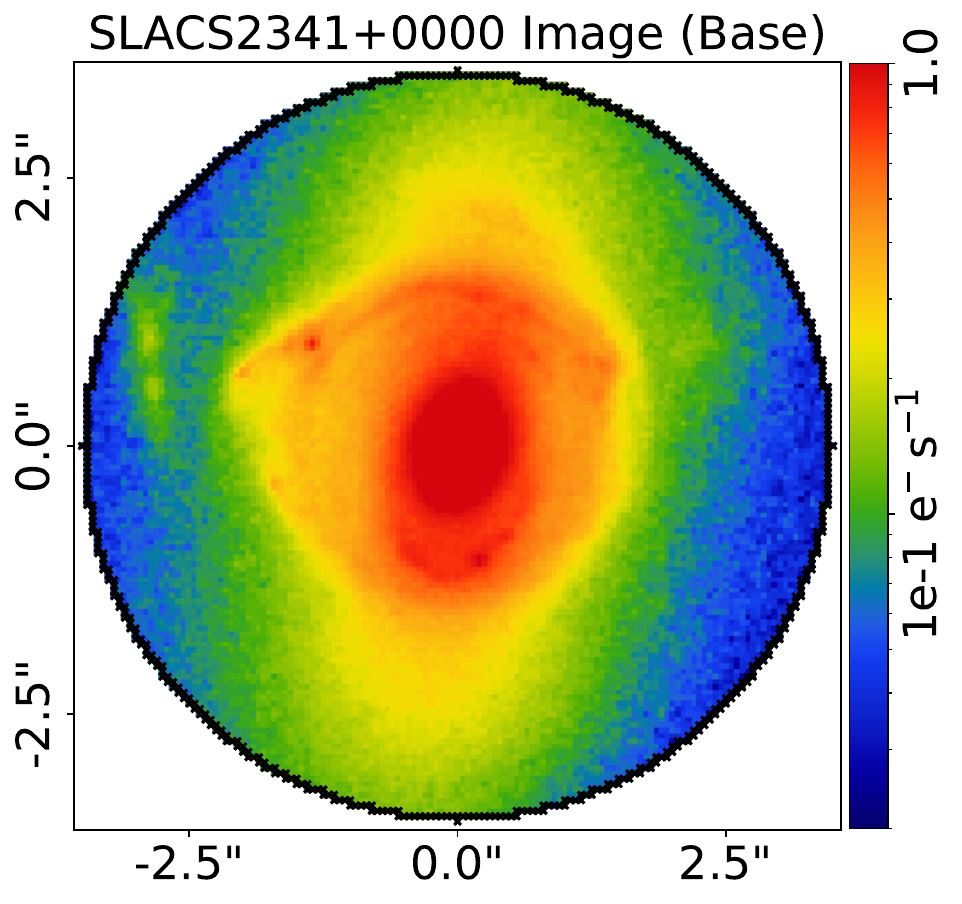}
\includegraphics[width=0.24\textwidth]{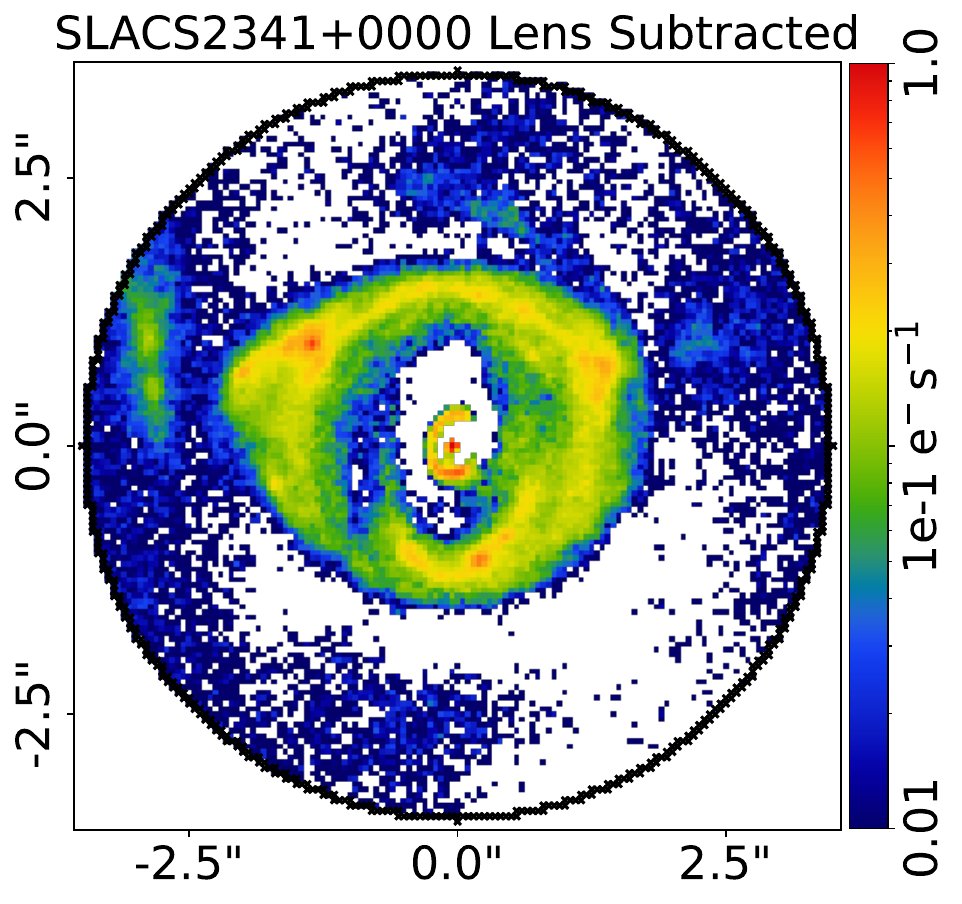}
\includegraphics[width=0.24\textwidth]{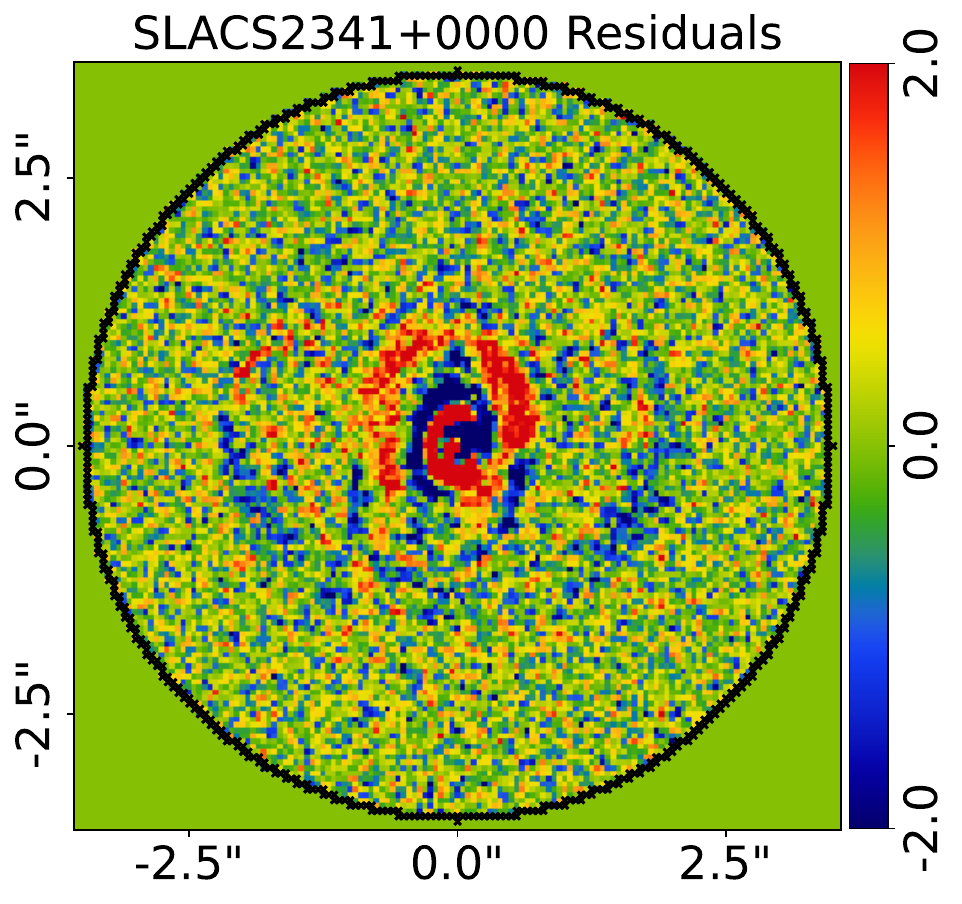}
\includegraphics[width=0.24\textwidth]{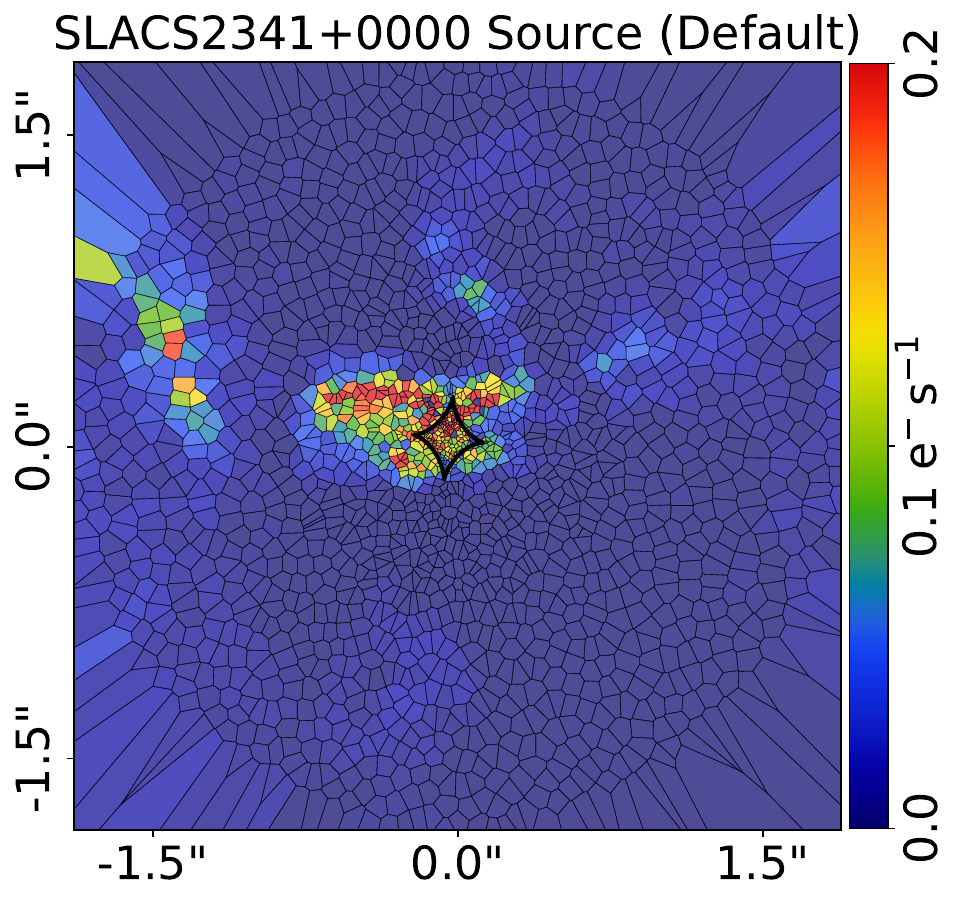}
\includegraphics[width=0.24\textwidth]{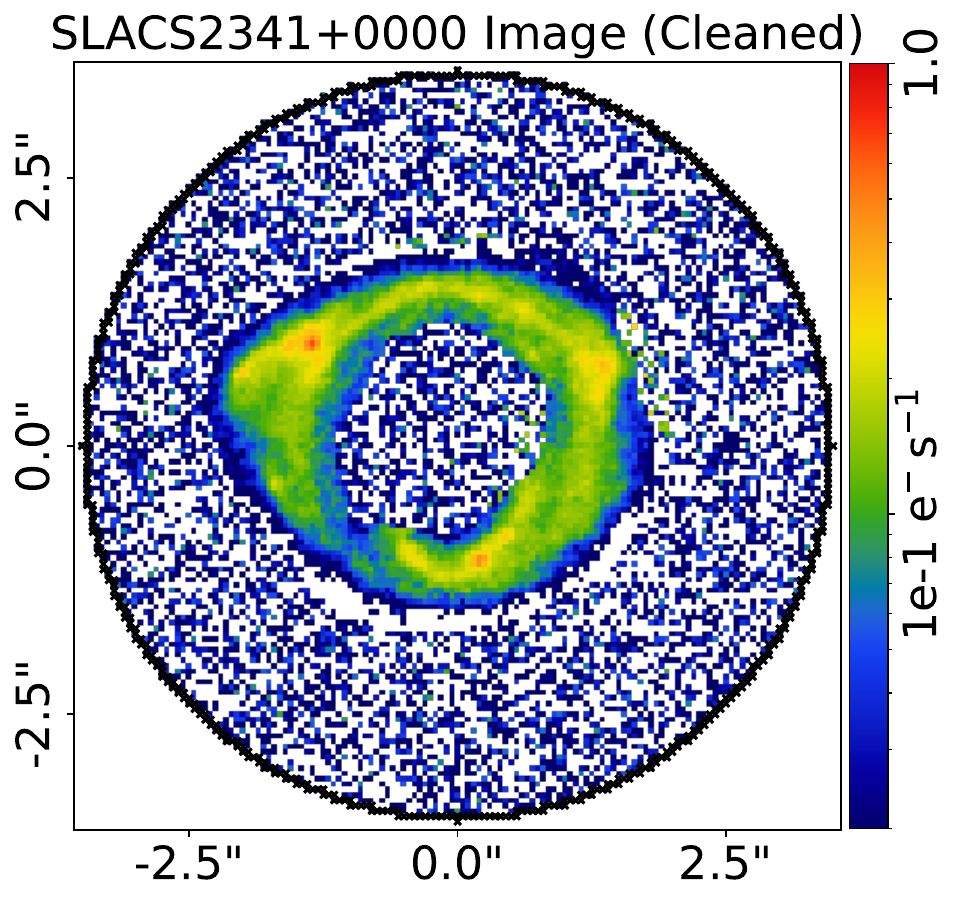}
\includegraphics[width=0.24\textwidth]{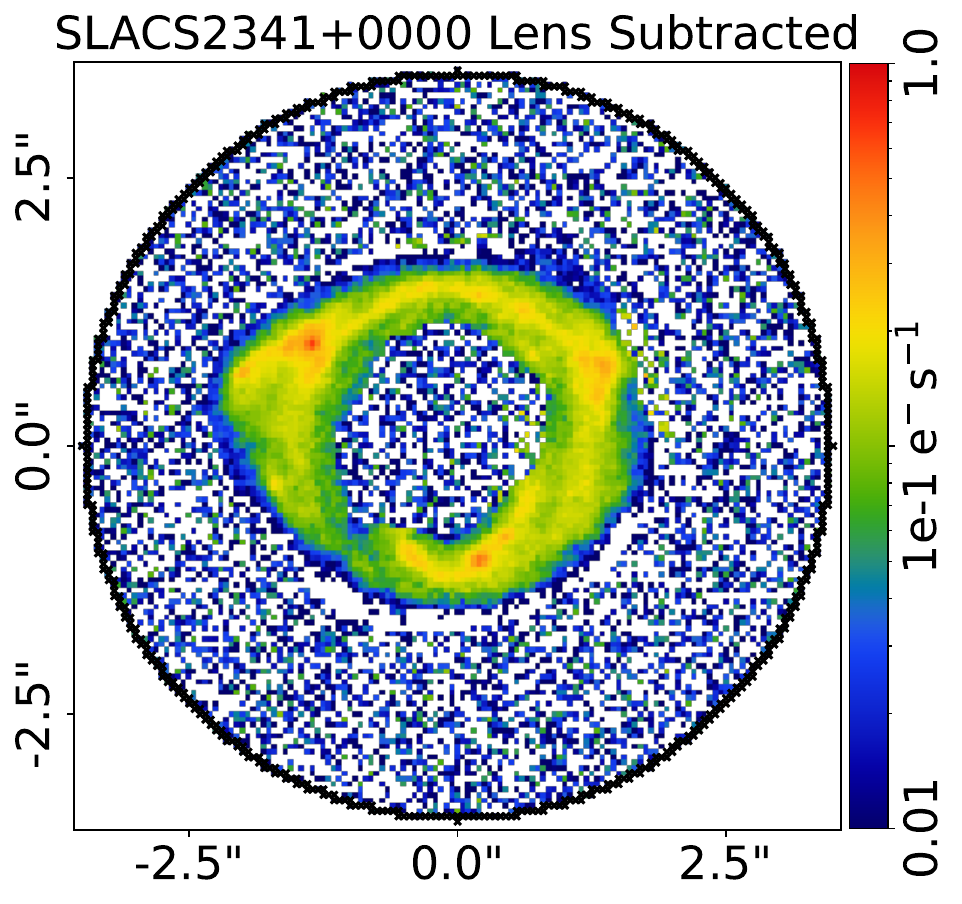}
\includegraphics[width=0.24\textwidth]{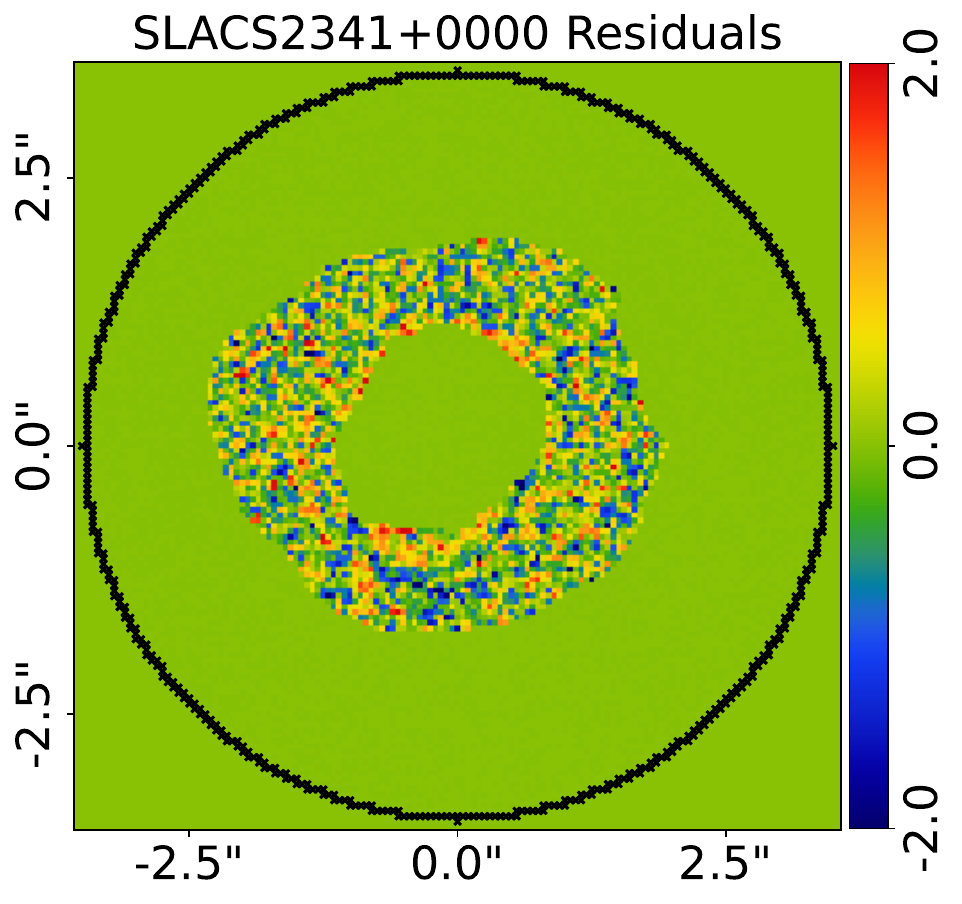}
\includegraphics[width=0.24\textwidth]{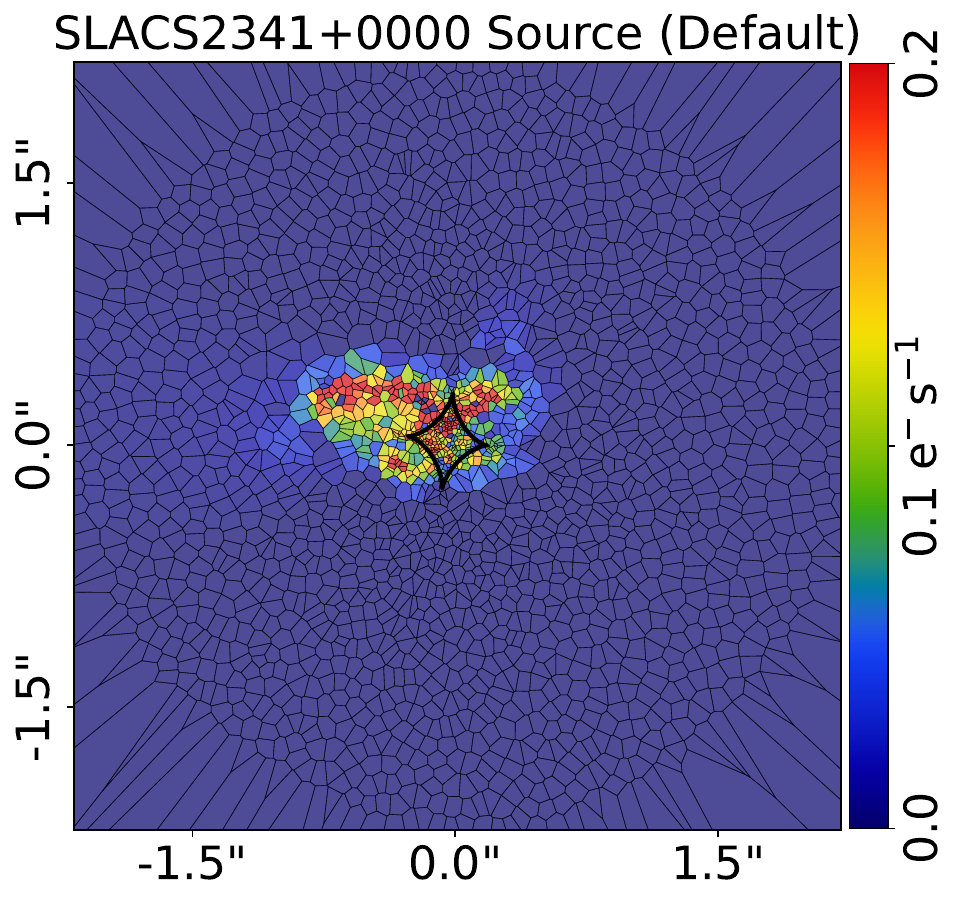}
\caption{
How removing lens light subtraction residuals (by fitting lens light cleaned data) can remove DM subhalo candidates, illustrated using SLACS2341+0000. The images (from left to right) are the observed image, lens subtracted image, normalized residuals (after subtracting the lens and lensed source models from the data) and source reconstruction. The top row shows fits to the original data and bottom row the lens light cleaned data. For SLACS2341+0000, the top row shows that the double S\'ersic lens light subtraction leaves significant residuals in the center and to the east, north and west of the image, because of non-symmetric emission indicative of a post-merger. The source reconstruction (right panel) fits these lens light residuals, as it cannot distinguish lens and source emission. This is why a DM subhalo is incorrectly favoured. The lens light cleaned data removes these lens light residuals, such that the source reconstruction only reconstructs the source. The evidence for a DM subhalo in turn reduces significantly. In other lenses where a DM subhalo candidate is removed by fitting the lens light cleaned data, similar residuals are seen if a double S\'ersic lens light model is used to subtract the lens light, but to a lesser degree. These are typically not due to post-merger features like in SLACS2341+0000, but other irregular morphological features (e.g a central bulge or disk-like structure).}
\label{figure:FPLight}
\end{figure*}

To investigate whether an inaccurate lens light subtraction produces false positives we fit lens light cleaned datasets. These are produced using a GUI which replaces the observed flux counts in the image data with Gaussian noise and increases the variances in all image pixels which -- from visual inspection -- appear to predominately contain lens light subtraction residuals. 
The pixels therefore do not contribute to the likelihood function given by equation \ref{eqn:evidence2}. An example is shown in \cref{figure:MaskScaled}. The log evidence increases by including a DM subhalo for fits using the lens light cleaned dataset is defined as $\Delta\,\mathrm{ln}\,\mathcal{Z}^{\rm Light}$, which is compared to $\Delta\,\mathrm{ln}\,\mathcal{Z}^{\rm Base}$ to isolate the dependence on the lens light subtraction.

The second and fourth columns of \cref{table:DetectSLACS} show $\Delta\,\mathrm{ln}\,\mathcal{Z}^{\rm Base}$ and $\Delta\,\mathrm{ln}\,\mathcal{Z}^{\rm Light}$ values for SLACS, where for nine lenses (ticks in column five) fitting the lens light cleaned data decreases $\Delta\,\mathrm{ln}\,\mathcal{Z}$ by more than $10$ ($\Delta\,\mathrm{ln}\,\mathcal{Z}^{\rm Base} - \Delta\,\mathrm{ln}\,\mathcal{Z}^{\rm Light} > 10$). This includes $7$ lenses which switch from candidate DM subhalos to non-detections, because $\Delta\,\mathrm{ln}\,\mathcal{Z}^{\rm Light} < 10$ and $\Delta\,\mathrm{ln}\,\mathcal{Z}^{\rm Base} > 10$. \cref{table:DetectBells} shows the same values for BELLS-GALLERY, where for $5$ lenses $\Delta\,\mathrm{ln}\,\mathcal{Z}$ decreases by more than $10$ and two lenses switch from favouring a DM subhalo to not. There are also $3$ BELLS-GALLERY lenses where $\Delta\,\mathrm{ln}\,\mathcal{Z}^{\rm Light} > 10$ and $\Delta\,\mathrm{ln}\,\mathcal{Z}^{\rm Base} <10$, meaning that fitting the lens light cleaned data means a DM subhalo is favoured when it was not for the baseline model.

\cref{figure:FPLight} shows the observed image (left column), lens subtracted image (left-centre column), normalized residuals (right-centre column) and source reconstructions (right column) of SLACS2341+0000, for fits to the original data (top row) and lens light cleaned data (bottom row). This is the SLACS lens with the largest decrease of $\Delta\,\mathrm{ln}\,\mathcal{Z}^{\rm Light}$ compared to $\Delta\,\mathrm{ln}\,\mathcal{Z}^{\rm Base}$. There is evidence that the lens galaxy has undergone a recent merger, with the residuals showing tidal stream features to the left, above and right of the lensed source. The lens light subtraction also shows a central dipole feature indicating the galaxy has not yet dynamically settled post-merger. The source reconstruction shown in the top right panel  reconstructs these lens light features towards the left, top and right of the source-plane. The bottom right panel shows these are not present in the source reconstruction of the lens light cleaned data, because the lens light residuals have been removed. The incorrect reconstruction of lens light features is responsible for the large decrease in $\Delta\,\mathrm{ln}\,\mathcal{Z}^{\rm Light}$.

Visual inspection of other lenses which show a large reduction in $\Delta\,\mathrm{ln}\,\mathcal{Z}$ when fitting lens light cleaned data indicates similar residuals are often present, which are therefore responsible for a DM subhalo being incorrectly being favoured. However, they are typically not post-merger features like in SLACS2341+0000 but fainter lens galaxy morphological features like a central bulge or bar. 

There will also be a more a subtle interplay between the lens subtraction and leftover lensed source emission, which to some degree will impact the DM subhalo inference. However, this is not responsible for the large changes of $\Delta\,\mathrm{ln}\,\mathcal{Z} > 10$ considered here. We note also that variance scaling (see \cref{VarScale}) was intended to mitigate these false positives, but is clearly insufficient in many lenses.

By removing lens light residuals via a GUI this source of DM subhalo false positives is successfully mitigated against. All remaining systematic tests therefore fit data which has been treated in this way.

\subsection{Subhalo Scanning With Different Source Resolution}\label{sec:source_only_results}

\begin{figure*}
\centering
\includegraphics[width=0.24\textwidth]{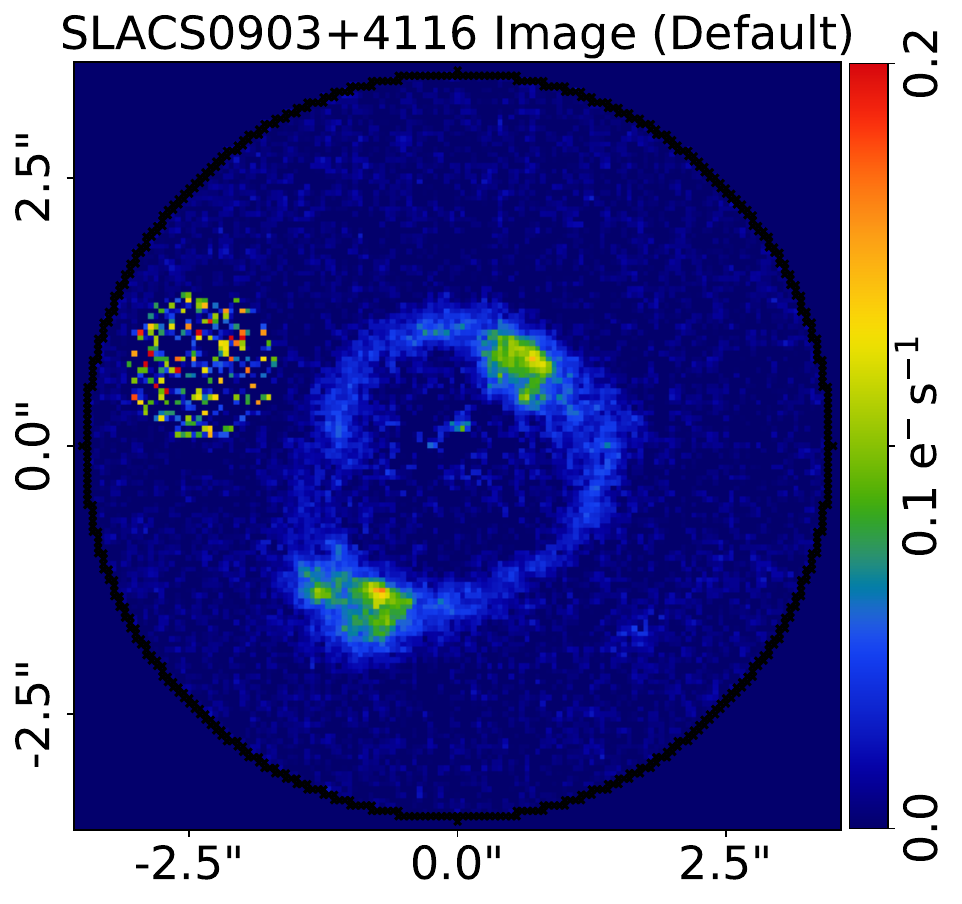}
\includegraphics[width=0.24\textwidth]{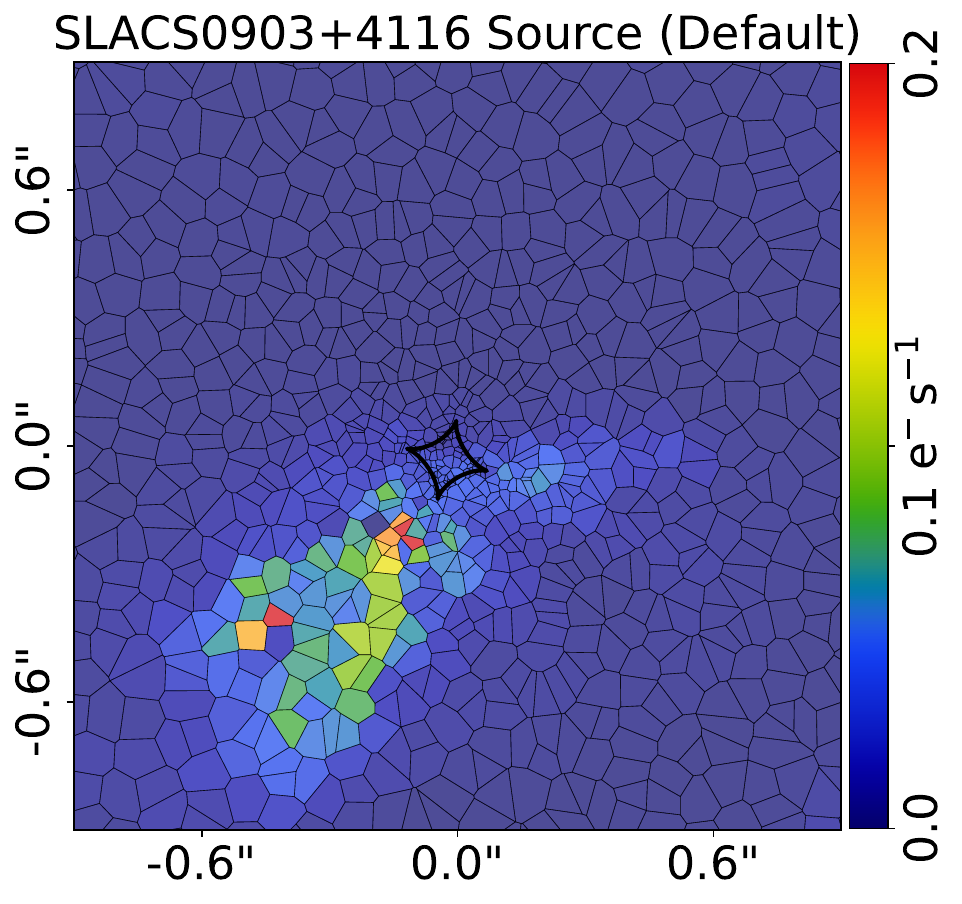}
\includegraphics[width=0.24\textwidth]{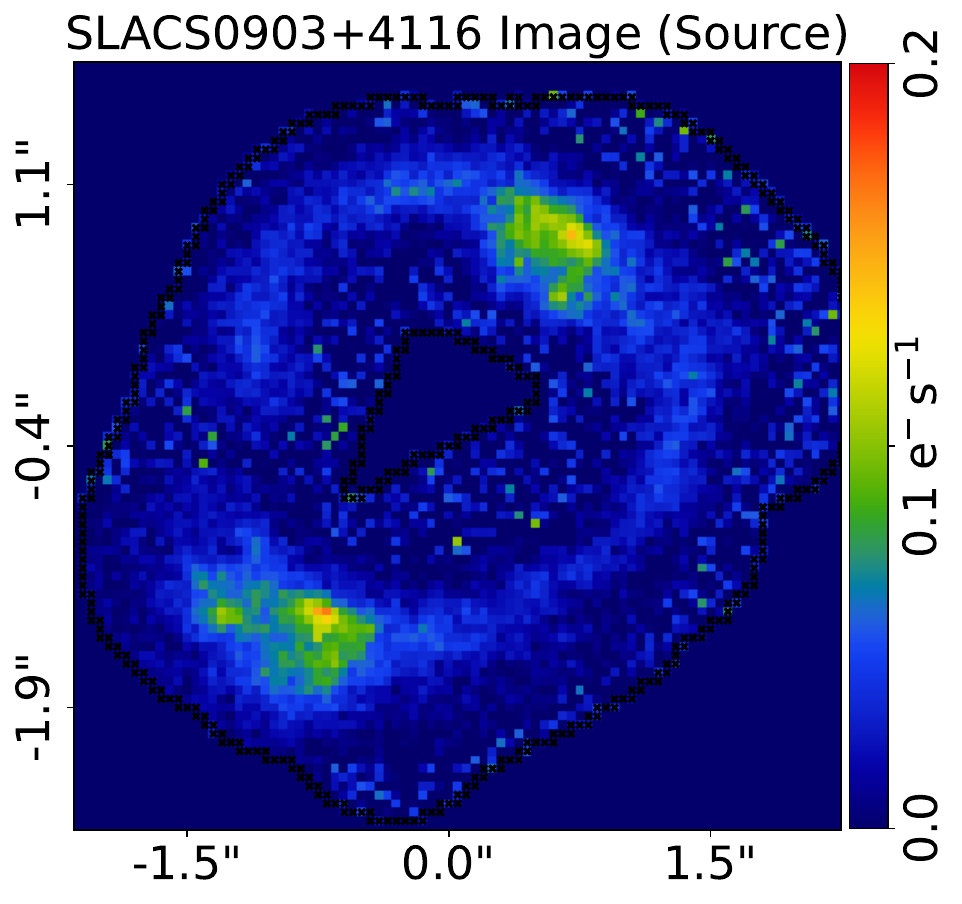}
\includegraphics[width=0.24\textwidth]{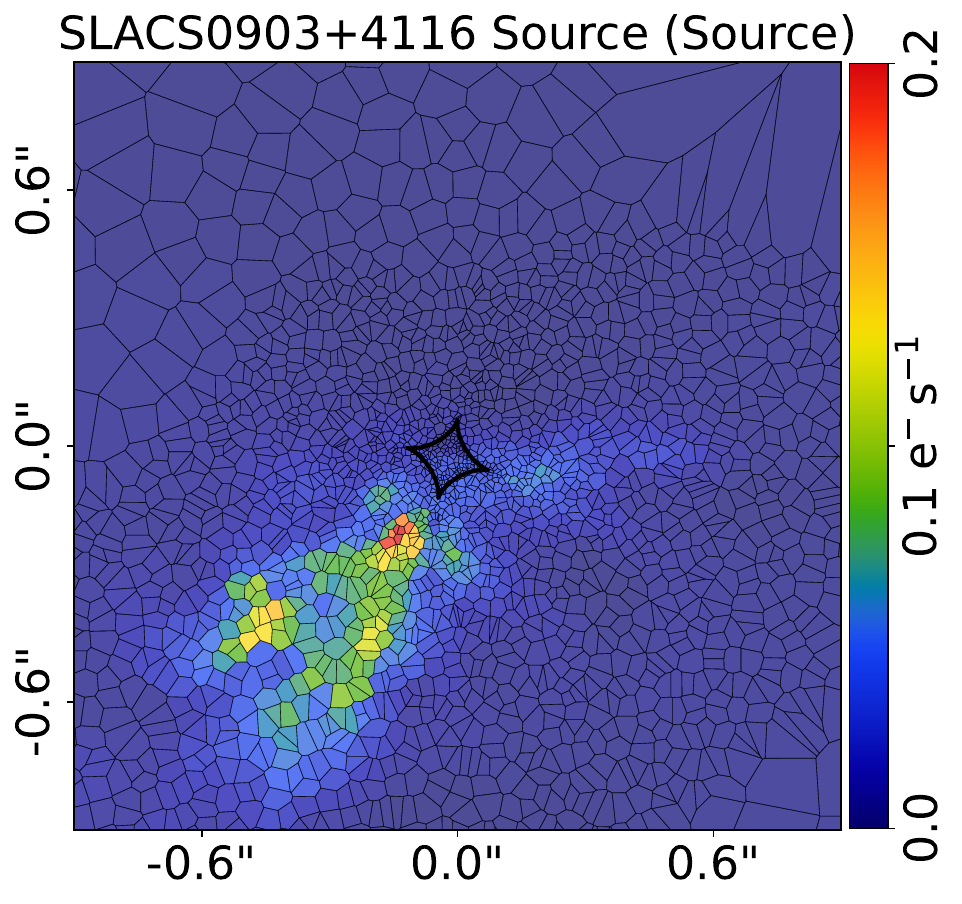}
\caption{
How making the source reconstruction higher resolution (by using a source-only mask) may remove a DM subhalo candidate, using the lens SLACS0903+4116. For this lens the default $3.5\arcsec$ mask gives $\Delta\,\mathrm{ln}\,\mathcal{Z}^{\rm Light} = 15.35$, whereas the source-only mask gives $\Delta\,\mathrm{ln}\,\mathcal{Z}^{\rm SO} = 3.91$. Fits using the default $3.5\arcsec$ mask (left panels) and source-only mask (right panels) are shown, showing the lens subtracted image and Voronoi source mesh. Using a source only mask reconstructs the source using more Voronoi pixels, resolving source structure such that the lens model no longer favours a DM subhalo.}
\label{figure:FPSource}
\end{figure*}

To investigate whether insufficient resolution of the source reconstruction leads to false positives we perform fits using source-only masks. These masks retain only image pixels with significant lensed source emission, an example of which is shown in \cref{figure:MaskScaled}. A GUI is used to mask the specific regions of the data which contain lensed source emission. All pixels outside of this custom mask are not ray-traced to the source plane and therefore are not used to construct the Voronoi mesh and reconstruct the source. This is in contrast to the lens light cleaned data above, which retained the $3.5\arcsec$ circular mask. The pixels outside the custom mask therefore also do not contribute to the $\chi^2$ given in \cref{eqn:evidence2}. A Voronoi mesh using a source-only mask therefore dedicates a larger fraction of Voronoi cells to reconstructing the source's brighter central regions. The number of source pixels is also fixed to 2500, the upper limit of the prior for previous fits using a circular mask. The log evidence increase by including a DM subhalo for fits using a source-only mask is defined as $\Delta\,\mathrm{ln}\,\mathcal{Z}^{\rm Source}$, which is compared to $\Delta\,\mathrm{ln}\,\mathcal{Z}^{\rm Light}$ to isolate the dependence on the source resolution.


The fourth and sixth columns of \cref{table:DetectSLACS} show the $\Delta\,\mathrm{ln}\,\mathcal{Z}^{\rm Light}$ and $\Delta\,\mathrm{ln}\,\mathcal{Z}^{\rm Source}$ values for SLACS. For four lenses (ticks in column 7) a higher resolution source model decreases $\Delta\mathrm{ln}\,\mathcal{Z}$ by more than $10$ ($\Delta\,\mathrm{ln}\,\mathcal{Z}^{\rm Light} - \Delta\,\mathrm{ln}\,\mathcal{Z}^{\rm Source} > 10$). This includes three lenses which go from candidate DM subhalos to non-detections ($\Delta\,\mathrm{ln}\,\mathcal{Z}^{\rm Light} > 10$ and $\Delta\,\mathrm{ln}\,\mathcal{Z}^{\rm Source} < 10$). \cref{table:DetectBells} shows the same values for BELLS-GALLERY, where for five lenses $\Delta\mathrm{ln}\,\mathcal{Z}$ decreases by more than $10$ and none switch from DM subhalo candidates to non detections. The lens BELLS1201+4743, marked with an asterix in column 6 of \cref{table:DetectBells}, switches from a DM non-detection ($\Delta\,\mathrm{ln}\,\mathcal{Z}^{\rm Light} < 10$) to a candidate DM subhalo ($\Delta\,\mathrm{ln}\,\mathcal{Z}^{\rm Source} > 10$).

Fig.~\ref{figure:FPSource} shows the lens subtracted images and source reconstructions of SLACS0903+4116, a lens where $\Delta\,\mathrm{ln}\,\mathcal{Z}^{\rm Light} = 18.54$ and $\Delta\,\mathrm{ln}\,\mathcal{Z}^{\rm Source} = 3.83$. The higher resolution source reconstruction produced using the source-only mask reconstructs more structure, improving the overall lens analysis such that a DM subhalo is no longer favoured. 

The DM subhalo results therefore depend on the source resolution. All remaining systematic tests therefore use source-only masks.

\subsection{Catastrophic Failures}\label{Catastrophic}

\begin{figure*}
\centering
\includegraphics[width=0.24\textwidth]{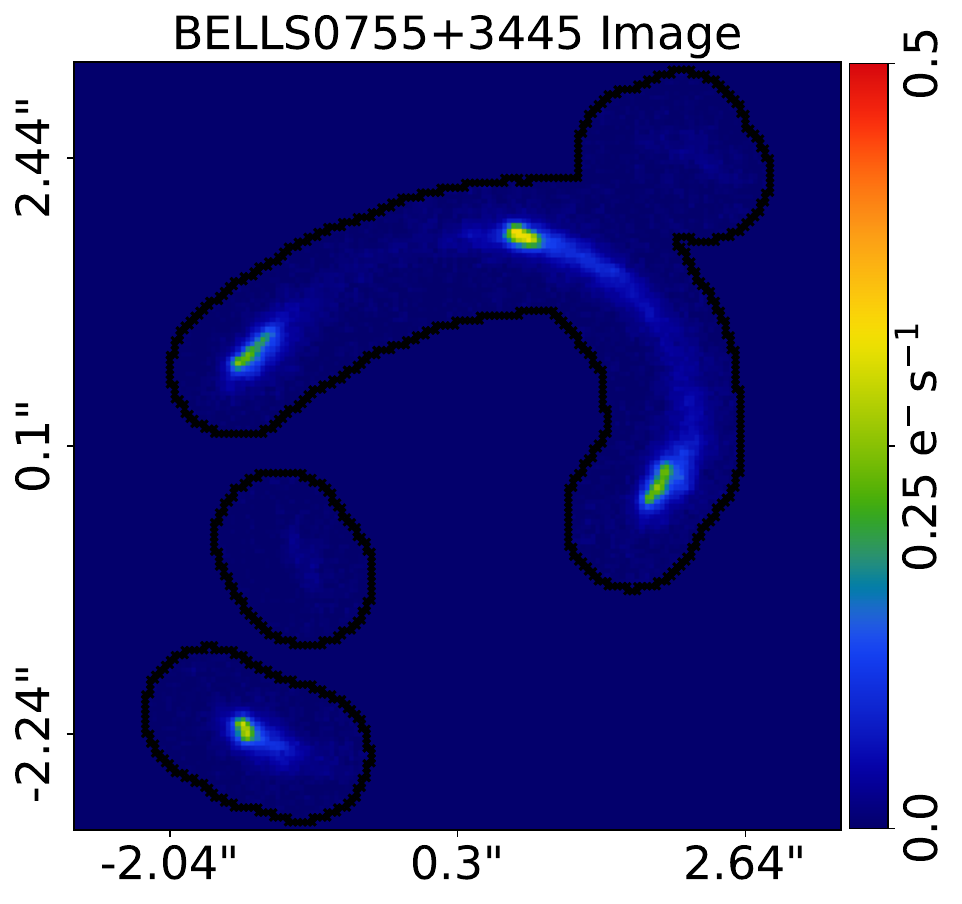}
\includegraphics[width=0.24\textwidth]{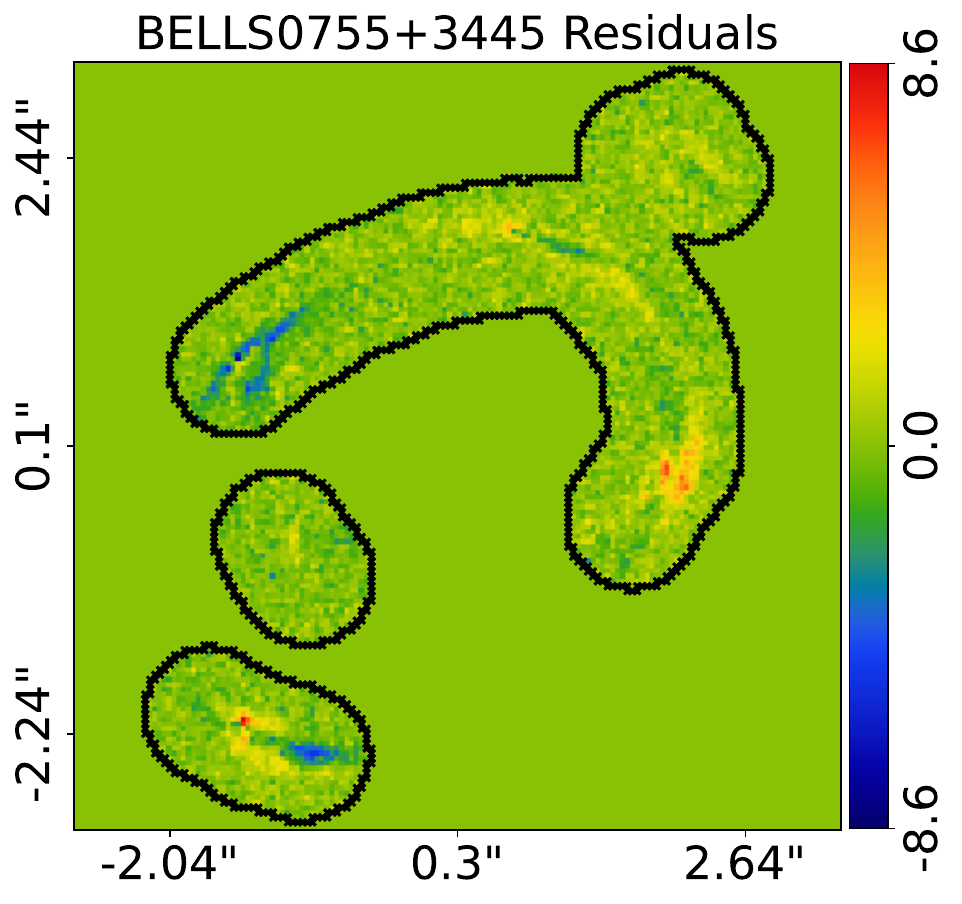}
\includegraphics[width=0.24\textwidth]{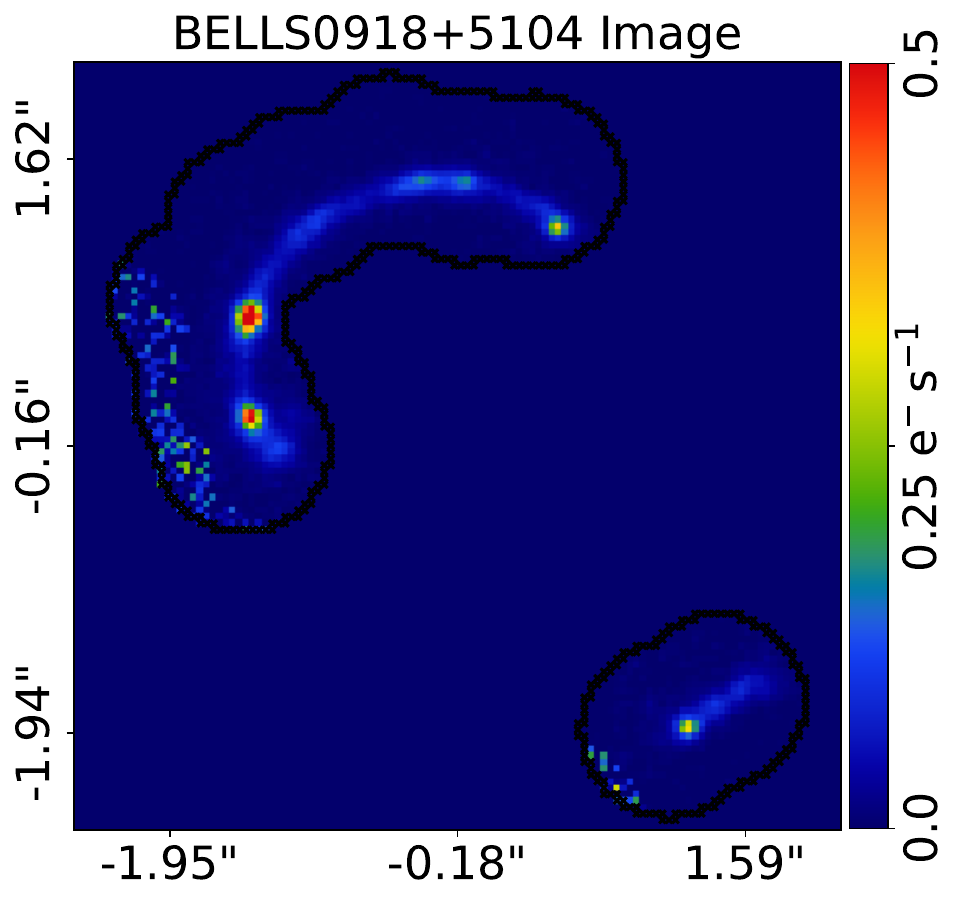}
\includegraphics[width=0.24\textwidth]{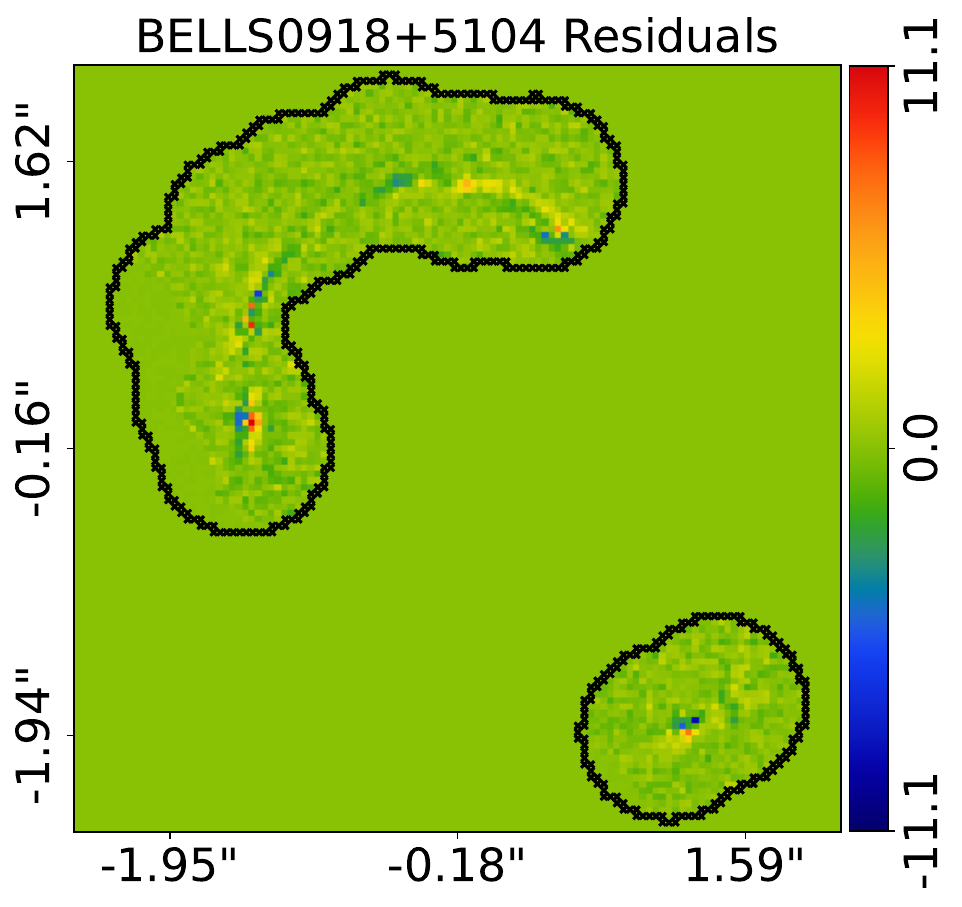}
\includegraphics[width=0.24\textwidth]{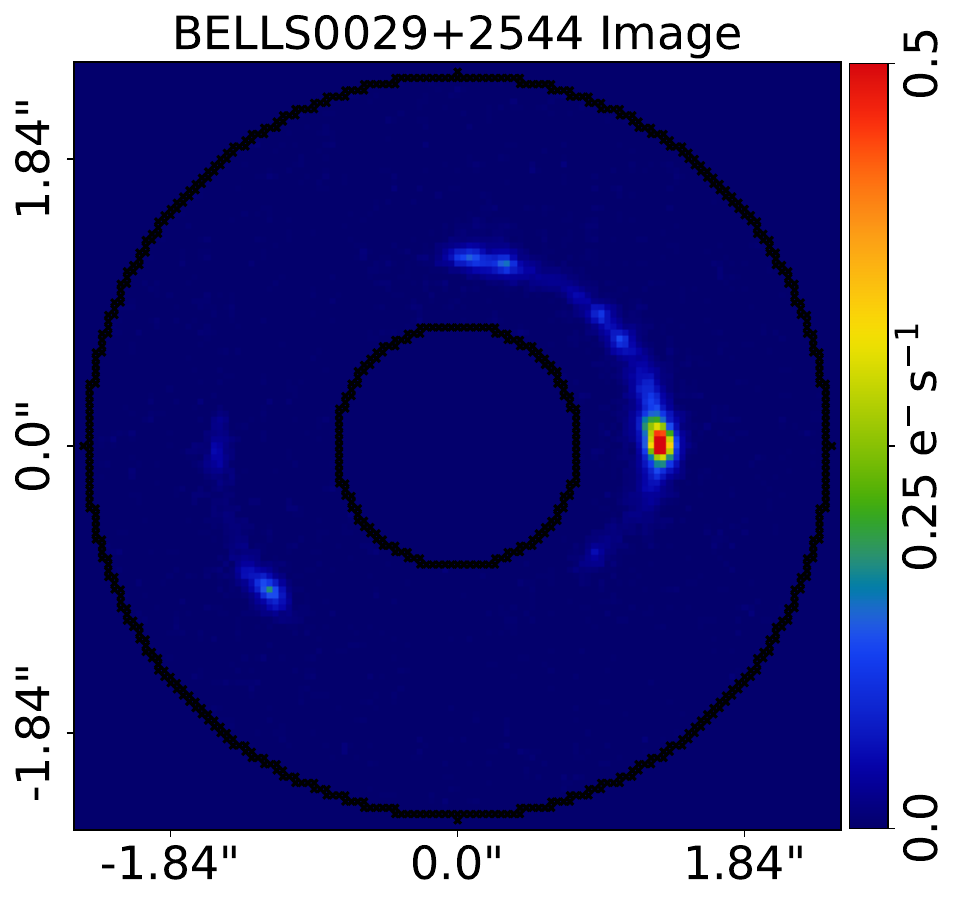}
\includegraphics[width=0.24\textwidth]{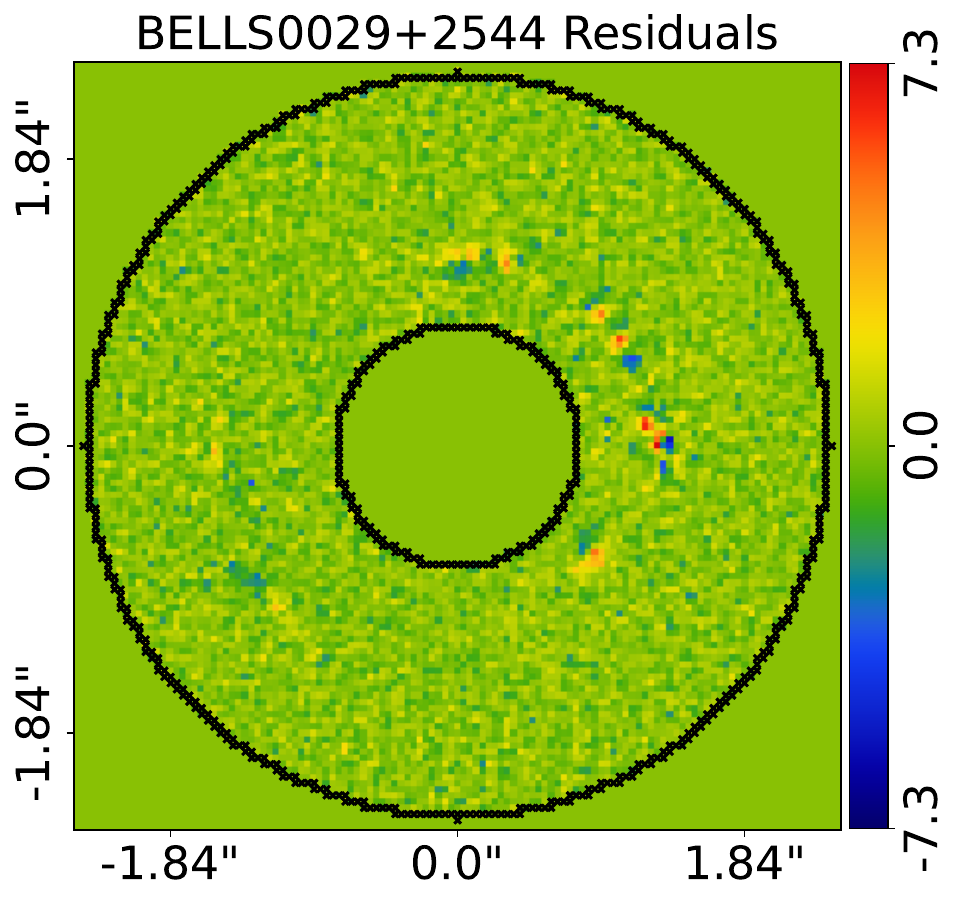}
\includegraphics[width=0.24\textwidth]{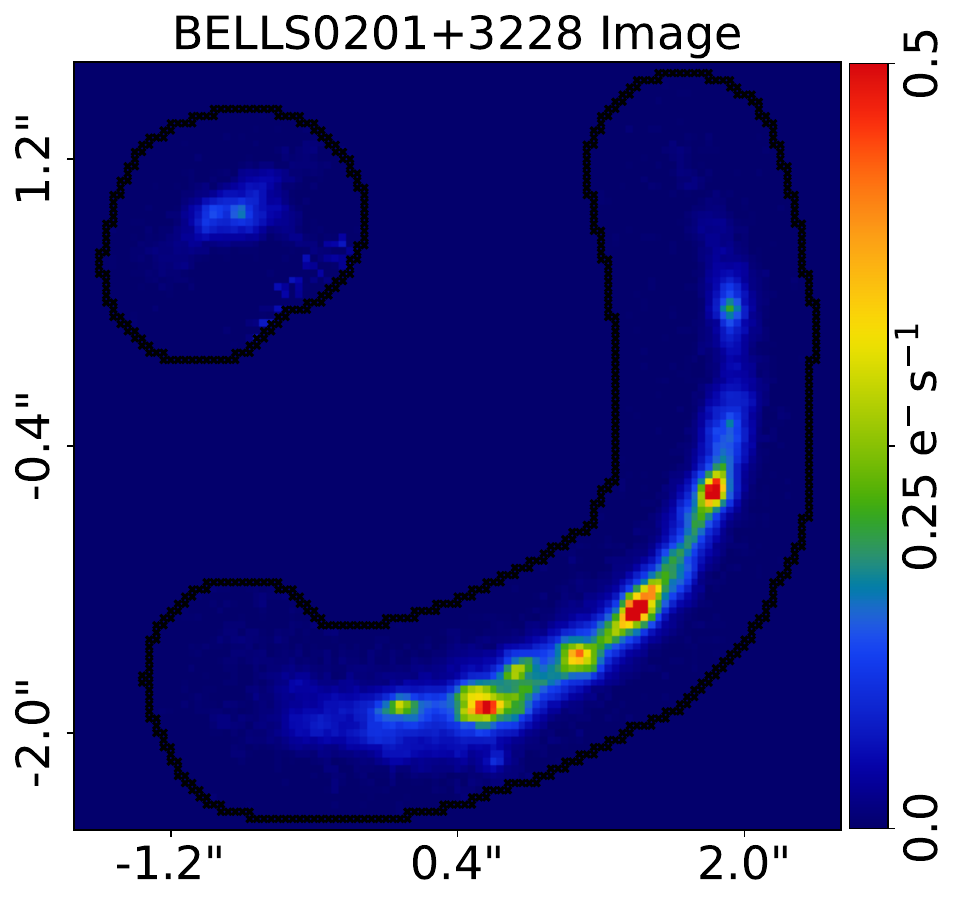}
\includegraphics[width=0.24\textwidth]{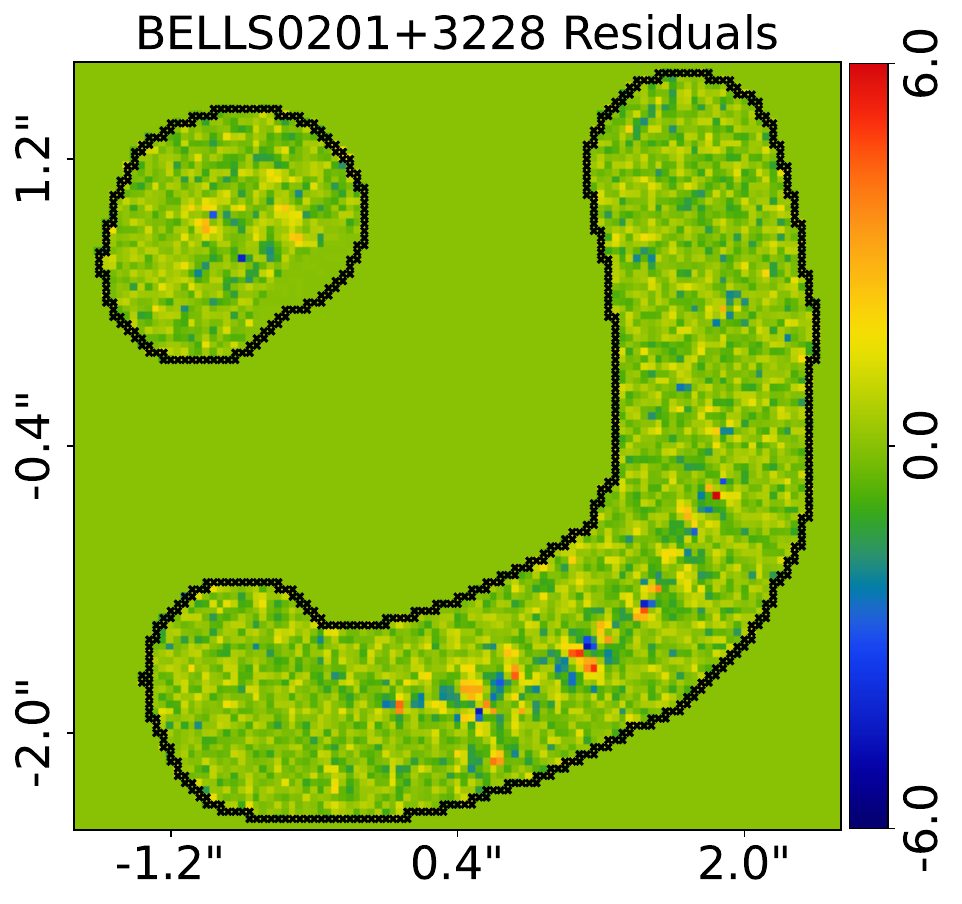}
\caption{
The four lenses which are catastrophic failures, based on their residuals exceeding of $3\sigma$ outliers across many image-pixels in two or more lensed source images. The lensed subtracted images and normalized residuals of each lens are shown (a normalized residual value indicates how much of a $\sigma$ outlier the fit is). 
}
\label{figure:FPFailures}
\end{figure*}

Before considering results for different mass models, we highlight four lenses where no mass model produces a satisfactory fit: BELLS0755+3445, BELLS0918+5104, BELLS0029+2544 and BELLS0201+3228. \cref{figure:FPFailures} shows the four lenses, where the residuals exceed $6\sigma$ in a large fraction of image-pixels containing the lensed source's emission. For the remaining 50 out of 54 lenses in our sample, the residuals of the lensed source are within $\sim 3\sigma$. These four lenses are catastrophic failures -- the significant residuals indicate that none of the lens models fitted in this work can attain a good quality of fit. We assign them to the category X for catastrophic failure and discard them from subsequent sections (noting that a DM subhalo is favoured in three of these lenses). \citet{Ritondale2019a} discuss BELLS0755+3445 as a lens where their fit produced significant residuals.

\subsection{Overall Subhalo Scanning Results}\label{Mass}

Using source-only masks, we compare the DM subhalo inferences after fitting all five different mass models: the PL, BPL, PL with multipoles, decomposed mass model and PL with line-of-sight galaxies. Before comparing how different mass models change the DM subhalo inference, we first consider $\Delta\,\mathrm{ln}\,\mathcal{Z}^{\rm Final}$, which is the the highest $\mathrm{ln}\,\mathcal{Z}$ value inferred assuming any of the five mass models with a DM subhalo minus the highest $\mathrm{ln}\,\mathcal{Z}$ value inferred for any mass model without a DM subhalo. $\Delta\,\mathrm{ln}\,\mathcal{Z}^{\rm Final}$ is given in the second last column of tables \cref{table:DetectSLACS} and \cref{table:DetectBells}. 

There are eight lenses which meet our criteria $\Delta\,\mathrm{ln}\,\mathcal{Z}^{\rm Final} > 10$. However, we assign 3 of these lenses as non-detections, because they have line-of-sight galaxies or post-merger features visible in their residuals, suggesting the model favouring a DM subhalo is likely spurious. These lenses have the category tag "ND / Los" for non-detection due to line-of-sight in the final column of tables \cref{table:DetectSLACS} and \cref{table:DetectBells}. We are therefore left with 5 DM subhalo candidates, which are assigned the category `Cand' for candidate and a total of 45 non-detections, which are assigned the category `ND'.


A small subset of model-fits do not produce a physically plausible lens model, instead inferring the demagnified solutions described by \citet{Maresca2021}. Their $\Delta \ln \mathcal{Z}$ values are omitted from the results and their corresponding results table entries have the entry `Demag'. This occurred for 11 fits in total: 8 out of 54 fits for the BPL, 2 out of 54 fits for the PL with multipoles and 1 out of 54 fits for a decomposed mass model.


\subsection{Subhalo Scanning Using Different Mass Models}\label{Mass}

\begin{figure}
\centering
\includegraphics[width=0.235\textwidth]{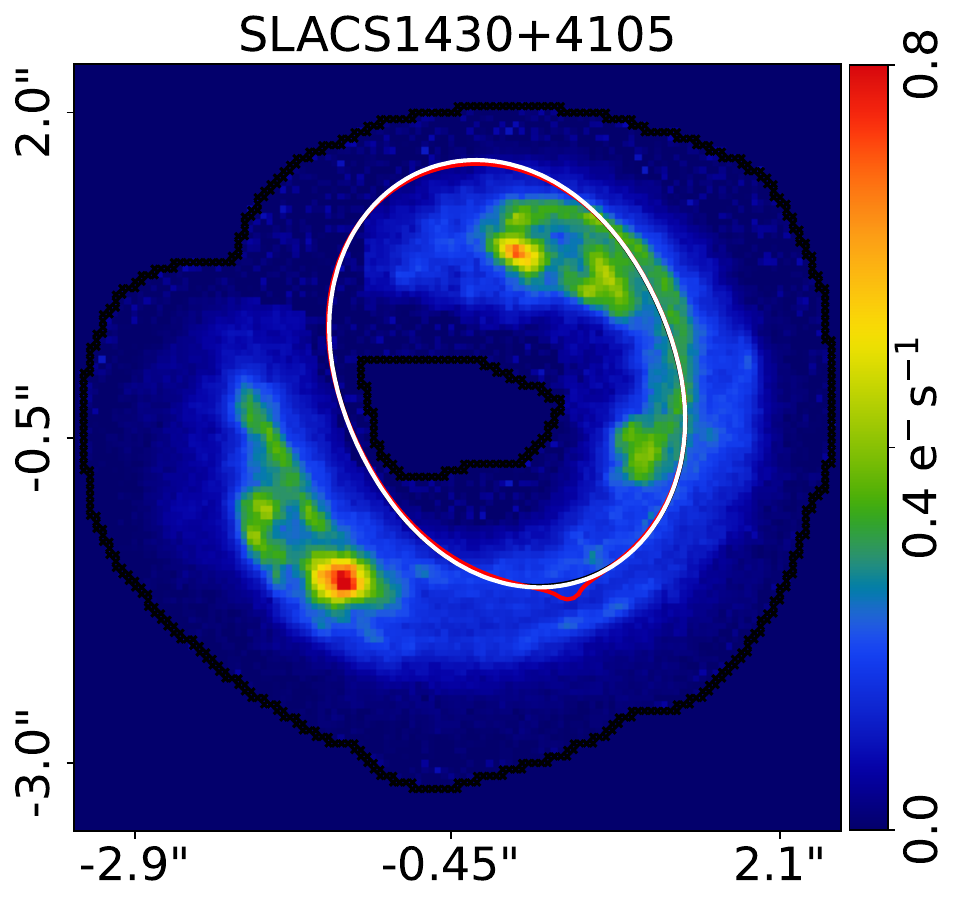}
\includegraphics[width=0.235\textwidth]{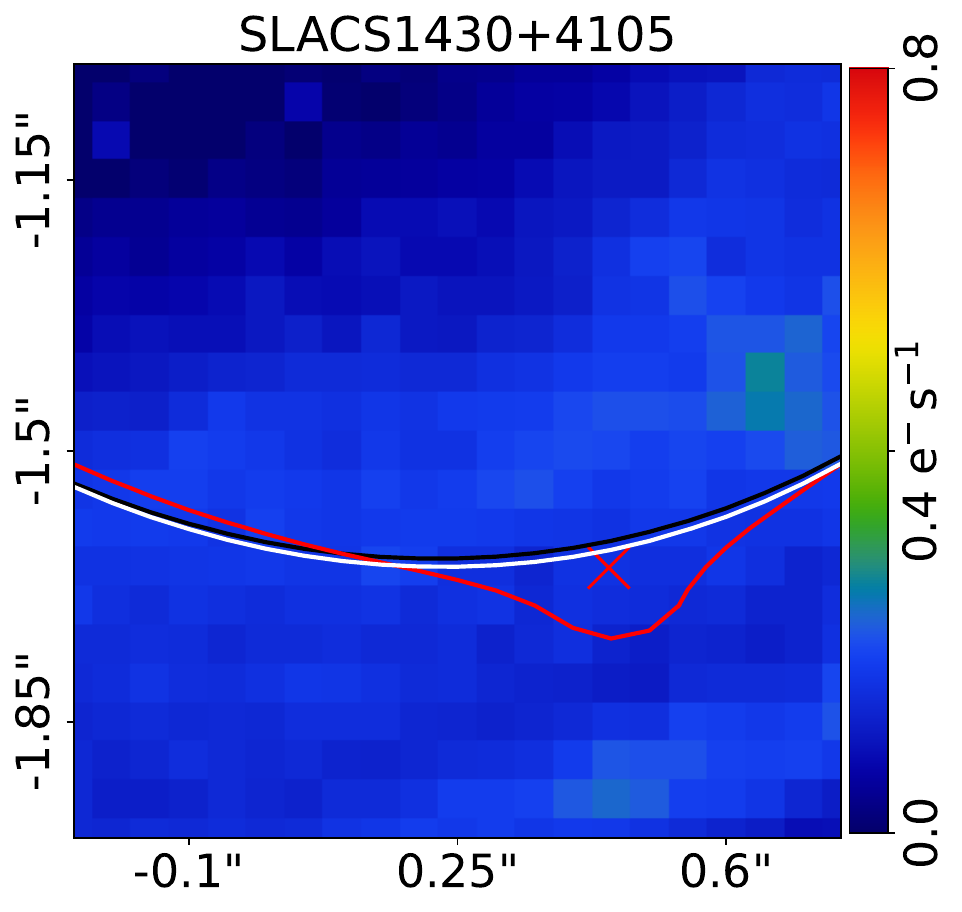}
\caption{
An attempt to visualize how making the lens mass model more complex may remove a DM subhalo candidate, illustrated using the lens SLACS1430+4105. The PL plus shear mass model gives $\Delta\,\mathrm{ln}\,\mathcal{Z}^{\rm Source} = 13.40$, compared to $\Delta\,\mathrm{ln}\,\mathcal{Z}^{\rm Decomp} = 6.54$ for a decomposed mass model which models separately the stellar and dark components. Both panels show the observed image, including the critical curves of the PL plus shear mass model without a DM subhalo (black), with a DM subhalo (red), and the decomposed mass model without a DM subhalo (white). The right panel zooms in on the location where the PL plus shear model favours a DM subhalo (red cross). Including a DM subhalo or fitting a decomposed mass model expands the critical curves outwards in the same direction, albeit the decomposed model expands to a much lesser degree}. 
\label{figure:FPMass2}
\end{figure}

We now consider what impact assuming a different mass model has. The log evidence increase for the BPL, PL with multipoles and decomposed mass models, including a DM subhalo, are denoted $\Delta\,\mathrm{ln}\,\mathcal{Z}^{\rm BPL}$, $\Delta\,\mathrm{ln}\,\mathcal{Z}^{\rm Multipole}$ and $\Delta\,\mathrm{ln}\,\mathcal{Z}^{\rm Decomp}$, respectively. 


There are four lenses where a PL mass model favours a DM subhalo ($\Delta\,\mathrm{ln}\,\mathcal{Z}^{\rm Source} > 10$) but at least one of the more complex mass models does not and the final DM subhalo inference disfavours a DM subhalo ($\Delta\,\mathrm{ln}\,\mathcal{Z}^{\rm Final} < 10$). This occurs in one lens for the BPL mass model ($\Delta\,\mathrm{ln}\,\mathcal{Z}^{\rm Source} > 10$ and $\Delta\,\mathrm{ln}\,\mathcal{Z}^{\rm BPL} < 10$), in two lenses for the PL with internal multipoles ($\Delta\,\mathrm{ln}\,\mathcal{Z}^{\rm Source} > 10$ and $\Delta\,\mathrm{ln}\,\mathcal{Z}^{\rm Multipole} < 10$) and two lenses for the decomposed mass model ($\Delta\,\mathrm{ln}\,\mathcal{Z}^{\rm Source} > 10$ and $\Delta\,\mathrm{ln}\,\mathcal{Z}^{\rm Decomp} < 10$). These values sum to five because this occurs for two different mass models in the same lens.

\begin{table}
\begin{adjustbox}{max width=\textwidth}
\centering
\begin{tabular}{ l | l | l | l | l} 
\multicolumn{1}{p{1.5cm}|}{Lens Name} 
& \multicolumn{1}{p{1.4cm}|}{Broken Power-Law} 
& \multicolumn{1}{p{1.4cm}|}{Power-Law with Multipoles} 
& \multicolumn{1}{p{1.4cm}|}{Decomposed} 
\\ \hline
SLACS1250+0523 & & 31.84 &\\[2pt] 
SLACS0959+0410 & & 44.52 &\\[2pt]
SLACS1430+4105 & 4.56 & &  15.69\\[2pt]
BELLS1110+3649 & & & 11.81 \\[2pt]
\end{tabular}
\end{adjustbox}
\caption{
In four lenses the PL mass model favoured a DM subhalo but at least one of the three more complex mass models (BPL, PL with multipoles, decomposed mass model) did not. This table shows the inferred $\Delta\,\mathrm{ln}\,\mathcal{Z}$ values for the complex mass model minus $\Delta\,\mathrm{ln}\,\mathcal{Z}$ inferred for the simpler PL (both models without a subhalo). The values are inferred in the \textbf{SH}$^1$ stage of the SLaM pipeline. Values of $\Delta\,\mathrm{ln}\,\mathcal{Z} > 10$ indicate that the more complex mass model is favoured over the PL, which occurs for all 4 lenses.
}
\label{table:MassFPPL}
\end{table}

For these four lenses, \cref{table:MassFPPL} shows the Bayesian evidence increase of the more complex mass models compared to the simpler PL, before a DM subhalo is added to both. For all four lenses this value is above $10$, confirming that the more complex mass model fits the lens better. The DM subhalo favoured in these four lenses when assuming a PL mass model were therefore false positive, which fitting a more accurate lens mass model removed. They are labeled FP-PL, for "False Positive Power-Law", in the final column of tables \cref{table:DetectSLACS} and \cref{table:DetectBells}.

We looked for a visual indicator to explain why the more complex mass models removes the DM subhalo detection. \cref{figure:FPMass2} shows an attempt to do this using the lens SLACS1430+4105, where the simpler PL mass model favours a DM subhalo ($\Delta\,\mathrm{ln}\,\mathcal{Z}^{\rm Source} = 13.40$) but the more complex decomposed model does not ($\Delta\,\mathrm{ln}\,\mathcal{Z}^{\rm Decomp} = 6.54$). \cref{figure:FPMass2} shows that the decomposed mass model infers a tangential critical curve (white line) which is slightly extended outwards compared to the PL mass model (black line). When a DM subhalo is included with the PL (red line), the tangential critical curve extends outwards in the same direction as the decomposed model, albeit to a much larger degree. Adding a DM subhalo to the decomposed model has a negligible impact on the tangential critical curve (not shown for visual clarity). Adding a DM subhalo to the simpler PL model and fitting a decomposed mass model (which is favoured by the Bayesian evidence overall) therefore produce stretching of the tangential critical curve in the same direction. They therefore both change the ray-tracing around the location the DM subhalo is detected, possibly explaining why it produces a a false positive for the PL, but it is certainly not conclusive.

There are two lenses where the PL mass model disfavoured a DM subhalo ($\Delta\,\mathrm{ln}\,\mathcal{Z}^{\rm Source} < 10$) but the decomposed mass model favoured one ($\Delta\,\mathrm{ln}\,\mathcal{Z}^{\rm Decomp} > 10$) and a DM subhalo was favoured overall across all mass models ($\Delta\,\mathrm{ln}\,\mathcal{Z}^{\rm Final} > 10$). These lenses are SLACS0029-0055 and SLACS1029+0420 are both assigned as candidate DM subhalos 
and we discuss these lenses in detail in \cref{Discussion}.

There are 8 lenses where the PL model disfavoured a DM subhalo ($\Delta\,\mathrm{ln}\,\mathcal{Z}^{\rm Source} < 10$) but at least one of the more complex mass models favoured one. However, in these eight lenses, a DM subhalo was not favoured overall ($\Delta\,\mathrm{ln}\,\mathcal{Z}^{\rm Final} < 10$). This occurred once for the BPL, five times for the PL with internal multipoles and four times for the decomposed mass model. For these lenses, the PL without a DM subhalo had higher values of $\mathrm{ln}\,\mathcal{Z}$ than the more complex mass models with or without a DM subhalo. These eight lenses highlight that fitting a mass model which is too complex to be justified given the data quality may also produce false positive DM subhalo detections.

\begin{table}
\begin{adjustbox}{max width=\textwidth}
\centering
\begin{tabular}{ l | l | l | l } 
\multicolumn{1}{p{1.5cm}|}{Lens Name} 
& \multicolumn{1}{p{1.4cm}|}{Broken Power-Law} 
& \multicolumn{1}{p{1.4cm}|}{Power-Law with Multipoles} 
& \multicolumn{1}{p{1.4cm}}{Decomposed} 
\\ \hline
 & & & \\[-2pt]
SLACS0946+1006 &  $8.62^{+2.15}_{-2.54}$   & $10.37^{+0.80}_{-0.51}$ & $10.37^{+0.80}_{-0.51}$ \\[2pt]
SLACS0029-0055 & $10.03^{+0.03}_{-0.03}$  &  & $10.03^{+0.03}_{-0.03}$ \\[2pt]
SLACS1029+0420 & &  & $9.05^{+2.57}_{-3.04}$ \\[2pt]
BELLS1226+5457 & $11.67^{+0.18}_{-0.37}$ & $11.48^{+0.16}_{-0.35}$ & $10.96^{+0.27}_{-0.15}$ \\[2pt]
BELLS1201+4743 & $11.23^{+0.12}_{-0.20}$ & & \\[2pt]

\end{tabular}
\end{adjustbox}
\caption{
The inferred DM subhalo masses, $\log_{10}[M^{\rm sub}_{\rm 200}$\,/\,M$_{\rm \odot}]$, for the five lenses in the "Cand" category. Masses are shown for the broken power-law, power-law with multipoles and decomposed mass models, which all include an external shear. Masses are shown for the models where a DM subhalo is favoured above our criteria of $\Delta\,\mathrm{ln}\,\mathcal{Z} > 10$. Errors quoted on $\log_{10}[M^{\rm sub}_{\rm 200}$\,/\,M$_{\rm \odot}]$ are at $3\sigma$ confidence.
}
\label{table:CandidateSLACS}
\end{table}

The inferred DM subhalo masses for the candidate strong lenses are given in \cref{table:CandidateSLACS}.

\subsection{Line-of-Sight Galaxies}\label{ResultLos}

\begin{figure}
\centering
\includegraphics[width=0.15\textwidth]{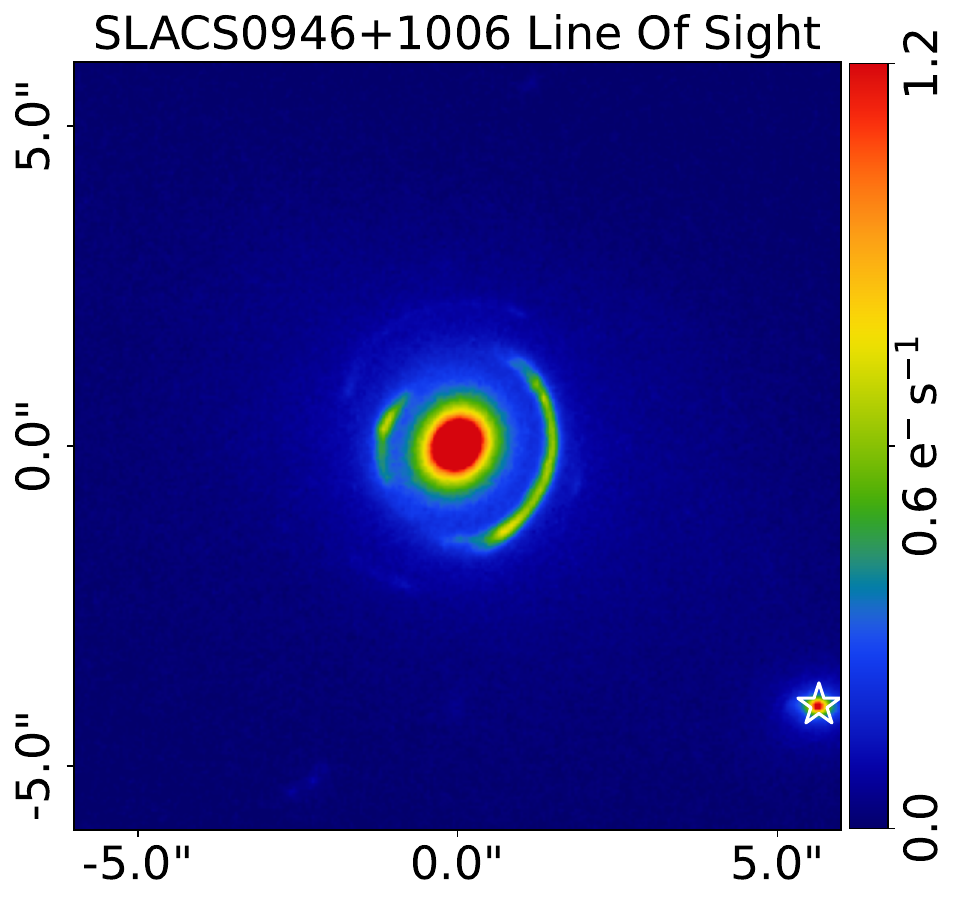}
\includegraphics[width=0.15\textwidth]{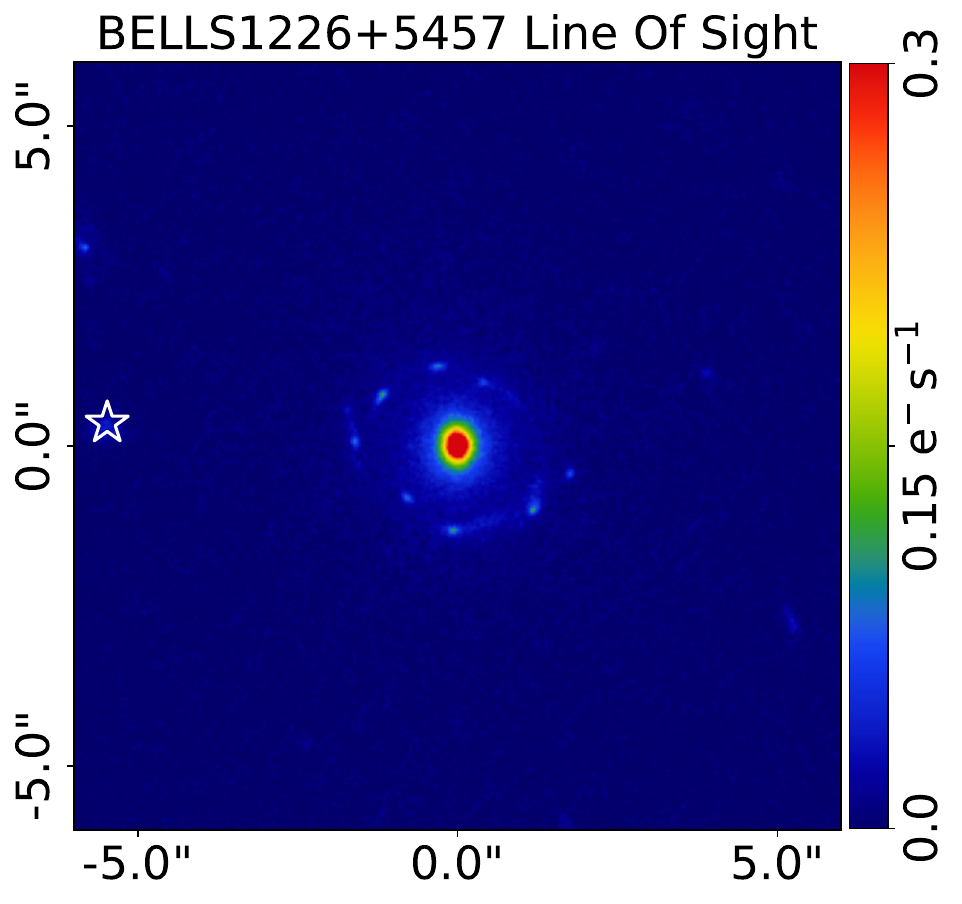}
\includegraphics[width=0.15\textwidth]{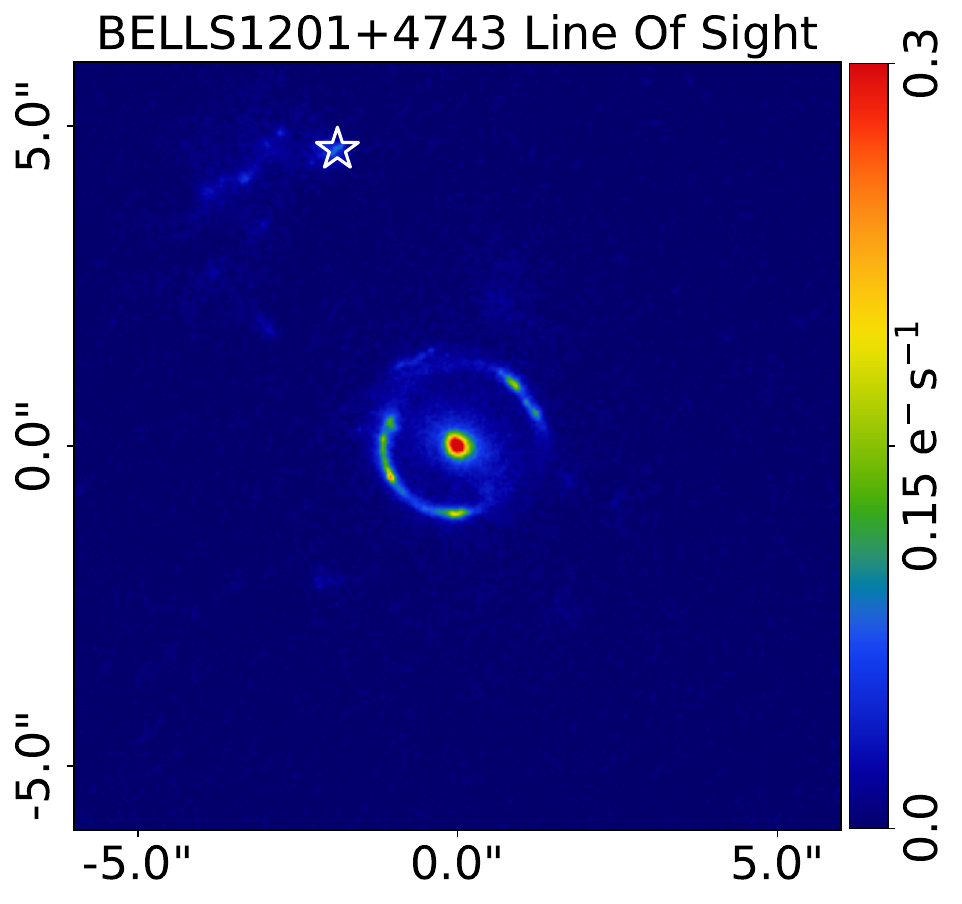}
\caption{
Three lenses where adding a line-of-sight galaxy to the lens model leads to a change of $|\Delta\,\mathrm{ln}\,\mathcal{Z}^{\rm Los} - \Delta\,\mathrm{ln}\,\mathcal{Z}^{\rm Source}| > 10$. These lenses were judged via visual inspection to not have a line-of-sight galaxy close enough to the lensed source that they should have a noticeable impact on the lens model. Catastrophic failures (see \cref{Catastrophic}) are not included. Black stars mark the centre of each line-of-sight galaxy, which are modeled with SIS mass profiles fixed to those centers. 
}
\label{figure:FPClump1}
\end{figure}

We now consider whether including line-of-sight galaxies in the lens model changes the DM subhalo inference, by inspecting results for the PL plus shear mass model with line-of-sight galaxies included (see \cref{ModelLos}). We first isolate all lenses where the PL with line-of-sight galaxies was favoured over all other mass models, by finding those where their $\mathrm{ln}\,\mathcal{Z}$ value is above $10$ any of the other four mass models. There are three lenses where this is the case: BELLS0113+0250, BELLS2342-0120 and BELLS1226+5457. 

BELLS0113+0250 and BELLS2342-0120 are two of the lenses shown in \cref{figure:FPClump0}, which were judged to have nearby line-of-sight galaxies just outside the lensed source. For both lenses, including line-of-sight galaxies notably impacts the DM subhalo inference. For BELLS2342-0120, the highest evidence mass model not including line-of-sight galaxies is the BPL. It gives $\mathrm{ln}\,\mathcal{Z} = 4588.00$ without a DM subhalo and $\mathrm{ln}\,\mathcal{Z} = 4601.19$ with a DM subhalo, meaning we would favour a DM subhalo. However, the PL with line-of-sight galaxies not including a DM subhalo has an even higher evidence ($\mathrm{ln}\,\mathcal{Z} = 4601.68$), meaning that we ultimately disfavour a DM subhalo. For BELLS0113+0250, ignoring the model with line-of-galaxies means we would infer $\Delta \mathrm{ln}\,\mathcal{Z} = 8.55$, which reduces to $\Delta \mathrm{ln}\,\mathcal{Z} = -0.17$ when considering the model with line-of-sight galaxies.


There are nine lenses where $\Delta\,\mathrm{ln}\,\mathcal{Z}^{\rm Los} > 10$, meaning that the PL with line-of-sight galaxies favours a DM subhalo. Four of these lenses are candidate subhalos. Two lenses belong to the FP-PL category, meaning the model favouring a DM subhalo is likely a false positive due to the PL being too simple. The remaining three lenses are examples where fitting a mass model which is too complex to be justified given the data quality can give a spurious DM detection. There are 42 lenses remaining where the inclusion of line-of-sight galaxies had no impact on the DM subhalo inference. This includes three lenses in \cref{figure:FPClump0}, which were judged to have nearby line-of-sight galaxies just outside the lensed source.

\section{Discussion}\label{Discussion}

\subsection{Expected Detections}\label{DMExpect}

We now estimate upper limits on the number of expected DM subhalo detections for a $\Lambda$CDM Universe, via the sensitivity analysis performed in \citet{Amorisco2022} and \citet{He2022b}. Using \texttt{PyAutoLens}, these works simulated realizations of strong lens images which included subhalos, with varying image-plane positions and masses, and quantified how detectable they are. Both works assumed parametric (cored) S\'ersic sources, whereas this work uses Voronoi mesh source reconstructions. Using parametric sources makes subhalos more detectable, therefore the expectations provided by these sensitivity maps are upper limits and this work should detect fewer DM subhalos in a CDM universe. We quote values from their work where the threshold for a detection is consistent with ours, a log Bayesian evidence difference of 10.

We first consider what is the lowest detectable DM subhalo mass our fits are sensitive to. \citet{Amorisco2022} find that for HST-like data at a lensed source S/N of $\sim 50$ we are sensitive to DM subhalos of at least $10^{9.0}$\,\,M$_{\rm \odot}$. Extrapolations of forecasts in \citet{He2022b} indicate that DM subhalos of masses $10^{9.0}$\,\,M$_{\rm \odot}$ are detectable. These are consistent with sensitivity mapping performed by \citet{Despali2022} using a different lens modelling code (for a threshold $\Delta\,\mathrm{ln}\,\mathcal{Z} = 10.0$). The lowest detectable mass depends critically on the source S/N, and for many lenses our source S/N is below $50$, meaning their lowest detectable mass will be above $10^{9.0}$\,\,M$_{\rm \odot}$. However, the majority of detections listed in \cref{table:DetectSLACS} and \cref{table:DetectBells} are above masses of $10^{10.0}$\,\,M$_{\rm \odot}$. The masses of the candidate DM subhalos are therefore feasible for our HST data. 


We now consider upper limits on the expected number of detections for subhalos between masses of $10^{10.0}$\,\,M$_{\rm \odot}$ and $10^{11.0}$\,\,M$_{\rm \odot}$ in a CDM universe. At higher masses, their reduced number counts means that the random chance of alignment drives the probability of detection, as opposed to data quality. For a sample with lens redshift $z_{\rm lens} = 0.2$  and source redshift $z_{\rm src} = 0.6$, \citet{Amorisco2022} predicts that there should be 0.025 detections per lens for subhalos in the mass range $10^{10-11}$\,\,M$_{\rm \odot}$ in the CDM case. For higher lens and source redshifts, the expected number of detections rises up to 0.1. They do not provide forecasts for masses above $10^{11.0}$\,\,M$_{\rm \odot}$, but the rarity of these objects means they would not significantly change expectations. 

For the simple lens model fitted in \cref{FitBase}, which do not address false positives due to the lens light and source resolution, if we consider every candidate detection with an inferred mass above $10^{10}$\,\,M$_{\rm \odot}$ this gives a rate of 13 out of 37 or $0.351$ detections per lens for SLACS and 7 out of 16 or $0.4375$ detections per lens for BELLS-GALLERY. Systematic associated with the double Sersic lens light model, low resolution source and PL plus shear mass model therefore led us to detect many more higher mass DM subhalos than expected in CDM. After improving the lens and source models we were left with five DM subhalo candidates, a number which does not exceed CDM expectations.

\subsection{Are any DM Subhalo Candidates Genuine?}\label{Genuine}

A key result of this paper is that changing the lens galaxy mass model changes the DM subhalo inference. For example, we identified 4 out of 54 lenses where a PL mass model produces a false positive removed by a more complex mass model (category `FP-PL') and two lenses where a decomposed mass model favours a DM subhalo when other models did not (category `Decomp'). Only 2 out of 54 lenses favoured a DM subhalo for all five mass models fitted. We cannot ascertain whether any DM subhalo candidate is genuine -- even for these two lenses, we cannot be certain whether another hypothetical mass model not fitted in this work would remove the DM subhalo detection.

To determine if they are genuine we must apply the technique used in other studies \citep{Koopmans2005, Vegetti2009, Ritondale2019a, Vernardos2022}, where free-form pixelized corrections are added to the lens's gravitational potential. This confirms a DM subhalo candidate is genuine by requiring that these corrections reconstruct a \textit{local} 2D overdensity in the lens's convergence, that is consistent with the parametric DM subhalo inferred via lens modeling. In many lenses, the corrections produce \textit{global} changes to the convergence, indicating that a DM subhalo candidate is actually accounting for a systematic in the lens model. Future work will assess our DM subhalo candidates using this technique.

Whilst our study cannot determine if any DM subhalo candidates are genuine, by considering the DM subhalo inferences for different mass models we can gain insight on DM subhalo strong lens analysis. We therefore now consider in detail the different assumptions made by the different mass models and relate this to how it changes our DM subhalo results.

\subsection{Removing DM Subhalo Candidates with a More Complex Mass Model}\label{Remove}

We showed evidence of four lenses (see \cref{table:MassFPPL}) where fitting a more complex mass model (either a PL with multipoles or decomposed mass model) did not favour a DM subhalo when the simpler PL model did. In all four lenses, the inferred Bayesian evidence for the more complex mass model was above that of the PL (both without a DM subhalo) by over $10$. These four lenses make up the category FP-PL for `False Positive Power-Law'.

The cause of this behaviour is illustrated in \citet[][hereafter H23]{He2023} using HST-like strong lens images simulated via a high resolution zoom-in cosmological simulation \citep{Richings2021} of a massive elliptical galaxy. H23 showed that a mismatch between the assumed lens mass model and the simulated lens galaxy's more complex underlying mass distribution could create a signal that resembles the perturbing effect of a DM subhalo. In certain simulated datasets, where a DM subhalo was not truly present, the lens model favoured a DM subhalo with an increase of log evidence of up to $30$. The PL favoured a DM subhalo for all five lenses in the FP-PL category with $\Delta\,\mathrm{ln}\,\mathcal{Z}^{\rm Source} \simeq 30$ or less.

Two of these lenses, SLACS1250-0523 and SLACS1430+4105, were in the sample of three lenses studied by \citet{Nightingale2019}. The authors showed that the stellar mass distribution of both lenses are composed of two elliptical components with unique axis ratios and position angles. When the authors fitted an SIE mass model to SLACS1430+4105 its inferred position angle went to a value between those inferred for each S\'ersic. They argued that the SIE model therefore adjusted its orientation to try and capture the lens's true complexity, which is captured by the decomposed mass model. Their study supports the argument that these two lenses have the type of complex features in their mass distribution which H23 showed cause false positive DM subhalo detections. Work by \citet{Vegetti2014} also did not favour a DM subhalo in SLACS1430+4105, supporting the false positive interpretation.

\textbf{Fitting more complex mass models can therefore remove false positives by adding complexity that is present in the lens's true mass distribution. Therefore, in 4 out of 54 lenses, or $7.4\%$} of our sample, the PL mass model produces false positive DM subhalo detections. Amongst these four lenses, the BPL removes one false positive, the PL plus multipoles removes two and the decomposed mass model removes two.

\subsection{Creating DM Subhalo Candidates with a Decomposed Mass Model}\label{Create}

In \cref{ResultSim} we showed that the PL lens mass model `absorbed' genuine DM subhalo signals by adjusting the inferred mass model parameters away from their true input values. For the decomposed mass model, the centres, axis ratios and position angles of the two S\'ersic components representing the decomposed model's stellar mass are tied to that of the lens galaxy's light (each S\'ersic component has mass-to-light ratio and gradient parameters that are free to vary). The restrictions this puts on the lens's two dimensional stellar mass distribution therefore may reduce this subhalo absorption effect and make DM subhalos not detected with the PL model detectable. 

This is a plausible interpretation of the results for the lenses SLACS0029-0055 and SLACS1029+0420, where a PL mass model did not favour a DM subhalo but the decomposed mass model did, with values of $\Delta \mathrm{ln}\,\mathcal{Z}^{Final} = 23.69$ and $\Delta \mathrm{ln}\,\mathcal{Z}^{Final} = 10.57$ respectively. However, we cannot be certain that the linking of light to mass in the decomposed mass model is a robust assumption. The decomposed mass model could be creating a false positive due to some form of missing complexity, in a similar fashion seen for the four lenses discussed above. \textbf{Fitting a decomposed mass model, which better captures the lens's true mass distribution, may therefore make DM subhalos detectable which are not detectable when fitting other lens mass models}. Future work will test this hypothesis by applying the potential corrections described in \cref{Genuine} to these two lenses.

\subsection{What Mass Model Complexity Is Missing?}\label{Missing}

The FP-PL category consists of four lenses where a DM subhalo was favoured for the PL model and disfavoured for the PL with multipoles or decomposed mass model, and the latter had a higher overall evidence. When comparing to the BPL instead of the PL, the results do not change for any of these lenses. The multipoles and decomposed mass models are therefore adding a form of complexity, not present in the PL or BPL, which removes DM subhalo candidates. These models add complexity to the mass distribution's angular structure, e.g., allowing azimuthal variations in the projected density that vary with radius. In contrast, the BPL only adds freedom radially. This is evidence that it is missing complexity in the angular structure of lens mass models which creates false positives DM subhalo candidates. This is consistent with H23 and is discussed by \citet{Kochanek2021} in the context of measuring the Hubble Constant with strong lenses. 

Of these four lenses, there are two where the PL with multipoles changed the DM inference and two where it was the decomposed mass model. The PL with multipoles and decomposed mass models therefore do not always give consistent DM subhalo results, because they add angular structure to the mass distribution in different ways. The fourth order multipole fitted in this work adds boxiness / diskiness to the mass distribution \citep{VanDeVyvere2022a}, whereas the decomposed mass model allows for mass twists and departures from a single axis-ratio \cite{Nightingale2019}.

Studies of local massive elliptical galaxies have revealed a diversity of complex structures, including kinematically distinct cores \citep{Krajnovic2011}, boxy / disky isophotes \citep{Emsellem2011}, isophotal twists and centre shifts \citep{Goullaud2018}. Recent works have investigated what impact these have on lens models \citep{Cao2021, VanDeVyvere2022a, VanDeVyvere2022b, Etherington2023}. Edge-on disks have also been shown to cause false positive DM signals  \citep{Hsueh2016, Hsueh2017, Hsueh2018}). These forms of complexity add smoothly changing radial and azimuthal features to the mass distribution, which the BPL, PL with multipoles and decomposed models add in different ways. 

Our results motivate the development of more complex mass models that add azimuthal freedom (and to a lesser degree, radial freedom) in a way that captures the true complexity of all lens galaxies. However, it is unclear how. Evaluating a mass model's deflection angles typically relies on it conforming to elliptical symmetry (e.g. that all iso-convergence contours correspond to a single position angle and axis ratio). Even if we are able to determine what complexity is missing from the mass model, it remains to be seen whether one can practically fit it as a parameterized lens mass model. Future work will build-on the results of this study in order to better understand what mass model complexity is missing. 

\subsection{Potential Corrections}

The potential corrections technique \citep{Koopmans2005, Vegetti2009, Suyu2010, Vernardos2022} can provide key insight on the missing mass model complexity \citep{Powell2022}. Performing the potential corrections analysis (see \cref{Genuine}) for different mass models and comparing the results will facilitate progress, because the complexity included and omitted should be reflected in the potential corrections themselves. 

This raises an important question, how well do the potential corrections perform in a regime where a DM subhalo is present, but there is also missing complexity in the lens mass model?  In this scenario, the lensing signal produced by a DM subhalo will be superimposed with the signal produced by missing complexity in the lens mass model. Would the potential corrections reproduce the local DM subhalo signal and simultaneously correct the mass model on a global scale, or would a degenerate solution be inferred such that the DM subhalo is rejected? This scenario is considered by \citet{Galan2022} who use wavelets to perform a multi-scale potential correction on simulated lenses. Their analysis indicates the signals are separable because they operate on different physical scales.

\subsection{What about Line of Sight Galaxies?}

There are two lenses where including line-of-sight galaxies had a meaningful impact on the DM subhalo inference, both of which had bright galaxies within $\sim 1.0"$ of the lensed source. There are three more lenses which had bright galaxies this close, but their inclusion did not change the DM subhalo inference. For the remaining 49 lenses, models including line-of-sight galaxies were fitted, but these objects were typically $\sim 3.0"$ or more from the lens and relatively faint. Provided line-of-sight galaxies are sufficiently far from the lens they therefore do not impact the DM subhalo inference, at least for HST quality data. Future work could quantify this more precisely, by estimating the masses of the line-of-sight galaxies from their luminous emission.

\subsection{Subhalo Masses}

H23 show that an overly simplistic mass model can lead to overestimates of $M^{\rm sub}_{\rm 200}$ by a factor of $\sim 4$. Given the uncertainty surrounding whether our DM subhalo candidates are genuine, interpreting their inferred masses, which are given in \cref{table:CandidateSLACS}, is difficult. We therefore focus on SLACS0946+1006, a confirmed DM subhalo \citep{Vegetti2010}, which passed our detection criteria for all mass models (category A). Our $3\sigma$ confidence intervals for $M_{\rm 200}^{\rm sub}$ -- with each of the different lens mass models (which all include an external shear) -- are: 

\begin{itemize}
    \item PL: $M^{\rm sub}_{\rm 200} = 1.00^{+0.32}_{-0.25} \times 10^{11}$M$_{\rm \odot}$,
    \item BPL: $M^{\rm sub}_{\rm 200} = 3.42^{+2.61}_{-1.93} \times 10^{11}$M$_{\rm \odot}$
    \item PL with multipoles: $M^{\rm sub}_{\rm 200} = 2.51^{+1.71}_{-1.1} \times 10^{11}$M$_{\rm \odot}$
    \item Decomposed mass model: $M^{\rm sub}_{\rm 200} = 0.85^{+0.87}_{-0.59} \times 10^{11}$M$_{\rm \odot}$
    \item PL with line-of-sight galaxies: $M^{\rm sub}_{\rm 200} = 2.53^{+3.72}_{-1.77} \times 10^{11}$M$_{\rm \odot}$.
\end{itemize}

The $M^{\rm sub}_{\rm 200}$ estimates therefore vary depending on the mass model, with the BPL value inconsistent with the PL. We anticipate that attempts to constrain more subtle DM properties like the subhalo's concentration will be more impacted by this degeneracy with the lens mass model \citep{Minor2021}. Understanding the missing complexity in strong lens mass models is important for ensuring that DM subhalo mass measurements are accurate. 
 
Even if our mass models were perfect, the mass estimates quoted in this work for any genuine DM subhalo have additional potential systematics. Our DM subhalo model assumes they lie on the mass-concentration relation from \citet{Ludlow2016} and we will overestimate the mass of any genuine DM subhalo which is more concentrated than this relation (because more concentrated NFW halos have a higher central density, making their perturbations to the lensing more prominent, see \citealt{Amorisco2022}). This is also shown by \citet{Minor2021, Minor2021a}. DM subhalos may also be at a different redshift to the lens, which can also lead to an incorrect mass estimate \citep{Li2016a, Despali2018, He2022b, Amorisco2022, Despali2022}.

\subsection{Improving Other Aspects of Lens Models}

The evidence favouring a DM subhalo decreased by more than $10$ when: (i) residuals from an inadequate lens light subtraction were removed in 12 out of 54 lenses; (ii) the source reconstruction resolution was increased in 7 out of 54 lenses. We identified this by refitting each lens with simple changes to the imaging data and masks, similar to those used by other studies \citep{Vegetti2014, Ritondale2019a}. Improving {\tt PyAutoLens} to mitigate these systematics is also straight forward, for example using more flexible lens light models (e.g. basis functions \citealt{Tagore2016}) and optimizing the code to reconstruct sources at higher resolution.

\subsection{Comparison with Other Works}

We now compare to studies by \citet{Vegetti2010}, \citet{Vegetti2014} and \citet{Ritondale2019a} who search for subhalos in the SLACS and BELLS-GALLERY samples. These works use lens light subtracted images, source-only masks and a PL mass model, thus our $\Delta\,\mathrm{ln}\,\mathcal{Z}^{\rm Source}$ values are the most suitable to compare. In certain lenses these studies include line-of-sight galaxies, meaning that we compare $\Delta\,\mathrm{ln}\,\mathcal{Z}^{\rm Los}$ values. 


\citet{Vegetti2010} present the detection of a DM subhalo in the lens system SLACS0946+1006, for which we infer $\Delta\,\mathrm{ln}\,\mathcal{Z}^{\rm Source} = 72.36$ and assign it as a DM subhalo candidate. Our inferred values of $(x,y) = (-1.22, -1.28)$ are consistent with the values presented in \citet{Vegetti2010}. Comparing subhalo mass is less straightforward, because \citet{Vegetti2010} assume a psuedo-Jaffe density profile whereas we assume an NFW. The pseudo-Jaffe parameterization is more centrally dense than the NFW, such that a factor of $\sim 10$ difference is expected between their inferred masses \citep{Vegetti2018}. The mass of $M^{\rm sub}_{\rm 200} = 1.00^{+0.32}_{-0.25} \times 10^{11}$M$_{\rm \odot}$ for our NFW subhalo model is therefore qualitatively above what one would have predicted by converting their pseudo-Jaffe inferred value of $3.51 \times 10^{9}$M$_{\rm \odot}$ to an NFW. Our results therefore agree with \citet{Vegetti2010}.

\citet{Vegetti2014} analyse the following $11$ SLACS lenses: SLACS0252+0039, SLACS0737+3216, SLACS0956+5100, SLACS0959+4416, SLACS1023+4230, SLACS1205+4910, SLACS1430+4105, SLACS1627-0053, SLACS2238-0754, SLACS2300+0022. They report no DM substructure detection for every system. All of these lenses are in our SLACS sample except SLACSJ0959+4416, which we removed due to a poor lens light subtraction. Our highest $\Delta\,\mathrm{ln}\,\mathcal{Z}^{\rm Source}$ value for a lens in common with this sample (omitting SLACS0946+1006) is SLACS0956+5100 with a value of $\Delta\,\mathrm{ln}\,\mathcal{Z}^{\rm Source} = 23.35$. We infer $\Delta\,\mathrm{ln}\,\mathcal{Z}^{\rm Source} > 10$ for one more shared lens, SLACS1430+4105. To claim a DM subhalo detection, \citet{Vegetti2014} require that the Bayesian evidence increases by 50. Therefore, for all $10$ overlapping lenses we are in agreement.

\citet{Ritondale2019a} analyse 17 lenses from the BELLS-GALLERY sample, of which 16 are shared with our sample (we removed a system with two lens galaxies). In 3 lenses they find that the addition of a subhalo in the lens model increases the Bayesian evidence by more than $100$; BELLS0742+3341, BELLS0755+3445 and BELLS1110+3649. For these three lenses we infer $\Delta\,\mathrm{ln}\,\mathcal{Z}^{\rm Source}$ values of $-3.59$, $1268.78$ and $12.65$ respectively. We attribute BELLS0755+3445 as a catastrophic failure and \citet{Ritondale2019a} specifically discuss this as a lens with an inaccurate mass model that causes a spurious DM subhalo inference. We find $\Delta\,\mathrm{ln}\,\mathcal{Z}^{\rm Source} > 100.0$ in three more lenses which we class as catastrophic failures, BELLS0918+5104, BELLS0029+2544 and BELLS0201+32284, which are not mentioned specifically by \citet{Ritondale2019a}. In the lens BELLS1226+5457 we infer $\Delta\,\mathrm{ln}\,\mathcal{Z}^{\rm Source} = 105.90$, which is reported below 100 in \citet{Ritondale2019a}. 

There are differences between our results and those of \citet{Ritondale2019a}. Assessing the cause for discrepancy is difficult. BELLS-GALLERY source galaxies are compact Lyman-alpha emitters \citep{Shu2016, Ritondale2019} which for fits to simulated lenses with similar source properties highlighted the need for higher source resolution (see \cref{A_Pre_Subhalo}). Therefore the differences are likely due to how each work approaches the source analysis. Although {\tt PyAutoLens} and the method of \citet{Ritondale2019a} are similar, there are differences in their implementation and the regularization schemes that are applied. More detailed study is warranted, especially in light of the systematics highlighted by \citet{Etherington2022} where stochasticity in the construction of the source can produce large spikes in the log likelihood. 

\section{Summary}\label{Summary}

In this work, we scan for DM subhalos in 54 strong lenses imaged by the Hubble Space Telescope (HST): twice as many as have been previously attempted \citep{Vegetti2014, Ritondale2019a}. To achieve this, we successfully developed a predominantly automated data processing pipeline, based on open-source lens modeling software \texttt{PyAutoLens}. By comparing lens models with and without DM subhalos, we infer the probability that each lens contains a DM substructure. Tested on idealized mock HST images of 16 lenses, our method correctly identifies DM substructures of mass $>10^{9.5}$M$_{\rm \odot}$ \citep[the expected sensitivity of HST][]{Amorisco2022, He2022b, Despali2022} without false positives, provided that the source galaxy reconstruction has sufficiently high resolution. 

We identify five DM subhalo candidates, including one previously identified in the lens SLACS0946+1006 \citep{Vegetti2010}. For two candidates fits using simpler models for the lens's mass did not favour a DM subhalo, but more complex mass models which use separate components for the stars and dark matter do. Future work will extend these fits using a pixel-grid based technique for the lens's gravitational potential \citep{Koopmans2005, Vegetti2009}, in order to definitively determine whether any of these candidates are genuine DM subhalo detections. We identify a total of $45$ non-detections, which are vital for overcoming Poisson statistics when constraining DM models \citep{Despali2022}.

We demonstrate that changing the complexity of the lens galaxy's mass model has a dramatic impact on the DM subhalo inference. Because our software is highly automated, we have been able to fit five different parametric forms for the lens's mass which are used in the literature: (i) power-law \citep{Tessore2015}; (ii) broken power-law \citep{Oriordan2019, Oriordan2020, Oriordan2021}; (iii) power-law including internal multipoles \citep{Chu2013}; (iv) decomposition of the lens into stellar and dark components \citep{Nightingale2019} and; (v) a power-law where the mass of nearby line-of-sight galaxies is also accounted for. An external shear term is included in all models. 

We demonstrate that fits assuming a more complex model for the lens's mass distribution may: (i) favour the inclusion of a DM subhalo when fits assuming a simpler lens mass model do not (2 out of 54 lenses) and; (ii) remove false positive DM subhalo detections found when assuming a simpler lens mass model (6 out of 54 lenses). The inferred DM subhalo masses also depend on the mass model that we assume. 

We believe that the main form of complexity missing in our lens mass models was in their azimuthal structure, and that effort must be placed on developing lens models that add this. If done correctly, the pay-off could be huge -- \textbf{enabling studies that are more sensitive to DM subhalos of lower masses than previously forecasted and which are devoid of false positive detections}.

The importance of automating strong lensing analysis will increase in future surveys. Several hundred lenses will be required for competitive constraints on DM physics \citep{Vegetti2018, Amorisco2022, He2022b, Despali2022}; and thousands (or tens of thousands) of lenses will soon be discovered by the James Webb Space Telescope, Euclid, and Roman Space Telescopes \citep{Collett2015}. These exquisite datasets and large lens samples will allow us to test $\Lambda$CDM on smaller scales of the Universe than ever before.

\section*{Software Citations}

This work uses the following software packages:

\begin{itemize}

\item
\href{https://github.com/rhayes777/PyAutoFit}{{PyAutoFit}}
\citep{pyautofit}

\item
\href{https://github.com/Jammy2211/PyAutoGalaxy}{{PyAutoGalaxy}}
\citep{pyautogalaxy}

\item
\href{https://github.com/Jammy2211/PyAutoLens}{{PyAutoLens}}
\citep{Nightingale2015, Nightingale2018, Nightingale2021}

\item
\href{https://github.com/astropy/astropy}{{Astropy}}
\citep{astropy1, astropy2}

\item
\href{https://bitbucket.org/bdiemer/colossus/src/master/}{{Colossus}}
\citep{colossus}

\item
\href{https://github.com/dfm/corner.py}{{Corner.py}}
\citep{corner}

\item
\href{https://github.com/joshspeagle/dynesty}{{Dynesty}}
\citep{Speagle2020}

\item
\href{https://github.com/matplotlib/matplotlib}{{Matplotlib}}
\citep{matplotlib}

\item
\href{numba` https://github.com/numba/numba}{{Numba}}
\citep{numba}

\item
\href{https://github.com/numpy/numpy}{{NumPy}}
\citep{numpy}

\item
\href{https://www.python.org/}{{Python}}
\citep{python}

\item
\href{https://github.com/scikit-image/scikit-image}{{Scikit-image}}
\citep{scikit-image}

\item
\href{https://github.com/scikit-learn/scikit-learn}{{Scikit-learn}}
\citep{scikit-learn}

\item
\href{https://github.com/scipy/scipy}{{Scipy}}
\citep{scipy}

\item
\href{https://www.sqlite.org/index.html}{{SQLite}}
\citep{sqlite}

\end{itemize}

\section*{Data Availability}

Analysis results are publically available at \url{https://github.com/Jammy2211/autolens_subhalo}.

\section*{Acknowledgements}

JWN is supported by the UK Space Agency, through grant ST/N001494/1, and a Royal Society Short Industry Fellowship.
AE is supported by STFC via grants ST/R504725/1 and ST/T506047/1. AA, SC, CSF and QH acknowledge support from the European Research Council (ERC) through Advanced Investigator grant DMIDAS (GA 786910). XYC and RL acknowledge the support of the National Nature Science Foundation of China (Nos 11988101,11773032,12022306), the support from the Ministry of Science and Technology of China (grant Nos. 2020SKA0110100), the science research grants from the China Manned Space Project (Nos. CMS-CSST-2021-B01, CMS-CSST-2021-A01), CAS Project for Young Scientists in Basic Research(No. YSBR-062), and the support from K.C.Wong Education Foundation. 

This work used both the Cambridge Service for Data Driven Discovery (CSD3) and the DiRAC Data-Centric system, project code dp195, which are operated by the University of Cambridge and Durham University on behalf of the STFC DiRAC HPC Facility (www.dirac.ac.uk). These were funded by BIS capital grant ST/K00042X/1, STFC capital grants ST/P002307/1, ST/R002452/1, ST/H008519/1, ST/K00087X/1, STFC Operations grants ST/K003267/1, ST/K003267/1, and Durham University. DiRAC is part of the UK National E-Infrastructure.


\appendix
\section{Regularization Formalism}\label{RegFormal}

The linear regularization matrix $\tens{H}$ used in \citealt{Warren2003} and \citet{Nightingale2015} is derived following the formalism given in \citet{Press2001}, where $\tens{H} = \tens{B}^T \tens{B}$ and the matrix $\tens{B}$ stores the regularization pattern of source pixels with one another. For example, to regularize a source pixel with its neighbor, assuming that pixel one is a neighbour of pixel two, and two of three, etc., the matrix $\tens{B_x}$ is given as
\begin{equation}
\begin{bmatrix}
-1  &  1 & 0  & 0 & ...\\ 
 0  & -1 & 1  & 0 & ...\\ 
 0  &  0 & -1 & 1 & ... \\ 
... & ... & ... & ... & ...
\end{bmatrix}
\, \, .
\end{equation}
For gradient regularization on an $N \times N$ square grid, this matrix gives the regularization of source pixels across the x-direction, where every N elements will be a row of zeros. This matrix then gives a regularization matrix $\tens{H_x} = \tens{B_x}^T \tens{B_x}$. For regularization in the y-direction, a second $\tens{B_y}$ matrix is generated, where the values of negative one are again across the diagonal and the values of positive ones are every N elements across from this, with the final N rows all zeros. $\tens{B_y}$ is then used to compute a second regularization matrix $\tens{H_y} = \tens{B_y}^T \tens{B_y}$, which is added to the first to give the overall regularization matrix $\tens{H} = \tens{H_x} + \tens{H_y}$. Gradient regularization used in this work follows the same pattern, but computes around 5-10 $\tens{H}$ matrices corresponding to regularization across all neighboring Voronoi vertex indices.

DM subhalo results use a scheme which adapts the degree of smoothing to the reconstructed source's luminous emission. First, an estimate of the flux that will be reconstructed by each Voronoi cell is computed using a previous model of the lensed source emission, $\Xi_{j}$, as
\begin{equation}
\label{eqn:LumReg}
v_{\rm  i} = \frac{\sum^{K}_{\rm  k=1} w \, \Xi_{\rm  Src,k}}{K}, 
\end{equation}
where the summation is over the K image-pixels allocated to each Voronoi source pixel and $w$ is given by \cref{eqn:NaturalNeighbor}. Each element in $\vec{v}$ is divided by $K$ to normalize for the number of allocated image pixels, thereby ensuring that the source pixels which (by chance) are allocated more image pixels do not receive a higher value of $v_{\rm  i}$ than those which are allocated fewer. The vector $\vec{V}$ is then computed, where each element is given by
\begin{equation}
V_{\rm  i} = \bigg[ \frac{v_{\rm  i}}{v_{\rm  max}} \bigg]^{L_{\rm  Lum}} .
\label{eqn:LumReg2}
\end{equation}
Each element is divided by the maximum value of $\vec{v}$ to scale all values between zero and one and raised to the power of the hyper-parameter $L_{\rm  Lum}$. $\vec{V}$ is then used to compute the luminosity-weighted regularization value of each source pixel as
\begin{equation}
\Lambda_{\rm  i} = \lambda_{\rm  Src} V_{\rm  i} + \lambda_{\rm  BG} (1-V_{\rm  i}) ,
\label{eqn:LumReg3}
\end{equation}
therefore leading to two regularization coefficients $\lambda_{\rm  Src}$ and $\lambda_{\rm  BG}$, which are both free parameters.

To perform luminosity weighted regularization, the 1D vector of regularization coefficients $\vec{\rm \Lambda}$ (see \cref{eqn:LumReg3}) is folded into the computation of $\tens{H}$. The $\vec{B}$ matrices above are redefined to include each pixel's effective regularization coefficient, $\lambda_{\rm eff}$, as $\tens{B}_{\rm  \Lambda} = \vec{\rm \Lambda}\tens{B}$, where $\vec{\Lambda}$ is given by \cref{eqn:LumReg3}. The corresponding regularization matrix is then $\tens{H}_{\rm  \Lambda} = \tens{B}_{\rm  \Lambda}^T \tens{B}_{\rm  \Lambda}$.

We use the \texttt{PyAutoLens} regularization scheme \texttt{AdaptiveBrightnessSplit}. This scheme also regularizes the source pixel values by interpolating values at a cross of surrounding points, which depends on the size of each source pixel and is independent of the number of connecting neighbours between source pixels, which can be unstable. Explicitly, the regularization term of the \texttt{AdaptiveBrightnessSplit} is given by (following the expression of \citealt{Warren2003})
\begin{equation}
    \begin{split}
    G_{L} = \sum_{j=1}^{J}\{ & \left[s_{j} - \tilde{s}\left(x_{j} + l_{j},\ y_{j} + l_{j}\right)\right]^2 \\
    + & \left[ s_{j} - \tilde{s}\left(x_{j} - l_{j},\ y_{j} + l_{j}\right)\right]^2 \\
    + & \left[ s_{j} - \tilde{s}\left(x_{j} + l_{j},\ y_{j} - l_{j}\right)\right]^2 \\
    + & \left[ s_{j} - \tilde{s}\left(x_{j} - l_{j},\ y_{j} - l_{j}\right)\right]^2 \},
    \end{split}
\end{equation}
where $s_{j}$ is the value of the source pixel $j$ (at position $\left(x_j,\ y_j\right)$). $\tilde{s}\left(x,\ y\right)$ is the natural neighbor interpolating function given by $\{s_j\}_{j=0}^{J}$. $l_j$ is the Voronoi ``length" of the $j$-th source pixel which is defined as the square root of the area of $j$-th source pixel (Voronoi cell), $A_{j}$. With the expression of $G_L$, the regularization matrix $H$ is then derived as \citep[see Eq.~13 of][]{Warren2003}
\begin{equation}
    H_{i,j} = \frac{1}{2}\frac{\partial G_L}{\partial s_i \partial s_j}.
\end{equation}
This regularization scheme is similar to what is used in the work of \citet{Vegetti2009}. The difference is that \citet{Vegetti2009} compute the difference between the value at a Delaunay vertex and the (barycentrically) interpolated values at associated intersecting points of horizontal (vertical) lines and Delaunay edges, while here we compute the value at a Voronoi vertex and the (natural neighboring) interpolated values at positions separated by the associated Voronoi ``length''. 
\section{Simulated Data Results}\label{ResultSim}

\subsection{Simulations}

\begin{figure*}
\centering
\includegraphics[width=0.24\textwidth]{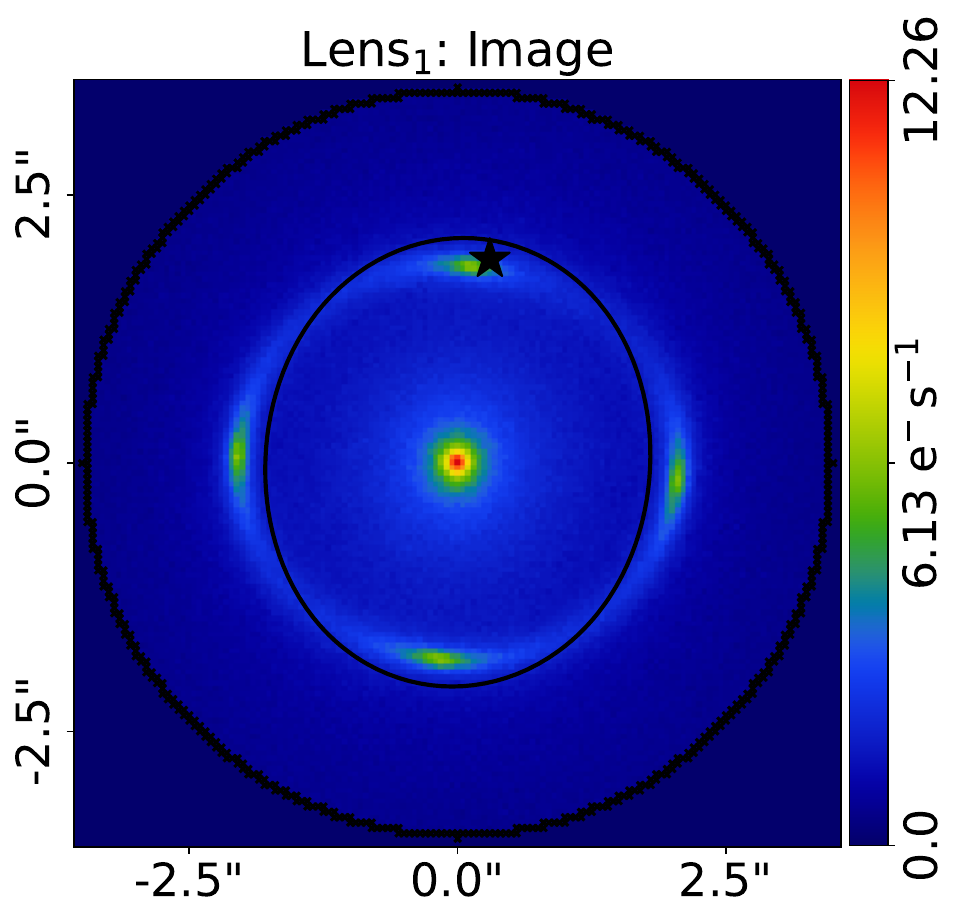}
\includegraphics[width=0.24\textwidth]{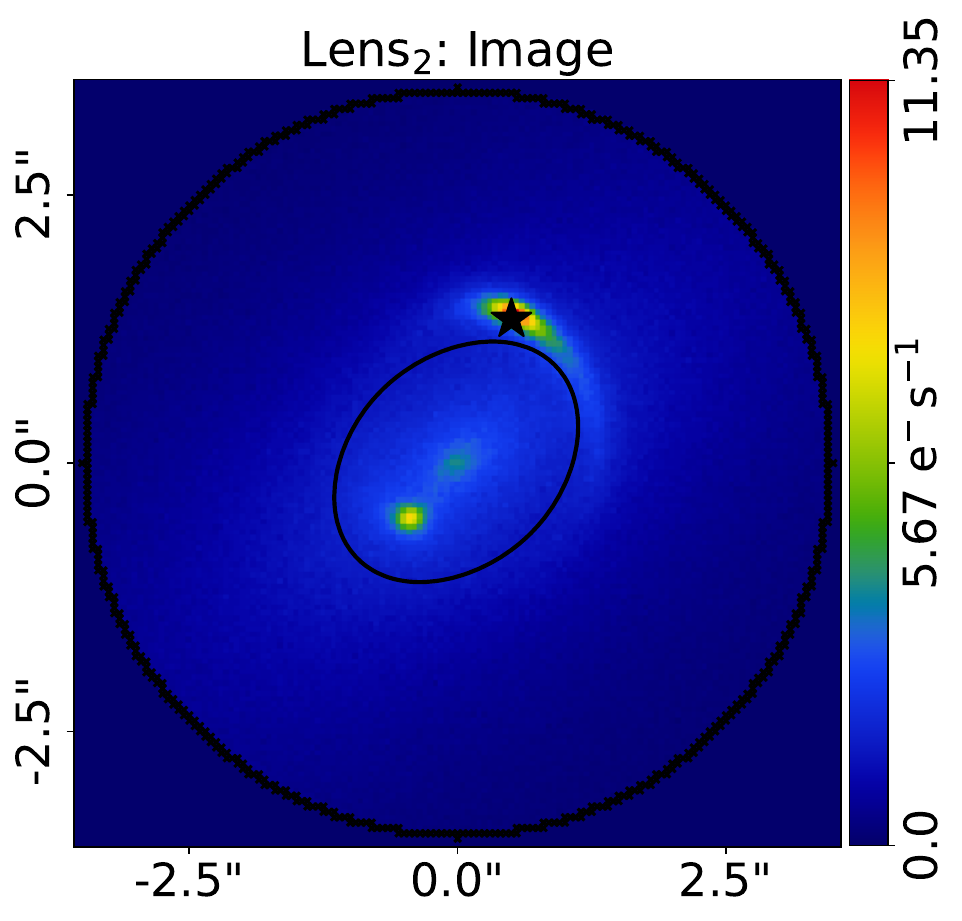}
\includegraphics[width=0.24\textwidth]{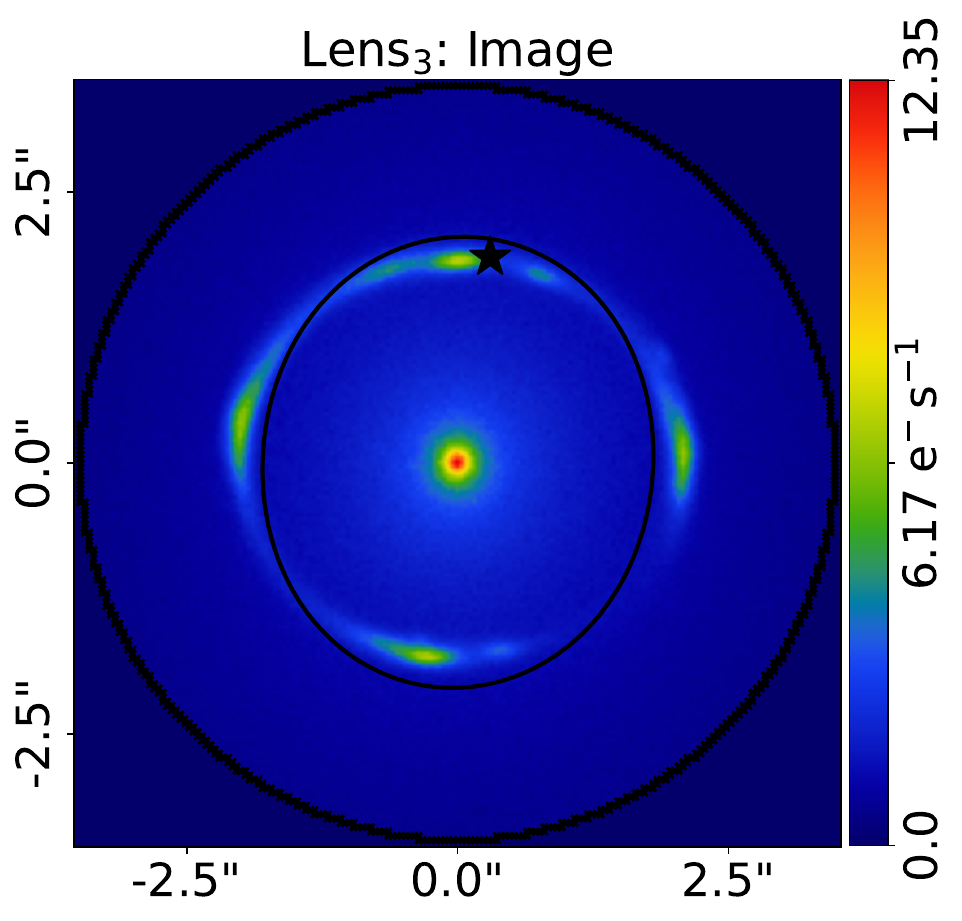}
\includegraphics[width=0.24\textwidth]{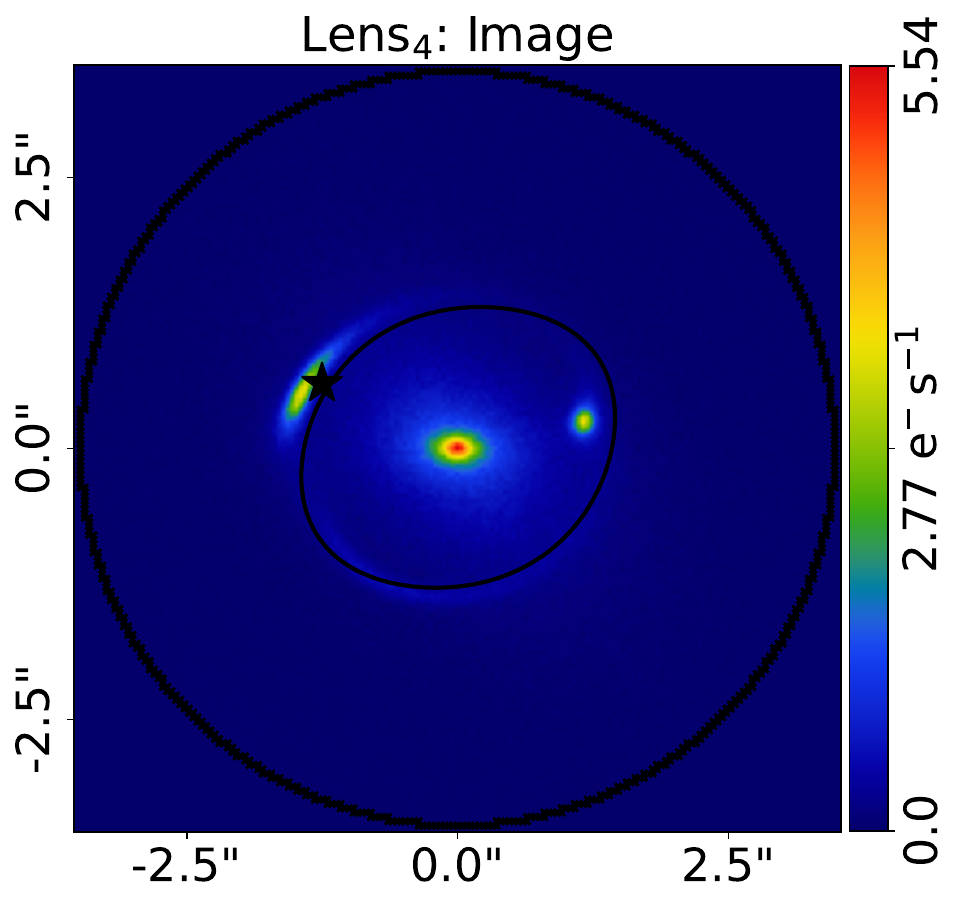}
\caption{
The simulated strong lens images we use to test {\tt PyAutoLens}'s subhalo analysis. Strong lens configurations include a quadruply imaged Einstein ring (Lens$_1$ and Lens$_3$) and two doubly imaged sources (Lens$_2$ and Lens$_4$). The source is either a single S\'ersic profile (Lens$_1$ and Lens$_2$) or between one and six \citet{eff87} light profiles (Lens$_3$ and Lens$_4$). Each simulated image includes a $10^{10.5}$M$_{\odot}$ subhalo at the location marked by the black star. Images including a subhalo at these locations with masses of $10^{10.0}$M$_{\odot}$ and $10^{9.5}$M$_{\odot}$ and without a subhalo are also fitted in this work.
} 
\label{figure:Simulations}
\end{figure*}

\begin{table*}
\resizebox{\linewidth}{!}{
\Large
\begin{tabular}{ l | l | l | l | l | l | l | l | l | l | l} 
\multicolumn{1}{p{2.0cm}|}{\centering Lens Name} 
& \multicolumn{1}{p{1.3cm}|}{$x^{\rm{mass}}$ $(\arcsec)$} 
& \multicolumn{1}{p{1.3cm}|}{$y^{\rm{mass}}$ $(\arcsec)$} 
& \multicolumn{1}{p{0.7cm}|}{$\epsilon_{\rm 1}^{\rm{mass}}$}
& \multicolumn{1}{p{0.7cm}|}{$\epsilon_{\rm 2}^{\rm{mass}}$}
& \multicolumn{1}{p{1.3cm}|}{$\theta_{\rm Ein}^{\rm{mass}}$ $(\arcsec)$} 
& \multicolumn{1}{p{0.7cm}|}{$\gamma^{\rm{mass}}$} 
& \multicolumn{1}{p{0.7cm}|}{$\gamma_{\rm 1}^{\rm{ext}}$} 
& \multicolumn{1}{p{0.7cm}|}{$\gamma_{\rm 2}^{\rm{ext}}$}
& \multicolumn{1}{p{1.3cm}|}{$x^{\rm sub}$ $(\arcsec)$} 
& \multicolumn{1}{p{1.3cm}}{$y^{\rm sub}$ $(\arcsec)$} 
\\ \hline
$\mathrm{Lens_{\rm 1}}$ & 0.0 & 0.0 & -0.023 & 0.0115 & 1.926 & 2.073 & 0.05 & 0.0 & 0.3 & 1.9 \\[0pt]
$\mathrm{Lens_{\rm 2}}$ & 0.0 & 0.0 & 0.019 & 0.109 & 1.1 & 1.9 & -0.007 & 0.006 & 0.5 & 1.34 \\[0pt]
$\mathrm{Lens_{\rm 3}}$ & 0.0 & 0.0 & -0.023 & 0.0115 & 1.926 & 2.073 & 0.05 & 0.0 & 0.3 & 1.9 \\[0pt]
$\mathrm{Lens_{\rm 4}}$ & 0.0 & 0.0 & 0.05 & 0.15 & 1.4 & 2.1 & 0.0 & 0.05 & -1.25 & 0.6 \\[0pt]
\end{tabular}
}
\caption{The parameters of the mass models of the simulated lenses.}
\label{table:MassModelSim}
\end{table*}

\begin{table*}
\resizebox{\linewidth}{!}{
\large
\begin{tabular}{ l | l | l | l | l | l | l | l } 
\multicolumn{1}{p{2.0cm}|}{\centering Lens Name} 
& \multicolumn{1}{p{1.3cm}|}{$x$ $(\arcsec)$} 
& \multicolumn{1}{p{1.3cm}|}{$y$ $(\arcsec)$} 
& \multicolumn{1}{p{0.7cm}|}{$\epsilon_{\rm 1}$}
& \multicolumn{1}{p{0.7cm}|}{$\epsilon_{\rm 2}$}
& \multicolumn{1}{p{1.3cm}|}{$I$} 
& \multicolumn{1}{p{0.7cm}|}{$R$ / $r_{\rm c}$} 
& \multicolumn{1}{p{0.7cm}}{$n$} 
\\ \hline
$\mathrm{Lens_{\rm 1}}$ & 0.01 & 0.01 & -0.05555 & 0.096225 & 0.3 & 0.3 & 2.5 \\[0pt]
$\mathrm{Lens_{\rm 2}}$ & 0.3 & 0.2 & -0.060356 & -0.165828 & 0.8 & 0.15 & 2.5 \\[0pt]
$\mathrm{Lens_{\rm 3}}$ & 0.0285 & 0.0404 & -0.24647 & -0.20769 & 11.585 & 0.03899 & \\[-2pt]
                        & 0.0711 & 0.1947 & 0.07803 & 0.00936 & 8.911 & 0.01015 & \\[-2pt]
                        & -0.0792 & 0.0760 & -0.35274 & 0.18159 & 6.3651 & 0.02594 & \\[-2pt]                        
                        & 0.0977 & 0.0726  & -0.23621 & -0.25923 & 5.5818 & 0.04237 & \\[-2pt]
                        & -0.0020 & 0.0020 & 0.51173 & -0.36744 & 2.254 & 0.02441 & \\[-2pt]                        
                        & 0.1131 & 0.0515 & 0.53965 & 0.05310 & 1.029 & 0.0336 & \\[0pt]
$\mathrm{Lens_{\rm 4}}$ & -0.2 & 0.02 & 0.53965 & 0.05310 & 1.029 & 0.0336 & \\[0pt]                        
\end{tabular}
}
\caption{The parameters of the source parameters of the simulated lenses. $\mathrm{Lens_{\rm 1}}$ and $\mathrm{Lens_{\rm 2}}$ correspond to Sersic parameters, whereas $\mathrm{Lens_{\rm 3}}$ and $\mathrm{Lens_{\rm 4}}$ are Elson, Free and Fall parameters.}
\label{table:SourceModelSim}
\end{table*}

We test our lens modelling and subhalo scanning pipelines on a sample of $16$ simulated lenses. For every lens we assume a lens galaxy and subhalo (when included) at $z=0.5$, and a source at $z=1.0$. Each lens galaxy is simulated using two centrally aligned S\'ersic profiles and a PL mass profile. Sources are simulated using either a single S\'ersic profile or a superposition of between 1 and 6 elliptical \citet[][hereafter EFF]{eff87} 
profiles,
\begin{equation}
\label{eqn:eff}
I_{\rm  EFF} (\xi) = I_{\rm  c} ( 1 + r^{\rm 2} / r^{\rm 2}_{\rm c} )^{-\eta} ,
\end{equation}
where $r_{\rm c}$ is a scale factor that relates the profile to the circular half light radius and $\eta$ controls the intensity gradient of the profile. The EFF profile is used by \citet{Cornachione2018} to model the BELLS-GALLERY source galaxies, and we adopt parameter values representative of their results. We assume $\eta = 1.5$ for all simulated sources.

We simulate four baseline models of lens light, lens mass and source light (see \cref{table:MassModelSim} and \cref{table:SourceModelSim}). The same mass model is used to simulate lenses one and three, which forms a nearly complete Einstein ring but with either a single S\'ersic source or six EFF profiles. For each lens, we create $4$ mock images with subhalos of mass $10^{10.5}$M$_{\odot}$, $10^{10.0}$M$_{\odot}$, $10^{9.5}$M$_{\odot}$ or where the subhalo is omitted. The $(x^{\rm sub},\,y^{\rm sub})$ coordinates of each subhalo are chosen to be near or on top of the lensed source's emission, to ensure they perturb the source's emission significantly enough to be detectable and thus able to test our subhalo scanning analysis.

The $4$ images including a $10^{10.5}$M$_{\odot}$ subhalo are shown in \cref{figure:Simulations}, where the subhalo locations are shown by a black star. Each simulated image has a resolution and signal-to-noise representative of the HST imaging of strong lenses we fit in this work. For Lens$_1$ and Lens$_2$ (which assume a single S\'ersic source) we assume a pixel-scale of $0.05\arcsec$. For Lens$_3$ and Lens$_4$ (which assume EFF profiles for the source) we assume $0.04\arcsec$. These pixel-scales are the resolution of the SLACS and BELLS-GALLERY HST images respectively. To create the images of the lens and source emission an iterative grid is used. This first evaluates each light profile (including ray-tracing if it is in the source-plane) at the centre of each image pixel. It then evaluates the profiles using higher resolution sub-grids within each image-pixel, in increments of $11 \times 11$, $21 \times 21$, $31 \times 31$, up to $301 \times 301$. A pixel intensity is computed until it reaches a fractional accuracy of $99.9999\%$ compared to the value computed using the previous sub-grid. This image has a uniform background sky added to it, is blurred with a Gaussian PSF with $\sigma=0.05 \arcsec$, has Poisson noise added, and then the input background sky subtracted.

This mock sample is idealized, in that the same lens light model (a double Sersic) and mass model (a PL plus shear) used to simulate each lens is assumed when we fit it. The primary purpose of this exercise is to build confidence that our method does not produce false positives in this idealized setup, and determine to what masses it can accurately recover input DM subhalos.

\subsection{Results}

\begin{table*}
\Large
\resizebox{\linewidth}{!}{
\begin{tabular}{ l | l | l | l | l | l | l | l | l | l | l | l | l | l } 
\multicolumn{1}{p{1.3cm}|}{Lens Name} 
& \multicolumn{1}{p{1.5cm}|}{Subhalo Mass}  
& \multicolumn{1}{p{1.7cm}|}{$\Delta\,\mathrm{ln}\,\mathcal{Z}^{\rm Base}$ (Circular Mask)}  
& \multicolumn{1}{p{1.7cm}|}{$\Delta\,\mathrm{ln}\,\mathcal{Z}^{\rm Base}$ (Annular Mask)}  
& \multicolumn{1}{p{2.0cm}|}{$x^{{\rm sub}}$\,[$\textrm{arcsec}$]}  
& \multicolumn{1}{p{1.9cm}|}{$y^{{\rm sub}}$\,[$\textrm{arcsec}$]}  
& \multicolumn{1}{p{2.0cm}|}{$\log_{10}$ \, $[M^{\rm sub}_{\rm 200}$\,/\,M$_{\rm \odot}]$}  
& \multicolumn{1}{p{2.1cm}|}{$\Delta\,\mathrm{ln}\,\mathcal{L}^{\rm Base}$ \,\,\, (No \,\,\,\, \,\, \, \, Interpolation)}  
& \multicolumn{1}{p{2.1cm}|}{$\Delta\,\mathrm{ln}\,\mathcal{L}^{\rm Base}$ \, \, \, (Linear \, \, \, \, \, Regularization)}  
& \multicolumn{1}{p{1.8cm}|}{$\Delta\,\mathrm{ln}\,\mathcal{L}^{\rm Base}$ (Both)}  
& \multicolumn{1}{p{1.9cm}|}{$\mathrm{ln}\,\mathcal{Z}^{\rm Base}$ \, \, \, (Annular Mask)}  
& \multicolumn{1}{p{1.8cm}|}{$\mathrm{ln}\,\mathcal{L}^{\rm Base}$ \, \, \, (Both)}  
& \multicolumn{1}{p{2.1cm}|}{$\mathrm{ln}\,\mathcal{L}^{\rm Base}$ \, \, \, (Linear \, \, \, \, \, Regularization)}  
& \multicolumn{1}{p{2.1cm}}{$\mathrm{ln}\,\mathcal{L}^{\rm Base}$ \,\,\, \, \, \, \, (No \,\,\,\, \,\, \, \, \, \, \, \, Interpolation)}  
\\ \hline
& & & & & & & & & & & & & \\[-6pt]
$\mathrm{Lens}_{1}$ & No Subhalo & -1.57 & 1.7 & $0.92^{+1.84}_{-1.32}$ & $1.24^{+1.42}_{-1.76}$ & $9.34^{+0.69}_{-3.11}$  & -0.18 & -1.11  & -2.1  & 8491.03 & 7782.59 & 7862.74  & 8468.02  \\[2pt]
$\mathrm{Lens}_{1}$ & $10^{9.5}M_{\odot}$ & \textbf{12.91} & \textbf{17.67} & $1.88^{+0.65}_{-0.15}$ & $0.21^{+0.18}_{-0.20}$ & $9.59^{+0.27}_{-0.34}$  & 3.35 & \textbf{28.26}  & \textbf{16.75}  & 8565.95 & 7897.7 & 7975.78  & 8543.27  \\[2pt]
$\mathrm{Lens}_{1}$ & $10^{10.0}M_{\odot}$ & \textbf{30.42} & \textbf{39.89} & $1.78^{+0.17}_{-0.13}$ & $0.27^{+0.07}_{-0.06}$ & $9.90^{+0.21}_{-0.28}$  & \textbf{33.78} & \textbf{33.13}  & \textbf{33.24}  & 8451.39 & 7788.9 & 7873.59  & 8434.49  \\[2pt]
$\mathrm{Lens}_{1}$ & $10^{10.5}M_{\odot}$ & \textbf{181.04} & \textbf{186.58} & $1.89^{+0.11}_{-0.08}$ & $0.31^{+0.05}_{-0.04}$ & $10.43^{+0.10}_{-0.13}$  & \textbf{119.8} & \textbf{202.16}  & \textbf{177.6}  & 8610.81 & 7961.62 & 8038.45  & 8581.56  \\[2pt]
$\mathrm{Lens}_{2}$ & No Subhalo & 0.76 & 0.45 & $0.89^{+1.32}_{-2.55}$ & $2.26^{+2.30}_{-2.57}$ & $8.47^{+2.47}_{-2.41}$  & -0.19 & \textbf{13.35}  & -0.22  & 10319.48 & 8659.13 & 8737.79  & 10276.96  \\[2pt]
$\mathrm{Lens}_{2}$ & $10^{9.5}M_{\odot}$ & \textbf{12.12} & 3.71 & $-0.53^{+1.22}_{-1.22}$ & $1.72^{+1.73}_{-2.46}$ & $10.66^{+0.33}_{-4.05}$  & 3.56 & \textbf{17.52}  & 1.28  & 10245.35 & 8525.47 & 8690.09  & 10199.79  \\[2pt]
$\mathrm{Lens}_{2}$ & $10^{10.0}M_{\odot}$ & \textbf{15.6} & 7.57 & $1.48^{+1.16}_{-0.40}$ & $0.44^{+0.24}_{-0.70}$ & $9.95^{+0.48}_{-1.44}$  & 5.05 & 3.87  & \textbf{12.25}  & 10277.55 & 8599.48 & 8731.97  & 10234.11  \\[2pt]
$\mathrm{Lens}_{2}$ & $10^{10.5}M_{\odot}$ & \textbf{12.83} & \textbf{15.3} & $1.50^{+0.17}_{-0.20}$ & $0.50^{+0.14}_{-0.11}$ & $10.35^{+0.38}_{-0.60}$  & \textbf{16.95} & \textbf{14.25}  & \textbf{56.05}  & 10309.66 & 8522.58 & 8556.9  & 10278.37  \\[2pt]
$\mathrm{Lens}_{3}$ & No Subhalo & 5.68 & 4.62 & $-1.81^{+1.82}_{-2.68}$ & $0.10^{+2.20}_{-1.68}$ & $8.04^{+1.52}_{-2.03}$  & 0.06 & -6.14  & -1.39  & 13846.96 & 12640.44 & 12833.41  & 13817.07  \\[2pt]
$\mathrm{Lens}_{3}$ & $10^{9.5}M_{\odot}$ & \textbf{58.99} & \textbf{33.75} & $1.95^{+0.21}_{-0.31}$ & $0.41^{+0.25}_{-0.12}$ & $9.43^{+0.38}_{-0.32}$  & \textbf{13.51} & \textbf{37.86}  & \textbf{38.01}  & 13895.6 & 12756.99 & 12908.08  & 13882.19  \\[2pt]
$\mathrm{Lens}_{3}$ & $10^{10.0}M_{\odot}$ & \textbf{128.92} & \textbf{108.11} & $1.90^{+0.09}_{-0.09}$ & $0.27^{+0.06}_{-0.05}$ & $9.87^{+0.18}_{-0.13}$  & \textbf{57.36} & \textbf{123.86}  & \textbf{107.06}  & 13872.69 & 12697.12 & 12893.02  & 13846.19  \\[2pt]
$\mathrm{Lens}_{3}$ & $10^{10.5}M_{\odot}$ & \textbf{431.22} & \textbf{638.2} & $1.95^{+0.07}_{-0.04}$ & $0.33^{+0.04}_{-0.03}$ & $10.70^{+0.09}_{-0.08}$  & \textbf{274.93} & \textbf{595.63}  & \textbf{531.53}  & 13867.75 & 12788.37 & 12918.48  & 13865.25  \\[2pt]
$\mathrm{Lens}_{4}$ & No Subhalo & \textbf{15.25} & 2.34 & $-0.16^{+1.71}_{-2.12}$ & $1.17^{+2.17}_{-2.52}$ & $8.09^{+2.45}_{-1.76}$  & -0.08 & 3.93  & -1.1  & 15447.45 & 14188.7 & 14285.39  & 15416.03  \\[2pt]
$\mathrm{Lens}_{4}$ & $10^{9.5}M_{\odot}$ & \textbf{34.71} & 3.61 & $0.71^{+0.85}_{-1.38}$ & $-1.30^{+1.54}_{-1.58}$ & $9.05^{+1.52}_{-2.56}$  & 0.96 & \textbf{32.49}  & \textbf{12.93}  & 15370.66 & 14119.52 & 14191.89  & 15343.7  \\[2pt]
$\mathrm{Lens}_{4}$ & $10^{10.0}M_{\odot}$ & \textbf{53.36} & \textbf{18.24} & $0.70^{+0.05}_{-0.18}$ & $-1.28^{+0.12}_{-0.10}$ & $9.79^{+0.29}_{-0.32}$  & \textbf{10.8} & \textbf{10.9}  & \textbf{29.02}  & 15394.6 & 14146.35 & 14201.78  & 15372.61  \\[2pt]
$\mathrm{Lens}_{4}$ & $10^{10.5}M_{\odot}$ & \textbf{186.26} & \textbf{89.54} & $0.63^{+0.05}_{-0.04}$ & $-1.43^{+0.11}_{-0.16}$ & $10.38^{+0.15}_{-0.24}$  & \textbf{70.11} & \textbf{105.94}  & \textbf{111.36}  & 15195.02 & 14003.25 & 14080.38  & 15216.3  \\[2pt]

\end{tabular}
}
\caption{
The results for fits to the simulated strong lens sample, including Bayesian evidence increases $\Delta\,\mathrm{ln}\,\mathcal{Z}$ of the subhalo scanning analysis where log evidence increases are computed compared to the lens model fitted without a subhalo. All mass models assume a PL plus shear and a lens light subtraction consisting of two S\'ersic profiles. Columns 3 and 4 show results using the baseline Voronoi source analysis, which includes natural neighbor interpolation and brightness based regularization. Column 3 shows $\Delta\,\mathrm{ln}\,\mathcal{Z}^{\rm Base}$ using a $3.5\arcsec$ circular mask and column 4 for an annular mask. Columns 5 to 7 show the inferred subhalo location $x^{\rm sub}$, $y^{\rm sub}$ and mass $\log_{10}[M^{\rm sub}_{\rm 200}$\,/\,M$_{\rm \odot}]$ for the fit using an annular mask, where errors are quoted at $3\sigma$ credible regions. The true input subhalo coordinates $(x^{\rm sub}$,\ $y^{\rm sub})$ for each lens are: $\mathrm{Lens}_{1} = (0.3\arcsec,\,1.9\arcsec)$, $\mathrm{Lens}_{2} = (0.5\arcsec,\,1.34\arcsec)$, $\mathrm{Lens}_{3} = (0.3\arcsec,\,1.9\arcsec)$ and $\mathrm{Lens}_{4} = (-1.25\arcsec,\,0.6\arcsec)$. Columns 8, 9 and 10 show results for three source analysis variants (all using annular masks), which respectively: (i) switch off Voronoi natural neighbor interpolation such that every image sub-pixel maps to only one source-pixel; (ii) use linear regularization (see \citet{Warren2003} and \citet{Nightingale2015}) instead of brightness based adaptive regularization and; (iii) make both changes simultaneously. The final 4 columns show the overall Bayesian evidence values, $\mathrm{ln}\,\mathcal{Z}$, inferred for all four source analysis variants in the final model fit which includes a DM subhalo. These can be compared to choose the optimal source analysis.
}
\label{table:DetectSim}
\end{table*}

\begin{figure*}
\centering
\includegraphics[width=0.23\textwidth]{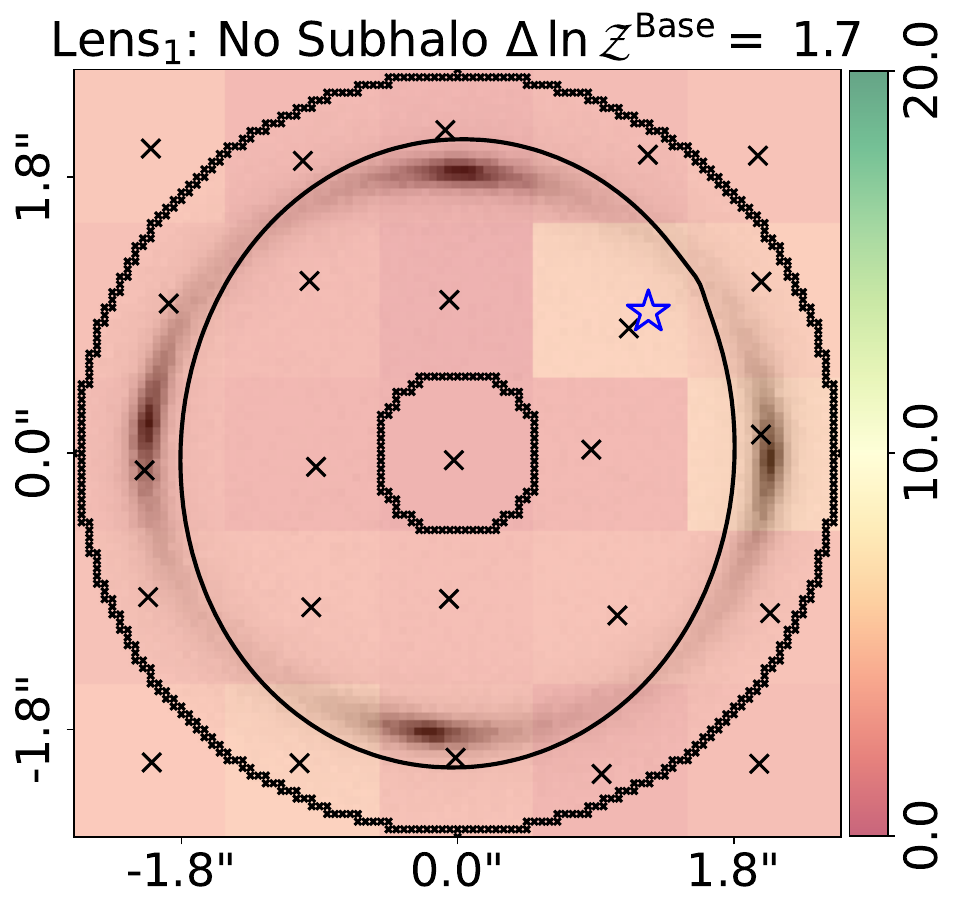}
\includegraphics[width=0.23\textwidth]{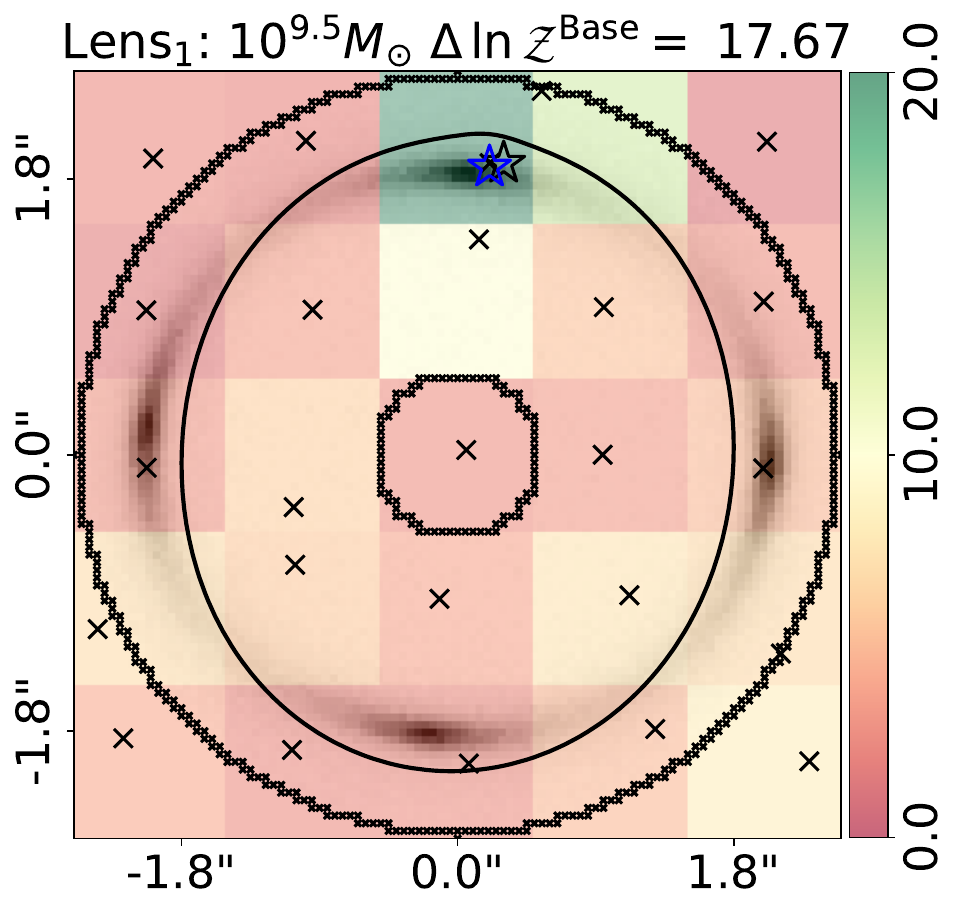}
\includegraphics[width=0.23\textwidth]{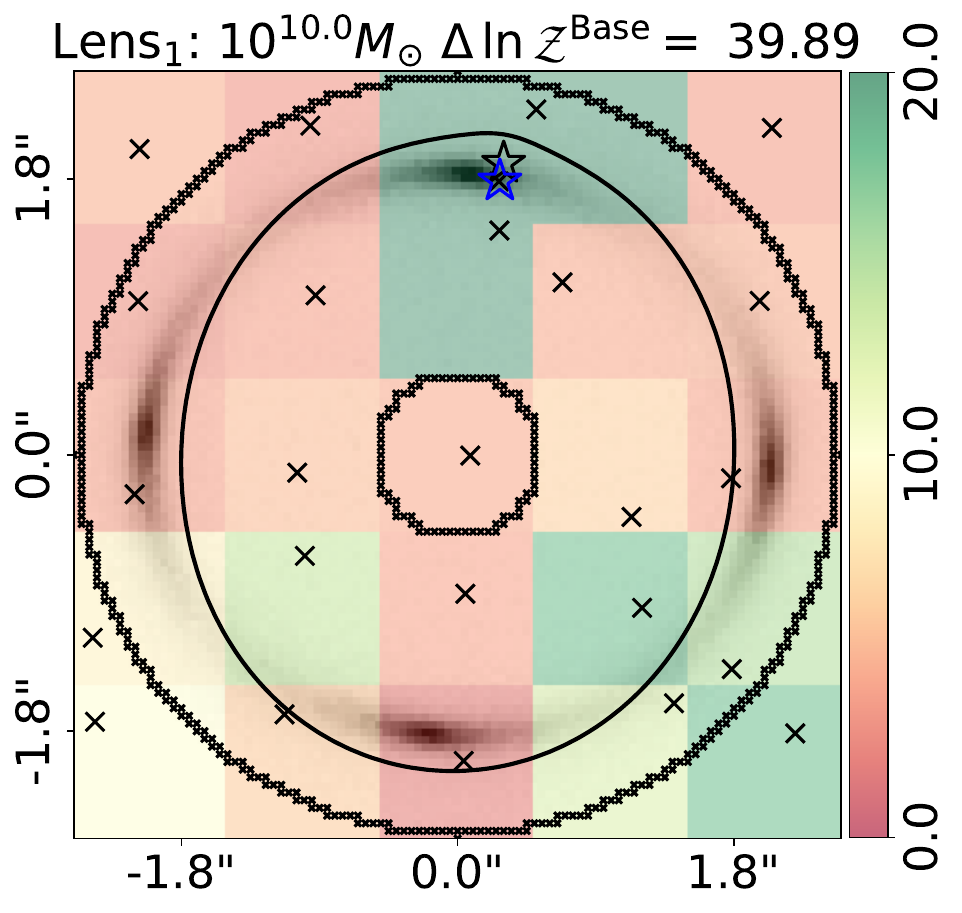}
\includegraphics[width=0.23\textwidth]{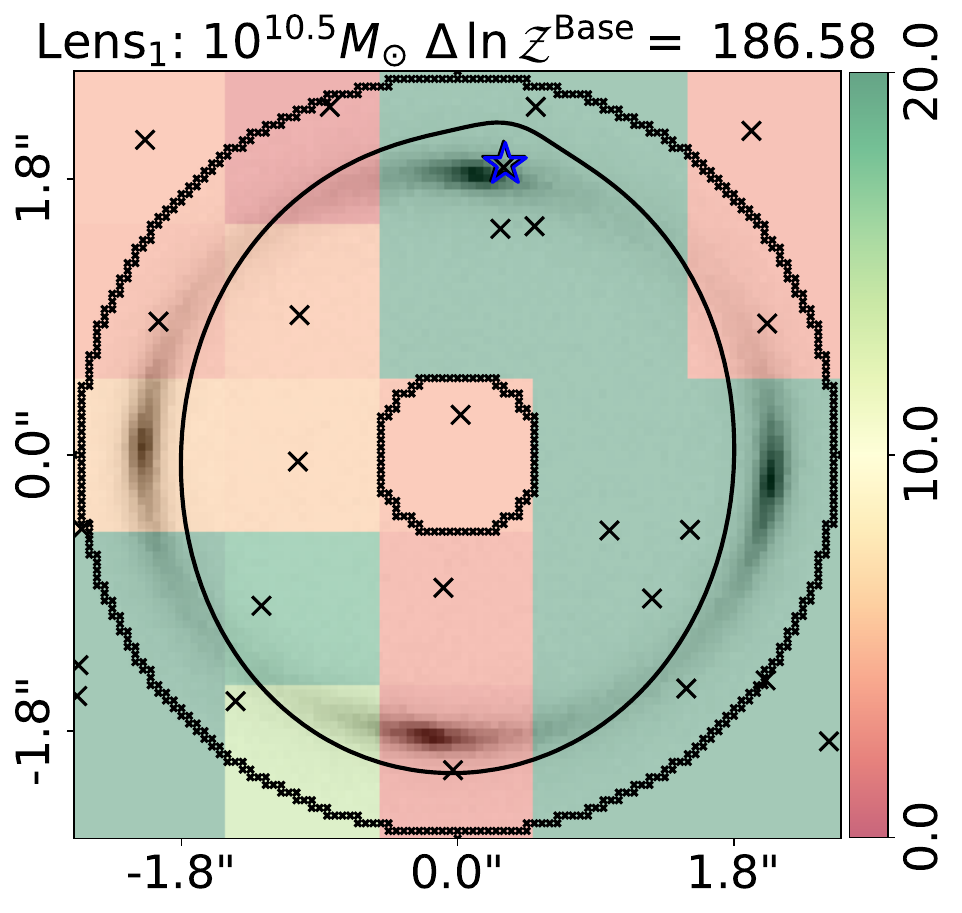}

\includegraphics[width=0.23\textwidth]{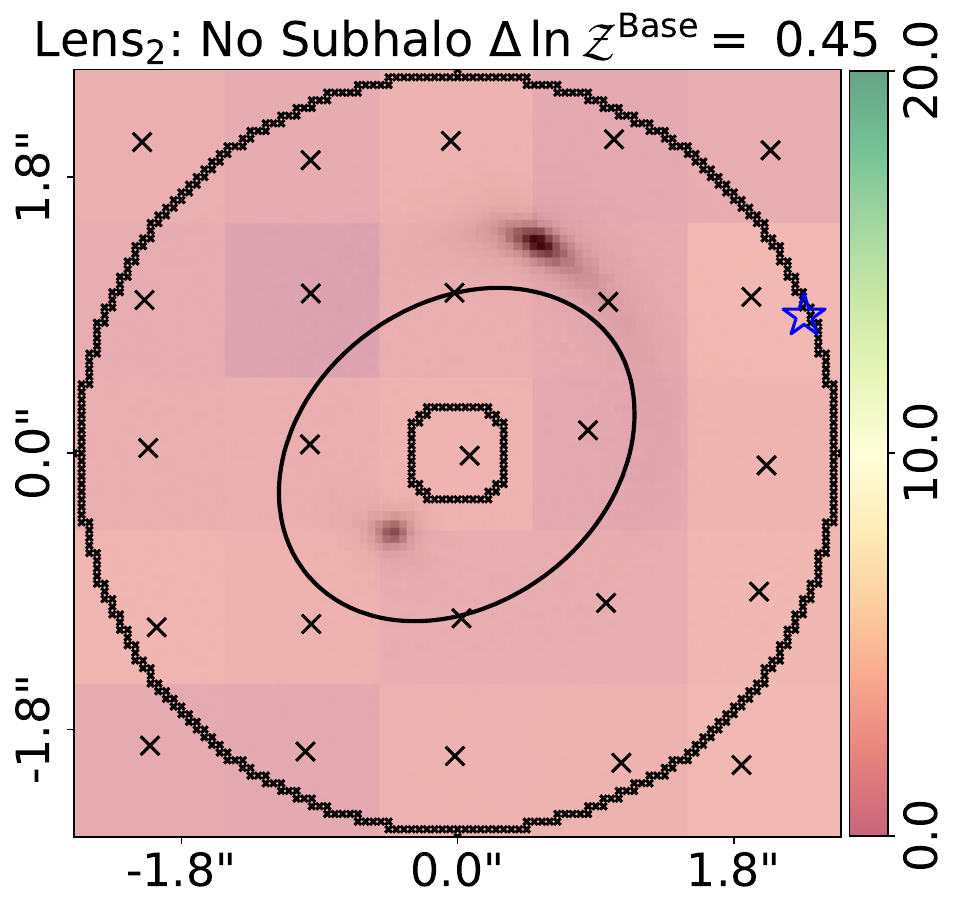}
\includegraphics[width=0.23\textwidth]{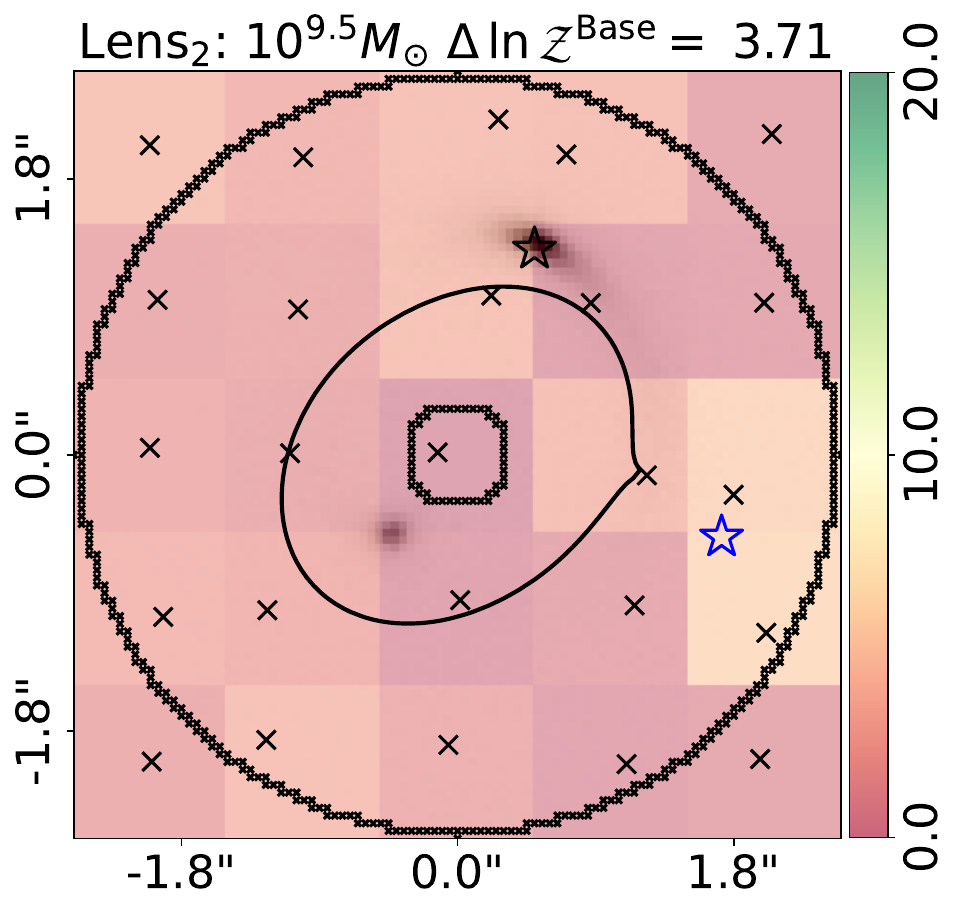}
\includegraphics[width=0.23\textwidth]{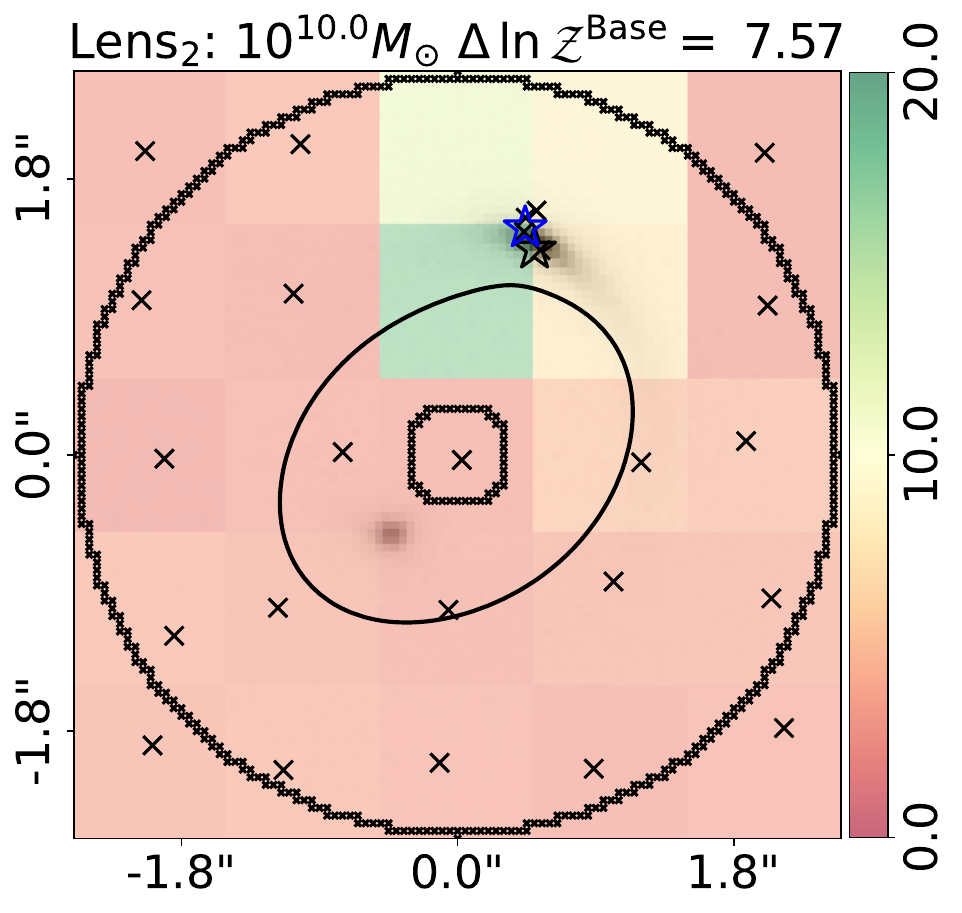}
\includegraphics[width=0.23\textwidth]{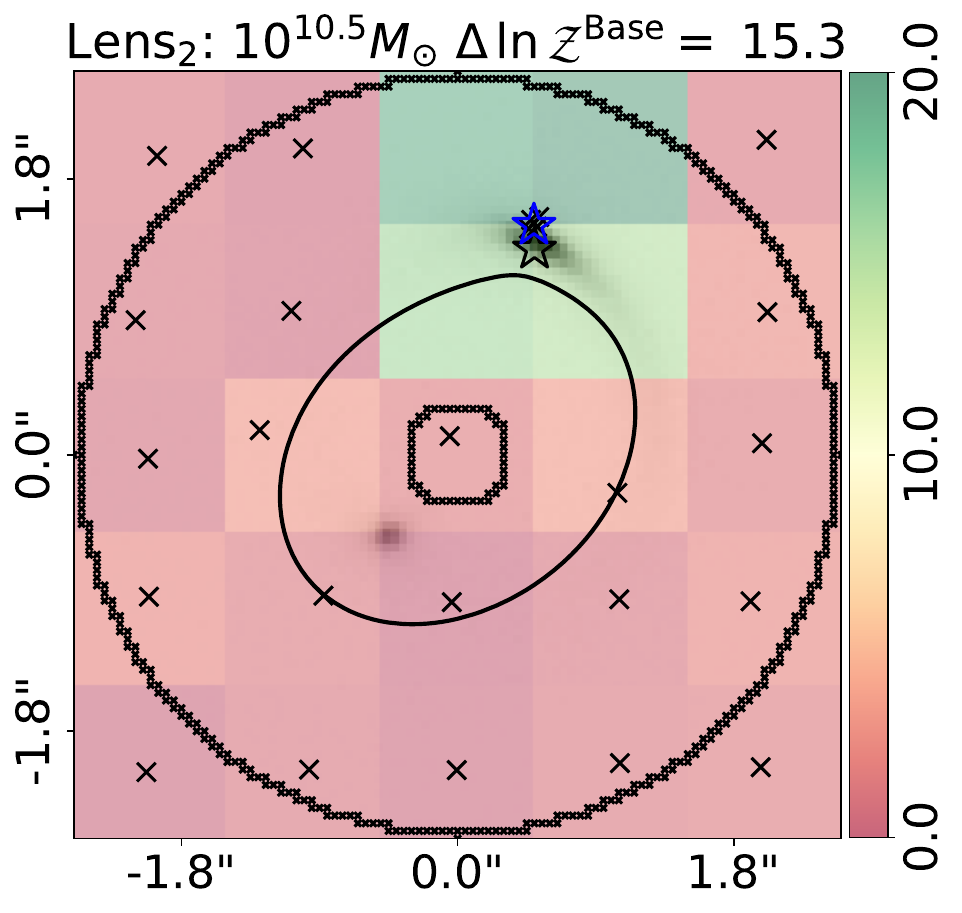}

\includegraphics[width=0.23\textwidth]{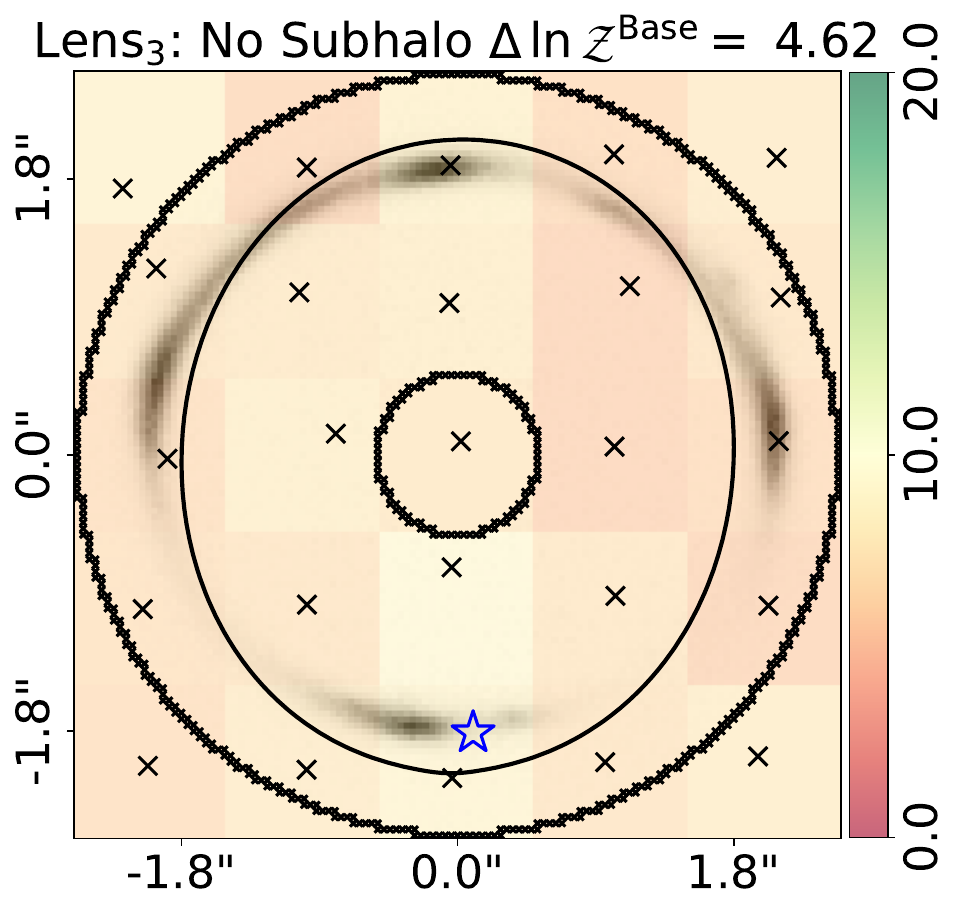}
\includegraphics[width=0.23\textwidth]{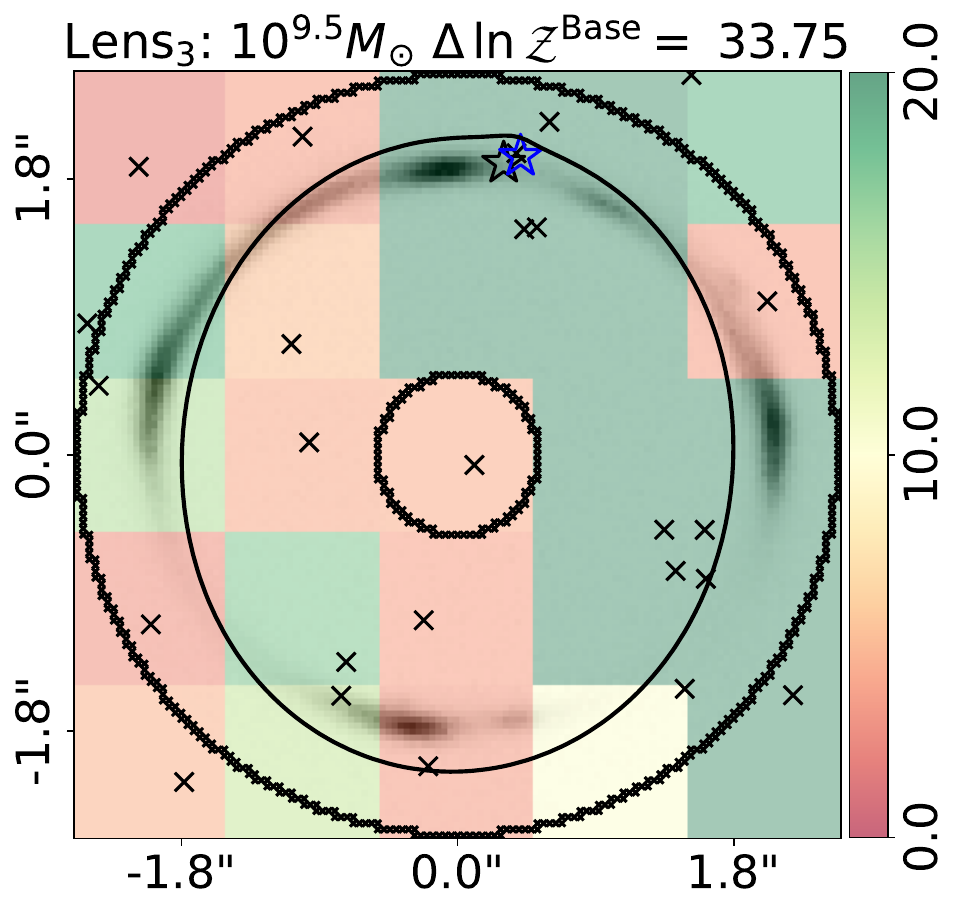}
\includegraphics[width=0.23\textwidth]{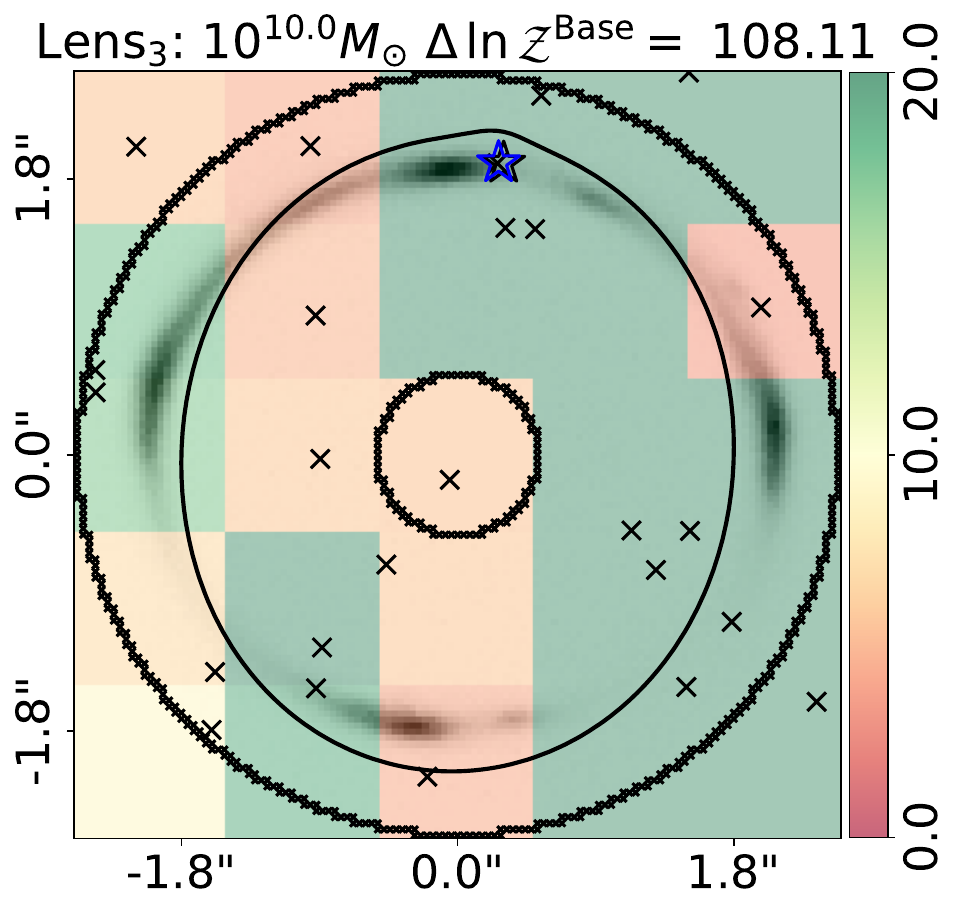}
\includegraphics[width=0.23\textwidth]{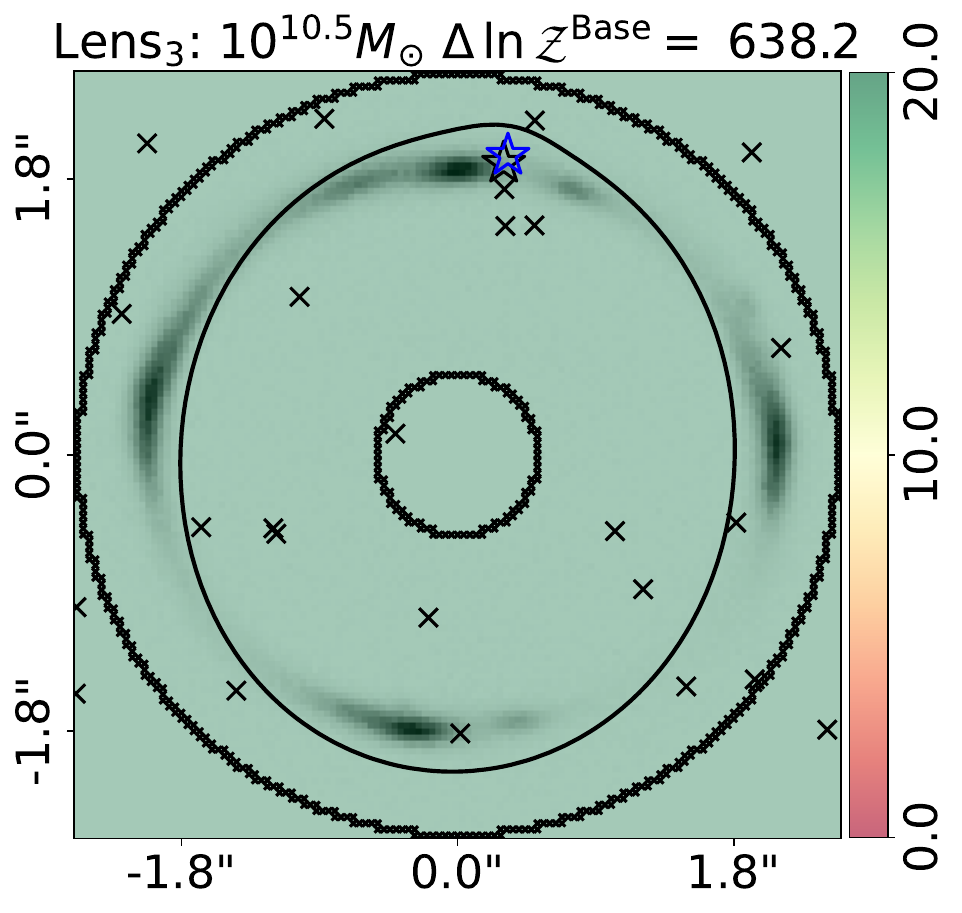}

\includegraphics[width=0.23\textwidth]{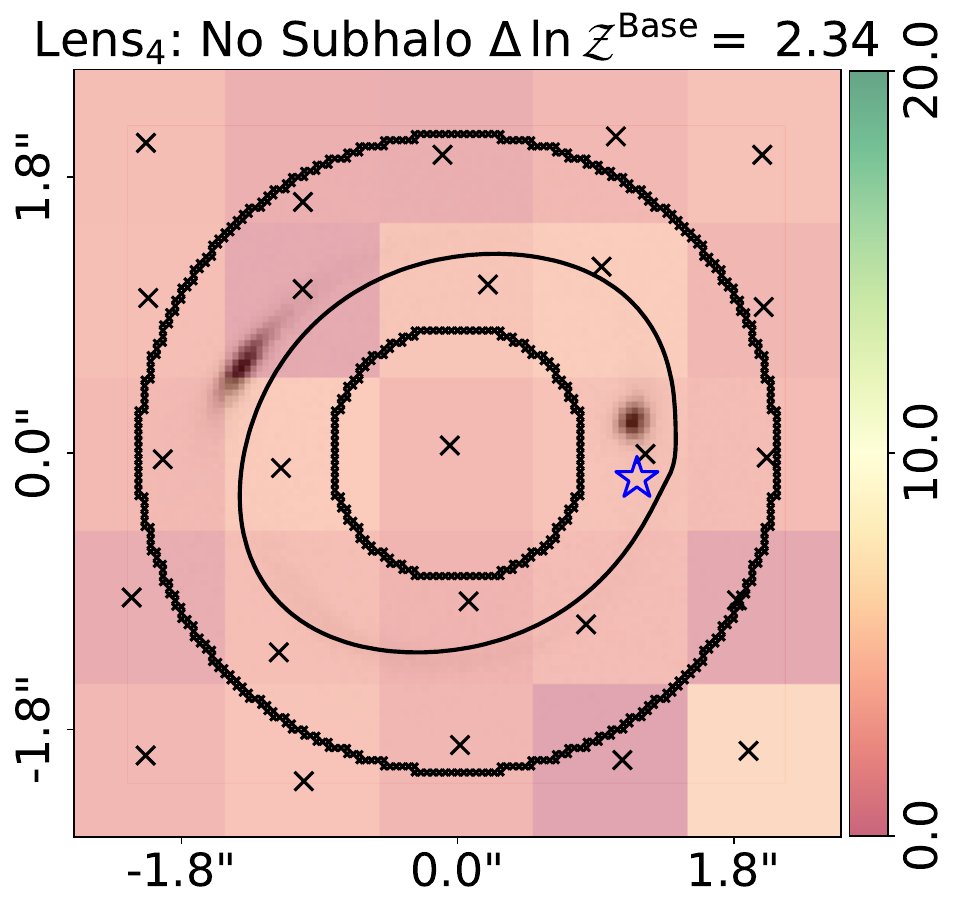}
\includegraphics[width=0.23\textwidth]{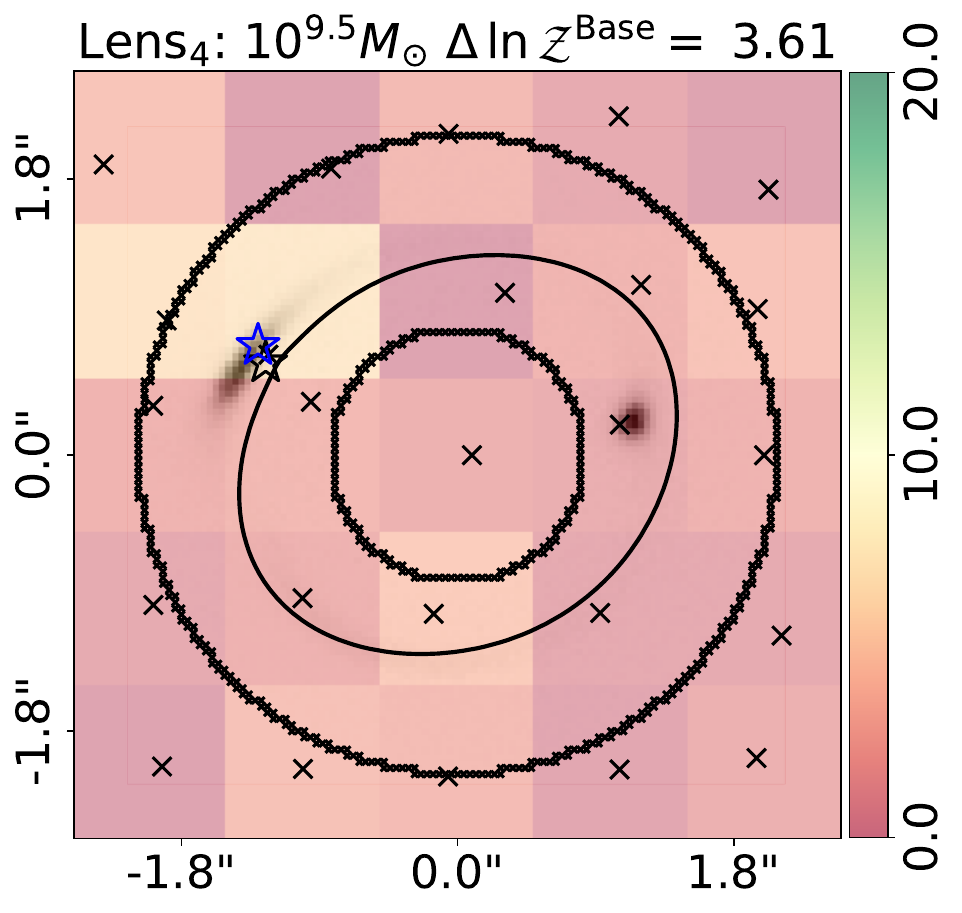}
\includegraphics[width=0.23\textwidth]{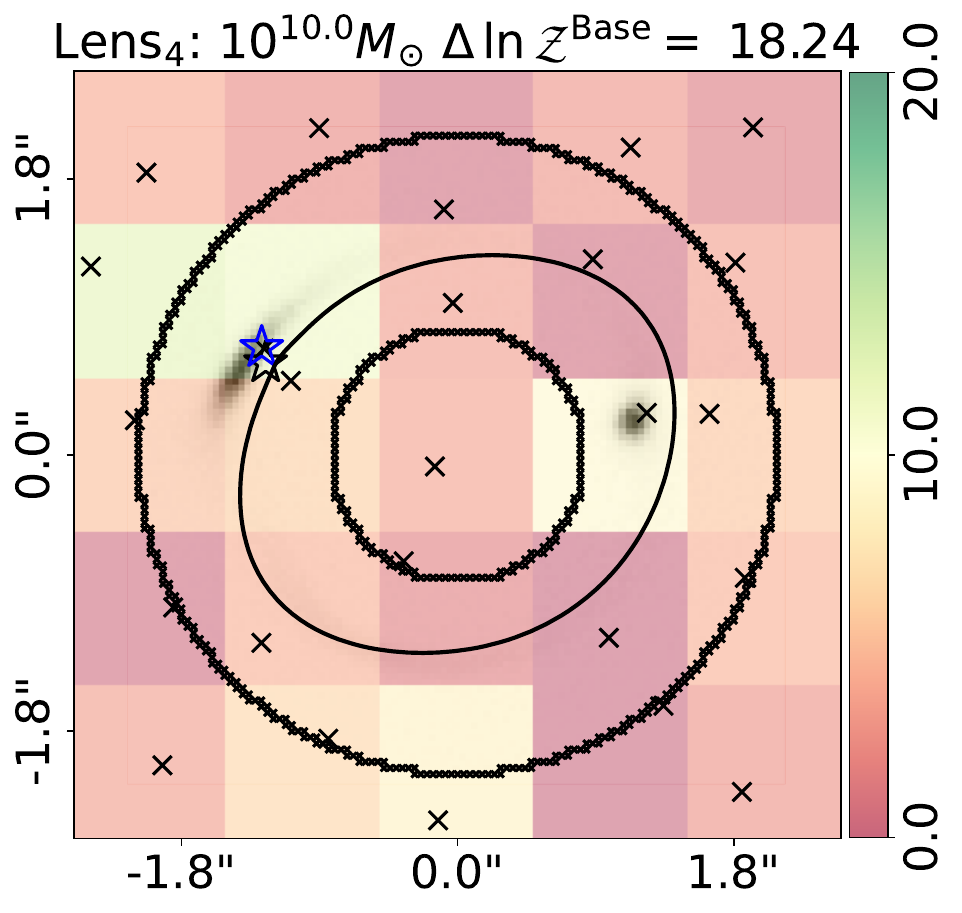}
\includegraphics[width=0.23\textwidth]{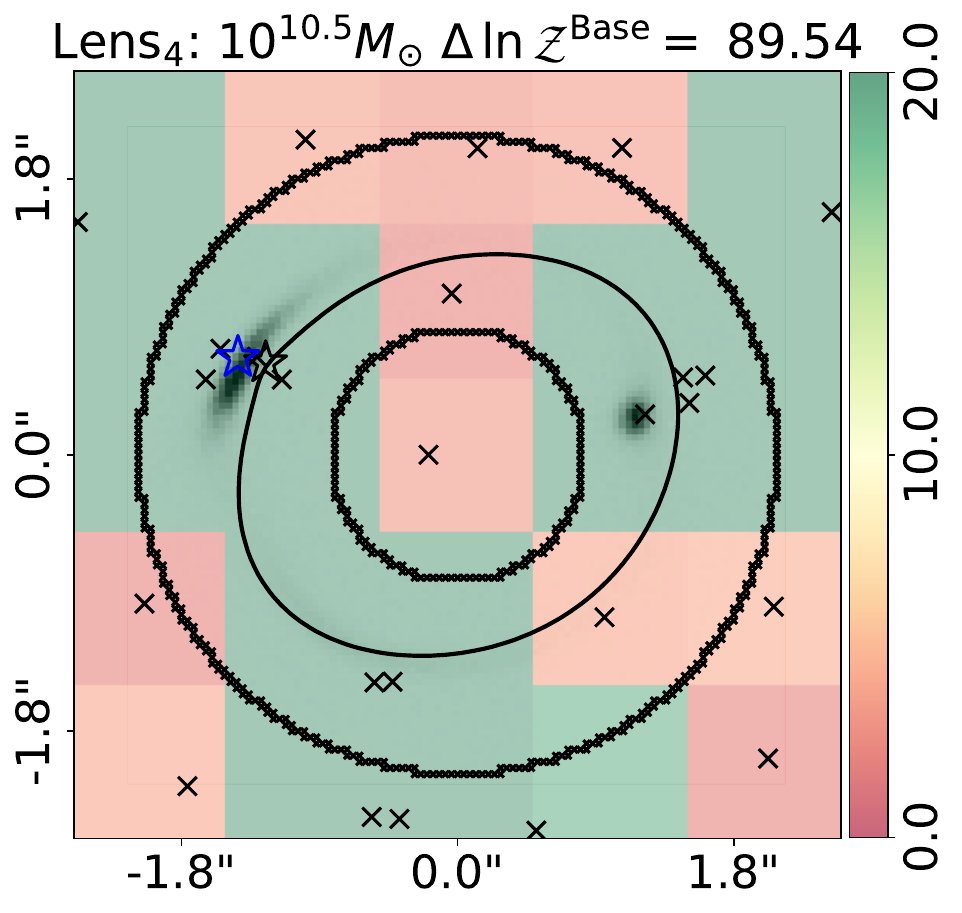}

\caption{
The Bayesian evidence increases $\Delta\,\mathrm{ln}\,\mathcal{Z}^{\rm Base}$ inferred by the subhalo scan for every simulated lens. Fits use a Voronoi mesh source reconstruction, annular masks and a power-law plus shear model with and without a subhalo. Each panel shows a $5 \times 5$ grid of {\tt dynesty} non-linear searches where the green and red overlay shows $\Delta\,\mathrm{ln}\,\mathcal{Z}^{\rm Base}$ values. The input subhalo mass increases from left to right and different lenses are shown across the rows. The (y, x) coordinates of each subhalo are confined to the 2D grid-cell via uniform priors and the inferred values are shown by black crosses. The true subhalo location is marked with a black star. The blue star shows the DM subhalo's maximum a posteriori inferred location for this fit. The colorbar ranges between 0 and 20 so that candidate DM subhalos ($\Delta\,\mathrm{ln}\,\mathcal{Z}^{\rm Base} > 10$ are colored green whereas non-candidates are red. 
}
\label{figure:SimDetect}
\end{figure*}

This section presents the results of fitting the $16$ simulated lenses with our subhalo scanning pipeline. We perform $2$ independent fits to every lens using a Voronoi mesh with the default $3.5\arcsec$ circular mask and an annular mask, where the outer and inner radii are chosen to be small whilst still containing all lensed source emission and the lens light subtraction is fixed to the maximum likelihood lens light model inferred using a circular mask. \cref{table:DetectSim} shows the inferred $\Delta\,\mathrm{ln}\,\mathcal{Z}$ values for every lens, and the inferred subhalo locations and masses for the annular mask fits, with errors quoted at $3\sigma$ confidence intervals.

We begin with the subhalo scanning results for the $4$ simulated lenses which do not contain a DM subhalo, to verify that our analysis does not produce false positives. The first, fifth, ninth and thirteenth rows of \cref{table:DetectSim} show the inferred $\Delta\,\mathrm{ln}\,\mathcal{Z}$ values for each fit. For Lens$_1$, Lens$_2$ and Lens$_3$ fits using either a circular or annular mask produce $\Delta\,\mathrm{ln}\,\mathcal{Z}^{\rm Base} < 10.0$ and therefore correctly do not favour the inclusion of a DM subhalo in the lens model. However, for Lens$_4$, the subhalo scan using the $3.5\arcsec$ circular mask infers $\Delta\,\mathrm{ln}\,\mathcal{Z}^{\rm Base} = 15.25$, incorrectly favouring a DM subhalo even though one is not present in the data. The corresponding annular mask result infers $\Delta\,\mathrm{ln}\,\mathcal{Z}^{\rm Vor} = 2.34$, which is correctly below $10$. Complex and compact sources (similar to those in BELLS-GALLERY) therefore may produce false positives when there is insufficient resolution in the source mesh to resolve it. In the main paper, fits are therefore performed to the HST data using source-only masks which dedicate higher resolution to the source reconstruction (annular masks are not used in the main paper due to lens light residuals requiring a more bespoke masking scheme to remove them). For the simulated lens results, we discuss hereafter only fits using annular masks. 

The $\Delta\,\mathrm{ln}\,\mathcal{Z}$ values for the $12$ other lenses which include subhalos of mass $10^{9.5}$M$_{\odot}$, $10^{10.0}$M$_{\odot}$ or $10^{10.5}$M$_{\odot}$ are also shown in \cref{table:DetectSim}. \cref{figure:SimDetect} shows their corresponding $5 \times 5$ subhalo grid searches for the Voronoi mesh with an annular mask. For all four lenses in our sample including a $10^{10.5}$M$_{\odot}$ subhalo, a model including a subhalo is favoured, with $\Delta\,\mathrm{ln}\,\mathcal{Z}^{\rm Base}$ values of 186.58, 15.3, 638.2 and 89.54. Three $10^{10.0}$M$_{\odot}$ subhalos are also recovered with $\Delta\,\mathrm{ln}\,\mathcal{Z}^{\rm Base}$ values of 39.89, 7.57, 108.11 and 18.24. The $10^{9.5}$M$_{\odot}$ subhalos are recovered in Lens$_1$ and Lens$_3$ with $\Delta\,\mathrm{ln}\,\mathcal{Z}^{\rm Base}$ values of 17.67 and 33.75. Provided the detection criteria of $\Delta\,\mathrm{ln}\,\mathcal{Z} > 10.0$ is met, the subhalo positions are recovered in all but one lens and masses are recovered -- in all but three lenses -- within the 3$\sigma$ credible regions of the posteriors. For these three lenses, the masses are close to the input, but are offset due to a slightly inaccurate lens mass model (see \cref{A_Pre_Subhalo}).

When a subhalo is present in the data, cells away from the subhalo's true location may show smaller increases in evidence, both in the cells neighbouring the true subhalo and further away but in proximity to the lensed source's emission. For example, in the top-right panel of \cref{figure:SimDetect} (Lens$_{\rm 1}$ with a $10^{10.5}$M$_{\odot}$ DM subhalo) values of $\Delta\,\mathrm{ln}\,\mathcal{Z}^{\rm Base} > 150$ are seen surrounding the true subhalo (black star), but values of $\Delta\,\mathrm{ln}\,\mathcal{Z}^{\rm Base} > 50$ are also seen towards the bottom-right and bottom-left of the grid. These models infer subhalo masses above or below the true input value in order to produce a lensing signal that is similar to the signal produced by the true DM subhalo. These are not false positives, because they are due to a subhalo being truly present somewhere in the data. Models which assume a subhalo offset from a true subhalo can therefore mimic its perturbing effect by rescaling its mass. The same behaviour is discussed in \citet{He2023} and therefore must be considered when interpreting the SLACS and BELLS-GALLERY lenses. This also highlights the multimodality of the lens model parameter space and demonstrates why our grid-search of \texttt{dynesty} fits removes it.

Our subhalo analysis therefore successfully detect DM subhalos down to masses of at least $10^{9.5}$M$_{\odot}$ for HST-quality data. It does not infer false positives, provided there is sufficient resolution in the source reconstruction. These conclusions are only valid for simulated data where the parametric lens light and mass models used to simulate the images are the same as those used to fit them. 

\subsection{Justification of Source Analysis}

Columns 8, 9, and 10 of \cref{table:DetectSim} show subhalo scanning results for fits using annular masks and the following source analysis variants: (i) switching off Voronoi natural neighbor interpolation such that each image sub-pixel maps to one Voronoi cell; (ii) using the Voronoi mesh equivalent of gradient regularization (see WD03 and \citealt{Nightingale2015}) and; (iii) doing both simultaneously. The subhalo inferences are as good as before. For all lenses not including a subhalo the different source variants correctly do not favour a subhalo, and when a subhalo is included it is recovered in the majority of lenses. Our DM subhalo results are therefore not sensitive to the specifics of the source analysis. This is because a Bayesian evidence increase of $\Delta\,\mathrm{ln}\,\mathcal{Z} > 10$ corresponds to a $\gtrsim 5\sigma$ result. Changing such a large Bayesian evidence increase via the source regularization or interpolation would require a much more radical change in the priors about how smooth galaxies are.

The final four columns of \cref{table:DetectSim} show the overall log Bayesian evidence values, $\mathrm{ln}\,\mathcal{Z}$, using the different source variants for the lens model including a DM subhalo. These can be compared in order to determine the optimal source analysis. In 15 out of 16 lenses, the highest evidence source analysis uses both natural neighbor Voronoi interpolation and adaptive luminosity based regularization, justifying our choice to use it in the main paper. \cref{table:DetectSim} shows that using adaptive regularization is more important than interpolation, a result that is consistent with the findings of \citealt{Nightingale2018}, in particular figures 6, 7, 8 and section 5, who showed that for compact sources different regions of the source reconstruction require different levels of regularization in order to ensure a clean reconstruction, a result also discussed by \citet{Vegetti2014}. Recently, \citet{Vernardos2022} performed a similar comparison of different source methods and argued in favour of `observationally motivated priors' for the source regularization. Detailed comparison is beyond the scope of this work. 


\subsection{Mass Models Adjust Parameters To Absorb Subhalo Perturbation}\label{A_Pre_Subhalo}

\begin{table*}
\Large
\resizebox{\linewidth}{!}{
\begin{tabular}{ l | l | l | l | l | l | l | l | l | l} 
\multicolumn{1}{p{2.0cm}|}{\centering Lens Name} 
& \multicolumn{1}{p{1.0cm}|}{\centering Subhalo Mass} 
& \multicolumn{1}{p{0.7cm}|}{$\epsilon_{\rm 1}^{\rm{mass}}$}
& \multicolumn{1}{p{0.7cm}|}{$\epsilon_{\rm 2}^{\rm{mass}}$}
& \multicolumn{1}{p{0.7cm}|}{$\theta_{\rm Ein}^{\rm{mass}}$ $(\arcsec)$} 
& \multicolumn{1}{p{0.7cm}|}{$\gamma_{\rm 1}^{\rm{ext}}$} 
& \multicolumn{1}{p{0.7cm}|}{$\gamma_{\rm 2}^{\rm{ext}}$}
& \multicolumn{1}{p{0.7cm}|}{$y^{\rm{mass}}$ $(\arcsec)$} 
& \multicolumn{1}{p{0.7cm}|}{$x^{\rm{mass}}$ $(\arcsec)$} 
& \multicolumn{1}{p{0.7cm}}{$\gamma^{\rm{mass}}$} 
\\ \hline

& & & & & & & & & \\[-6pt]

$\mathrm{Lens}_{1}$ & No Subhalo & $0.0106^{+0.0033}_{-0.0029}$ & $-0.0159^{+0.0054}_{-0.0051}$ & $1.9244^{+0.0031}_{-0.0021}$ & $-0.0005^{+0.0015}_{-0.0022}$ & $0.0524^{+0.0080}_{-0.0040}$ & $-0.0000^{+0.0013}_{-0.0015}$ & $0.0005^{+0.0015}_{-0.0013}$ & $2.0725^{+0.1129}_{-0.0636}$ \\[2pt]
$\mathrm{Lens}_{1}$ & $10^{9.5}M_{\odot}$ & $0.0143^{+0.0036}_{-0.0028}$ & $-0.0250^{+0.0049}_{-0.0047}$ & $1.9273^{+0.0044}_{-0.0040}$ & $0.0013^{+0.0019}_{-0.0017}$ & $0.0485^{+0.0066}_{-0.0066}$ & $0.0005^{+0.0016}_{-0.0017}$ & $0.0003^{+0.0010}_{-0.0011}$ & $2.0509^{+0.0997}_{-0.0937}$ \\[2pt]
$\mathrm{Lens}_{1}$ & $10^{10.0}M_{\odot}$ & $0.0132^{+0.0038}_{-0.0029}$ & $-0.0245^{+0.0046}_{-0.0049}$ & $1.9276^{+0.0046}_{-0.0040}$ & $0.0010^{+0.0022}_{-0.0019}$ & $0.0467^{+0.0066}_{-0.0055}$ & $0.0020^{+0.0017}_{-0.0014}$ & $0.0010^{+0.0012}_{-0.0012}$ & $2.0180^{+0.0972}_{-0.0871}$ \\[2pt]
$\mathrm{Lens}_{1}$ & $10^{10.5}M_{\odot}$ & $0.0192^{+0.0038}_{-0.0036}$ & $-0.0336^{+0.0043}_{-0.0049}$ & $1.9414^{+0.0042}_{-0.0056}$ & $0.0024^{+0.0021}_{-0.0023}$ & $0.0507^{+0.0060}_{-0.0067}$ & $0.0055^{+0.0015}_{-0.0012}$ & $0.0015^{+0.0010}_{-0.0012}$ & $2.0913^{+0.0792}_{-0.0963}$ \\[2pt]

$\mathrm{Lens}_{2}$ & No Subhalo & $0.1179^{+0.0257}_{-0.0268}$ & $0.0309^{+0.0204}_{-0.0217}$ & $1.1002^{+0.0059}_{-0.0078}$ & $0.0118^{+0.0184}_{-0.0193}$ & $0.0003^{+0.0136}_{-0.0154}$ & $-0.0089^{+0.0187}_{-0.0184}$ & $-0.0089^{+0.0130}_{-0.0161}$ & $1.9036^{+0.0404}_{-0.0467}$ \\[2pt]
$\mathrm{Lens}_{2}$ & $10^{9.5}M_{\odot}$ & $0.1387^{+0.0327}_{-0.0211}$ & $0.0202^{+0.0181}_{-0.0176}$ & $1.1093^{+0.0047}_{-0.0052}$ & $0.0144^{+0.0225}_{-0.0147}$ & $-0.0067^{+0.0105}_{-0.0136}$ & $-0.0054^{+0.0169}_{-0.0181}$ & $-0.0054^{+0.0093}_{-0.0124}$ & $1.9498^{+0.0423}_{-0.0382}$ \\[2pt]
$\mathrm{Lens}_{2}$ & $10^{10.0}M_{\odot}$ & $0.1848^{+0.0372}_{-0.0316}$ & $0.0150^{+0.0218}_{-0.0180}$ & $1.1182^{+0.0089}_{-0.0082}$ & $0.0388^{+0.0260}_{-0.0211}$ & $-0.0055^{+0.0140}_{-0.0138}$ & $0.0007^{+0.0166}_{-0.0194}$ & $-0.0103^{+0.0146}_{-0.0165}$ & $1.9904^{+0.0590}_{-0.0429}$ \\[2pt]
$\mathrm{Lens}_{2}$ & $10^{10.5}M_{\odot}$ & $0.1643^{+0.0317}_{-0.0258}$ & $-0.0019^{+0.0170}_{-0.0203}$ & $1.1402^{+0.0065}_{-0.0059}$ & $0.0098^{+0.0199}_{-0.0169}$ & $-0.0070^{+0.0109}_{-0.0129}$ & $0.0204^{+0.0130}_{-0.0141}$ & $-0.0087^{+0.0112}_{-0.0116}$ & $2.0662^{+0.0445}_{-0.0400}$ \\[2pt]

$\mathrm{Lens}_{3}$ & No Subhalo & $0.0109^{+0.0022}_{-0.0021}$ & $-0.0227^{+0.0030}_{-0.0030}$ & $1.9251^{+0.0028}_{-0.0027}$ & $-0.0002^{+0.0013}_{-0.0012}$ & $0.0494^{+0.0045}_{-0.0043}$ & $-0.0012^{+0.0013}_{-0.0010}$ & $0.0000^{+0.0009}_{-0.0010}$ & $2.0616^{+0.0767}_{-0.0736}$ \\[2pt]
$\mathrm{Lens}_{3}$ & $10^{9.5}M_{\odot}$ & $0.0145^{+0.0018}_{-0.0021}$ & $-0.0253^{+0.0030}_{-0.0022}$ & $1.9278^{+0.0032}_{-0.0035}$ & $0.0015^{+0.0014}_{-0.0012}$ & $0.0479^{+0.0037}_{-0.0052}$ & $0.0010^{+0.0009}_{-0.0009}$ & $0.0007^{+0.0010}_{-0.0009}$ & $2.0431^{+0.0642}_{-0.0795}$ \\[2pt]
$\mathrm{Lens}_{3}$ & $10^{10.0}M_{\odot}$ & $0.0158^{+0.0021}_{-0.0015}$ & $-0.0212^{+0.0024}_{-0.0024}$ & $1.9284^{+0.0027}_{-0.0033}$ & $0.0022^{+0.0013}_{-0.0010}$ & $0.0492^{+0.0037}_{-0.0047}$ & $0.0017^{+0.0008}_{-0.0007}$ & $0.0013^{+0.0009}_{-0.0008}$ & $2.0356^{+0.0597}_{-0.0749}$ \\[2pt]
$\mathrm{Lens}_{3}$ & $10^{10.5}M_{\odot}$ & $0.0264^{+0.0189}_{-0.0016}$ & $-0.0380^{+0.0038}_{-0.0057}$ & $1.9685^{+0.0110}_{-0.0026}$ & $-0.0051^{+0.0052}_{-0.0005}$ & $0.0882^{+0.0029}_{-0.0007}$ & $0.0059^{+0.0024}_{-0.0004}$ & $0.0031^{+0.0006}_{-0.0005}$ & $2.6520^{+0.0391}_{-0.0244}$ \\[2pt]

$\mathrm{Lens}_{4}$ & No Subhalo & $0.1640^{+0.0167}_{-0.0132}$ & $0.0486^{+0.0095}_{-0.0098}$ & $1.4134^{+0.0224}_{-0.0195}$ & $0.0523^{+0.0070}_{-0.0105}$ & $-0.0037^{+0.0077}_{-0.0082}$ & $0.0001^{+0.0061}_{-0.0063}$ & $0.0019^{+0.0055}_{-0.0081}$ & $2.1476^{+0.1228}_{-0.0913}$ \\[2pt]
$\mathrm{Lens}_{4}$ & $10^{9.5}M_{\odot}$ & $0.1622^{+0.0178}_{-0.0161}$ & $0.0561^{+0.0101}_{-0.0087}$ & $1.4325^{+0.0181}_{-0.0197}$ & $0.0415^{+0.0078}_{-0.0085}$ & $-0.0074^{+0.0057}_{-0.0066}$ & $0.0042^{+0.0046}_{-0.0050}$ & $0.0017^{+0.0081}_{-0.0042}$ & $2.2550^{+0.0868}_{-0.1007}$ \\[2pt]
$\mathrm{Lens}_{4}$ & $10^{10.0}M_{\odot}$ & $0.1937^{+0.0291}_{-0.0139}$ & $0.0541^{+0.0112}_{-0.0118}$ & $1.4713^{+0.0309}_{-0.0196}$ & $0.0413^{+0.0060}_{-0.0054}$ & $-0.0181^{+0.0064}_{-0.0078}$ & $0.0084^{+0.0053}_{-0.0041}$ & $0.0018^{+0.0063}_{-0.0041}$ & $2.3810^{+0.0796}_{-0.0822}$ \\[2pt]
$\mathrm{Lens}_{4}$ & $10^{10.5}M_{\odot}$ & $0.2317^{+0.0251}_{-0.0270}$ & $0.0733^{+0.0152}_{-0.0129}$ & $1.5363^{+0.0339}_{-0.0295}$ & $0.0343^{+0.0059}_{-0.0054}$ & $-0.0286^{+0.0033}_{-0.0046}$ & $0.0119^{+0.0040}_{-0.0041}$ & $0.0015^{+0.0042}_{-0.0042}$ & $2.5390^{+0.0642}_{-0.0622}$ \\[2pt]

\end{tabular}
}
\caption{The median PDF parameter estimates with $3\sigma$ confidence intervals for the power-law + shear model fits to each simulated lens dataset. Fits use a Voronoi mesh and annular mask and are performed directly before the subhalo detection grid-search.}
\label{table:MassModelBefore}
\end{table*}

\begin{figure*}
\centering
\includegraphics[width=0.19\textwidth]{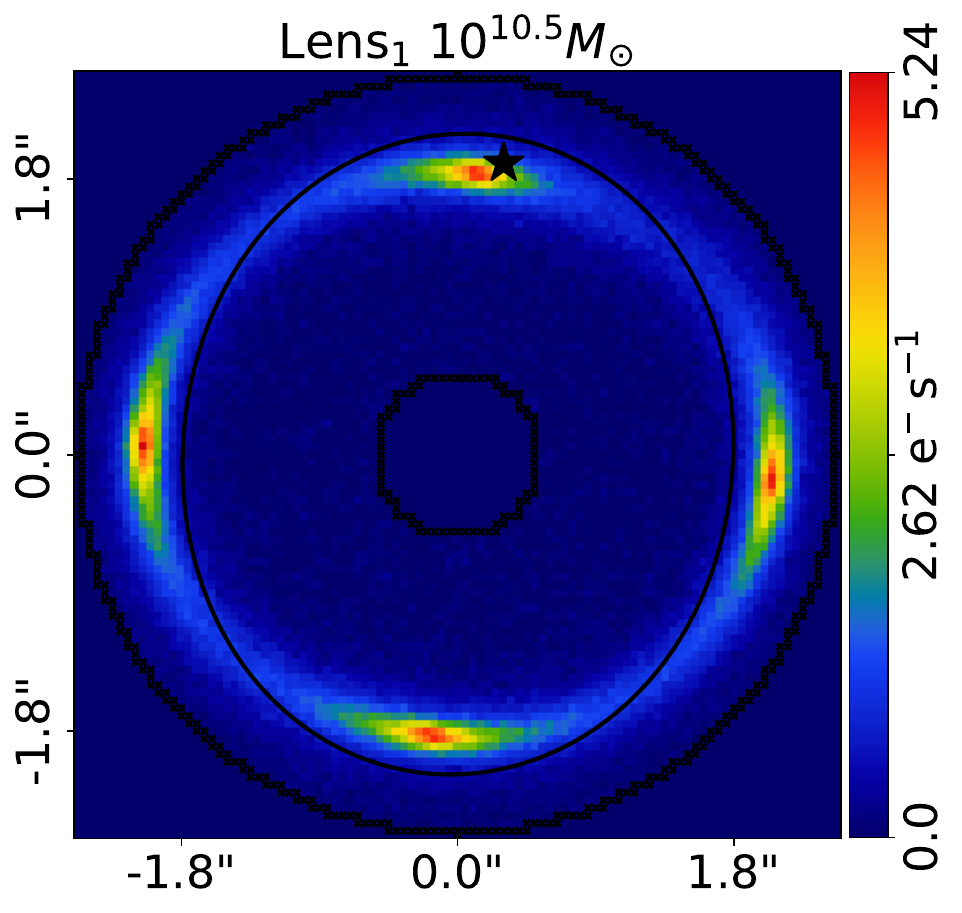}
\includegraphics[width=0.19\textwidth]{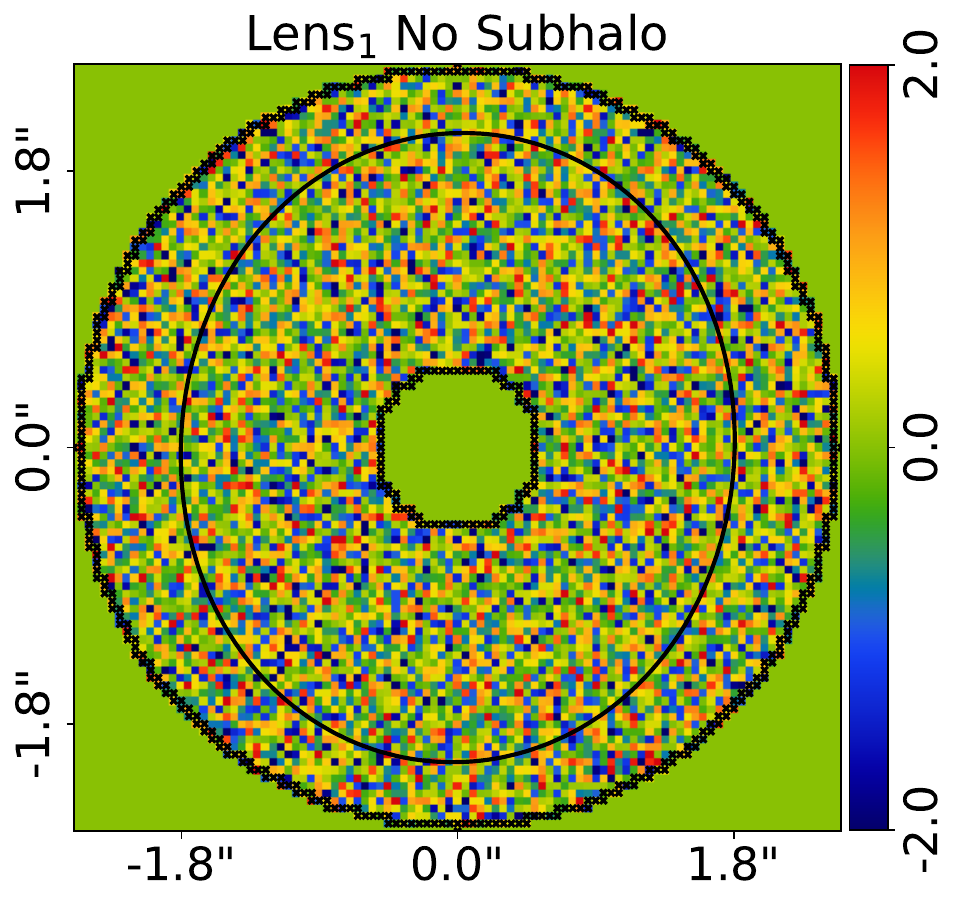}
\includegraphics[width=0.19\textwidth]{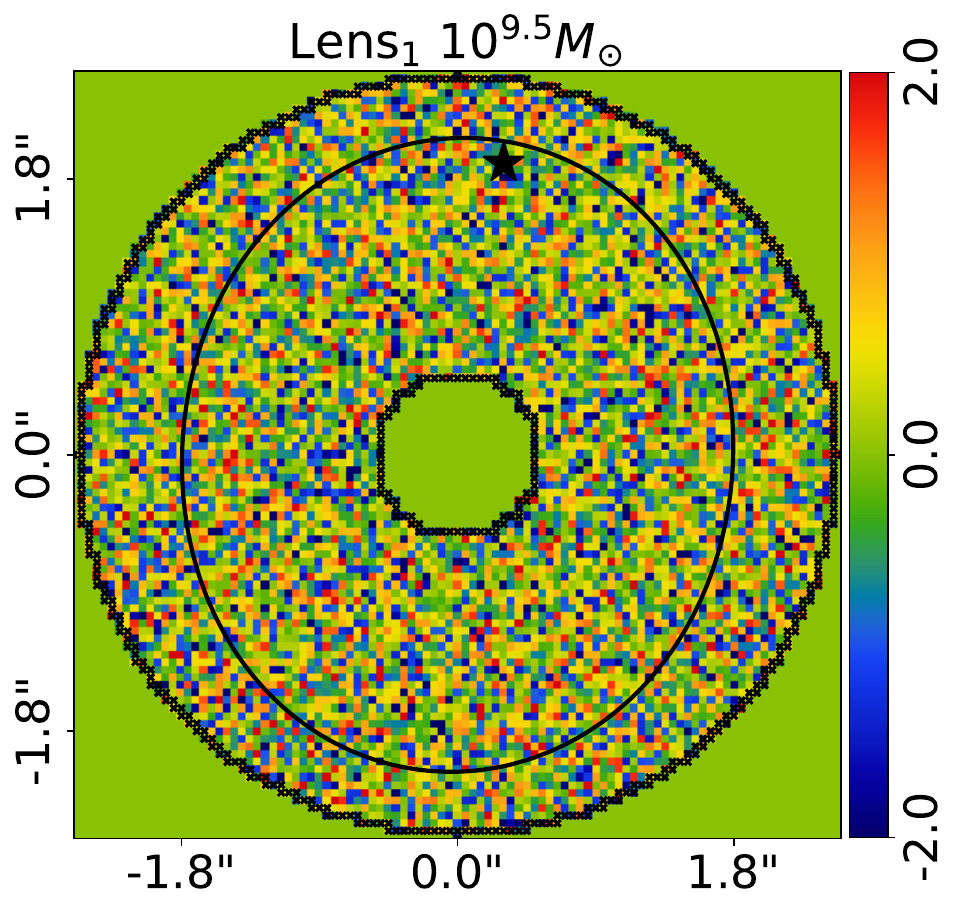}
\includegraphics[width=0.19\textwidth]{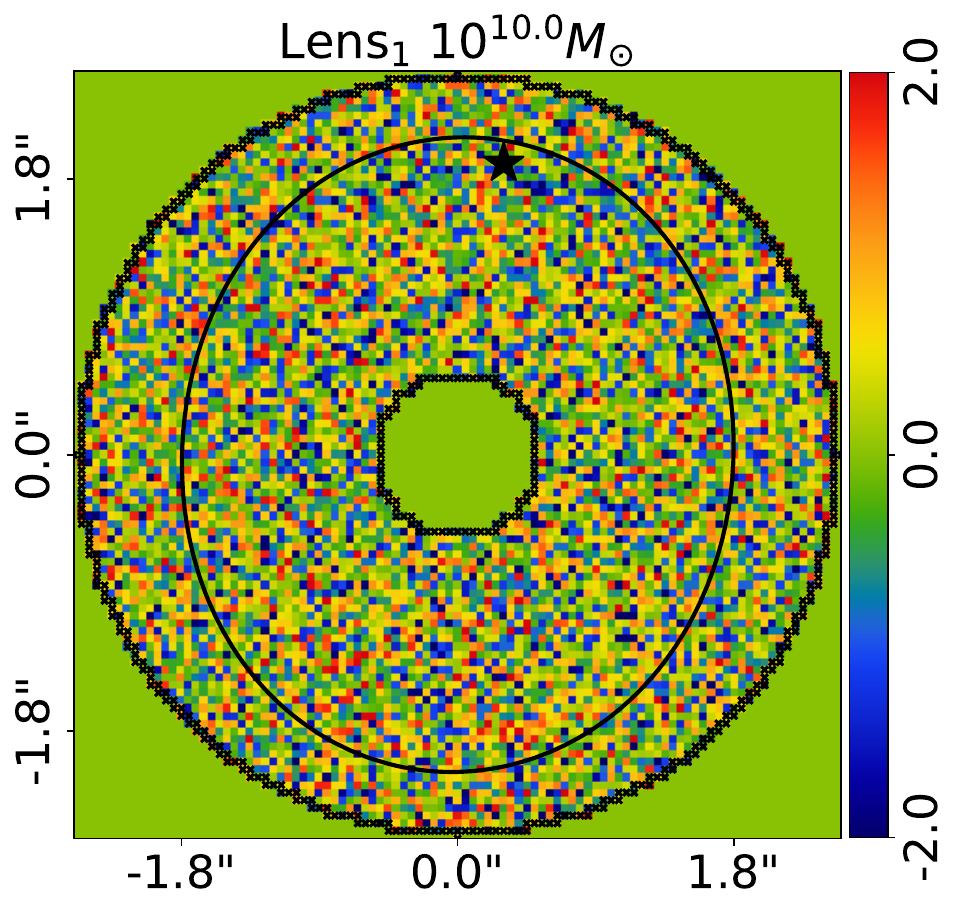}
\includegraphics[width=0.19\textwidth]{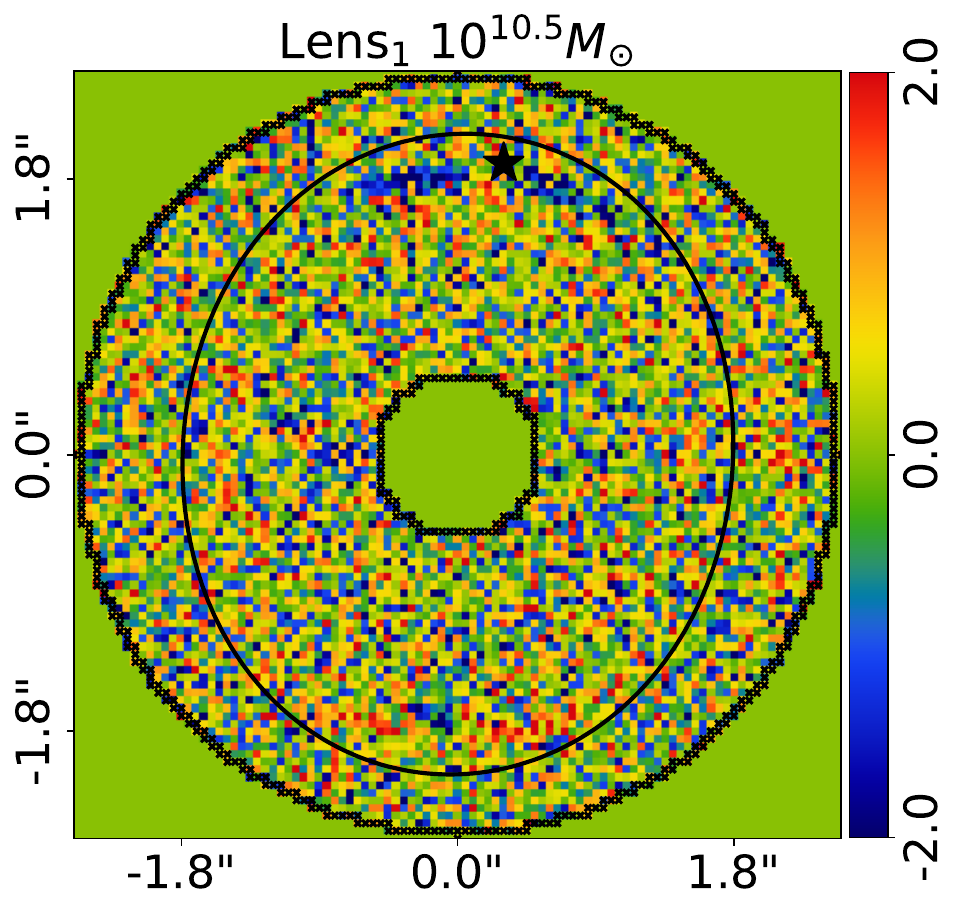}

\includegraphics[width=0.19\textwidth]{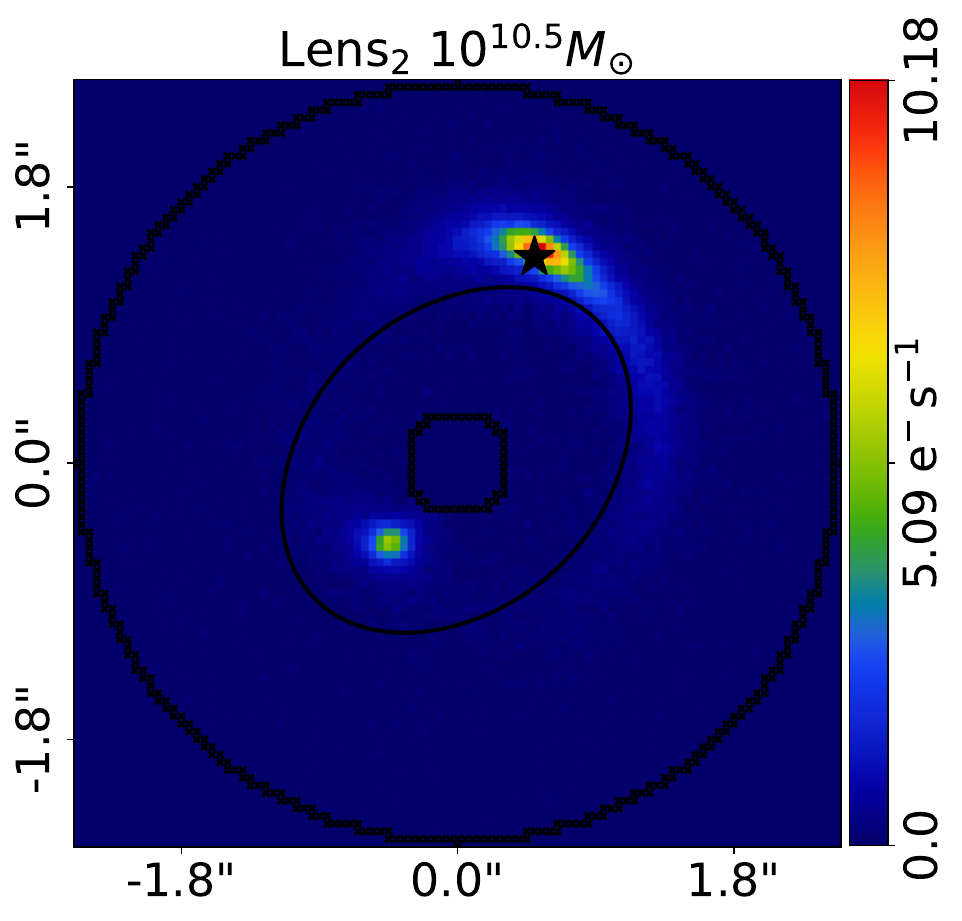}
\includegraphics[width=0.19\textwidth]{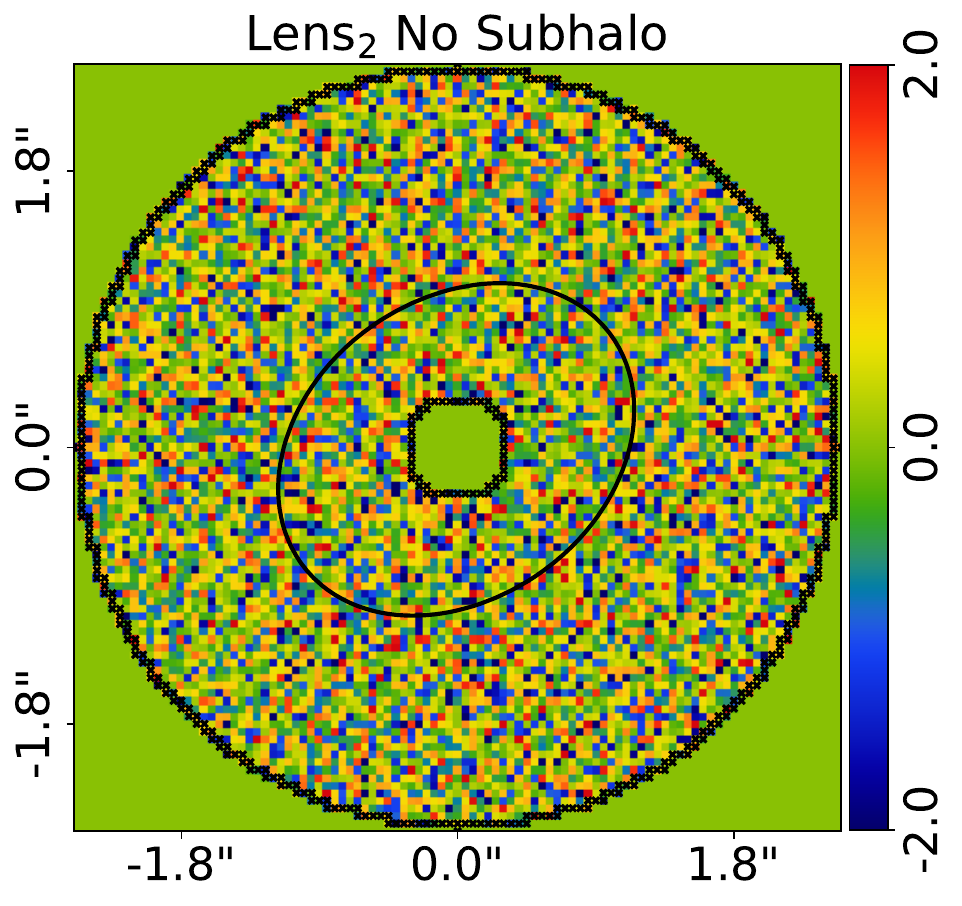}
\includegraphics[width=0.19\textwidth]{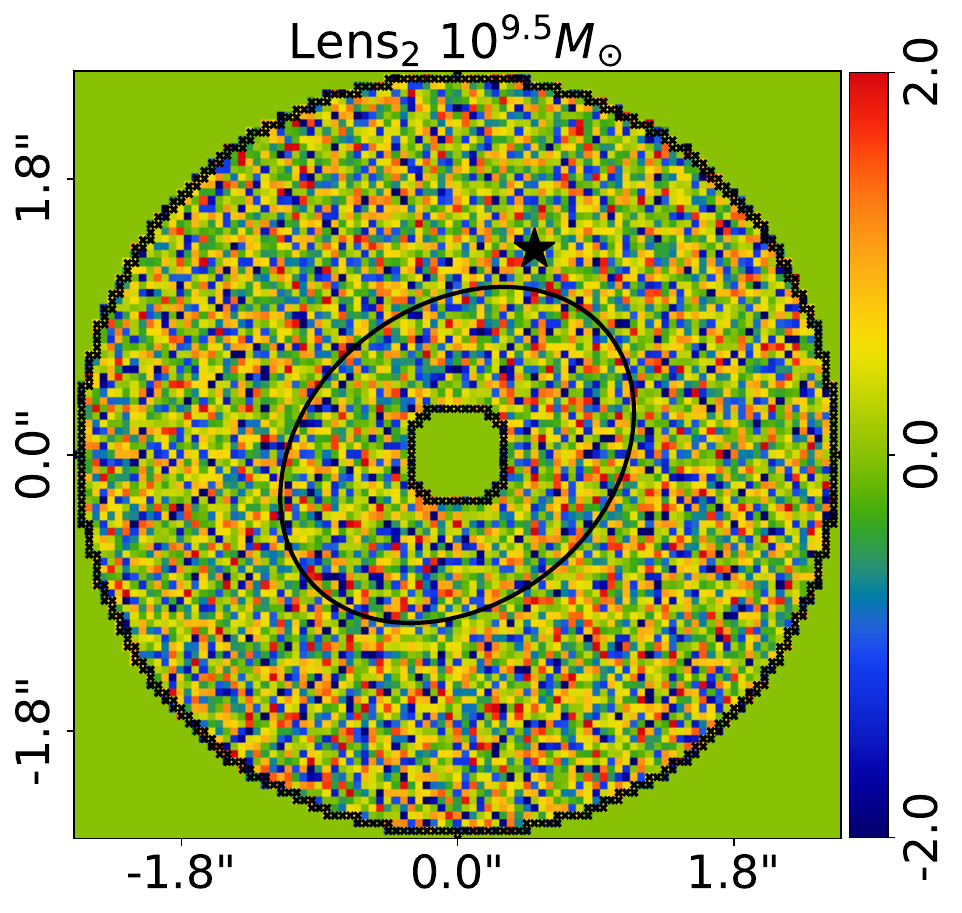}
\includegraphics[width=0.19\textwidth]{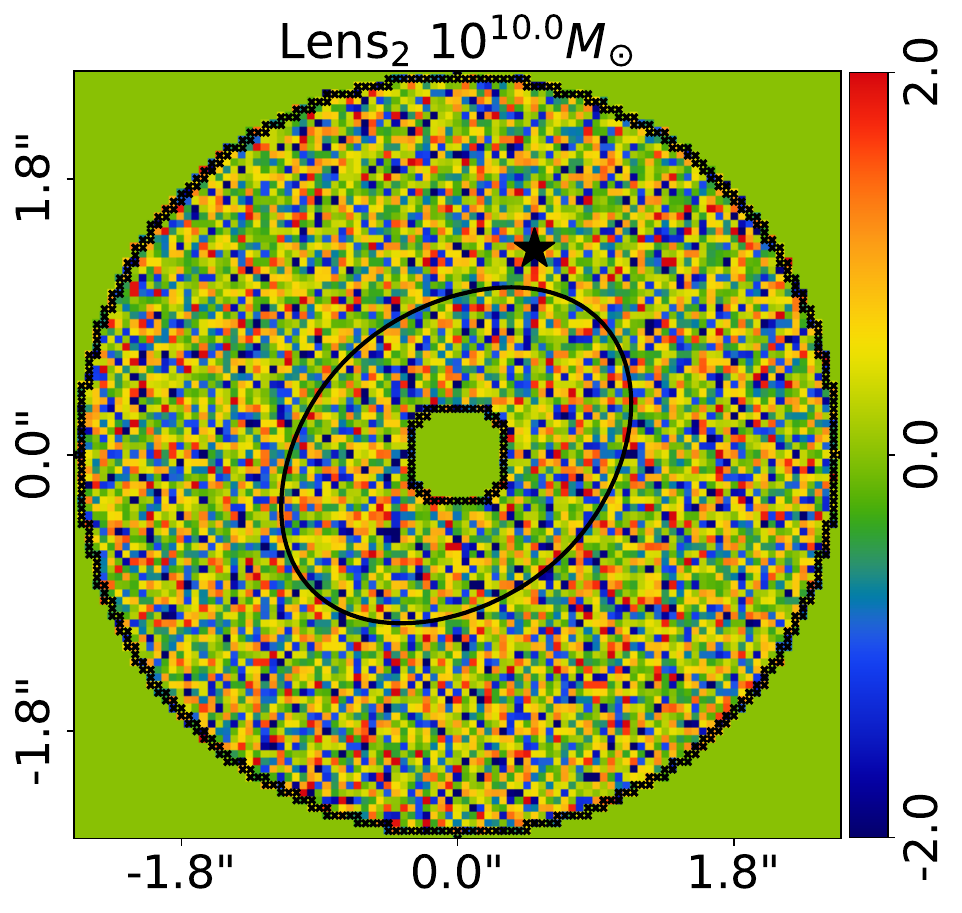}
\includegraphics[width=0.19\textwidth]{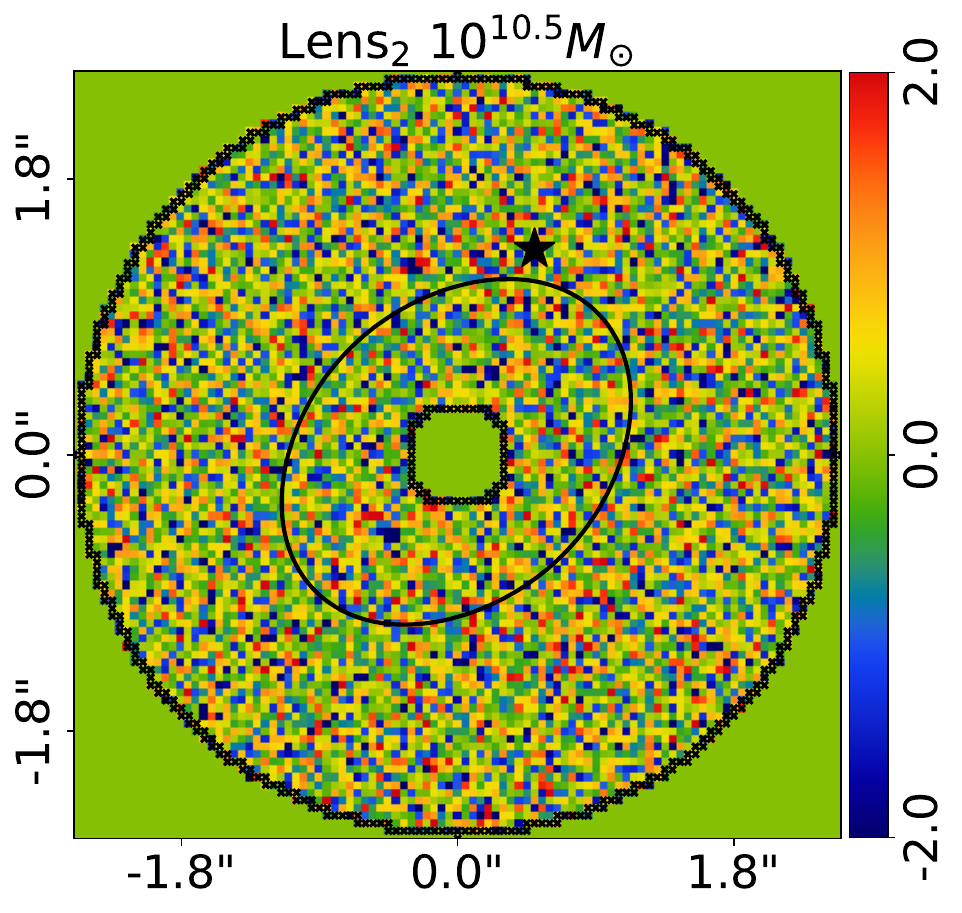}

\includegraphics[width=0.19\textwidth]{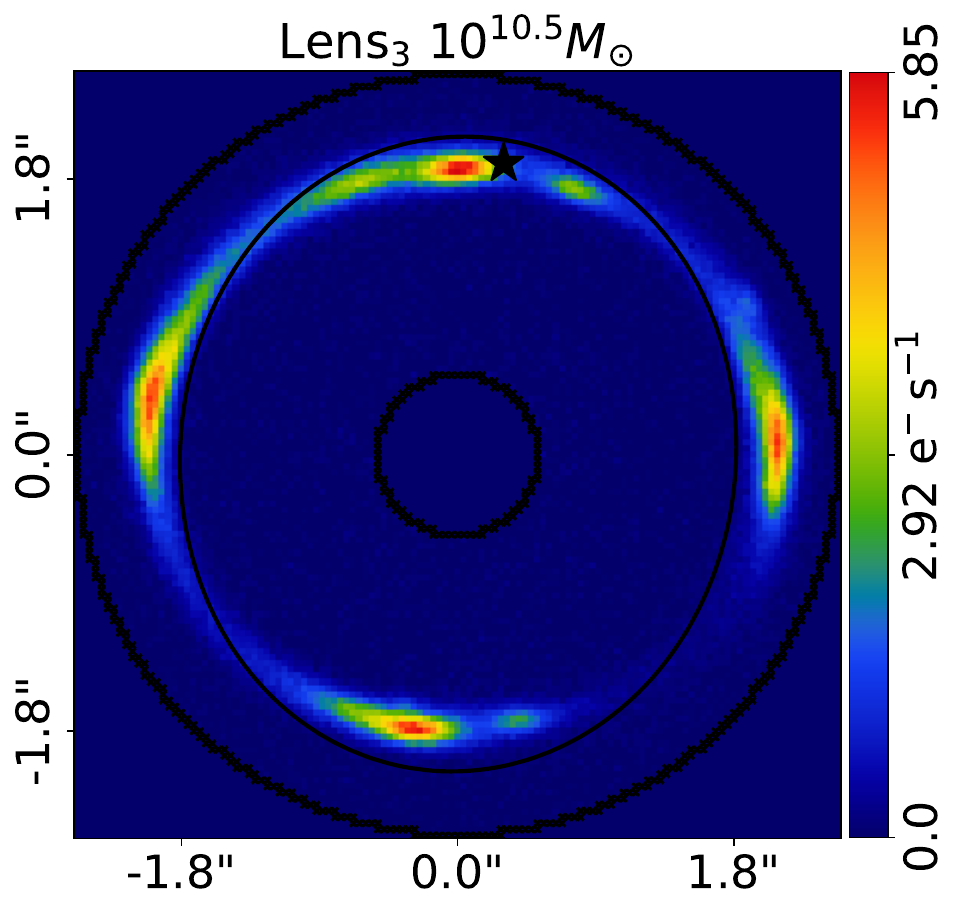}
\includegraphics[width=0.19\textwidth]{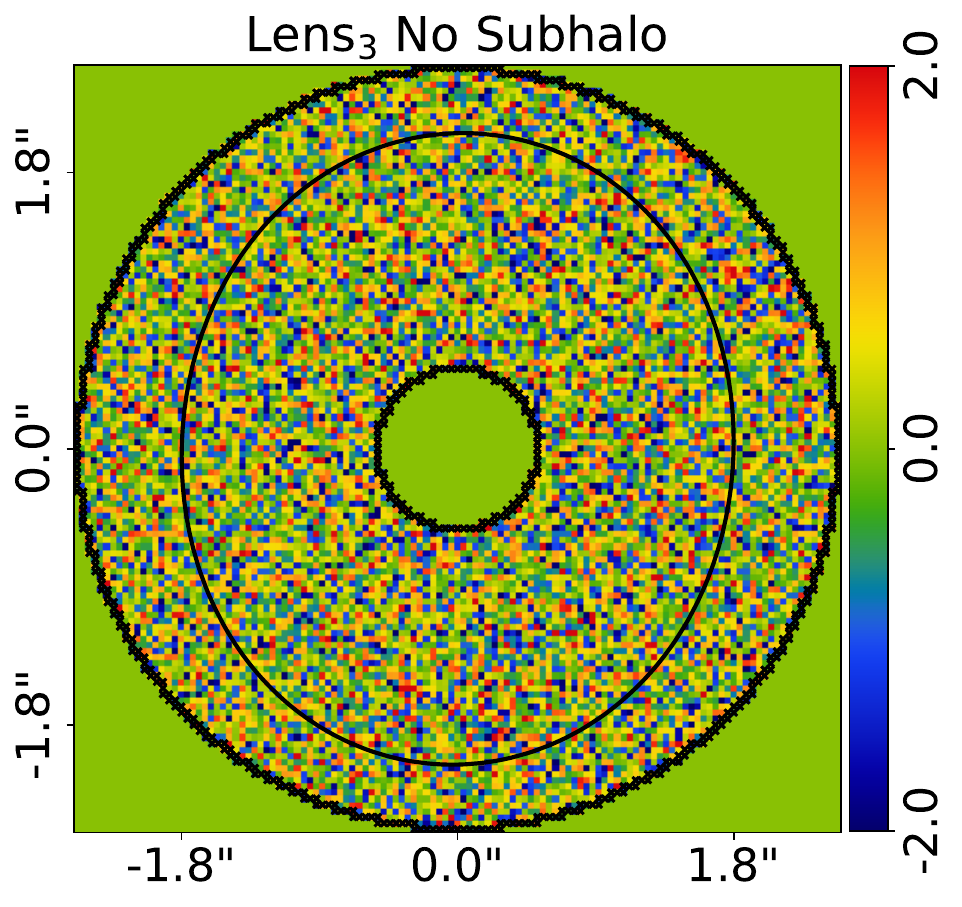}
\includegraphics[width=0.19\textwidth]{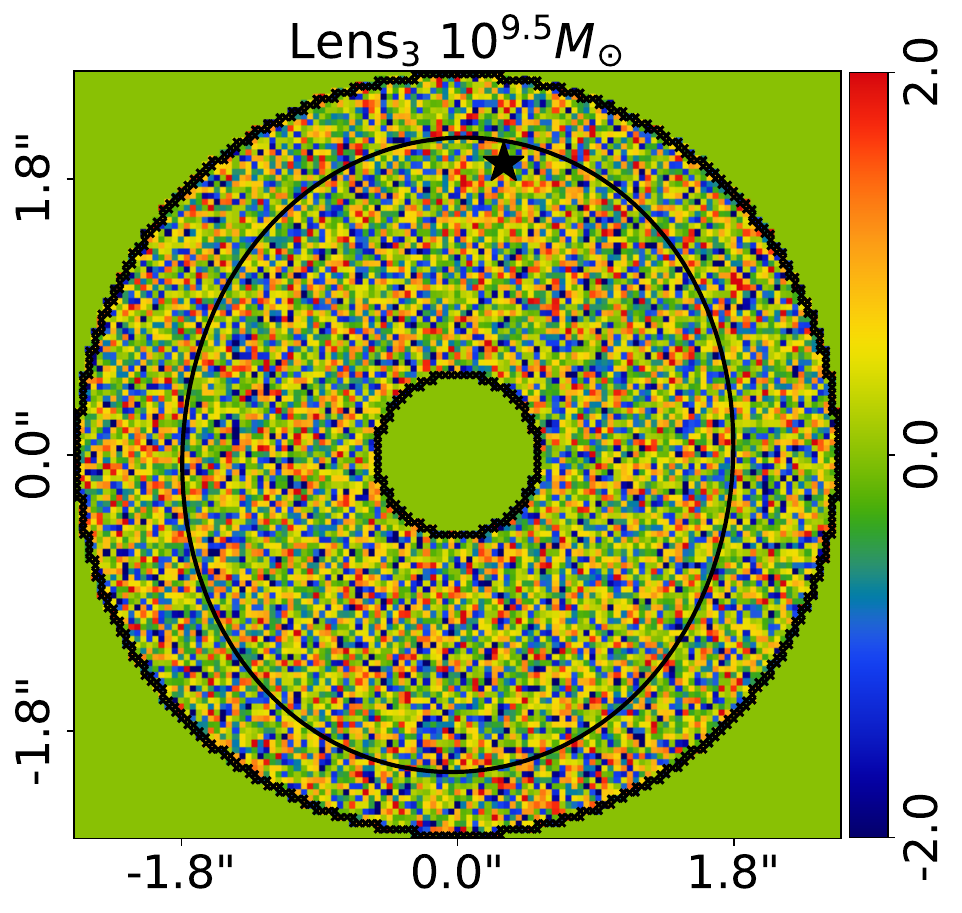}
\includegraphics[width=0.19\textwidth]{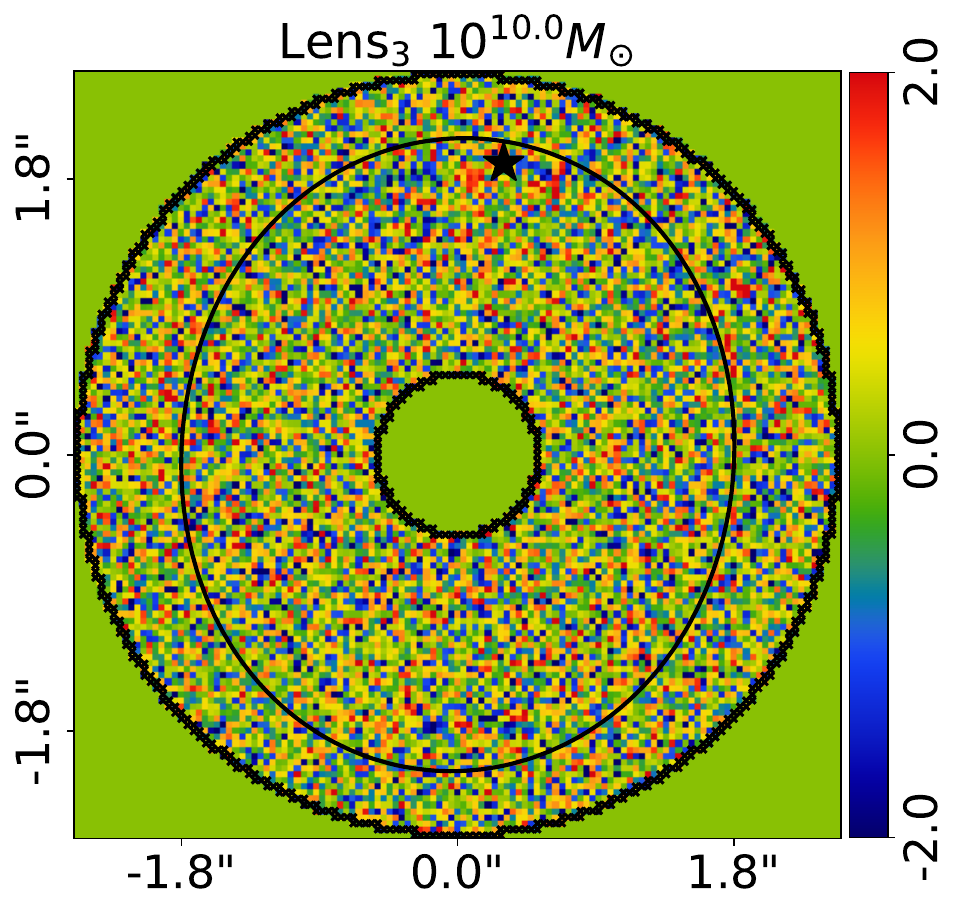}
\includegraphics[width=0.19\textwidth]{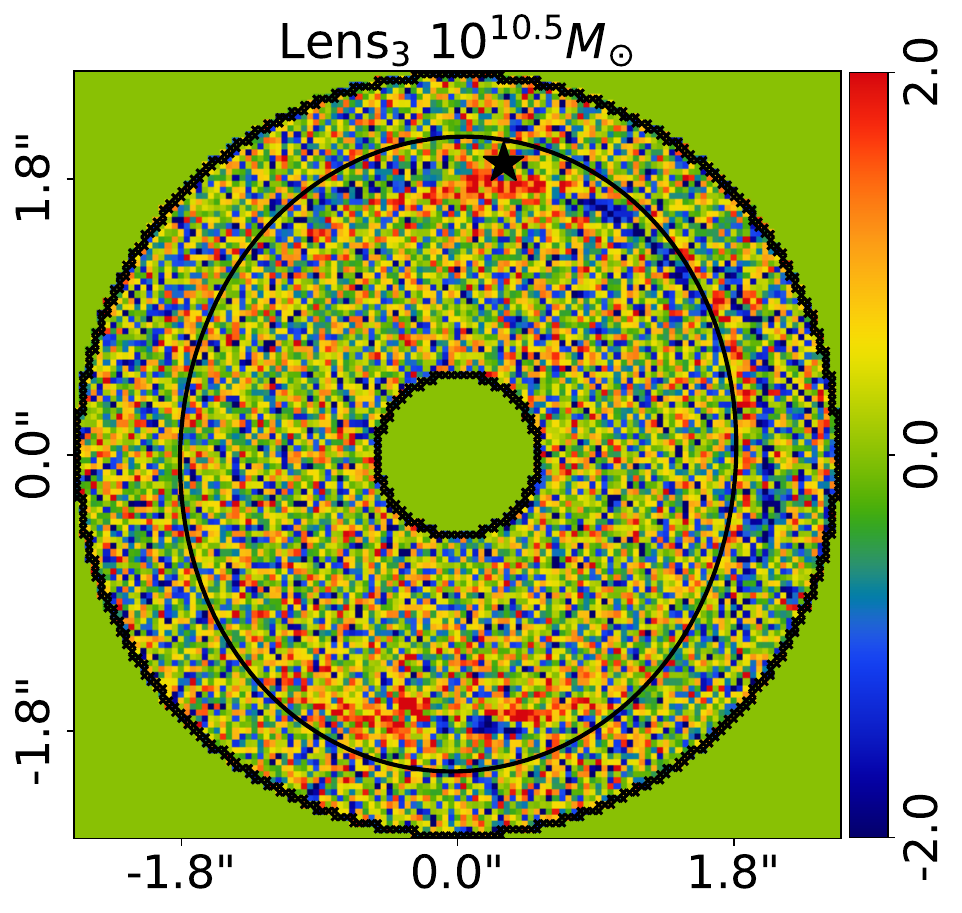}

\includegraphics[width=0.19\textwidth]{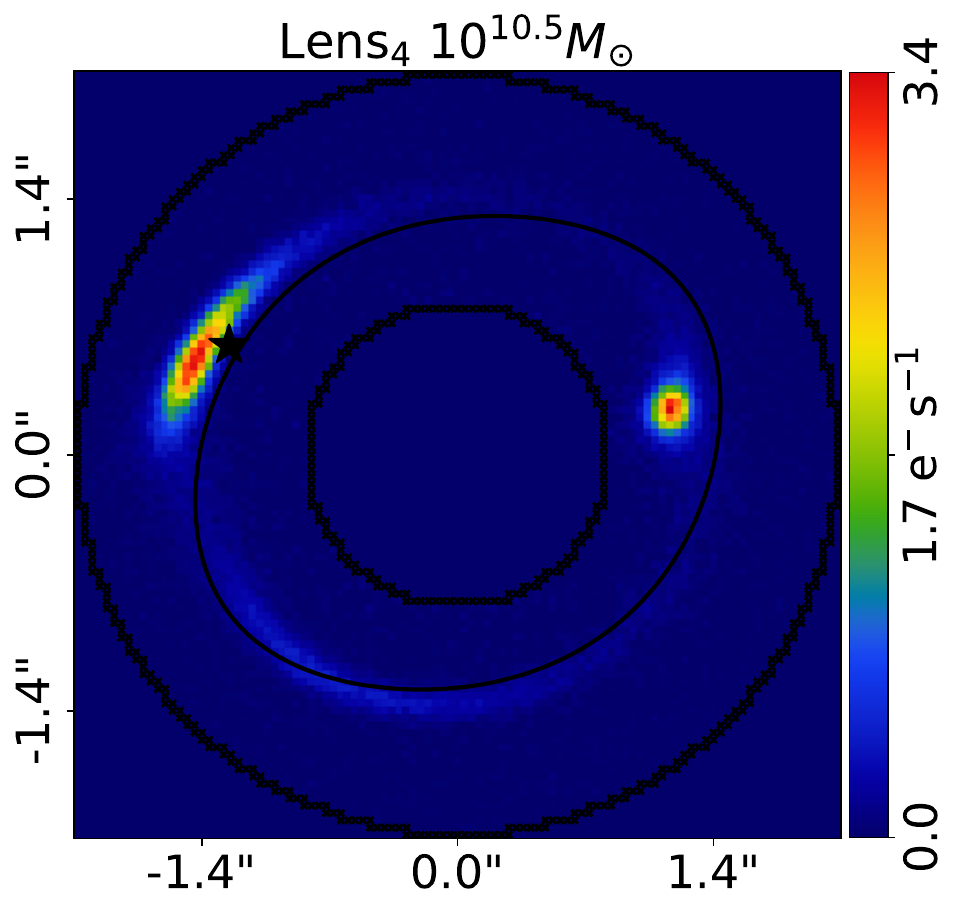}
\includegraphics[width=0.19\textwidth]{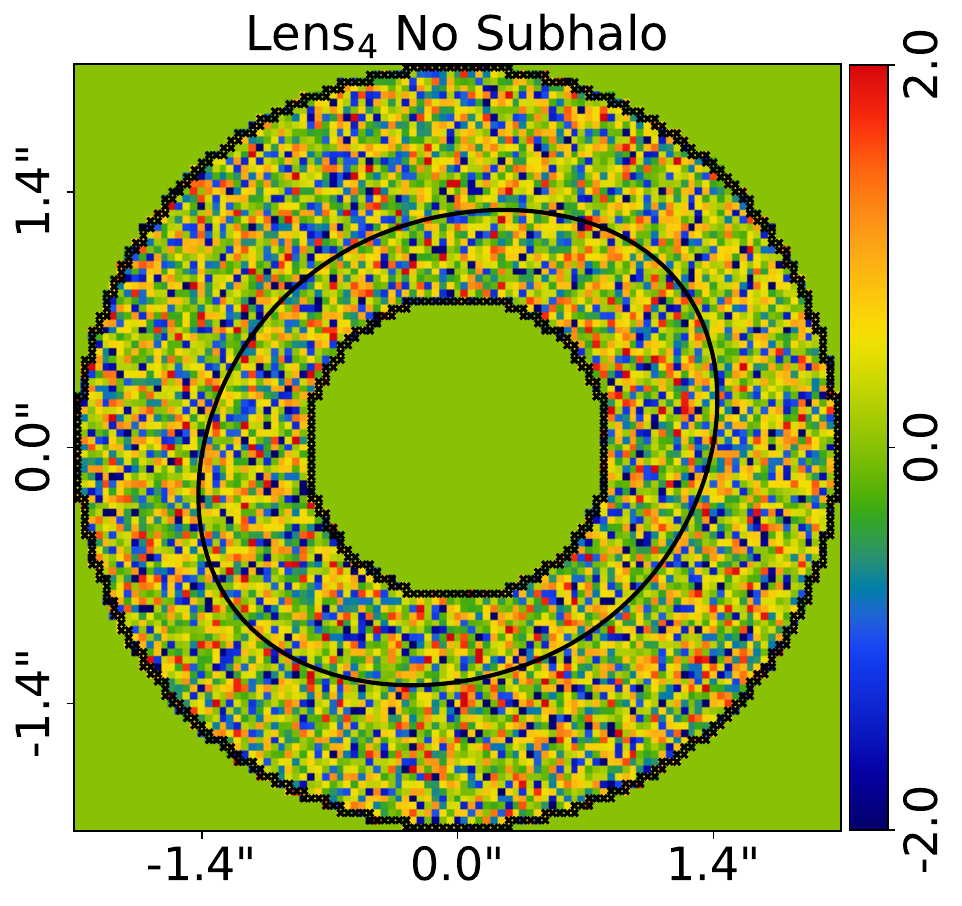}
\includegraphics[width=0.19\textwidth]{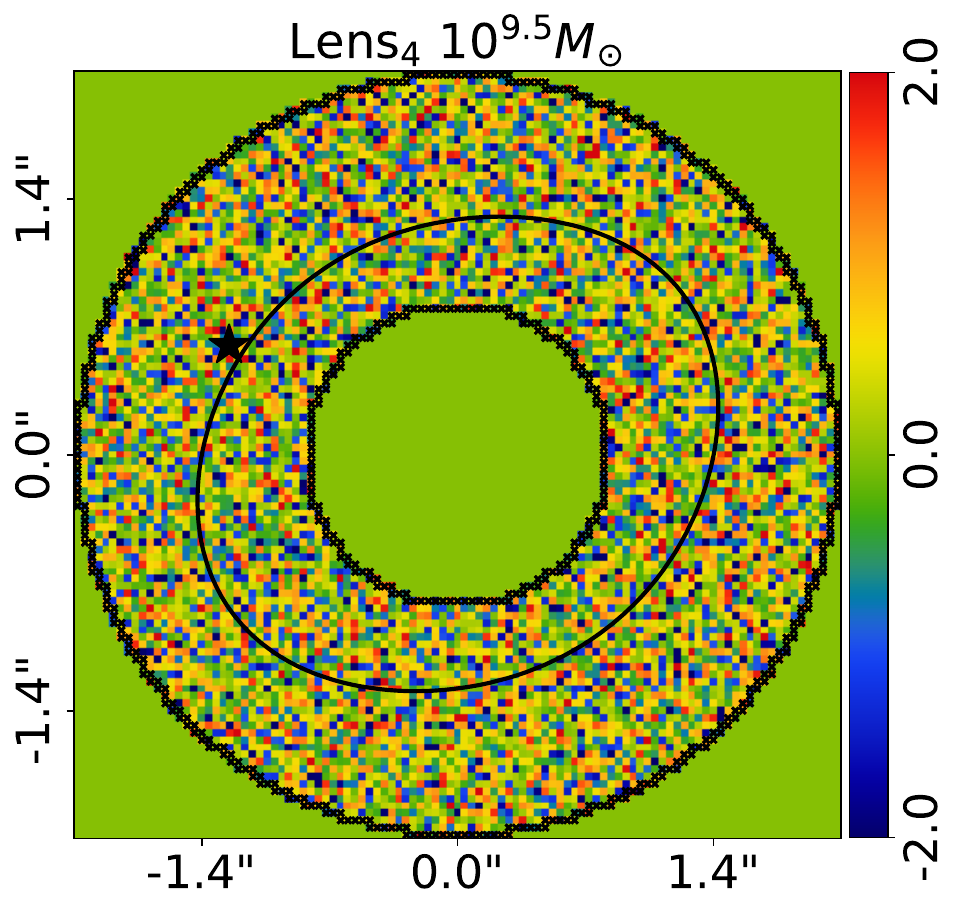}
\includegraphics[width=0.19\textwidth]{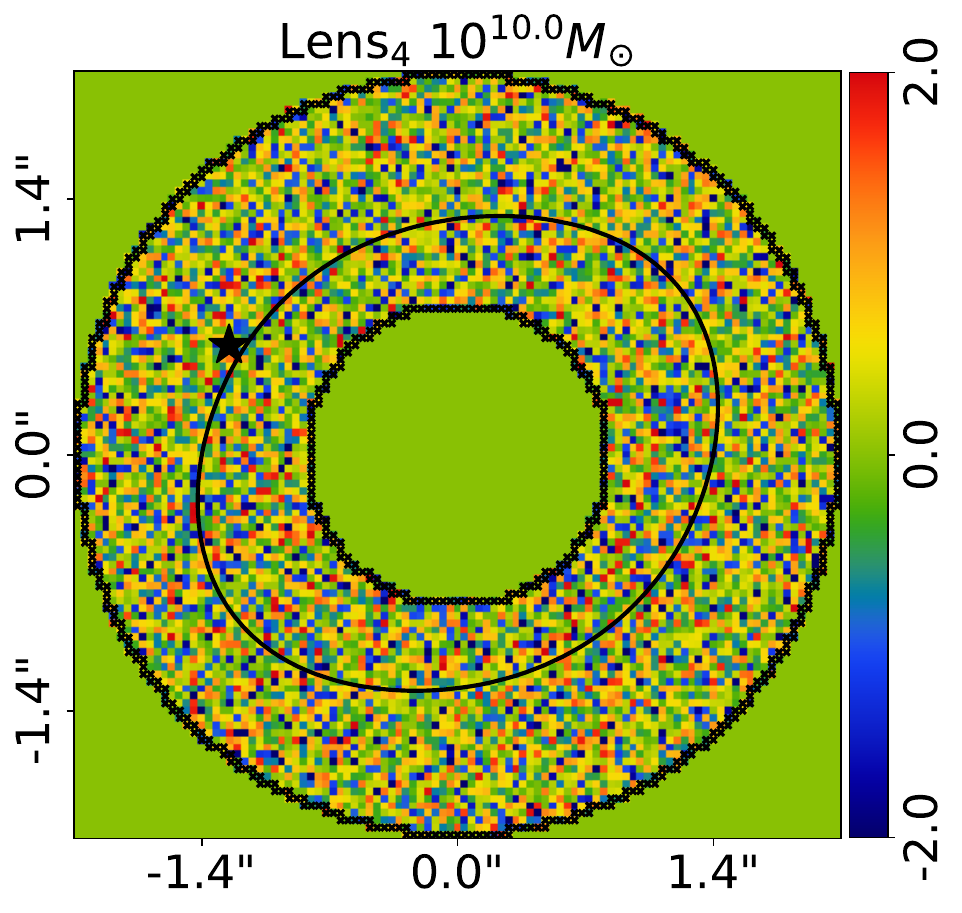}
\includegraphics[width=0.19\textwidth]{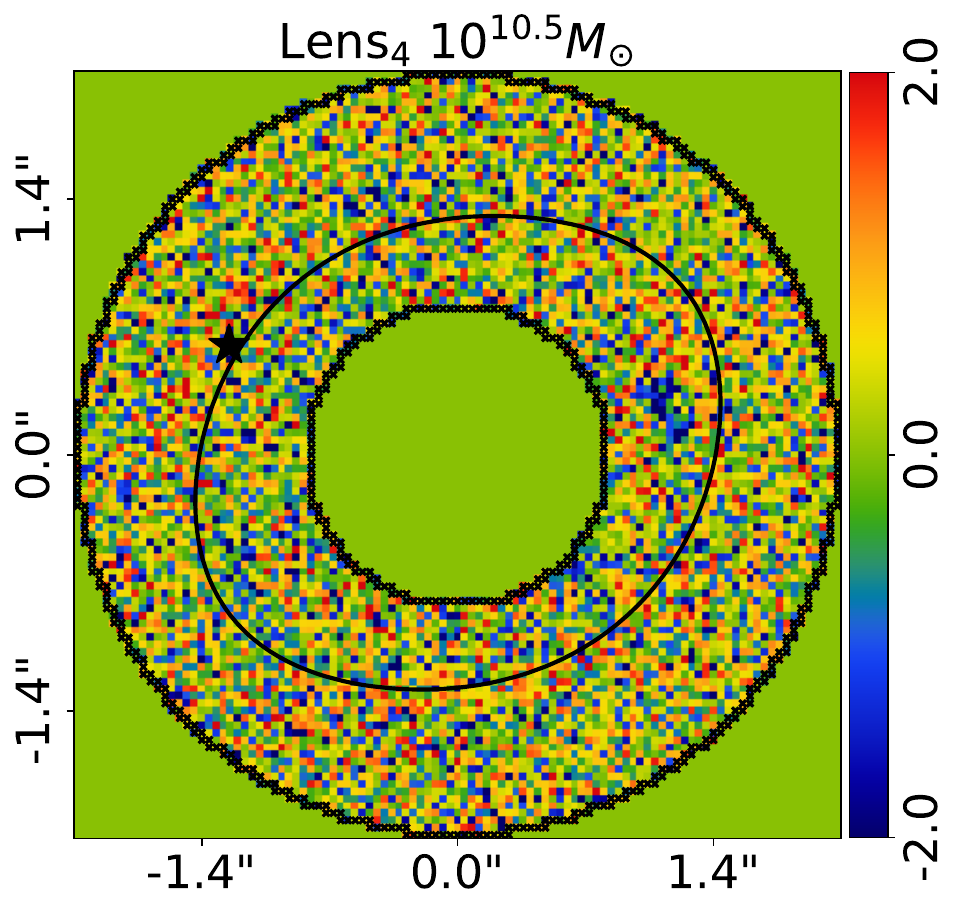}

\caption{The lens subtracted images (left panel) and normalized residuals (data minus model divided by noise) of a power-law plus shear model-fit without a subhalo to each simulated dataset. The lens subtracted images are for each dataset with a $10^{10.5}M_{\odot}$ subhalo. Panels to the right show the residuals for datasets where the input subhalo mass increases from left to right and different lenses are shown across the rows. Fits are shown for a Voronoi mesh source reconstruction and annular masks. The true subhalo location is marked with a black star. For certain lenses and source models, the power-law plus shear model leaves residuals in the vicinity of the subhalo.}
\label{figure:SimResi}
\end{figure*}

Table \ref{table:MassModelBefore} shows the power law plus shear mass models inferred directly before the subhalo search. For all datasets without a subhalo, the inferred parameters are accurate and consistent with the input values given in table \ref{table:MassModelSim}. However, for certain datasets including a high mass subhalo, offsets from the true parameter values are seen. The most extreme example is for Lens$_{\rm 3}$ with a $10^{10.5}$~M$_{\odot}$ subhalo. A density slope of $\gamma^{\rm mass} = 2.6520^{+0.0391}_{-0.0244}$ is inferred compared to $\gamma^{\rm mass} = 2.0616^{+0.0767}_{-0.0736}$ for data without a subhalo (the true value is $\gamma^{\rm mass} = 2.073$). The magnitude of the offsets vary across the other datasets and across different mass model parameters (but not the external shear parameters). Furthermore, the offsets decrease as the input subhalo mass is reduced. This is evidence that the PL mass model is adjusting its parameters to `absorb' the perturbing effect of the subhalo in certain lenses, impacting our ability to detect the DM subhalo.

The normalized residuals of the PL model-fits to the simulated datasets using a Voronoi mesh and annular mask are shown in \cref{figure:SimResi}. The four simulated lenses run from top to bottom, with the input subhalo mass decreasing from left to right (the right most panel shows data without a subhalo). The presence of a $10^{10.5}$M$_{\odot}$ subhalo creates residuals in the majority of fits. These show characteristic features of subhalo residuals that are localized primarily around a single image of the lensed source near the subhalo's true location, marked on \cref{figure:SimResi} as a black star \citep{Vegetti2009, Vegetti2010}. There are lenses where a $10^{10.5}$M$_{\odot}$ subhalo does not produce visible residuals, which overlap with the lenses whose PL model is offset from the true input model, reaffirming the notion that changes to the mass model may absorb the subhalo signal. For lenses where no subhalo is included (right columns), the residuals are consistent with Gaussian noise. A lack of visible residuals does not necessarily mean that a subhalo is undetectable, because the source reconstruction has the flexibility to adapt its reconstruction to account for the subhalo perturbation. One may still ultimately infer an evidence increase in the subhalo scanning analysis because the inclusion of the subhalo improves the likelihood via the regularization terms (see \citealt{Suyu2006}).

The largest offsets of mass model parameters for data with high mass subhalos are seen in Lens$_{3}$ and Lens$_{4}$, which are the simulated lenses with complex and compact sources (e.g. BELLS-GALLERY like). For more compact sources the mass model therefore appears more able to absorb the subhalo signal. This is consistent with the results of \citep{Ritondale2019a} who performed sensitivity mapping of the BELLS-GALLERY lenses and noted reduced sensitivity due to weaker constraints on the mass model parameters for more compact sources.

The centre of PL mass models which absorb high mass DM subhalos are also offset from their true values of $(0.0\arcsec,\ 0.0\arcsec)$. For example, for Lens$_{3}$ with a $10^{10.5}$M$_{\odot}$ subhalo the inferred centre is $(x^{\rm mass},\ y^{\rm mass}) = (0.005\arcsec,\ 0.003\arcsec)$ and for ${\rm Lens}_{4}$ it is $(x^{\rm mass},\ y^{\rm mass}) = (0.002\arcsec,\ 0.012\arcsec)$. Both these centres are offset from $(0.0\arcsec,\ 0.0\arcsec)$ at $3\sigma$ confidence. The decomposed mass models fitted in this work tie the two dimensional stellar mass distribution to the emission of the lens galaxy's light, and therefore put strong constraints on the stellar mass profile centre (as well as the ellipticity components). This may reduce a decomposed mass model's ability to absorb a DM subhalo signal.






\bibliography{library, manual, citations}            

\label{lastpage}

\end{document}